\documentclass[12pt,a4paper,twoside]{report}
\PassOptionsToPackage{dvipsnames}{xcolor}
\usepackage[a4paper, twoside, textwidth=6.5in, textheight=8.75in, inner=27mm]{geometry}

\usepackage{amssymb}
\usepackage{amsmath}

\usepackage[british]{babel}
\usepackage[T1]{fontenc}
\usepackage[utf8]{inputenc}
\usepackage{amsmath,amssymb,amstext,amsfonts}

\usepackage[dvipsnames]{xcolor}
\usepackage{graphicx}
\usepackage{hyperref}
\hypersetup{bookmarksnumbered=true,
  pdftitle={High precision black hole scattering},
  pdfauthor={Benjamin Johannes Sauer},
  pdfkeywords={black hole scattering, post-Minkowskian, WQFT, Feynman integrals}}
\usepackage{cancel}
\usepackage{float}
\usepackage{bbm}
\usepackage{mathtools}
\usepackage{comment}
\usepackage{dsfont}
\usepackage[sort&compress,square,numbers]{natbib} 
\usepackage{lscape}
\usepackage{siunitx}
\usepackage{mathdots}
\usepackage{multirow}

\usepackage{dsfont}
\usepackage{slashed,cancel,lscape,caption,array,graphicx,subcaption}
\usepackage{amsmath}
\usepackage{nicefrac}
\usepackage{mathtools}
\usepackage{mathrsfs}
\usepackage{multirow}
\usepackage{booktabs}
\usepackage{chngcntr}
\usepackage{nicefrac}
\usepackage{colortbl}
\usepackage{graphicx}
\usepackage{times}
\usepackage{float,slashed}

\usepackage{tikz}
\usetikzlibrary{decorations.pathmorphing}
\usetikzlibrary{decorations.markings}
\usetikzlibrary{positioning, shapes, snakes, arrows}

\usetikzlibrary{shapes.misc}

\tikzset{ 
	graviton/.style={line width=.8pt, -latex,decorate, decoration={snake, segment length=4pt,amplitude=1.8pt, pre length=.1cm, post length=.25cm}},
	worldline/.style={gray, line width=1pt},
	worldlineBold/.style={black, line width=.6pt},
        background/.style={black,dotted,line width=1pt},
	zUndirected/.style={line width=1pt},
	zParticle/.style={line width=1pt,postaction={decorate},decoration={markings,mark=at position .6 with {\arrow[#1]{latex}}}},
        righleft/.style={line width=1pt,postaction={decorate},decoration={markings,mark=at position .999 with {\arrow[#1]{latex}},mark=at position .01 with {\arrowreversed[#1]{latex}}}},
	zParticleF/.style={line width=1pt,postaction={decorate}},
	cscalar/.style={line width=1pt,postaction={decorate},decoration={markings,mark=at position .6 with {\arrow[#1]{latex}}}},
	bscalar/.style={dashed, line width=1pt,postaction={decorate},decoration={markings,mark=at position .6 with {\arrow[#1]{latex}}}},
	cscalar2/.style={line width=1pt,postaction={decorate},decoration={markings,mark=at position .8 with {\arrow[#1]{latex}}}},
	photon/.style={line width =.8pt, decorate, decoration={snake, segment length=3pt, amplitude=1.8pt,  pre length=.1cm, post length=.1cm}},
	 mid arrow/.style={postaction={decorate,decoration={
        markings,
        mark=at position .5 with {\arrow[#1]{latex}}}}} ,
        worlddot/.style={dotted, line width=.8pt},
	worlddot2/.style={dotted, line width=1pt},
        cross/.style={cross out, draw=black, minimum size=2*(#1-\pgflinewidth), inner sep=0pt, outer sep=0pt},
cross/.default={5pt}   }
 \usepackage{tikz-feynman}

 \newcommand{\red}[1]{\textcolor{red}{#1}}
 \newcommand{\blue}[1]{\textcolor{blue}{#1}}
 \newcommand{\white}[1]{\textcolor{white}{#1}}

\newcommand{\zenodoOne}{\href{https://doi.org/10.5281/zenodo.18340333}{\tt zenodo.org} \cite{zenodo1} }
\newcommand{\zenodoTwo}{\href{https://doi.org/10.5281/zenodo.14604438}{\tt zenodo.org} \cite{zenodo2} }
\DeclareMathAlphabet{\mathpzc}{OT1}{pzc}{m}{it}

\def\cO{\mathcal{O}}

\def\eps{\epsilon}

\def\d{\mathrm{d}}

\def\dd{{\delta\!\!\!{}^-}}

\def\d{\mathrm{d}}
\def\eps{\epsilon}

\def\customsizeg#1{{\hbox{$\left#1\vbox to20.5\p@{}\right.\n@space$}}}
\newcommand*\Bell{\ensuremath{\boldsymbol\ell}}

\newcommand{\I}{\mathrm{i}}
\renewcommand{\d}{\ensuremath\mathrm{d}}

\def\nn{\nonumber}

\def\i0{i0^{+}}

\newcommand{\vev}[1]{\langle #1\rangle}
\newcommand{\wev}[1]{\langle #1\rangle_{\rm WQFT}}

\newcommand{\xd}{{\dot x}}

\newcommand{\be}{\begin{equation}}
\newcommand{\ee}{\end{equation}}
\newcommand{\ba}{\begin{align}}
\newcommand{\ea}{\end{align}}
\ifx\genfrac\sdflkaj\else\fi
\newcommand{\sfrac}[2]{{\textstyle\frac{#1}{#2}}}

\newcommand{\ord}{\mathcal{O}}
\newcommand{\uz}{\underline{z}}
\DeclareMathOperator{\K}{K}
\DeclareMathOperator{\sign}{sign}

\newcommand{\rd}{\mathrm{d}}
\begin{document}

\begin{titlepage}
        \begin{center}
            \vspace*{1cm}
     
            {
              \LARGE
              \textbf{High precision black hole scattering}
            }
     
            \vspace{1.3cm}
     
            \textbf{Dissertation}
            \\
            zur Erlangung des akademischen Grades
            \\
            doctor rerum naturalium
            \\
            (Dr. rer. nat.)
            \\
            im Fach Physik
            \\
            Spezialisierung: Theoretische Physik
            \\
            \vspace{1.3cm}
            eingereicht an der
            \\
            Mathematisch-Naturwissenschaftlichen Fakult\"at
            \\
            der Humboldt-Universit\"at zu Berlin
            \\
            \vspace{1.3cm}
            von
            \\
            \textbf{M. Sc. Benjamin Johannes Sauer}
            \\
           
            \vfill
                   {
            Pr\"asidentin der Humboldt-Universit\"at zu Berlin:
            \\
            Prof. Dr. Julia von Blumenthal
            \\
            \vspace{.2cm}
            Dekan der Mathematisch-Naturwissenschaftlichen Fakult\"at:
            \\
            Prof. Dr. Emil List-Kratochvil
            \\
            \vspace{1.3cm}
            }   \vfill
            {\raggedright   
       Gutachter:
        1. Prof. Dr. Jan Plefka
       \\
       \hspace{1.8cm}
       2. Prof. Dr. Pierre Vanhove
       \\
       \hspace{1.8cm}
       3. Prof. Dr. Mao Zeng
       \\
       \vspace{.2cm}
       Tag der m\"undlichen Pr\"ufung: 15.09.2026
       \\
       }         
        \end{center}
     \end{titlepage}
 
     \chapter*{Abstract}

     In this thesis we compute the impulse of a classical black hole scattering encounter in the Post-Minkowskian (PM) expansion - a weak-field expansion in Newton's constant $G$ - up to the fifth order. We describe the black holes as point particles within an effective field theory framework and couple them through gravity. The classical equations of motion are solved diagrammatically using the worldline quantum field theory (WQFT) formalism, in which the solution with causal boundary conditions is encoded in tree-level Feynman diagrams built from graviton and worldline fields. The Feynman integrals arising at each order are reduced by advanced IBP reduction techniques and evaluated with the method of differential equations. We find that, in the solution of these integrals, periods over Calabi--Yau manifolds - familiar from string theory - appear and contribute to the classical impulse at the fifth Post-Minkowskian order. The central new results of this work are the first analytic results for the scattering observables at fifth Post-Minkowskian order, together with the appearance of Calabi--Yau threefold periods in a classical physics observable.
     
     We begin by reviewing the effective field theory description of black holes and the Post-Minkowskian expansion, followed by the WQFT formalism, in which the classical solution of the equations of motion is given by tree-level one-point functions. We then analyse the general structure of the impulse integrand and show how tensor and integration-by-parts (IBP) reduction, together with symmetry relations, simplify it enough to make the computation at high PM order feasible in the first place.
     
     The reduced integrand is integrated using the method of differential equations, which we review and apply at third Post-Minkowskian order. We also introduce how periods over non-trivial geometries enter Feynman integrals. We then discuss in detail the integrals contributing at fourth and fifth Post-Minkowskian order and how their computation is realised.
     
     Finally, we use the computed impulse to derive the scattering angle and the radiated energy. We discuss their analytic structure at the various Post-Minkowskian orders and analyse how tail and memory effects contribute to each order. In addition, we address the split into conservative and dissipative contributions and the subtleties that arise from it.

     \chapter*{Zusammenfassung}

     In dieser Arbeit berechnen wir den Impuls eines klassischen Schwarzloch-Streuprozesses in der Post-Minkowski-Entwicklung (PM), eine Entwicklung in der Newtonschen Konstante $G$, bis zur fünften Ordnung. Wir beschreiben die Schwarzen Löcher im Rahmen einer effektiven Feldtheorie als Punktteilchen und koppeln sie über die Gravitation. Die klassischen Bewegungsgleichungen werden diagrammatisch mithilfe des Formalismus der Weltlinien Quantenfeldtheorie (WQFT) gelöst, in dem die Lösung mit kausalen Randbedingungen durch Feynman-Diagramme auf Baumniveau aus Graviton- und Weltlinien-Feldern gegeben ist. Die in jeder Ordnung auftretenden Feyn\-man-Integrale werden mit der Methode der Differentialgleichungen ausgewertet. Wir stellen fest, dass in der Lösung dieser Integrale Perioden über Calabi--Yau Mannigfaltigkeiten, bekannt aus der Stringtheorie, auftreten und in fünfter Post-Minkowski-Ordnung zum klassischen Impuls beitragen. Die zentralen neuen Ergebnisse dieser Arbeit sind die ersten analytischen Resultate für die Streuobservablen in fünfter Post-Minkowski-Ordnung sowie das Auftreten von Perioden über Cala\-bi-Yau-Dreifaltigkeiten (CY3) in einer klassischen physikalischen Observable.
     
     Zunächst geben wir einen Überblick über die effektive Feldtheoriebeschreibung Schwarzer Löcher und die Post-Minkowski-Entwicklung, gefolgt vom WQFT-Formalismus, in dem die klas\-sische Lösung der Bewegungsgleichungen durch Einpunktfunktionen auf Baumniveau gegeben ist. Anschließend analysieren wir die allgemeine Struktur des Integranden des Impulses und zeigen, wie Tensor- und Integration-by-parts-(IBP-)Reduktion zusammen mit Symmetrierelationen ihn so weit vereinfachen, dass die Berechnung in hoher PM-Ordnung überhaupt erst durchführbar wird.
     
     Der reduzierte Integrand wird mit der Methode der Differentialgleichungen integriert, die wir in dritter Post-Minkowski-Ordnung vorstellen und anwenden. Danach führen wir auch ein, wie Perioden über nichttriviale Geometrien hier auftreten. Danach diskutieren wir ausführlich die in vierter und fünfter Post-Minkowski-Ordnung beitragenden Integrale und wie sich ihre Berechnung umsetzen lässt.
     
     Schließlich nutzen wir den berechneten Impuls, um den Streuwinkel und die abgestrahlte Energie abzuleiten. Wir erörtern ihre analytische Struktur in den verschiedenen Post-Minkowski-Ordnungen und identifizieren, welche physikalischen Effekte in jeder Ordnung neu hinzutreten. Insbesondere behandeln wir die Aufspaltung in konservative und dissipative Beiträge und die damit verbundenen Schwierigkeiten.

     \chapter*{List of Publications}

     This thesis is mainly based on the content of the following peer-reviewed publications:
     
     \vspace{.4cm}
     \noindent
     \cite{Jakobsen:2022psy}
     G. U. Jakobsen, G. Mogull, J. Plefka and B. Sauer,
     ``\textit{All Things Retarded: Radiation-Reaction in Worldline Quantum Field Theory}'',
     JHEP 10 (2022) 128
     arXiv: 2207.00569

     \vspace{.4cm}
     \noindent
     \cite{Jakobsen:2023ndj}
     G. U. Jakobsen, G. Mogull, J. Plefka, B. Sauer and Y. Xu,
     ``\textit{Conservative Scattering of Spinning Black Holes at Fourth Post-Minkowskian Order}'',
     Phys.Rev.Lett. 131 (2023) 15, 151401
     arXiv: 2306.01714

     \vspace{.4cm}
     \noindent
     \cite{Jakobsen:2023hig}
     G. U. Jakobsen, G. Mogull, J. Plefka and B. Sauer,
     ``\textit{Dissipative Scattering of Spinning Black Holes at Fourth Post-Minkowskian Order     }'',
     Phys.Rev.Lett. 131 (2023) 24, 241402 \\
     arXiv: 2308.11514

     \vspace{.4cm}
     \noindent
     \cite{Jakobsen:2023pvx}
     G. U. Jakobsen, G. Mogull, J. Plefka and B. Sauer,
     ``\textit{Tidal effects and renormalization at fourth Post-Minkowskian order}'',
     Phys.Rev.D 109 (2024) 4, L041504
     arXiv: 2312.00719

     \vspace{.4cm}
     \noindent
     \cite{Klemm:2024wtd}
     A. Klemm, C. Nega, B. Sauer and J. Plefka,
     ``\textit{Calabi-Yau periods for black hole scattering in classical general relativity     }'',
     Phys.Rev.D 109 (2024) 12, 124046
     arXiv: 2401.07899

     \vspace{.4cm}
     \noindent
     \cite{Driesse:2024xad}
     M. Driesse, G. U. Jakobsen, G. Mogull, J. Plefka, B. Sauer and J. Usovitsch,
     ``\textit{Conservative Black Hole Scattering at Fifth Post-Minkowskian and First Self-Force Order}'',
     Phys.Rev.Lett. 132 (2024) 24, 241402
     arXiv: 2403.07781

     \vspace{.4cm}
     \noindent
     \cite{Driesse:2024feo}
     M. Driesse, G. U. Jakobsen, A. Klemm, G. Mogull, C. Nega, J. Plefka, B. Sauer and J. Usovitsch,
     ``\textit{Emergence of Calabi-Yau manifolds in high-precision black-hole scattering}'',
     Nature 641 (2025) 8063, 603-607
     arXiv: 2411.11846

     \vspace{.4cm}
     \noindent
     \cite{Duhr:2025lbz}
     C. Duhr, S. Maggio, C. Nega, B. Sauer, L. Tancredi, F. J. Wagner,
     ``\textit{Aspects of canonical differential equations for Calabi-Yau geometries and beyond     }'',
     JHEP 06 (2025) 128
     arXiv: 2503.20655

     \vspace{.4cm}
     \noindent
     \cite{Driesse:2026qiz}
     M. Driesse, G. U. Jakobsen, G. Mogull, C. Nega, J. Plefka, B. Sauer and J. Usovitsch,
     ``\textit{Conservative Black Hole Scattering at Fifth Post-Minkowskian and Second Self-Force Order}'',\\
     Phys.Rev.Lett. 137 (2026) 8, 081402, arXiv:2601.16256

     \vfill
     \noindent 
     \newpage
     \textbf{Selbständigkeitserklärung:}
     \\
     Ich erkläre, dass ich die Dissertation selbständig und nur unter Verwendung der von mir gemäß § 7 Abs. 3 der Promotionsordnung der Mathematisch-Naturwissenschaftlichen Fakultät, veröffentlicht im Amtlichen Mitteilungsblatt der Humboldt-Universität zu Berlin Nr. 42/2018 am 11.07.2018 angegebenen Hilfsmittel angefertigt habe.

\begin{figure}[h!]
        \centering
        \includegraphics[width=0.2\textwidth]{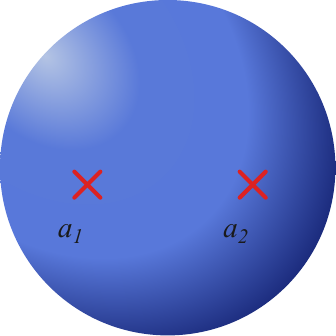}
         \caption{The punctured Riemann-sphere on which the integration of $I_1$ is defined.}
        \label{fig:puncRiemann}
    \end{figure}

\begin{figure}[h!]
        \centering
        \includegraphics[width=0.6\textwidth]{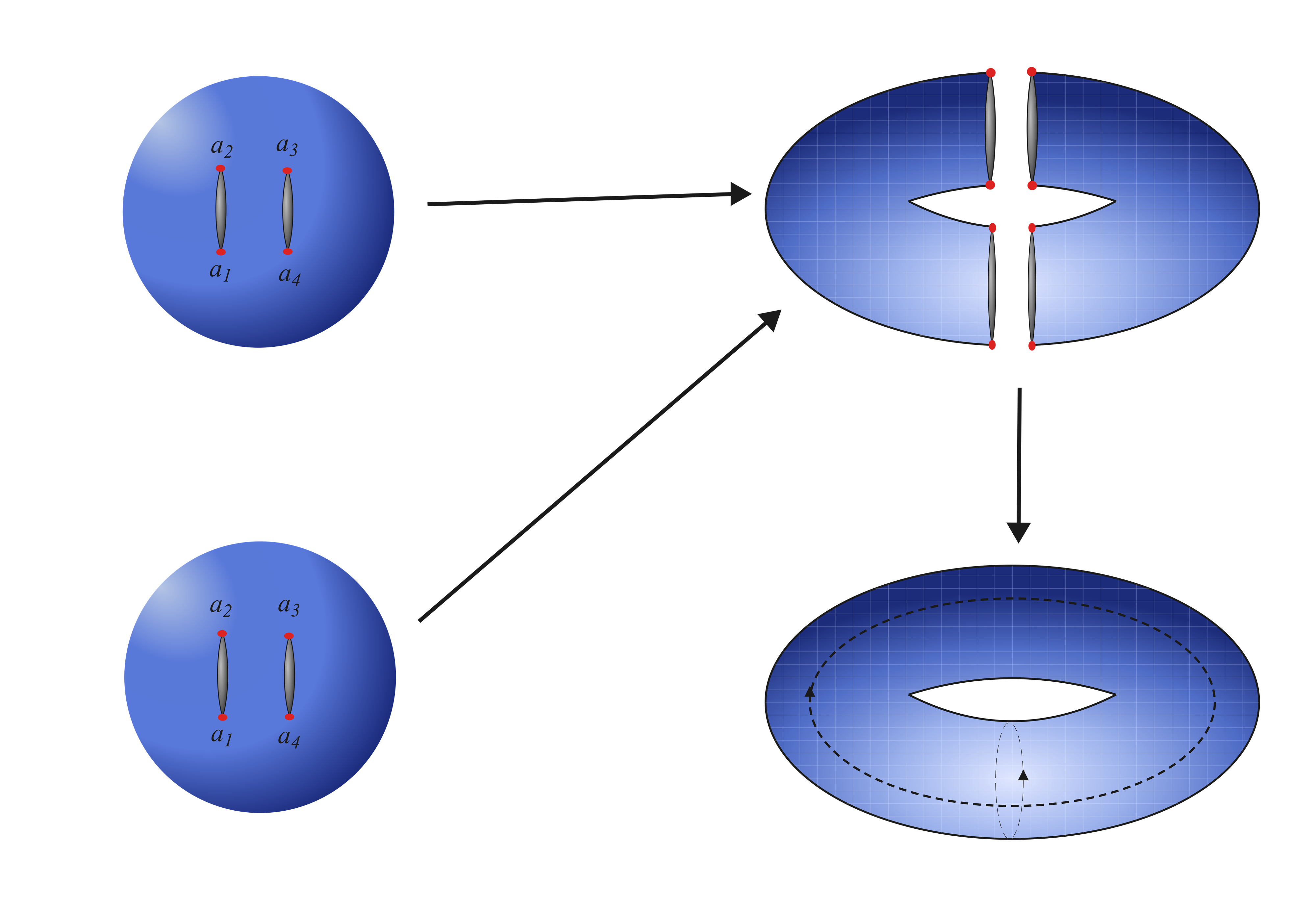}
         \caption{The construction of the torus as the Riemann surface of the integrand. Two copies of the Riemann sphere are glued together along their branch cuts.}
        \label{fig:torusconstruction}
    \end{figure}

     In dieser Dissertation wurde "Claude AI, Opus 4.8" zur sprachlichen Verbesserung des Textes dieser Dissertation und zur Hilfe zur Erstellung der Fig.~\ref{fig:scatter}, Fig.~\ref{fig:puncRiemann}, Fig.~\ref{fig:contoursEll}, Fig.~\ref{fig:contoursLog}, Fig.~\ref{fig:torusconstruction}, Fig.~\ref{fig:torusPunct} verwendet.
\newpage

\chapter*{Acknowledgements}
I want to thank Jan Plefka for being my supervisor during this PhD and for supporting and believing in this research. I greatly enjoyed my time in his group, and the balance of guidance and independence that I was given throughout my PhD.

Furthermore, I want to thank Mathias Driesse, Claude Duhr, Gustav Jakobsen, Albrecht Klemm, Sara Maggio, Gustav Mogull, Christoph Nega, Lorenzo Tancredi, Johann Usovitsch and Fabian Wagner for the exciting collaborations on many projects during my PhD.

In particular, I want to thank Gustav Jakobsen for many enlightening discussions about gravity, Christoph Nega for many discussions and explanations concerning Feynman integrals and Calabi--Yau periods, and Johann Usovitsch for many discussions about IBP reductions.

I would also like to thank everyone in the RTG and in Jan's group for a friendly and productive research environment. I am very grateful that I could take part in so many conferences, for the people I met on these trips, and for the many discussions that took place along the way.

I also want to thank Pierre Vanhove and Mao Zeng for agreeing to act as referees for this thesis, and Pasquale Pavone and Peter Uwer for being part of my defence committee.

I am also grateful to everyone I met during my Bachelor's and Master's studies, in particular my friends Vincent Stegmaier, Jakob Neef, Nele Stetzuhn, Anna Reckwitz and Nils Niggemann. They made studying physics great fun, and it would not have been the same without them.

Zuletzt möchte ich auch meinen Eltern, meinen Gro{\ss}eltern und Phips daf\"ur danken, dass sie mich immer unterst\"utzt haben und immer f\"ur mich da waren.

This work was funded by the Deutsche Forschungsgemeinschaft (DFG, German Research Foundation) Projektnummer 417533893/GRK2575 ``Rethinking Quantum Field Theory''.
\tableofcontents

\clearpage

\chapter{Introduction}
\label{sec:intro}

The two-body problem is one of the most thoroughly studied questions in the history of physics, be it the Kepler problem describing the motion of planets around the sun or the hydrogen atom in quantum mechanics. In the quantum-mechanical case the solutions are given by spherical harmonics and Laguerre polynomials, and their understanding marked the very beginning of atomic physics and all that has followed from it. In Newtonian mechanics the trajectories of the planets are given by elliptic orbits, providing not only the foundation of astronomy but also some of the first evidence for the heliocentric model.

In recent years the two-body problem has experienced a revival of interest in a different corner of physics, that of general relativity~\cite{Einstein:1915by}. The first detection of gravitational waves in 2015~\cite{LIGOScientific:2016aoc,LIGOScientific:2016wyt} not only led to a Nobel prize but also provided yet another confirmation of Einstein's theory and opened up a completely new window onto the largest structures in the universe~\cite{LIGOScientific:2018mvr,LIGOScientific:2020ibl,LIGOScientific:2021usb,LIGOScientific:2021djp,Antelis:2016icm}. The sources of these gravitational waves are two-body systems of very massive objects, most commonly pairs of black holes or neutron stars. In such systems the two bodies spiral into each other, emitting a gravitational-wave signal, until they finally merge into a single body. These merger events are the main source of detectable gravitational-wave signals, and their theoretical description is key to a better understanding of general relativity and the universe~\cite{Will:2014kxa,Yunes:2016jcc,Shankaranarayanan:2022wbx}.

The complexity and non-linearity of general relativity make the two-body problem considerably harder to solve than the hydrogen atom in quantum mechanics or the Kepler problem in Newtonian gravity. To learn from gravitational-wave observations we nevertheless need theoretical predictions of high accuracy. One modern strategy is to rely on numerical relativity~\cite{Pretorius:2005gq, Campanelli:2005dd, Baker:2005vv}, which provides essentially exact solutions of Einstein's equations up to controllable numerical errors but is computationally expensive and fails in certain corners of parameter space. The complementary, and perhaps theorist-favourite, strategy is perturbation theory. Expanding the solution in a small parameter has led to successes across many areas of physics. In particle physics in particular, the framework of perturbative expansion in terms of Feynman diagrams has produced not only highly precise predictions but also a transparent intuitive picture of the underlying processes. Computing scattering amplitudes in powers of the coupling constant gives rise to loop integrals, and the difficulty resides in the evaluation of thousands or even millions of such integrals. While higher and higher corrections in the perturbative series become smaller, they also grow considerably more intricate. Methods such as integration-by-parts (IBP) reductions~\cite{Chetyrkin:1981qh,Laporta:2000dsw} and the method of differential equations~\cite{Kotikov:1990kg,Bern:1993kr,Remiddi:1997ny,Gehrmann:1999as,Henn:2013pwa,Henn:2013nsa} have, however, enabled tremendous progress in extending these computations to high orders.

In the bound two-body problem in general relativity the natural perturbative expansion is around Newtonian mechanics. A double expansion in the relative velocity and Newton's constant $G$ describes the inspiral of two black holes. This Post-Newtonian (PN)~\cite{Blanchet:2013haa} double expansion makes the computations tractable but also restricts the regime of validity, since it applies only to widely separated bodies moving with small relative velocities. As a well-established formalism the PN expansion is by now complete through fifth order~\cite{Porto:2026fsd,Blumlein:2020pyo,Bini:2025vuk}, with partial results available up to 7PN~\cite{Brunello:2025gpf,Brunello:2026anu}. These computations are surprisingly effective, especially when one does not aim at the radiated waveform directly. Resummation techniques such as the effective-one-body (EOB) formalism~\cite{Buonanno:1998gg,Buonanno:2000ef} combine PN results with known analytic limits to construct an effective one-body Hamiltonian. One resums thereby around the geodesic motion of a test particle in the metric of a single black hole, which allows for the generation of waveform templates of much higher precision.

The self-force (SF) expansion~\cite{Mino:1996nk,Pound:2021qin,Barack:2018yvs}, in contrast, expands in the mass ratio $m_1/m_2$ of the two black holes, treating the lighter body as a perturbation of the geodesic motion in the metric of the heavier one. It is applicable in a much larger region of parameter space, but it is also so intricate that solutions can in practice only be obtained numerically or through an additional expansion.

The Post-Minkowskian (PM) expansion is a single expansion in Newton's constant $G$---an expansion around flat space---and represents an ideal middle ground between the two previous approaches. It allows for analytic results while remaining fully relativistic and exact in the velocity. The downside is that it is naturally tailored to scattering processes rather than to bound systems. While the scattering process is theoretically as interesting as the bound problem, it is not a priori clear how it can help us make use of gravitational-wave signals. Here the mapping of unbound to bound quantities enters the picture. One can map, in part, the unbound potential or the radiated energy to its bound counterpart, and one can even map scattering observables such as the scattering angle to the bound periastron advance~\cite{Iwasaki:1971vb,Cheung:2018wkq,Cristofoli:2019neg,Kalin:2019rwq,Kalin:2019inp,Cho:2021arx,Saketh:2021sri,Gonzo:2023goe,Gonzo:2024zxo}. This mapping holds up to non-local-in-time contributions originating from so-called tail effects~\cite{Bern:2021yeh,Cho:2021arx}. Because of the non-linearity of Einstein's gravity, terms depending on the entire history of the trajectory appear and make a complete mapping to bound orbits impossible with present techniques. The scattering observables can in turn feed into EOB models and be used to generate waveform templates from purely PM data~\cite{Damour:2016gwp,Damour:2017zjx,Antonelli:2019ytb,Damgaard:2021rnk,Buonanno:2024byg}.

The PM scattering angle was first computed by Westpfahl in 1985~\cite{Westpfahl:1985tsl} up to next-to-leading order (2PM), and it took more than thirty years until the 3PM order was obtained in 2019 by Bern et al.~\cite{Bern:2019nnu}, after the detection of gravitational waves had made such computations drastically more relevant. The result was completed in 2021 by the inclusion of radiative effects~\cite{Herrmann:2021tct}. The key reason why this computation became feasible after such a long pause is that it was built on QFT scattering amplitudes~\cite{Bjerrum-Bohr:2022blt,Kosower:2022yvp,Buonanno:2022pgc} together with an effective field theory (EFT) description of black holes as point particles~\cite{Goldberger:2004jt}. The remaining complexity now resides almost entirely in the evaluation of Feynman integrals, the building blocks of the full analytic answer. The same techniques led in 2021 to the 4PM scattering angle~\cite{Bern:2021dqo,Bern:2021yeh,Dlapa:2021npj,Dlapa:2021vgp}, which was completed one year later by the inclusion of radiative effects~\cite{Dlapa:2022lmu}. The latter group employs an EFT formalism that is not based on amplitudes but nevertheless makes heavy use of QFT techniques.

The most established frameworks today, all of which include radiative effects, are the worldline effective field theory (WEFT)~\cite{Porto:2016pyg,Levi:2018nxp,Goldberger:2022ebt,Goldberger:2022rqf}, the amplitudes-based KMOC formalism~\cite{Kosower:2018adc} that builds observables from (cut and uncut) amplitudes, and the worldline quantum field theory (WQFT)~\cite{Mogull:2020sak,Jakobsen:2022psy}, which sits in between. Within these formalisms the 3PM and 4PM results have by now been confirmed several times~\cite{Bern:2019crd,Kalin:2020fhe,Parra-Martinez:2020dzs,Cheung:2020gyp,Herrmann:2021tct,Ruf:2021egk,Bjerrum-Bohr:2021vuf,Bjerrum-Bohr:2021din,Brandhuber:2021eyq,DiVecchia:2021bdo,Riva:2021vnj,Mougiakakos:2022sic,Jakobsen:2022zsx,Jakobsen:2022psy,Jakobsen:2022fcj,DiVecchia:2022piu,Heissenberg:2022tsn,Riva:2022fru,FebresCordero:2022jts,Damgaard:2023ttc,Jakobsen:2023ndj,Jakobsen:2023hig,Bern:2021dqo,Bern:2021yeh,Dlapa:2021npj,Dlapa:2021vgp}. The motivation to push these expansions to ever higher orders comes from the next generation of gravitational-wave detectors~\cite{Punturo:2010zz, ET:2025xjr, LISA:2017pwj, Reitze:2019iox}, whose sensitivity will exceed that of the current ones by orders of magnitude and which demand theoretical data of correspondingly high precision. The new detectors will also operate over a much broader frequency band, probing a far larger region of parameter space and calling in particular for analytic results, since numerical relativity does not deliver data across wide regions of phase space and across all relevant scales. Current rough estimates suggest that data through 7PM order~\cite{Buonanno:2022pgc,ET:2025xjr,Antonelli:2019ytb} will be required to fully exploit this next generation of detectors. 
To reach these precision goals one has to go in two directions. The inclusion of higher order spin effects ($S^\mu\sim G$) and tidal effects (starting at $G^6$) leaves the loop order of the computation constant but increases the Post-Minkowskian order. This was done using both worldline~\cite{Kalin:2020mvi, Kalin:2020fhe, Kalin:2020lmz, Mogull:2020sak, Jakobsen:2021smu, Dlapa:2021npj, Dlapa:2021vgp, Mougiakakos:2021ckm, Riva:2021vnj, Dlapa:2022lmu, Dlapa:2023hsl, Liu:2021zxr, Mougiakakos:2022sic, Riva:2022fru, Jakobsen:2021lvp, Jakobsen:2021zvh, Jakobsen:2022fcj, Jakobsen:2022zsx, Jakobsen:2022psy, Shi:2021qsb,Bastianelli:2021nbs, Comberiati:2022cpm,Wang:2022ntx,Ben-Shahar:2023djm, Bhattacharyya:2024aeq, Jakobsen:2023ndj, Jakobsen:2023hig, Jakobsen:2023pvx,Haddad:2024ebn,Bonocore:2024uxk, Bonocore:2025stf,Mogull:2025cfn,Hoogeveen:2025tew} and amplitudes approaches~\cite{Neill:2013wsa, Luna:2017dtq, Kosower:2018adc, Cristofoli:2021vyo, Bjerrum-Bohr:2013bxa, Bjerrum-Bohr:2018xdl, Bern:2019nnu, Bern:2019crd, Bjerrum-Bohr:2021wwt, Cheung:2020gyp, Bjerrum-Bohr:2021din, DiVecchia:2020ymx, DiVecchia:2021bdo, DiVecchia:2021ndb, DiVecchia:2022piu, Heissenberg:2022tsn, Damour:2020tta, Herrmann:2021tct, Damgaard:2019lfh, Damgaard:2021ipf, Damgaard:2023vnx, Aoude:2020onz, AccettulliHuber:2020dal, Brandhuber:2021eyq, Bern:2021dqo, Bern:2021yeh, Bern:2022kto, Bern:2023ity, Damgaard:2023ttc, Brandhuber:2023hhy, Brandhuber:2023hhl, DeAngelis:2023lvf, Herderschee:2023fxh, Caron-Huot:2023vxl, FebresCordero:2022jts, Bohnenblust:2023qmy,Akpinar:2024meg, Akpinar:2025bkt,Gonzo:2024zxo, Gonzo:2023goe,Alaverdian:2024spu}.

On the other hand one can also increase the loop order of the computations to increase the Post-Minkowskian order. 

The main obstacle to doing this is the evaluation of the underlying Feynman integrals. Notably, the integrals that appear in KMOC, the worldline EFT, and WQFT take the same form in the end~\cite{Capatti:2024bid,Damgaard:2023vnx}, which underlines their role as the truly framework-independent building blocks of the problem.
The state-of-the-art using worldline approaches is 5PM and first self-force order results including radiative effects~\cite{Driesse:2024feo,Driesse:2024xad} and conservative results at second self-force order~\cite{Driesse:2026qiz}, work to which the present author contributed. In the amplitude approach the state of the art is the computation of the potential contributions to the scattering angle at first and second self-force order~\cite{Bern:2025wyd}.

The goal of this thesis is therefore to describe the necessary steps that allow these state-of-the-art computations of the fifth Post-Minkowskian order of the impulse with a focus on modern integration techniques, which might find a broader application than just classical gravity computations. We discuss in Ch.~\ref{sec:wqft} the effective description and physical foundations of the black hole scattering process. In Ch.~\ref{sec:inin} we review the WQFT formalism that we employ to perform this computation. In Ch.~\ref{sec:intred} we review the techniques used to reduce our integrand, and we exploit them in Ch.~\ref{sec:struct} to analyse its general structure. We continue with the method of differential equations for polylogarithmic integrals and apply it at 3PM order to illustrate the strategy used at higher orders in Ch.~\ref{sec:methdiff}. In Ch.~\ref{sec:geom} we generalise this method and analyse how non-trivial geometries arise in Feynman integrals and how one can compute the associated integrals. The key geometry encountered will turn out to be a Calabi--Yau $n$-fold, familiar from string theory as a higher-dimensional generalisation of the elliptic curve. We then use these methods to reproduce known integrals at 4PM order in Ch.~\ref{sec:4pm}, and to compute the integrals needed at 5PM order in Ch.~\ref{sec:5pm1sf} and Ch.~\ref{sec:5pm2sf}. These results are finally used in Ch.~\ref{sec:obs} to derive observables up to fifth Post-Minkowskian order.

\chapter{Black hole scattering}
\label{sec:wqft}

 \section{General relativity and black holes}

 General relativity was formulated by Einstein in 1915~\cite{Einstein:1915by}, exactly 100 years prior to the first detection of gravitational waves. The \emph{equivalence principle} is the guiding principle of the theory. The gravitational field is represented by the spacetime metric $g_{\mu\nu}$, which encodes the structure of spacetime and the coupling of gravity to energy and momentum.
 
 The dynamics can be described by the Einstein-Hilbert action, supplemented by the matter action $S_m[\phi_i,g]$ for fields collectively denoted $\phi_i$,
 \begin{equation}
 S = \frac{1}{16\pi G}\int \d^4x\,\sqrt{-g}\,R \;+\; S_m[\phi_i,g_{\mu\nu}],
 \end{equation}
 where $g \equiv \det g_{\mu\nu}$ and $R = g^{\mu\nu}R_{\mu\nu}$ is the Ricci scalar constructed from the Ricci tensor $R_{\mu\nu} = R^{\rho}{}_{\mu\rho\nu}$, which again can be constructed from the Riemann curvature tensor
\begin{align}
        R^{\lambda}{}_{\mu\nu\kappa} =\partial_{\nu}\Gamma^{\lambda}{}_{\kappa\mu}- \partial_{\kappa}\Gamma^{\lambda}{}_{\nu\mu} 
        +\Gamma^{\lambda}{}_{\eta \kappa}\Gamma^{\eta}{}_{\nu\mu} -\Gamma^{\lambda}{}_{\eta \nu}\Gamma^{\eta}{}_{\kappa\mu}\, ,
\end{align}
 where the \emph{Christoffel symbols} are given as
\begin{align}
        \Gamma^{\mu}_{\nu\rho} \;=\; \tfrac{1}{2}g^{\mu\sigma}\bigl(\partial_\nu g_{\rho\sigma}+\partial_\rho g_{\nu\sigma}-\partial_\sigma g_{\nu\rho}\bigr).
\end{align}
 Variation with respect to $g^{\mu\nu}$ yields the \emph{Einstein field equations},
 \begin{equation}
 G_{\mu\nu} \equiv R_{\mu\nu} - \tfrac{1}{2}g_{\mu\nu}R \;=\;- 8\pi G\, T_{\mu\nu},\qquad
 T_{\mu\nu} =- \frac{2}{\sqrt{-g}}\frac{\delta S_m}{\delta g^{\mu\nu}}.
 \end{equation}
 These field equations relate the Einstein tensor $G_{\mu\nu}$ to the stress--energy tensor $T_{\mu\nu}$ of the matter content.
The contracted Bianchi identity $\nabla^\mu G_{\mu\nu}=0$ ensures the energy-momentum conservation $\nabla^\mu T_{\mu\nu}=0$.
The covariant derivative $\nabla_\mu$ acts on a tensor as
\begin{align}
        \nabla_\mu T^{\nu}{}_{\rho} \;=\; \partial_\mu T^{\nu}{}_{\rho} + \Gamma^{\nu}_{\mu\sigma}\,T^{\sigma}{}_{\rho} - \Gamma^{\sigma}_{\mu\rho}\,T^{\nu}{}_{\sigma}.
\end{align}

One of the most simple but also most fascinating solutions of the Einstein field equations in vacuum is the Schwarzschild metric, found by Karl Schwarzschild in 1916~\cite{Schwarzschild:1916uq}. It describes the gravitational field outside a non-spinning, non-charged spherical mass:
\begin{align}
        ds^2 = \left(1 - \frac{2GM}{r}\right)dt^2 - \left(1 - \frac{2GM}{r}\right)^{-1}dr^2 - r^2\,d\theta^2- r^2\,\sin^2{(\theta)}\,d\phi^2,
 \end{align}
where we use a mostly minus metric. In these coordinates we see an apparent singularity at the \emph{event horizon} $r = r_S \equiv 2GM$. If the radius of the mass is smaller than this event horizon, the solution describes a Schwarzschild \emph{black hole} and no causal signal, including light, can escape the black hole once it passes the event horizon. Spacetime itself becomes singular in the middle of the black hole ($r=0$).

\begin{figure}[h!]
        \includegraphics[width=0.9\textwidth]{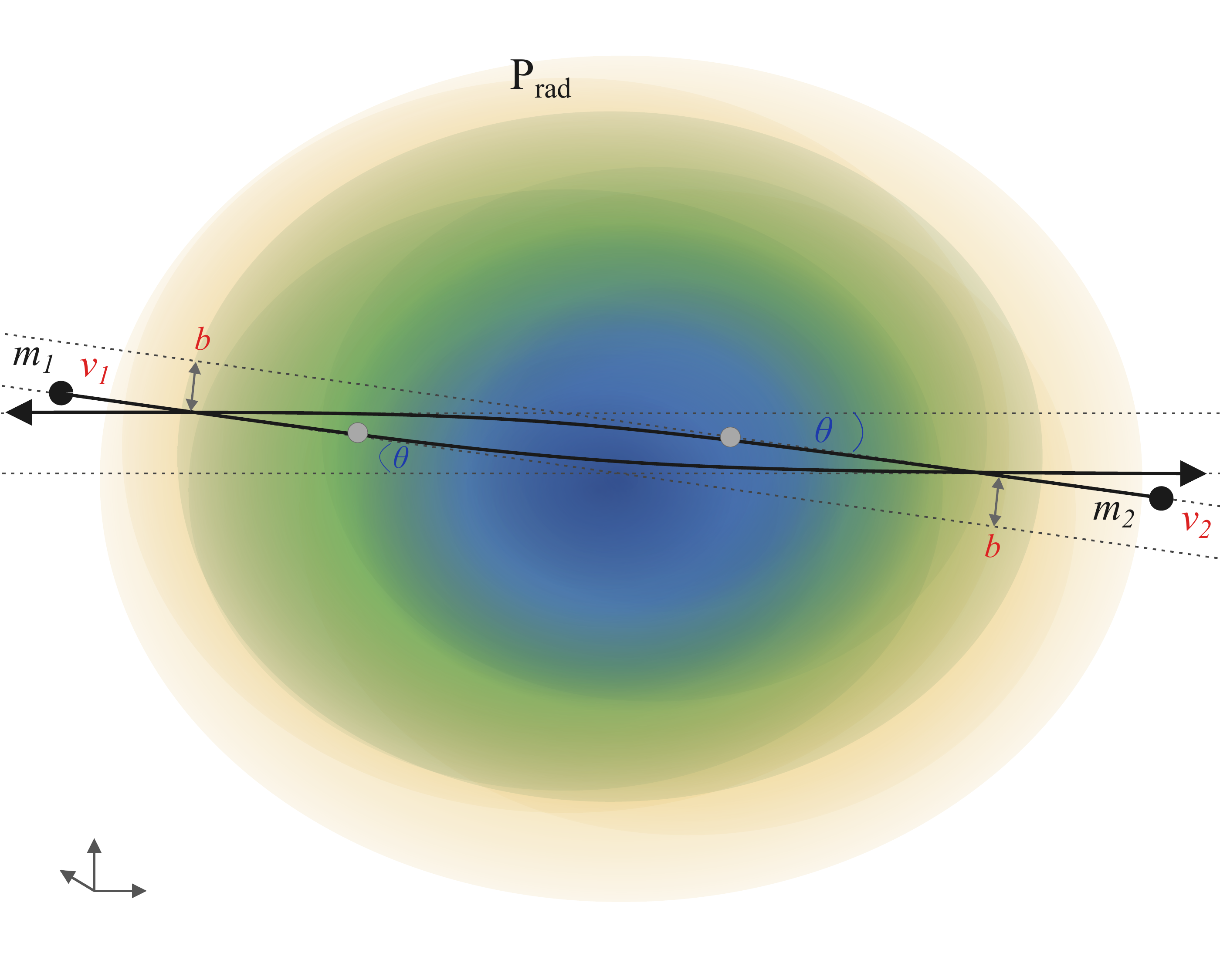}
        \centering
        \caption{The scattering process of two black holes. The scattering process is determined by the impact parameter $b$ and the initial velocities $v_i$ and the masses of the black holes $m_i$. During the process the momentum $P_{\text{rad}}$ gets radiated.}
        \label{fig:scatter}
        \end{figure}
Astrophysical black holes also carry angular momentum or spin and are described by the \emph{Kerr metric}~\cite{Kerr:1963ud}, which is characterised by the two parameters mass $M$ and the spin $S$.
The \emph{no-hair theorem}~\cite{Ruffini:1971bza} states that all stationary black holes can be described fully by their mass, spin and charge, very similar to elementary particles. Since charge is very often negligible for black holes, we can look at black holes as compact objects governed solely by their mass and spin.

The physical process, which we will be describing and for which we will try to do high precision calculations in this thesis, is the scattering event of two Schwarzschild black holes, depicted in Fig.~\ref{fig:scatter}. In the absence of spin the process is governed solely by the two black hole masses $m_i$, their initial velocity $v_i$ and the impact parameter $b$. The goal of this thesis will be to determine the radiated momentum $P_{\text{rad}}$ and the scattering angle $\theta$ to very high precision.

To reach this goal we will use the similarity of black holes and elementary particles to make use of techniques known from particle physics. 
 In the remainder of this chapter we therefore focus on the effective description of black hole scattering. We first discuss the simplest case of a single point particle coupled to gravity.

\section{The massive point particle}\label{sect:2.1}
We will follow for the rest of the chapter the discussion in~\cite{review}.
Consider a point particle of mass $m$ propagating in a background spacetime, flat or curved, equipped with a metric tensor $g_{\mu\nu}(x)$. The action for such a particle is proportional to the invariant length of its worldline and its mass,
\be\label{Sds}
S=-m  \int \d s = -m \int \d\tau \sqrt{g_{\mu\nu}(x)\,\xd^{\mu}(\tau) \, \xd^{\nu}(\tau)} \, ,
\ee
where the classical trajectory is parametrised by $x^{\mu}(\tau)$ with $\tau$ an arbitrary worldline coordinate and we set $c=1$. The invariance of $s$ under reparametrisations $\tau \mapsto \tilde\tau(\tau)$ encodes the geometric statement that the action depends on the worldline itself and not on its parametrisation. The square-root form is, however, not practical to handle in practice, and we linearise it by introducing an einbein $e(\tau)$,
\be\label{Spp0}
S_{\rm pp} = -\frac{m}{2} \int \d\tau \,( e^{-1}\, \xd^{2} + e)\, .
\ee
Treating $e(\tau)$ as an independent field, its algebraic equation of motion is
\be\label{massshell1}
0= \xd^{2} - e^{2}\,,
\ee
and substituting the solution $e=\sqrt{\xd^{2}}$ back into~\refeq{Spp0} recovers the original action~\refeq{Sds}. The linearised action $S_{\rm pp}$ remains reparametrisation invariant, with the infinitesimal transformation $\tau \mapsto \tau + \xi(\tau)$ acting as
\be
\delta x^{\mu} = \xi \xd^{\mu}\, , \qquad \delta e = \frac{d}{d\tau} (\xi e) \, ,
\ee
and we use this freedom to fix the gauge $e(\tau)=1$. The equation of motion~\refeq{massshell1} then reduces to $\xd^{2}=1$ and $\tau$ becomes the proper time measured in the rest frame of the particle. The gauge-fixed action takes the very compact form
\be\label{Spp}
S_{\rm pp} = - \frac{m}{2}\int \d\tau \, \, g_{\mu\nu}(x)\,\xd^{\mu}(\tau) \, \xd^{\nu}(\tau) \, ,
\ee
where an irrelevant constant has been dropped. The conjugate momentum reads
\be
p_{\mu} = m \, g_{\mu\nu}\xd^{\nu}\, ,
\ee
and obeys the mass-shell condition $p^{2}=m^{2} $ which follows directly from~\refeq{massshell1}, now interpreted as a constraint on the initial data. Variation of the gauge-fixed action with respect to $x^{\mu}$ yields the geodesic equation,
\be\label{geodesiceq}
0 = {\ddot x}^{\mu} + \Gamma^{\mu}{}_{\nu\rho} \xd^{\nu}\xd^{\rho}\, .
\ee
The geodesic equation in turn preserves the gauge condition $\xd^{2}=1$, since
\be
\frac{d}{d\tau} \left ( g_{\mu\nu}(x)\,\xd^{\mu}(\tau) \, \xd^{\nu}(\tau) \right ) =0
\ee
is itself implied by~\refeq{geodesiceq}. Imposing $\xd^{2}=1$ at one instant of $\tau$ is therefore sufficient to enforce it everywhere along the trajectory. In flat spacetime $g_{\mu\nu}=\eta_{\mu\nu}$ the geodesic equation reduces to the familiar straight-line motion
\be
x^{\mu}(\tau) = x_0^{\mu} + v^{\mu}\, \tau \label{eq:straighTr},
\ee
with $v^{2}=1$.
 
 \section{Worldline effective field theory description of black holes}

We have introduced the action of a single point particle coupled to gravity. Coupling two such worldlines through their joint interaction with the gravitational field yields the leading terms of an effective description of a black hole binary. 
The same description applies not only to black holes but to any sufficiently compact massive object, such as a neutron star. As usual for an EFT, the description is only valid in the presence of a large separation of scales. In our case the separation $|b|$ of the two bodies has to be large compared to their intrinsic size. For a black hole of mass $m$ that size is set by the Schwarzschild radius $r_S = 2Gm$, while for a neutron star it is set by its radius $r_{NS}$, which is of comparable magnitude to its associated Schwarzschild radius.

A compact body is not a perfect point particle---it has internal structure and can in general be deformed. To account for this we must include corrections to the worldline action in the form of higher-dimensional operators. These operators are constrained only by the underlying symmetries of the theory, namely general coordinate transformations in the bulk and reparametrisation invariance on the worldline, and they come with so-called Wilson coefficients. The Wilson coefficients are a priori undetermined and have to be fixed by a matching computation against an independent description of the compact body. They also encode information about its microscopic structure, and are therefore in principle measurable in gravitational-wave signals. This effective field theory framework for compact massive bodies was introduced in ref.~\cite{Goldberger:2004jt} and originally applied to the non-relativistic, Post-Newtonian, regime of the gravitational two-body problem; we refer also to ref.~\cite{Levi:2018nxp,Goldberger:2007hy} for a detailed review.

We can build higher dimensional operators by contracting the curvature tensor $R^{\mu}{}_{\nu\rho\kappa}$ and the velocities $\dot x^{\mu}(\tau)$ of the compact objects to scalar quantities and add them order by order in an energy expansion.
We also include factors of the einbein $e$ to ensure reparametrisation invariance on the worldline. 
We get up to linear order in the curvature coupling as the extension of the action $S_{\rm pp}$ in \refeq{Spp0} 
\begin{align}
S_{\rm pp}'=\int \d\tau \big( &-\sfrac{me(\tau)}{2}
-\sfrac{m}{2}e(\tau)^{-1} g_{\mu\nu}[x(\tau)]\, \dot x^{\mu}(\tau)\dot x^{\nu}(\tau)  + c_{R,1} e(\tau) \,R[x(\tau)]  \nn \\
& + {c_{R,2}}{e(\tau)}^{-1} R_{\mu\nu}[x(\tau)] \,\dot x^{\mu}(\tau)\dot x^{\nu}(\tau) \big) + \cO(R^{2})\, .
\end{align}
The couplings $c_{R,1}$ and $c_{R,2}$ are the Wilson coefficients characterising the internal structure of the compact body. The two operators linear in $R$ can be removed by a field redefinition of the metric,
\be
g_{\mu\nu}\Rightarrow g_{\mu\nu}\left (1 + \kappa_{1}\frac{e^{2}}{\dot x^{2}}R \right ) +\kappa_{2} R_{\mu\nu}\,,
\ee
with a suitable choice of $\kappa_{1,2}$. Observables are invariant under such field redefinitions, so that the Lagrangians are equivalent. This eliminates the linear-curvature terms at the cost of introducing higher-curvature ones, and the Einstein-Hilbert bulk action picks up higher-curvature corrections. By the same logic, we have not included any terms coupling to the acceleration $\ddot x^{\mu}(\tau)$, since these can be removed by a redefinition of the trajectory $x^{\mu}(\tau)$. Since we can omit the Ricci tensor in our action, the higher-dimensional operators should be built only from the traceless part of the Riemann curvature tensor, the Weyl tensor $C_{\mu\nu\rho\sigma}$, and the velocity $\dot x^{\mu}(\tau)$. The gravitational dynamics on the worldline are therefore encoded in the Weyl tensor $C_{\mu\nu\rho\kappa}$, which can conveniently be decomposed into the electric and magnetic curvature tensors
\be
  E_{\mu \nu}:= R_{\mu \alpha \nu \beta} \dot{x}^{ \alpha} \dot{x}^{ \beta}\,, \qquad
  B_{\mu \nu}:= R^{*}_{\mu \alpha \nu \beta} \dot{x}^{ \alpha} \dot{x}^{ \beta}\, ,
\ee
with the dual Riemann tensor $R^*_{{\mu\nu}\alpha\beta}=\frac12 \varepsilon_{\mu\nu}{}^{\rho\sigma} R_{\rho\sigma\alpha\beta}$.
These satisfy the relation
\begin{align}
    E_{\mu\nu}+\I
    B_{\mu\nu}
   &=
   (C_{\mu\alpha\nu\beta}+\I
   C^*_{\mu\alpha\nu\beta}) {\dot x}^{\alpha} {\dot x}^{\beta}
   =
   \frac18{G_{\mu\alpha}}^{\rho\sigma}{G_{\nu\beta}}^{\lambda\tau}C_{\rho\sigma\lambda\tau}{\dot x}^\alpha{\dot x}^\beta\,, \label{eq:EBextra}
\end{align}
which involves the dual Weyl tensor $C^*_{{\mu\nu}\alpha\beta}=\frac12 \varepsilon_{\mu\nu}{}^{\rho\sigma} C_{\rho\sigma\alpha\beta}$ and the tensor
\begin{align}
    G_{\mu\nu\rho\sigma}=2g_{\mu[\rho}g_{\sigma]\nu}+\I \varepsilon_{\mu\nu\rho\sigma}\,.
\end{align}
We now invert this relation to show that we can also express the Weyl tensor in terms of the electric and magnetic curvature tensor \cite{Vines:2016unv}:
\begin{align}
    C_{\mu\nu\rho\kappa} +
    \I
    C^{\ast}_{\mu\nu\rho\kappa}
    =
    G_{\mu\nu}{}^{\alpha\sigma}G_{\rho\kappa}{}^{\beta\epsilon}
    \left
     (
         E_{\sigma\epsilon}
        + \I
         B_{\sigma\epsilon}
     \right ) {\dot x}_{\alpha}{\dot x}_{\beta}\,. \label{eq:inverted}
\end{align}
To derive this we insert the relation \refeq{eq:EBextra} into the right-hand side of \refeq{eq:inverted}, making use of the additional identity (valid in four dimensions)
\begin{align}
    G_{\mu\nu\rho\sigma}={G_{\mu\nu\lambda}}^\alpha{G_{\rho\sigma}}^{\lambda\beta}{\dot x}_\alpha{\dot x}_\beta\,,
\end{align}
which follows from $\varepsilon_{[\mu\nu\rho\sigma}\pi_{\lambda]}=0$, with $\pi_\mu$ an arbitrary vector. Taking the real part of \refeq{eq:inverted} we arrive at the explicit relation
\begin{align} \label{poprel}
\begin{aligned}
 C_{\mu\nu\rho\kappa} =&
     - 2E_{\mu[\rho}(g_{\kappa]\nu}-2{\dot x}_{\kappa]} {\dot x}_{\nu})
     + 2E_{\nu[\rho}(g_{\kappa]\mu}-2{\dot x}_{\kappa]} {\dot x}_{\mu})
    \\ &
        + 2B_{[\mu  }{}^{\lambda} {\dot x}_{\nu]}{\dot x}^\alpha\varepsilon_{\alpha\rho \kappa  \lambda}
        + 2B_{[\rho }{}^{ \lambda} {\dot x}_{\kappa]}{\dot x}^\alpha\varepsilon_{\alpha\mu \nu  \lambda}
     \,,
\end{aligned}
\end{align}
which expresses the Weyl tensor in terms of the gravito-electromagnetic tensors. Hence in four dimensions the building blocks of the higher-dimensional operators are the velocity $\dot x$ and the gravito-electromagnetic tensors $E_{\mu\nu}$ and $B_{\mu\nu}$.

Restricting to parity-even interactions, that is, those with an even number of Levi-Civita tensors in the action, the first correction to the point particle action~\refeq{Spp0} reads~\cite{Kalin:2020mvi}
\begin{align}    S &\!=\!-m \int\!\d\tau \left[\frac{1}{2e} g_{\mu\nu}\dot x^{\mu}\dot x^{\nu} + \frac{e}{2}
       +
       \frac{c_{E^2}}{e^{3}} E_{\mu \nu} E^{ \mu \nu}
       \!+\!\frac{c_{B^2}}{e^{3}} B_{\mu \nu} B^{\mu \nu}\right]\!\!.
\end{align}
The coefficients $c_{E/B^2}$ are related to the static Love numbers~\cite{Hinderer:2007mb,Binnington:2009bb}. It can be shown~\cite{Parra-Martinez:2025bcu,Porto:2016zng,Ivanov:2022qqt} that the static Love numbers vanish for spinless black holes, and we therefore ignore these corrections in what follows. We emphasise, however, that they can be incorporated in the present framework with very little conceptual effort~\cite{Jakobsen:2023pvx,Heissenberg:2022tsn,AccettulliHuber:2020dal,Kalin:2020lmz,Mougiakakos:2022sic,Jakobsen:2022psy,Mandal:2023hqa,Bini:2012gu,Henry:2019xhg,Bini:2020flp,Cheung:2020sdj,Haddad:2020que,Bern:2020uwk}.

For the Schwarzschild black holes considered in this thesis, the relevant action therefore reduces to the bare point particle action,
\begin{align}    S_{\text{Schw.}} &\!=\!-m \int\!\d\tau \left[\frac{1}{2e} g_{\mu\nu}\dot x^{\mu}\dot x^{\nu} + \frac{e}{2}
     \right]\!\!.
\end{align}

 \chapter{Worldline quantum field theory}
 \label{sec:inin}
 
 In the previous chapter we derived an effective action that describes the scattering of two black holes within an EFT framework. In principle this action can be used to derive the classical equations of motion and to solve them, although, as is so often the case in physics, this is impossible to do without perturbation theory. In this chapter we develop an efficient way to obtain the equations of motion in a Post-Minkowskian expansion around flat spacetime, and we explain how to encode them diagrammatically through a corresponding set of Feynman rules. 
 
 We review the WQFT formalism~\cite{Mogull:2020sak} applied to the gravitational scattering of two massive bodies; in this chapter we follow closely~\cite{review} and see also~\cite{Jakobsen:2023oow}.
 \section{Classical dynamics from one-point functions}
 
Following our discussion in Sec.~\ref{sect:2.1}, the system we consider is obtained by coupling two massive point particles in curved spacetime to the Einstein-Hilbert action together with a gauge-fixing term,
 \be\label{eq:WQFTGR}
 S_{\text{WQFT}}= - \sum_{i=1}^{2} \int \d\tau \left (\,\frac{m_{i}}{2}\, g_{\mu\nu}[x_{i}(\tau)] \xd_{i}^{\mu}(\tau)
 \xd_{i}^{\nu}(\tau) \right )
 - \dfrac{1}{16\pi G} \int \d^{d}x \sqrt{-g}\, R[g]  + S_{\rm g.f.}\, ,
 \ee
 with $G$ the $d$-dimensional Newton's constant. The gauge-fixing term $S_{\rm g.f.}$ is defined in App.~\ref{sec:gaugefixing}. We work in flat space in a mostly-minus signature for the metric and our conventions are collected in App.~\ref{App:Conventions}. Since we will work in the weak-field approximation we expand the metric around flat Minkowski space,
 \be
 \label{eq:gravitondef}
 g_{\mu\nu}=\eta_{\mu\nu}+\kappa h_{\mu\nu}\, , \qquad \kappa= \sqrt{32\pi G} \, .
 \ee
 Expanding the bulk action to quadratic order in the graviton field reproduces the standard kinetic Lagrangian of linearised gravity, while higher orders give rise to the graviton self-interactions; we refer the reader to ref.~\cite{Badger:2023eqz} for further details.

 At the same time we expand $x_i(\tau)$ around the geodesic trajectory in flat space~\refeq{eq:straighTr}:
 \begin{align}
        x_i^\mu(\tau)=x^\mu_{0,i}+v_i^\mu \tau +z_i^\mu(\tau)\,. \label{eq:xexpansion}
 \end{align}
We will now work in the path-integral formalism and obtain the classical equations of motion as the classical limit of the quantised theory. The partition function reads
 \begin{align}
         Z[J_i]=\int
         D[\phi_i]
         \exp{ \left[ \dfrac{i}{\hbar} \left( S[\phi_i] +\sum_i \int \d x_i J_i(x_i) \phi_i(x_i) \right) \right]}\,,
 \end{align}
 where $x_i= x^\mu$ or $x_i =\tau$ depending on whether $\phi_i$ is a bulk or a worldline (WL) field.
 In particular we will have $\phi_i \in \{ h_{\mu\nu}, z^{\mu}_{i}(\tau)\}$ and $x_h=x^\mu$ and $x_z=\tau$.
 We then define one-point functions through
 \begin{align}
         \vev{\mathcal{O}}_{\rm WQFT} = \int D[\phi_i] \, \mathcal{O}[\phi_i]\, e^{\frac{\I}{\hbar}
 S[\phi_i]} \,,
 \end{align}
 with a normalised path-integral measure so that $\vev{1}_{\rm WQFT}=1$. By construction, the one-point functions $\vev{\mathcal{O}}_{\rm WQFT}$ depend only on the background parameters $\{v_{i},x_{0,i}\}$. The key property that we exploit in the WQFT approach is that the \emph{tree-level one-point functions solve the classical equations of motion}.
 
 To show this we consider the $\hbar$ dependence of a generic correlation function $\wev{\psi_{1}\ldots \psi_{n}}$ with $\psi_i \in \{ h_{\mu\nu}, z^{\mu}_{i}(\tau)\}$. 
 A one point function with a loop scales as $\hbar$:
 \begin{align}
        \begin{tikzpicture}[baseline={(current bounding box.center)}]
                \begin{feynman}
                \coordinate (a1) at (-0.5,0);
                \coordinate (a2) at (0.5,0);
                \draw [photon] (a1) -- (a2) ;
                \draw [fill] (a1) circle (.0) ;
                \draw [fill] (a2) circle (.0) node [right] {$\sim \hbar$};
                \end{feynman}
                \end{tikzpicture} ,\quad
                \begin{tikzpicture}[baseline={(current bounding box.center)}]
                        \begin{feynman}
                        \coordinate (a1) at (-0.5,0);
                        \coordinate (a2) at (0.5,0);
                        \draw [zUndirected] (a1) -- (a2) ;
                        \draw [fill] (a1) circle (.0)  ;
                        \draw [fill] (a2) circle (.0) node [right] {$\sim \hbar$};
                        \end{feynman}
                        \end{tikzpicture} ,\quad \begin{tikzpicture}[baseline={(current bounding box.center)}]
                                \coordinate (in) at (-1,0);
                                \coordinate (out1) at (1,0);
                                \coordinate (out2) at (1,0.5);
                                \coordinate (x) at (0,0);
                                \node (k) at (0,-1.3) {$$};
                                \draw (out1) ;
                                \draw (out2) node [right] {$$};
                                \draw [dotted] (in) -- (x);
                                \draw [zParticle] (x) -- (out1);
                                \draw [zParticle] (x) to[out=40,in=180] (out2);
                                \draw [graviton] (x) -- (k);
                                \draw [fill] (x) circle (.08);
                                \draw [fill] (1,-.3) circle (.0) node [right] {$\sim \dfrac{1}{\hbar}$};
                                \end{tikzpicture},\Rightarrow \begin{tikzpicture}[baseline={(current bounding box.center)}]
                                        \coordinate (in) at (-1,0);
                                        \coordinate (out1) at (2,0);
                                        \coordinate (x) at (.5,0);
                                        \draw (out1) ;
                                        \draw [zUndirected] (in) -- node [below] {\tiny $\sim \hbar$}(x);
                                        \draw [zParticle] (x) --node [below] {\tiny$\sim \hbar$} (out1);
                                        \draw [zUndirected] (in) to[out=45,in=135]node [above] {\tiny$\sim \hbar$} (x);
                                        \draw [fill] (x) circle (.08)node [below] {\tiny $\sim \hbar^{-1}$};
                                        \draw [fill] (in) circle (.08)node [below] {\tiny $\sim \hbar^{-1}$};
                                        \draw [] (out1) circle (.0) node [right] {$\sim \hbar$};
                                        \end{tikzpicture}.
 \end{align}
 This follows from the fact that the action $S_{\text{WQFT}}$ does not depend on $\hbar$, so the vertices that arise from expanding the interaction terms in $e^{\frac{\I}{\hbar} S_{\text{WQFT}}}$ scale as $\hbar^{-1}$, while the propagators, obtained from an exact treatment of the quadratic terms in the action, scale as $\hbar$. For a generic connected tree-level diagram with $n$ external legs it is then easy to see that the diagram scales as $\hbar^{n-1}$, with $n$ incoming or outgoing particles  and propagators $\sim \hbar^n$ and an internal structure $\sim \hbar^{-1}$.
 A loop diagram may be viewed as a tree in which one vertex of arbitrary multiplicity has been replaced by a loop, and every loop raises the $\hbar$-counting by one. A connected $n$-point correlator therefore scales as

 \begin{align}
         \vev{\psi_{1}\ldots \psi_{n}}_{\rm WQFT,conn}\sim \sum_{\rm L=0}^{\infty}\hbar^{n-1+L} \,(\text{$L$-loop connected $n$-point diagrams})\,.
 \end{align}
 The consequence is that only tree-level one-point functions contribute at the classical level.
 The expectation values of the form $\vev{\psi_{1}\ldots \psi_{n}}_{\rm WQFT}$ are thus given by the product of one point functions
 \begin{align}
         \vev{\psi_{1}\ldots \psi_{n}}_{\rm WQFT}=\vev{\psi_{1}}_{\rm WQFT}\ldots \vev{\psi_{n}}_{\rm WQFT}\;.
 \end{align}
 This also implies that $\vev{ \frac{\delta S[\phi_i]}{\delta \phi_i} }$ factorises in the classical limit, which reduces
 the Schwinger--Dyson equation to
 \begin{align}
         0=\lim_{\hbar\to 0}  \vev{ \frac{\delta S[\phi_i]}{\delta \phi_i} }= \frac{\delta S[\vev{\phi_i}^{\rm tree}_{\rm WQFT}]}{\delta \vev{\phi_i}^{\rm tree}_{\rm WQFT}}\,.
 \end{align}
 These are again the Euler-Lagrange equations or simply the classical equations of motion.

 We summarise our findings. We have a procedure to compute classical expectation values, which obey the classical equations of motion, in terms of tree-level one-point functions. One problem remains, however. In order to ensure convergence of the path integrals we have to assume vanishing boundary conditions for all fields at $\tau=\pm\infty$. The fields under consideration are $h_{\mu\nu}$ and $z_i^\mu$, which should certainly vanish at $\tau = -\infty$ but which are not required to vanish at $\tau = +\infty$, where the system carries the imprint of the scattering.
 
 This mismatch of boundary conditions is tied to the choice of propagators. In an in--out QFT computation, where the expectation value is taken between an incoming and an outgoing vacuum state, one uses Feynman propagators
 \begin{align}
         \dfrac{1}{k^2+ \i0}
 \end{align}
 to enforce vanishing boundary conditions at both $\tau=\pm\infty$. In a classical computation one uses instead retarded propagators
 \begin{align}
         \dfrac{1}{(k^0+\i0)^2-\boldsymbol{k}^2}
 \end{align}
 to enforce causal boundary conditions. As we will see, the resolution is simply to replace every Feynman propagator by a retarded one and to enforce a causality flow in our Feynman diagrams. This prescription follows directly from the classical equations of motion, and it can be put on formal footing using the in-in formalism, to which we now turn.
 
 \section{Causal boundary conditions through the in-in formalism}
 
 We have seen that in the in--out formalism we solve the equations of motion but with the wrong boundary conditions. If we were to solve the equations of motion in the standard way, we would simply use retarded boundary conditions and hence retarded Green's functions to obtain a classical solution. 
 We now discuss the in-in formalism, which derives the same conclusion in a formal manner~\cite{Galley:2009px,Jakobsen:2022psy,Kalin:2022hph,Schwinger:1960qe,Keldysh:1964ud}.

 In the standard in--out formalism, which we used naively in the previous chapter, one computes, via the path integral, expectation values between the initial vacuum state and the final vacuum state. As already discussed this makes little sense for us, since our fields do not vanish for $\tau=\infty$, so that there is no final vacuum state. Instead we want an expectation value which depends only on the initial vacuum state from which the system evolves obeying causality. The desired expectation value is basically the expectation value of an operator in the Heisenberg picture
 \begin{align}
         \vev{\hat{\phi_i}_H(t,\boldsymbol{x})}_{in-in}:=\vev{0 |\hat{\phi_i}_H(t,\boldsymbol{x})|0} =\vev{0 |\hat{U}(-\infty,t)\hat{\phi_i}_I(t,\boldsymbol{x})\hat{U}(t,-\infty)|0}  ,
 \end{align}
 where $\hat{\phi_i}_H$ and $\hat{\phi_i}_I$ are the operators in the Heisenberg and interaction pictures. 
 The vacuum states $|0\rangle=|0 \rangle_{in}$ are here the initial vacuum states, which give the formalism its name.
 We define a generalised generating functional $W[J_{i,1},J_{i,2}]$,
 \begin{align}
         \exp{\left( \dfrac{i}{\hbar} W[J_{i,1},J_{i,2}] \right)} =\vev{0|\hat{U}_{J_{i,2}}(-\infty,\infty)\hat{U}_{J_{i,1}}(\infty,-\infty)|0}\,.
 \end{align}
 The desired expectation value follows from
 \begin{align}
         \dfrac{\delta W[J_{i,1},J_{i,2}]}{\delta J_{i,1}}\bigg|_{J_{i,j}=0}&=\vev{0 |\hat{U}(-\infty,\infty)\hat{U}(\infty,t)\hat{\phi}_{i,1,I}(t,\boldsymbol{x})\hat{U}(t,-\infty)|0} \nn\\
        =- \dfrac{\delta W[J_{i,1},J_{i,2}]}{\delta J_{i,2}}\bigg|_{J_{i,j}=0}&= \vev{0 |\hat{U}(-\infty,t){\hat{\phi}_{i,2,I}}(t,\boldsymbol{x})\hat{U}(t,\infty)\hat{U}(\infty,-\infty)|0} \nn\\
       &=  \vev{0 |\hat{U}(-\infty,t)\hat{\phi_i}_I(t,\boldsymbol{x})\hat{U}(t,-\infty)|0} \,.
 \end{align}
 A path-integral representation of this functional can be obtained by doubling the fields, with $\phi_{i,1}$ propagating forward in time and $\phi_{i,2}$ propagating backwards in time,
 \begin{align}
         &\exp{\left( \dfrac{i}{\hbar} W[J_{i,1},J_{i,2}] \right)}= \nn\\
         &\int D[\phi_{i,1},\phi_{i,2}]\exp{\left[ \dfrac{i}{\hbar} \left( S[\phi_{i,1}]- S[\phi_{i,2}]+\int \d x_i (J_{i,1}(x_i)\phi_{i,1}(x_i)-J_{i,2}(x_i)\phi_{i,2}(x_i))\right)\right]}\,.
 \end{align}
 The two field copies are linked by the boundary conditions
 \begin{align}
         \phi_{i,1}(t=\infty,\boldsymbol{x})&=\phi_{i,2}(t=\infty,\boldsymbol{x})\,,\\
         \phi_{i,1}(t=-\infty,\boldsymbol{x})&=\phi_{i,2}(t=-\infty,\boldsymbol{x})=0\,.
 \end{align}
 After the field doubling, the simple propagator of the field $\phi_i$ becomes a propagator matrix~\cite{Jakobsen:2022psy,Jakobsen:2023oow},
 \begin{align}
        \underline D_{i,A,B}&=
        \left(
                 \begin{array}{cc}
                 D_{i,F}(x,y) &  -D_{i,-}(x,y)\\
                 -D_{i,+}(x,y) & -D_{i,D}(x,y) \\
                 \end{array}
                 \right),
 \end{align}
 where $A,B \in \{ 1,2\}$ and $\{F,D,+,-\}$ denotes the Feynman, Dyson, and the two Wightman propagator prescriptions; see~\cite{Jakobsen:2022psy} for a detailed derivation of the propagator matrix. It ensures the right time flow and boundary conditions of our fields. 
 For a scalar field we would have
 \begin{align}
        D_{F/D}= \dfrac{1}{k^2 \pm \i0},\quad D_{+/-}= i 2\, \pi \,\delta(k^2) \theta(\pm k^0).
 \end{align}
 
 \subsection*{Schwinger--Keldysh basis}
 
 It is convenient to switch to the Schwinger--Keldysh basis~\cite{Keldysh:1964ud} to simplify the propagator matrix,
 \begin{align}
         \phi_{i,+}=\frac{1}{2}(\phi_{i,1}+\phi_{i,2})\,,\quad \phi_{i,-}=\phi_{i,1}-\phi_{i,2}\,,\nn\\
         J_{i,+}=\frac{1}{2}(J_{i,1}+J_{i,2})\,,\quad J_{i,-}=J_{i,1}-J_{i,2}\,.
 \end{align}
 These new fields obey the boundary conditions
 \begin{align}
       &  \phi_{i,-}(t=\infty,\boldsymbol{x})=0\,,\\
       &  \phi_{i,+}(t=-\infty,\boldsymbol{x})=\phi_{i,-}(t=-\infty,\boldsymbol{x})=0\,.
 \end{align}
 The plus field obeys exactly the boundary conditions that we want for our physical field obeying retarded boundary conditions fixed only at $\tau=-\infty $. 

It is also useful to change the basis of propagators by using the relations
\begin{align}
        D_{i,F}(x,y)=D_{i,ret}(x,y)-D_{i,-}(x,y)=D_{i,adv}(x,y)-D_{i,+}(x,y)\;,\nn\\
        D_{i,D}(x,y)=D_{i,ret}(x,y)+D_{i,+}(x,y)=D_{i,adv}(x,y)+D_{i,-}(x,y)\;.
\end{align}
The propagator matrix for the Schwinger--Keldysh basis reads
 \begin{align}
        \underline D_{i,a,b}&
        =\left(
                \begin{array}{cc}
               \frac{1}{4}( D_{i,F}- D_{i,D}- D_{i,+}- D_{i,-}) & \frac{1}{2}( D_{i,F}+ D_{i,D}- D_{i,+}+D_{i,-})\\
               \frac{1}{2}( D_{i,F}+ D_{i,D}+ D_{i,+}-D_{i,-}) &  D_{i,F}- D_{i,D}+ D_{i,+}+D_{i,-} \\
                \end{array}
                \right)\\
                & =\left(
                 \begin{array}{cc}
                -\frac{1}{2} D_{i,H}(x,y) &  D_{i,ret}(x,y)\\
                 D_{i,adv}(x,y) & 0 \\
                 \end{array}
                 \right),
 \end{align}
 where $a,b \in \{+,-\}$.
 The new propagators in our matrix are the retarded and advanced propagators on the off-diagonal and the Hadamard propagator $D_{i,H}=(D_{i,+}+D_{i,-})$ on the diagonal.
 
 We will see shortly that only the retarded entry contributes at the classical level. In the new fields the generating functional reads
 \begin{align}
         \exp{\left( \dfrac{i}{\hbar} W[J_{i,+},J_{i,-}] \right)}=
         \int D[\phi_{i,+},\phi_{i,-}]&\exp{}\Bigg[ \dfrac{i}{\hbar} \Big( S[\phi_{i,+}+\phi_{i,-}/2]- S[\phi_{i,+}-\phi_{i,-}/2]\nn \\
         &+\int \d x_i (J_{i,+}(x)\phi_{i,-}(x)+J_{i,-}(x)\phi_{i,+}(x))\Big)\Bigg]\,,
 \end{align}
 and the desired expectation value is generated by
 \begin{align}
         \vev{\hat{\phi_i}_H(t,\boldsymbol{x})}_{in-in}&=\dfrac{\delta W[J_{i,+},J_{i,-}]}{\delta J_{i,-}}\bigg|_{J_{i,\pm}=0}\nn\\
         &=Z_0^{-1}\int D[\phi_{i,+},\phi_{i,-}]\phi_+(x)\exp{\left[ \dfrac{i}{\hbar} \left( S[\phi_{i,+}+\phi_{i,-}/2]- S[\phi_{i,+}-\phi_{i,-}/2]\right)\right]},
 \end{align}
 where 
 \begin{align}
        Z_0=
        &\int D[\phi_{i,+},\phi_{i,-}]\exp{\left[ \dfrac{i}{\hbar} \left( S[\phi_{i,+}+\phi_{i,-}/2]- S[\phi_{i,+}-\phi_{i,-}/2]\right)\right]}
 \end{align}
 We can therefore use the in-in formalism to derive the classical equations of motion diagrammatically through one-point functions with an outgoing $\phi_{i,+}$ leg.
 
 Expanding our action in $\phi_{i,-}$ we find only terms odd in $\phi_{i,-}$,
 \begin{align}
         S_{in-in,i, int}=\phi_{i,-}\left(\dfrac{\delta S_{int}[\phi_i]}{\delta \phi_i}\right)\bigg|_{\phi\rightarrow \phi_{i,+}}+\mathcal{O}(\phi_{i,-}^3)\,,
 \end{align}
 where crucially $S_{int,i}[\phi_i]$ is the standard in--out action.
This finding implies that all fields sourced from the worldline must be ``$-$''-fields.
It follows from this that only vertices with a single minus leg can contribute. All higher vertices would lead to a loop level graph, since there is no ``$--$''-propagator. Additionally only the retarded ``$+-$''-propagator can contribute to one-point functions with an outgoing ``$+$''-leg. We refer to ref.~\cite{Jakobsen:2022psy} for a detailed derivation.
The vertices themselves take the same form as in the in--out formalism. The source vertices must carry a ``$-$''-leg, while the outgoing field is a ``$+$''-field, so that the retarded propagators always point towards the outgoing line.
 From here on we may therefore forget about the in-in formalism and simply use the in--out Feynman rules, replacing all propagators by retarded propagators pointing towards the outgoing line and thereby ensuring a causality flow, as already anticipated in the previous section. Furthermore we will set from now on $\hbar =1$.

 \subsection*{A comment on the split into conservative and dissipative effects and on the choice of propagators}
 
 The computation we will perform is intended, among other things, to inform EOB models and ultimately the generation of waveform templates. For that purpose one needs to split the dynamics into conservative and dissipative pieces. We have just described how to compute the full physical process with retarded propagators, and the split between conservative and dissipative effects will then rely primarily on a different choice of propagator for the gravitons. We will therefore also discuss the use of Feynman propagators in the following chapters. The $\i0$ prescription can often be left implicit, since many of the methods we employ are independent of it. We will return to this issue in the results chapter, where we describe how to perform the split at each order in PM. Roughly speaking, the (admittedly imperfect) idea behind the use of Feynman propagators is to employ an in--out formalism in the bulk, which enforces the absence of outgoing radiation via Feynman boundary conditions, while keeping an in-in formalism on the worldline. 

 \section{Feynman rules}
 
 In this section we collect the Feynman rules that will be needed to compute observables.
 
 \subsection{Gravitons}
 
 To derive the graviton Feynman rules we expand the Einstein-Hilbert action around flat space using~\refeq{eq:gravitondef}. For the kinetic part of the action we find
 \be
 \mathcal{L}_{\rm EH}|_{h^{2}} + \mathcal{L}_{\rm g.f.}= -\sfrac12 h_{\mu\nu}
 \left [ \eta^{\mu(\rho}\eta^{\sigma)\nu} - \sfrac 12 \eta^{\mu\nu} \eta^{\rho\sigma}\right ] \partial^{2}h_{\rho\sigma}
 \, ,
 \ee
 from which we can compute the graviton propagator,
 \begin{equation}\label{eq:hPropagator}
 \begin{tikzpicture}[baseline={(current bounding box.center)}]
 \coordinate (x) at (-.5,0);
 \coordinate (y) at (1.5,0);
 \draw [graviton] (x) -- (y) node [midway, below] {$k$} node [midway, above] {$\rightarrow$};
 \draw [fill] (x) circle (.08) node [above] {$h_{\mu\nu}$};
 \draw [fill] (y) circle (.08) node [above] {$h_{\rho\sigma}$};
 \end{tikzpicture}=\frac{\I P_{\mu\nu;\rho\sigma}}{k^2}\,, \qquad \text{where} \quad P_{\mu\nu;\rho\sigma}=\eta_{\mu(\rho}\eta_{\sigma)\nu}-
 \frac{1}{d-2}\eta_{\mu\nu}\eta_{\rho\sigma}\,.
 \end{equation}
 For a retarded prescription this becomes
 \begin{align}
        \dfrac{\I P_{\mu \nu;\rho \sigma}}{(k^0+\i0)^2-\boldsymbol{k}^2}\,,
 \end{align}
 while the Feynman prescription reads
 \begin{align}
         \dfrac{\I P_{\mu \nu;\rho \sigma}}{k^2+\i0}\,.
  \end{align}
 The complexity of gravity arises from its self-interactions, which translate in the Feynman-diagram language into graviton vertices. They are generated by expanding the interaction part of the Einstein-Hilbert action, which produces an infinite tower of vertices, of which only a finite number contribute at any given order in perturbation theory. Schematically, the interaction part takes the form
 \be\label{EHint}
 \mathcal{L}_{\rm{EH, int}} \propto \sum_{n=1}^{\infty} G^{n/2}\left [ \partial^{2}h^{n+2}\right ]\, ,
 \ee
 where the bracket denotes the order in derivatives and graviton fields. Concretely, the computation of the impulse at $n$-th PM order ($G^{n}$) requires graviton vertices up to multiplicity $n+1$.
 
 The vertices have the schematic structure
 \be
 \begin{tikzpicture}[baseline={(current bounding box.center)}]
         \begin{feynman}
         \coordinate (a1) at (-0.5,0);
         \coordinate (a2) at (0.5,0);
         \coordinate (a3) at (0,-1.0);
         \coordinate (o) at (0,-0.5);
         \diagram*{ (o) -- [graviton] (a1), (o) -- [graviton] (a2), (o) -- [graviton] (a3) };
         \draw [fill] (a1) circle (.0) ;
         \draw [fill] (a2) circle (.0) ;
         \draw [fill] (a3) circle (.0) ;
         \draw [fill] (o) circle (.06);
         \end{feynman}
         \end{tikzpicture} \sim G^{1/2} k^{2}  \, , \qquad
 \begin{tikzpicture}[baseline={(current bounding box.center)}]
 \label{gravitonvertexscaling}
 \begin{feynman}
         \coordinate (a1) at (-0.5,0);
         \coordinate (a2) at (0.5,0);
 \coordinate (a3) at (0.5,-1);
 \coordinate (a4) at (-0.5,-1);
 \coordinate (o) at (0,-0.5);
 \diagram*{ (o) -- [graviton] (a1), (o) -- [graviton] (a2), (o) -- [graviton] (a3), (o) -- [graviton] (a4)};
 \draw [fill] (o) circle (.06);
 \end{feynman}
 \end{tikzpicture} \sim G k^{2}  \, , \qquad
 \begin{tikzpicture}[baseline={(current bounding box.center)}]
 \begin{feynman}
         \coordinate (a1) at (-0.5,-0.2);
         \coordinate (a2) at (0.5,-0.2);
 \coordinate (a3) at (0.5,-1);
 \coordinate (a4) at (-0.5,-1);
 \coordinate (a5) at (0,0);
 \coordinate (o) at (0,-0.5);
 \diagram*{ (o) -- [graviton] (a1), (o) -- [graviton] (a2), (o) -- [graviton] (a3), (o) -- [graviton] (a4), (o) -- [graviton] (a5)};
 \draw [fill] (o) circle (.06);
 \end{feynman}
 \end{tikzpicture} \sim G^{3/2} k^{2}  \,, \qquad
 \begin{tikzpicture}[baseline={(current bounding box.center)}]
 \begin{feynman}
         \coordinate (a1) at (-0.5,-0.2);
         \coordinate (a2) at (0.5,-0.2);
 \coordinate (a3) at (0.5,-0.8);
 \coordinate (a4) at (-0.5,-0.8);
 \coordinate (a5) at (0,0);
 \coordinate (a6) at (0,-1);
 \coordinate (o) at (0,-0.5);
 \diagram*{ (o) -- [graviton] (a1), (o) -- [graviton] (a2), (o) -- [graviton] (a3), (o) -- [graviton] (a4), (o) -- [graviton] (a5), (o) -- [graviton] (a6)};
 \draw [fill] (o) circle (.06);
 \end{feynman}
 \end{tikzpicture} \sim G^{2} k^{2}  \,, \qquad \ldots
 \ee
 as indicated in \refeq{EHint}, and they grow rapidly in size. Although this infinite tower of vertices may appear problematic at first sight, the highest complexity in any Feynman diagram in practice always arises from the (multiple) three-graviton vertices.
 
 \subsection{Worldline dynamics}
 We can derive the propagator of the worldline field $z_i^\mu(\omega)$ after expanding $S_{\rm WQFT}$ around flat space from
 \begin{align}
        \mathcal{L}_{\rm{WQFT}}|_{z_i^2}=-\dfrac{m_i}{2} \eta_{\mu \nu}  {\dot z_i}^{\mu} {\dot z_i}^{\nu}.
 \end{align}
After Fourier transformation this yields the retarded worldline propagator
 \begin{align}
          \begin{tikzpicture}[baseline={(current bounding box.center)}]
            \coordinate (in) at (-0.6,0);
            \coordinate (out) at (1.4,0);
            \coordinate (x) at (-.2,0);
            \coordinate (y) at (1.0,0);
            \draw [zParticle] (x) -- (y) node [midway, below] {$\omega$};
            \draw [background] (in) -- (x);
            \draw [background] (y) -- (out);
            \draw [fill] (x) circle (.08) node [above] {$\mu$};
            \draw [fill] (y) circle (.08) node [above] {$\nu$};
          \end{tikzpicture}=\frac{-\I\eta^{\mu\nu}}{m_i \,( \omega + \i0)^{2}}\,.
        \end{align}
 The worldline interaction term in \refeq{eq:WQFTGR} arises from expanding the metric as in~\refeq{eq:gravitondef} and expanding the trajectory around the flat space geodesic as in~\ref{eq:xexpansion} and couples a single graviton to the worldline field $z_i^\mu$:
 \begin{align}\label{eq:allOrderWL}
 S_{\rm WQFT}^{\rm int} =& -\frac{m\kappa}{2} h_{\mu\nu}[x_0+v\tau + z(\tau)] \,\left  (v^{\mu}v^{\nu}
 + 2 v^{(\mu} {\dot z}^{\nu)}(\tau) +{\dot z}^{\mu}(\tau){\dot z}^{\nu}(\tau)
  \right ) \nn\\
  =&
 -m  \kappa \sum_{n=0}^\infty\frac{i^n}{n!}
 \int_{k,\omega_1,\ldots,\omega_n}
 e^{ik\cdot x_0}\dd\bigg(k\cdot v+\sum_{i=1}^n\omega_i\bigg)
 h_{\mu\nu}(-k)\left(\prod_{i=1}^nz^{\rho_i}(-\omega_i)\right) \times \nn \\
 &\times \left(\frac12\left(\prod_{i=1}^nk_{\rho_i}\!\right)\!v^\mu v^\nu+
 \sum_{i=1}^n\omega_i\left(\prod_{j\neq i}^nk_{\rho_j}\right)\!
 v^{(\mu}\delta^{\nu)}_{\rho_i}+
 \sum_{i<j}^n\omega_i\omega_j\left(\prod_{l\neq i,j}^nk_{\rho_{\text{l}}}\right)
 \delta^{(\mu}_{\rho_i}\delta^{\nu)}_{\rho_j}\right) \, ,
 \end{align}
 which gives rise to an infinite tower of graviton--worldline vertices. We use here the notation $\int_q =\int \frac{\d^d q}{(2 \pi )^d}$, $d=4-2 \epsilon$, $\dd(x)= 2\pi \delta(x)$ and we dropped the worldline index $i$. The computation of the impulse at $n$-th PM order ($G^{n}$) requires worldline vertices up to multiplicity $n$.  At leading order the one-point vertex reads
 \begin{align}\label{eq:vertexH}
 \begin{tikzpicture}[baseline={(current bounding box.center)}]
 \coordinate (in) at (-1,0);
 \coordinate (out) at (1,0);
 \coordinate (x) at (0,0);
 \node (k) at (0,-1.3) {$h_{\mu\nu}(k)$};
 \draw [dotted] (in) -- (x);
 \draw [dotted] (x) -- (out);
 \draw [graviton] (x) -- (k);
 \draw [fill] (x) circle (.08);
 \end{tikzpicture}=
 -\I\frac{m \kappa}{2}e^{ik\cdot x_0}\dd(k\cdot v)v^\mu v^\nu\,,
 \end{align}
 with $k$ outgoing. The two-point vertex reads
 \begin{align}\label{eq:zH}
 \begin{tikzpicture}[baseline={(current bounding box.center)}]
 \coordinate (in) at (-1,0);
 \coordinate (out) at (1,0);
 \coordinate (x) at (0,0);
 \node (k) at (0,-1.3) {$h_{\mu\nu}(k)$};
 \draw (out) node [right] {$z^\rho(\omega)$};
 \draw [dotted] (in) -- (x);
 \draw [zParticle] (x) -- (out);
 \draw [graviton] (x) -- (k);
 \draw [fill] (x) circle (.08);
 \end{tikzpicture}=
 \frac{m \kappa}{2}e^{ik\cdot x_0}\dd(k\cdot v+\omega)
 \left(2\omega v^{(\mu}\delta^{\nu)}_\rho+v^\mu v^\nu k_\rho\right)\,,
 \end{align}
 where the energy $\omega$ is also taken as outgoing. The three-point vertex is
 \begin{align}
 \begin{tikzpicture}[baseline={(current bounding box.center)}]
 \coordinate (in) at (-1,0);
 \coordinate (out1) at (1,0);
 \coordinate (out2) at (1,0.5);
 \coordinate (x) at (0,0);
 \node (k) at (0,-1.3) {$h_{\mu\nu}(k)$};
 \draw (out1) node [right] {$z^{\rho_1}(\omega_1)$};
 \draw (out2) node [right] {$z^{\rho_2}(\omega_2)$};
 \draw [dotted] (in) -- (x);
 \draw [zParticle] (x) -- (out1);
 \draw [zParticle] (x) to[out=40,in=180] (out2);
 \draw [graviton] (x) -- (k);
 \draw [fill] (x) circle (.08);
 \end{tikzpicture}=&
 \I m \kappa e^{ik\cdot x_0}\dd(k\cdot v+\omega_1+\omega_2)\nn
 \qquad\qquad\qquad\qquad\qquad
 \end{align}
 \vspace{-3.5em}
 \begin{align}
 \qquad\qquad\qquad\qquad\qquad\quad
 \times \left(\frac12k_{\rho_1} k_{\rho_2}v^\mu v^\nu+
 \omega_1k_{\rho_2} v^{(\mu}\delta^{\nu)}_{\rho_1}+
 \omega_2k_{\rho_1} v^{(\mu}\delta^{\nu)}_{\rho_2}+
 \omega_1\omega_2\delta^{(\mu}_{\rho_1}\delta^{\nu)}_{\rho_2}\right).\label{eq:zzH}
 \end{align}
 The second worldline deflection mode is on the same worldline. The dots represent the unperturbed trajectory $x_0+v\tau$. 
 
 The higher-point vertices follow analogously and have the following iterative structure:
 \begin{subequations}
 \begin{align}
   \left[\quad\begin{tikzpicture}[baseline={(current bounding box.center)},scale=.7]
   \coordinate (in) at (-1,0);
   \coordinate (out1) at (1,0);
   \coordinate (out2) at (1,0.5);
   \coordinate (out3) at (1,0.9);
   \coordinate (out4) at (1,1.4);
   \coordinate (x) at (0,0);
    \node (k1) at (-.8,-1.3) {};
     \node (k2) at (0,-1.2) {};
     \node (k3) at (.8,-1.3) {};
   \draw (out2) node [right] {$\!\!\!\vdots$};
   \draw (out4) node [right] {$z^\mu(\omega)$};
   \draw [dotted] (in) -- (x);
   \draw [zParticle] (x) -- (out1);
   \draw [zParticle] (x) to[out=30,in=180] (out3);
   \draw [zParticle] (x) to[out=60,in=180] (out4);
    \draw [graviton] (x) -- (k2);
   \draw [fill] (x) circle (.08);
   \end{tikzpicture}\right]_{\omega=0}&=
   \frac{\partial}{\partial x_0^{\mu}}\quad\,
   \begin{tikzpicture}[baseline={(current bounding box.center)},scale=.7]
     \coordinate (in) at (-1,0);
     \coordinate (out1) at (1,0);
     \coordinate (out2) at (1,0.5);
     \coordinate (out3) at (1,0.9);
     \coordinate (x) at (0,0);
      \node (k1) at (-.8,-1.3) {};
     \node (k2) at (0,-1.2) {};
     \node (k3) at (.8,-1.3) {};
     \draw (out1) node [right] {};
     \draw (out2) node [right] {$\!\!\!\vdots$};
     \draw (out3) node [right] {};
     \draw [dotted] (in) -- (x);
     \draw [zParticle] (x) -- (out1);
     \draw [zParticle] (x) to[out=30,in=180] (out3);
      \draw [graviton] (x) -- (k2);
     \draw [fill] (x) circle (.08);
     \end{tikzpicture}\,, \\
   \left[-\I\frac{\partial}{\partial\omega}\quad\begin{tikzpicture}[baseline={(current bounding box.center)},scale=.7]
   \coordinate (in) at (-1,0);
   \coordinate (out1) at (1,0);
   \coordinate (out2) at (1,0.5);
   \coordinate (out3) at (1,0.9);
   \coordinate (out4) at (1,1.4);
   \coordinate (x) at (0,0);
    \node (k1) at (-.8,-1.3) {};
     \node (k2) at (0,-1.2) {};
     \node (k3) at (.8,-1.3) {};
   \draw (out2) node [right] {$\!\!\!\vdots$};
   \draw (out4) node [right] {$z^{\mu}(\omega)$};
   \draw [dotted] (in) -- (x);
   \draw [zParticle] (x) -- (out1);
   \draw [zParticle] (x) to[out=30,in=180] (out3);
   \draw [zParticle] (x) to[out=60,in=180] (out4);
   \draw [graviton] (x) -- (k2);
   \draw [fill] (x) circle (.08);
   \end{tikzpicture}\right]_{\omega=0}&=
   \frac{\partial}{\partial v^{\mu}}\quad
   \begin{tikzpicture}[baseline={(current bounding box.center)},scale=.7]
     \coordinate (in) at (-1,0);
     \coordinate (out1) at (1,0);
     \coordinate (out2) at (1,0.5);
     \coordinate (out3) at (1,0.9);
     \coordinate (x) at (0,0);
     \node (k1) at (-.8,-1.3) {};
     \node (k2) at (0,-1.2) {};
     \node (k3) at (.8,-1.3) {};
     \draw (out2) node [right] {$\!\!\!\vdots$};
     \draw [dotted] (in) -- (x);
     \draw [zParticle] (x) -- (out1);
     \draw [zParticle] (x) to[out=30,in=180] (out3);
     \draw [graviton] (x) -- (k2);
     \draw [fill] (x) circle (.08);
     \end{tikzpicture}\,.
 \end{align}
 \end{subequations}
 
 \section{The impulse}
 
 Having established the WQFT formalism and the use of one-point functions for the computation of observables, we now turn to the observable on which this thesis focuses, the impulse $\Delta p_i$ of the two bodies. It encodes both the scattering angle and the radiated momentum of the system, both of which can be used to inform waveform models. The impulse is given by
 \begin{align}
         \Delta p_i^\mu =& p_i^\mu(\tau\rightarrow \infty)-p_i^\mu(\tau \rightarrow -\infty)=m_i \int \d \tau_i \dfrac{d^2 z_i^\mu(\tau_i)}{d \tau_i^2} \nn\\
         =& -m_i \int_\omega \omega^2 \vev{z_i^\mu(\omega) \dd(\omega)}_{\rm WQFT}=- m_i \lim_{\omega \rightarrow 0} \omega^2 \vev{z_i^\mu(\omega)}_{\rm WQFT}\,,
 \end{align}
 and admits the PM expansion
 \begin{align}
         \Delta p_i = \sum_{n=1}^{\infty} G^n \,\Delta p_{i}^{(n)}\,.
 \end{align}
 The coefficient $\Delta p_{i}^{(n)}$ is computed as the sum of all tree-level Feynman diagrams with an outgoing $z_i^\mu$ line at order $G^n$.
 
 \subsection{Vectorial structure}
 
 In the physical setup of Fig.~\ref{fig:scatter} three vectors enter the problem, the impact parameter and the two initial velocities. The impact parameter is defined by $b^\mu=x_{0,2}^\mu-x_{0,1}^\mu$, where we make use of translation invariance to enforce $b\cdot v_i=0$. Lorentz symmetry then allows us to expand the impulse on the basis of these three vectors,
 \begin{align}
         \Delta p_i^\mu=a_b b^\mu+ a_{v_1} v_1^\mu+a_{v_2} v_2^\mu\,.\label{eq:bvecStructImp}
 \end{align}
 The velocities fulfil
 \begin{align}
        v_i^2=1,\; v_1 \cdot v_2 =\gamma.
 \end{align}
 In the absence of spin the scattering process is planar. In momentum-space this translates via a Fourier-transformation, see App.~\ref{App:Integrals}, into
 \begin{align}
         \Delta \tilde p_i^\mu=\tilde a_b q^\mu+ \tilde a_{v_1} v_1^\mu+\tilde a_{v_2} v_2^\mu \,. \label{eq:vecStructImp}
 \end{align}
 It is straightforward to see that $\tilde a_b$ must have an even mass dimension and that $\tilde a_{v_i}$ must have an odd mass dimension. 
 
 \subsection{Mass polynomial structure}
 The expressions for the impulse exhibit a polynomial structure in the masses at every PM order. From the Feynman rules one finds that the impulse at $n$-th PM order takes the form
 \begin{align}
         \Delta p_{i}^{(n)}=&\sum_{j=1}^{n} m_1^{j}m_2^{n-j+1} \,a_{j-1,n-j}\nn \\
         =&m_2^{n+1} \sum_{j=1}^{n} q_r^{j} a_{j-1,n-j}, \label{eq:selfforceexp}
 \end{align}
 with $q_r=m_1/m_2$, so that the PM expansion is supplemented by a self-force expansion in the mass ratio $q_r$. The same integral families contribute to $a_{i,j}$ and $a_{j,i}$, and in fact $a_{i,j}=a_{j,i}$ for conservative scattering. We will therefore refer to $a_{i,j}$ with $i \le j$ as the $(i)$-SF contribution and to $a_{j,i}$ with $i \le j$ as the $\overline{i}$-SF contribution, since the latter would be the $i$-th order in an expansion using $\bar{q}_r=m_2/m_1$ instead. The first 1SF contributions appear at 3PM order, while 2SF contributions appear for the first time at 5PM order.
 
 \section{Integrand construction}
 
 The integrand is now obtained simply by summing over all possible tree-level graphs. The challenge lies in generating these diagrams efficiently and in inserting the vertices in a controlled manner. We employ a Berends--Giele type recursion to enumerate all tree-level diagrams and their symmetry factors, which has the additional advantage of making the direction of the retarded propagators easy to track. Up to 5PM we find the following diagram counts,
 \begin{align}
         \begin{tabular}{l|c|c|c|c|c}
                 PM order & 1 & 2 & 3 & 4&5 \cr
                 \hline
                 \# Feyn. graphs & 1 & 4 & 27 & 201&1595
                 \end{tabular}\,.
 \end{align}
 The number of Feynman diagrams grows with the PM order, and an efficient implementation of the Feynman-rule insertion is therefore essential. We use a dedicated \texttt{FORM}~\cite{Davies:2026cci,Ruijl:2017dtg} code for this purpose. The dominant contribution to the size of the integrand does not actually come from the number of graphs but from single graphs containing a maximal number of three-graviton vertices. We will therefore work in an optimised gauge that minimises the number of terms in the three-graviton vertex; the details of the gauge are collected in App.~\ref{sec:gaugefixing}. With this choice the $n$-graviton vertex contains the following number of terms,
 \begin{align}
         \begin{tabular}{l|c|c|c|c}
                 n & 3 & 4 & 5 & 6 \cr
                 \hline
                 \# terms  & 33 & 366 & 4630 & 51360
                 \end{tabular}\,.
 \end{align}
 For example, at 5PM the graph containing four three-graviton vertices has naively $33^4\sim 10^6$ terms, which is roughly the size of twenty graphs involving a single six-graviton vertex.
 
 One could optimise the integrand generation further by using on-shell unitarity methods~\cite{Bern:1994zx,Bern:1994cg,Britto:2004nc,Bern:2007ct,Bern:2008pv,Bern:2010tq,Carrasco:2011hw, Bern:2024vqs,Haddad:2025cmw,He:2025how} or by exploiting a self-force EFT~\cite{Cheung:2024byb,Kosmopoulos:2023bwc, Mougiakakos:2024nku,Mougiakakos:2024lif,Hoogeveen:2025tew}. The size of the integrand will not be the dominant bottleneck for the present computation in comparison with the other challenges we will face, and we therefore use the present setup as our working framework.
 
 \section{Feynman-graph conventions}
 
 In the following we use Feynman graphs to denote both actual Feynman diagrams and Feynman integrals. They take the form
 \begin{align}\label{eq:leadingDefDiagram}
 \begin{tikzpicture}[baseline={(current bounding box.center)}]
 \coordinate (inA) at (-1,0.7);
 \coordinate (outA) at (1,0.7);
 \coordinate (inB) at (-1,-0.7);
 \coordinate (outB) at (1,-0.7);
 \coordinate (xA) at (0,0.7);
 \coordinate (xB) at (0,-0.7);
 \draw [dotted] (inA) -- (xA);
 \draw [dotted] (inB) -- (xB);
 \draw [dotted] (xB) -- (outB);
 \draw [zUndirected] (xA) -- (outA) node [midway, above] {$\quad\omega=0$};
 \draw [fill] (xA) circle (.08);
 \draw [fill] (xB) circle (.08);
 \draw [photon] (xA) -- (xB) node [midway, left] {$q\!\uparrow$};
 \draw (inA) node [left] {$1$};
 \draw (inB) node [left] {$2$};
 \end{tikzpicture}=\I m_1 m_2 8\pi G\!
 \int_q e^{iq\cdot b}\dd(q\cdot v_1)\dd(q\cdot v_2)
 (-v_1^\nu v_1^\rho q^\mu)
 \frac{P_{\nu\rho;\sigma\lambda}}{q^2}v_2^\sigma v_2^\lambda\,.
 \end{align}
 Three different kinds of lines appear in these graphs. The dotted lines represent the unperturbed worldlines of the black holes. The solid straight line denotes the outgoing $z^\mu$ field, that is, the perturbation of the trajectory. Whenever we compute the impulse we have an outgoing $z^\mu$ line on which we set $\omega=0$. Finally, the wiggly lines in the bulk between the worldlines are gravitons. We always implicitly assume a total momentum $q$ flowing from the lower worldline to the upper one.

 We have to integrate in this diagram over $q$. Usually tree-level Feynman diagrams imply that one does not need to integrate since momentum conservation on the vertices makes all integration trivial. Our Feynman rules though only imply energy conservation on the worldlines and only the bulk vertices have $d$-dimensional momentum conservation. This means that even though we have only tree-level graphs, we have to perform non-trivial integrations.
 
 We also note that the Feynman graphs are drawn in $q$-space, while the impulse is naturally an object in $b$-space, so that a Fourier transformation from $q$ to $b$ has to be performed at every PM order. We will also always find that $q\cdot v_i = 0$, enforced by two delta functions arising from the Feynman rules.
 
 At 2PM order we encounter, for example, graphs of the form
 \begin{align*}\label{eq:2PMdeflection1}
         \!\!\!\!\!\!\!\!\!\!\!\!\!\!\!\!\!\!\!\!\!\!\!\!
         \begin{tikzpicture}[baseline={(current bounding box.center)}]
         \coordinate (inA) at (-1,.7);
         \coordinate (outA) at (2,.7);
         \coordinate (inB) at (-1,-.7);
         \coordinate (outB) at (2,-.7);
         \coordinate (xA) at (0,.7);
         \coordinate (yA) at (1,.7);
         \coordinate (xB) at (0,-.7);
         \coordinate (yB) at (1,-.7);
         \draw [dotted] (inA) -- (xA);
         \draw [zUndirected] (xA) -- (yA) node [midway, above] {$\stackrel{\displaystyle k\cdot v_1}{\rightarrow}$};
         \draw [zUndirected] (yA) -- (outA) node [midway, above] {$\omega\!=\!0$};
         \draw [dotted] (inB) -- (xB);
         \draw [dotted] (xB) -- (yB);
         \draw [dotted] (yB) -- (outB);
         \draw [fill] (xA) circle (.08);
         \draw [fill] (yA) circle (.08);
         \draw [fill] (xB) circle (.08);
         \draw [fill] (yB) circle (.08);
         \draw [photon] (xA) -- (xB) node [midway, left] {$k\!\uparrow$};
         \draw [photon] (yA) -- (yB) node [midway, right] {$\uparrow\!q\!-\!k$};
         \draw (inA) node [left] {$1$};
         \draw (inB) node [left] {$2$};
         \end{tikzpicture}&=\I 16\pi^{2}G^{2} m_1m_2^2
         \int_{k,q}
         \frac{(q^\mu-k^\mu)\dd(q\cdot v_1)\dd(q\cdot v_2)\dd(k\cdot v_2)}
         {k^2(k-q)^2(k\cdot v_1+\i0)^2}e^{iq\cdot b}\times
         \end{align*}
         \vspace{-3.5em}
         \begin{align}
                 \qquad\qquad\qquad\qquad\qquad\qquad
                 \left[8\gamma^2(k\cdot v_1)^2 - (2\gamma^2-1)^2k\cdot(q-k)\right]\, .
                 \end{align}
 In addition to the Fourier transformation we now also have a loop integral over $k$ to perform. At $n$-PM order we will have to evaluate the Fourier transform of an $(n-1)$-loop integral. Since the Fourier transformation is always the same and is essentially trivial, see~\refeq{eq:FT}, the challenge resides entirely in the loop integrals, and a central focus of this thesis is to find efficient ways to solve them. Interpreting the dotted lines on the worldlines as delta functions, the graphs above already provide a diagrammatic representation of the corresponding loop integral.
 
 We will therefore use the same type of graphs for Feynman integrals, with the outgoing line omitted,
 \begin{align}
 \!\!\!\!\!\!\!\!\!\!\!\!\!\!\!\!\!\!\!\!\!\!\!\!
 \begin{tikzpicture}[baseline={(current bounding box.center)}]
 \coordinate (inA) at (-1,.7);
 \coordinate (outA) at (2,.7);
 \coordinate (inB) at (-1,-.7);
 \coordinate (outB) at (2,-.7);
 \coordinate (xA) at (0,.7);
 \coordinate (yA) at (1,.7);
 \coordinate (xB) at (0,-.7);
 \coordinate (yB) at (1,-.7);
 \draw [dotted] (inA) -- (xA);
 \draw [zUndirected] (xA) -- (yA) node [midway, above] {};
  \draw [dotted] (yA) -- (outA) ;
 \draw [dotted] (inB) -- (xB);
 \draw [dotted] (xB) -- (yB);
 \draw [dotted] (yB) -- (outB);
 \draw [fill] (xA) circle (.08);
 \draw [fill] (yA) circle (.08);
 \draw [fill] (xB) circle (.08);
 \draw [fill] (yB) circle (.08);
 \draw [photon] (xA) -- (xB) node [midway, left] {};
 \draw [photon] (yA) -- (yB) node [midway, right] {};
 \draw (inA) node [left] {};
 \draw (inB) node [left] {};
 \end{tikzpicture}&=
 \int_{k}
 \frac{\dd(k\cdot v_2)}
 {k^2(k-q)^2(k\cdot v_1+\i0)}\,.
 \end{align}
 In this case every solid line represents a denominator. The quadratic graviton propagators are drawn as wiggly lines and the linear worldline propagators as solid lines, while the dotted line that closes the loop is interpreted as a delta function. Note that we display only single powers and no numerators. If we want to square a propagator we mark this by a dot,
 \begin{align}
         \!\!\!\!\!\!\!\!\!\!\!\!\!\!\!\!\!\!\!\!\!\!\!\!
         \begin{tikzpicture}[baseline={(current bounding box.center)}]
         \coordinate (inA) at (-1,.7);
         \coordinate (outA) at (2,.7);
         \coordinate (inB) at (-1,-.7);
         \coordinate (outB) at (2,-.7);
         \coordinate (xA) at (0,.7);
         \coordinate (yA) at (1,.7);
         \coordinate (xB) at (0,-.7);
         \coordinate (yB) at (1,-.7);
         \draw [dotted] (inA) -- (xA);
         \draw [zUndirected] (xA) -- (yA) node [midway, above] {};
          \draw [dotted] (yA) -- (outA) ;
         \draw [dotted] (inB) -- (xB);
         \draw [dotted] (xB) -- (yB);
         \draw [dotted] (yB) -- (outB);
         \draw [fill] (xA) circle (.08);
         \draw [fill] (yA) circle (.08);
         \draw [fill] (xB) circle (.08);
         \draw [fill] (yB) circle (.08);
         \draw [fill,blue!70] (0.5,.7) circle (.08);
         \draw [photon] (xA) -- (xB) node [midway, left] {};
         \draw [photon] (yA) -- (yB) node [midway, right] {};
         \draw (inA) node [left] {};
         \draw (inB) node [left] {};
         \end{tikzpicture}&=
         \int_{k}
         \frac{\dd(k\cdot v_2)}
         {k^2(k-q)^2(k\cdot v_1+\i0)^{\blue{2}}}\,.
         \end{align}
 From now on we drop the Fourier transformation; the relevant Fourier transforms are collected in App.~\ref{App:Integrals}.
 
\chapter{Integrand reduction}
\label{sec:intred}
We have seen how we can generate an integrand for the impulse at a given PM order. Even though we only have tree-level diagrams we need to integrate due to the missing full four-momentum conservation on the worldline. We should also mention that we can derive the same integrand for the impulse from methods like WEFT~\cite{Kalin:2022hph,Kalin:2020mvi} or the amplitudes based KMOC~\cite{Kosower:2018adc} formalism. Furthermore, one can use amplitudes based methods to compute a potential~\cite{Cheung:2018wkq} which is equivalent to computing the conservative impulse in $b^\mu$ direction~\cite{Cristofoli:2019neg,Bjerrum-Bohr:2018xdl, Bern:2019nnu,Bern:2019crd,Cheung:2020gyp,Bjerrum-Bohr:2021din,Damgaard:2021ipf,Damgaard:2022jem,DiVecchia:2021bdo,Herrmann:2021lqe}.
At 1PM there is no loop integration to be done and at 2PM we can do the integration directly without further complications. The result is well established and can be found in~\cite{Westpfahl:1985tsl}. It was also reproduced using WQFT in~\cite{Mogull:2020sak} and we will not recompute it here. We will look directly at the 3PM calculation, which is also well established~\cite{Bern:2019nnu,Bern:2019crd,Cheung:2020gyp,Kalin:2020fhe,Damour:2019lcq,Mogull:2020sak,Bjerrum-Bohr:2021din,Herrmann:2021tct,Jakobsen:2022fcj,Damgaard:2021ipf,Damgaard:2022jem,Kalin:2022hph,Kalin:2020mvi,Damour:2020tta,DiVecchia:2021bdo,Herrmann:2021lqe,Bjerrum-Bohr:2018xdl,Bini:2021gat}.
We want to use this 3PM calculation to review the methods we will use to compute the 4PM and 5PM contributions. 

The first important step will be to reduce our integrand. We will face many integrals in our calculation and the first step will be to simplify those and reduce the number of integrals one has to compute. 
\section{Tensor reduction}
The first reduction one typically has to do in a loop calculation is the so-called tensor reduction. The idea is to reduce an integral, which is a tensor, to a scalar quantity. This task will be in our case very simple. In other calculations this can become a non-trivial task though. 
In our integrand we will encounter integrals, which have at most tensor rank 1 of the form
\begin{align}
        \int_{\ell_1,...,\ell_n} \ell_i^\mu \dfrac{N(\ell_1,...,\ell_n, q, v_1,v_2)}{D(\ell_1,...,\ell_n, q, v_1,v_2)},
\end{align}
where $N(\ell_1,...,\ell_n, q, v_1,v_2)$ and $D(\ell_1,...,\ell_n, q, v_1,v_2)$ are some scalar quantities. 
We can now use Lorentz symmetry to replace
\begin{align}
        \ell_i^\mu \rightarrow \sum_{j=1}^2 (\ell_i \cdot v_j) \check{v}_j^\mu- \dfrac{(\ell_i \cdot q)}{|q|^2} q^\mu,
\end{align}
with the dual velocities 
 \begin{align}
        \check v_1^\mu =\dfrac{\gamma v_2^\mu-v_1^\mu}{\gamma^2-1},\;\text{and}\;\check v_2^\mu =\dfrac{\gamma v_1^\mu-v_2^\mu}{\gamma^2-1},
 \end{align}
 which satisfy $\check v_i \cdot v_i=1$ and $\check v_i \cdot v_j=0$ for $i \neq j$.
The tensorial structure factorises and we find
\begin{align}
       & \int_{\ell_1,...,\ell_n} \ell_i^\mu \dfrac{N(\ell_1,...,\ell_n, q, v_1,v_2)}{D(\ell_1,...,\ell_n, q, v_1,v_2)}\\
       &=  \int_{\ell_1,...,\ell_n} \sum_{j=1}^2 \check{v}_j^\mu (\ell_i \cdot v_j) \dfrac{N(\ell_1,...,\ell_n, q, v_1,v_2)}{D(\ell_1,...,\ell_n, q, v_1,v_2)}-q^\mu \int_{\ell_1,...,\ell_n} \dfrac{(\ell_i \cdot q)}{|q|^2} \dfrac{N(\ell_1,...,\ell_n, q, v_1,v_2)}{D(\ell_1,...,\ell_n, q, v_1,v_2)}.
\end{align}
This way we can tensor reduce all integrals to scalar integrals. 
\section{Integral families}
Once tensor reduction has been carried out on the integrand, what remains is a large number of scalar integrals of the schematic form
\begin{align}
        \int_{\ell_1,\dots,\ell_n} \dd(\ell_1\cdot v_{i_1})\cdots \dd(\ell_n\cdot v_{i_n})\,
      &  \prod \frac{1}{\bigl(\sum c_{i,j}\,\ell_i + d_j\, q\bigr)^{2}}\,
        \prod \frac{1}{(\sum d_i\,\ell_i \cdot v_j)^{\nu_i}}\,\nn \\
     & \times  \prod (\ell_i\cdot\ell_j)^{\nu_{i+m}}\,
        \prod (\ell_i\cdot q)^{\nu_{i+n}}\,, \label{eq:scalarints}
\end{align}
where $v_1$ and $v_2$ are the asymptotic four-velocities of the two black holes, $q$ is the momentum transfer between them, and the $c_{i,j}, d_i$ are real constants. We refer to $v_1, v_2$ and $q$ collectively as the \emph{external vectors}, and to $\ell_1,\dots,\ell_n$ as the \emph{loop momenta}.

The integrand in \refeq{eq:scalarints} contains four distinct types of building blocks. 
\emph{(i)} The $n$ linear delta functions $\dd(\ell_i\cdot v_j)$ pair each loop momentum with one of the two asymptotic velocities. In the self-force decomposition of \refeq{eq:selfforceexp}, the $n$-SF contribution is the configuration in which $n$ loop momenta are constrained against $v_2$ and the remaining ones against $v_1$ (and vice versa for $\bar n$-SF).

 \emph{(ii)} The \emph{quadratic propagators} $1/(\sum c_{i,j}\,\ell_i+d_j\,q)^{2}$ originate from graviton exchanges in the bulk and depend on a linear combination of loop momenta and the momentum transfer; the Feynman rules of the worldline theory enforce that each such denominator appears with multiplicity one in the original integrand. 

\emph{(iii)} The \emph{linear propagators} $1/(\ell_i\cdot v_j)^{\nu_i}$ descend from the worldline fields. These can appear raised to a maximum power of 2, but can also appear in the numerator e.g. raised to a negative power.
The delta functions together with the two propagator families fix the \emph{topology} (or \emph{sector}) of the integral, which in turn determines the function space; this point will be revisited below. 

\emph{(iv)} Finally, the remaining numerator factors $(\ell_i\cdot \ell_j)$ and $(\ell_i\cdot q)$ are the \emph{irreducible scalar products} (ISPs): scalar products of momenta that cannot be re-expressed as linear combinations of the denominators present in the integral.
The representation \refeq{eq:scalarints} suffers from several ambiguities. First, the integral is invariant under linear shifts of the loop momenta,
\begin{align}
        \ell_i \;\longrightarrow\; \sum_j c_j\,\ell_j + d\,q,
\end{align}
with $c_j, d$ constants, and under permutations
\begin{align}
        \ell_i \;\longleftrightarrow\; \ell_j.
\end{align}
Both operations change the apparent form of the propagators and delta functions while leaving the value of the integral unchanged. Beyond these, the integrand also enjoys a discrete symmetry on the external kinematics: as a Lorentz scalar it depends only on $v_1\cdot v_2 = \gamma$ and $q^2 = -|\vec q\,|^2$ (the constraints $q\cdot v_i = 0$ and $v_i^2 = 1$ being assumed throughout), so it is invariant under
\begin{align}
        v_1 \;\longleftrightarrow\; v_2 \qquad\text{and}\qquad q \;\longrightarrow\; -q.
\end{align}
A second source of ambiguity is the freedom to rewrite ISPs as sums of other scalar products.

To remove (some of) these ambiguities, we organise the scalar integrals into a so-called \emph{integral family}. Depending on the topologies that appear, several integral families may be required to cover the full integrand; for gravitational scattering at zeroth and first self-force order, however, every integral that arises can be cast into a single family of the form
\begin{align}
        {\cal P}_{\{n\}}^{\{\sigma\}}
        =
        \int_{\ell_1,\dots,\ell_{\text{L}}}
        \frac{
          \prod_{i=1}^L\dd^{(n_i-1)}(\ell_i\cdot v_{\alpha_i})
          }{
          \prod_{i=1}^{L}D_{i}^{n_{i+L}}(\sigma_{i})
          \prod_{I<J}D_{IJ}^{n_{IJ}} \label{eq:generalFamily4d}
          }\,,
      \end{align}
      where
      \begin{align}
         D_{i}&=\ell_i\cdot v_{{\bar \alpha}_i}+\sigma_i \i0\,,
      \end{align}
    and the graviton propagators $D_{IJ}$ with $I,J \in \{q,0,i/j\}$ and $i,j \in [1,L]$ are
      \begin{align}\label{DIJs}
        D_{ij}&=(\ell_i-\ell_j)^2
        \,,
        D_{qi}=(\ell_i-q)^2
        \,,
        D_{0i}=\ell_i^2
        \,.
      \end{align}
and where $\dd^{(n)}$ denotes the $n$'th derivative of the delta function, $\alpha_i\in \{1,2\}$ and $\bar 1/ \bar 2=2/1$. The total number of propagators including the delta functions at a given loop order $L$ is in our case $\frac{L(L+1)}{2}+3 L$.

Bringing every integral at every loop order into this form generally requires the use of \emph{partial-fraction identities}, which we discuss later. At 3PM, shifts and relabellings already suffice.
Even after choosing an integral family, some residual ambiguity remains. Loop-momentum shifts and relabellings can map one integral of the family onto another within the same family.

\subsection{The 3PM-0SF family}
At two loops and 0SF order, the integral family takes the explicit form
\begin{align}
        {{\cal P}^{\text{3PM-0SF}}}_{\{n\}}^{\{\sigma\}}
        =
        \int_{\ell_1,\ell_2}
        \frac{
          \prod_{i=1}^2\dd^{(n_i-1)}(\ell_i\cdot v_{2})
          }{
          \prod_{i=1}^{7}D_{i}^{n_{i+2}}
          }\,,
      \end{align}
      where the 7 propagators are given as
      \begin{align}\label{prop0SF}
        D_{1/2}=\ell_{1/2}\cdot v_{1}+\sigma_{1/2} \i0\,,
        D_{3}=(\ell_1-\ell_2)^2
        \,,
        D_{4/5}=(\ell_{1/2}-q)^2
        \,,
        D_{6/7}=\ell_{1/2}^2.
      \end{align}
When $n_{1}=n_{2}=1$, the integral family admits a particularly transparent representation: passing to the rest frame of one of the worldlines, $v_2 = (1,\vec 0)$, the delta functions can be integrated out explicitly. At 0SF order this rewriting is especially valuable because it exposes the trivial $\gamma$ dependence of the integrals. One is left with the Euclidean integral
\begin{align}
{{\cal P}^{\text{3PM-0SF}}}_{\{n\}}^{\{\sigma\}}
        =\dfrac{(-1)^{\sum_{i=3}^9 n_i}}{{\sqrt{\gamma^2-1}}^{n_3+n_4}}  \int_{\boldsymbol{\ell_1}\boldsymbol{\ell_2}} \dfrac{1}{\prod_{i=1}^7 \hat{D}_i^{n_{i+2}}}, \label{eq:0sfgamma}
\end{align}
with
\begin{align}
        \hat{D}_{1/2}=\boldsymbol{\ell}_{1/2}\cdot \boldsymbol{e}, \hat{D}_{3}=(\boldsymbol{\ell}_{1}- \boldsymbol{\ell}_{2})^2,
        \hat{D}_{4/5}=(\boldsymbol{\ell}_{1/2}- \boldsymbol{q})^2,\hat{D}_{6/7}=(\boldsymbol{\ell}_{1/2})^2
\end{align}
where $\boldsymbol e^2 = 1$, $\boldsymbol e\cdot\boldsymbol q = 0$, $\boldsymbol q^2 = |q|^2$, and the spatial integrations are performed in $d = 3-2\epsilon$ dimensions. The remaining integral is manifestly $\gamma$-independent. The same construction extends to higher loop orders and demonstrates that the function space at 0SF is, trivially, that of algebraic functions of $\gamma$. The pole structure of the integrals can only originate from the linear propagators: the graviton propagators are unable to go on-shell and therefore cannot generate any more complex structure. This will no longer be true at 1SF

Because the rewriting~\refeq{eq:0sfgamma} is so natural at 0SF, we will also use this representation when drawing Feynman graphs for the 0SF integrals. Only one worldline survives in this picture: the upper line carries the auxiliary direction $\vec e$ ingoing and outgoing as before, while the lower line now carries a single outgoing momentum $\vec q$. The result is a three-point graph of the type
\begin{align*}
\begin{tikzpicture}[baseline={([yshift=-1ex]current bounding box.south)},scale=1.2]
        \coordinate (inA) at (0.4,.7);
        \coordinate (outA) at (3.4,.7);
        \coordinate (inB) at (0.4,-.7);
        \coordinate (outB) at (3.4,-.7);
        \coordinate (xA) at (0.9,.7);
        \coordinate (xB) at (0.9,-.7);
        \coordinate (xM) at (0.9,0);
        \coordinate (xAM) at (0.9,.35);
        \coordinate (xBM) at (0.9,-.35);
        \coordinate (yA) at (1.9,.7);
        \coordinate (yB) at (1.9,-.7);
        \coordinate (yC) at (1.9,-1.2);
        \coordinate (xC) at (0.9,-1.2);
        \coordinate (zC) at (2.9,-1.2);
        \coordinate (yM) at (1.9,0);
        \coordinate (yAM) at (1.9,.35);
        \coordinate (yBM) at (1.9,-.35);
        \coordinate (zA) at (2.9,.7);
        \coordinate (zB) at (2.9,-.7);
        \coordinate (zM) at (2.9,0);
        \coordinate (zAM) at (2.9,.35);
        \coordinate (zBM) at (2.9,-.35);
        \draw [dotted] (inA) --node [midway, above]{$\rightarrow \vec{e}$}  (outA);
        \draw [photon] (yB) --node [midway, right]{$\downarrow \vec{q}$} (yC);
        \draw [photon] (xA) to[out=-45,in=180] (yB);
        \draw [photon] (yA) -- (yB);
        \draw [zUndirected] (xA) -- (yA);
        \draw [zUndirected] (yA) -- (zA);
        \draw [photon] (zA) to[out=-90-45,in=0] (yB);
        \foreach \v in {yA,yB,xA,yC,zA}
        \draw[fill] (\v) circle (.08);
      \end{tikzpicture}.
\end{align*}
For all such three-point graphs the loop integration is to be understood in $d = 3-2\epsilon$ spatial dimensions.
It is also a good diagrammatic representation of what the solution at 0SF is: the geodesic motion of a test particle in the Schwarzschild metric in a Post-Minkowskian expansion.
\subsection{The 3PM-1SF family}
The integral family at two loops and 1SF order has a very similar shape; the only structural change concerns the delta functions and linear propagators, but this small modification is enough to introduce a non-trivial $\gamma$ dependence and a qualitatively new propagator structure. The integrals take the form
\begin{align}
        {{\cal P}^{\text{3PM-1SF}}}_{\{n\}}^{\{\sigma\}}
        =
        \int_{\ell_1\ell_2}
        \frac{
          \dd^{(n_1-1)}(\ell_1\cdot v_{1})\dd^{(n_2-1)}(\ell_2\cdot v_{2})
          }{
          \prod_{i=1}^{7}D_{i}^{n_{i+2}} 
          }\,,
      \end{align}
      with the seven propagators
      \begin{align}
        D_{1}=\ell_1\cdot v_{{2}}+\sigma_1 \i0\,,
        D_{2}=\ell_2\cdot  v_{{1}}+\sigma_2& \i0\,,
        D_{3}=(\ell_1-\ell_2)^2+\sigma_3 \sign{ (\ell_1^0-\ell_2^0)}\, \i0
        \,,\nn \\    
        D_{4/5}=(\ell_{1/2}-q)^2
        \,, &
        D_{6/7}=\ell_{1/2}^2
      \end{align}
As before, it is convenient to integrate out the delta functions to expose the structure of the integrals, following~\cite{Bjerrum-Bohr:2021vuf}. This will also make clear why we can neglect the $\i0$ prescription for all graviton propagators except $D_3$: 
\begin{align}
        {{\cal P}^{\text{3PM-1SF}}}_{\{1,1,n\}}^{\{\sigma\}}
                =\dfrac{(-1)^{\sum_{i=4}^9 n_i}}{{\sqrt{\gamma^2-1}}^{n_3+n_4}}  \int_{\boldsymbol{\ell_1}\boldsymbol{\ell_2}} \dfrac{1}{\prod_{i=1}^7 \hat{D}_i^{n_{i+2}}}, \label{eq:1sfgamma}
        \end{align}
        where the seven propagators now read
      \begin{align}
        \hat{D}_{1/2}=\boldsymbol{\ell}_{1/2}\cdot \boldsymbol{e}, \hat{D}_{3}=(\boldsymbol{\ell}_{1}- \boldsymbol{\ell}_{2})^2-2 (\gamma-1)(\boldsymbol{\ell}_{1}\cdot \boldsymbol{e})(\boldsymbol{\ell}_{2}\cdot \boldsymbol{e}),
        \hat{D}_{4/5}=(\boldsymbol{\ell}_{1/2}- \boldsymbol{q})^2,\hat{D}_{6/7}=(\boldsymbol{\ell}_{1/2})^2
\end{align}
As at 0SF, the $\gamma$ dependence stemming from the linear propagators factorises out cleanly. Unlike the 0SF case, however, the residual integral itself still depends on $\gamma$. The decisive difference lies in 
\begin{align}
\hat D_3 = (\boldsymbol{\ell}_1 - \boldsymbol{\ell}_2)^2 - 2(\gamma-1)\,(\boldsymbol{\ell}_1\cdot\boldsymbol e)(\boldsymbol{\ell}_2\cdot\boldsymbol e)\,.
\end{align} 
This denominator depends on $\gamma$ explicitly and, crucially, can go on-shell. The remaining graviton propagators stay strictly off-shell, so the $i0$ prescription, which we suppress in the equations above, is relevant only for $\hat D_3$ and the linear propagators. We refer to denominators of this type as \emph{active} propagators.
The active propagators are the reason that integrals will evaluate to a non-trivial function space, the determination of which is one of the main objectives of the following sections and this thesis in general.

The active propagators are very easy to detect just from the Feynman diagrams. An active graviton cannot touch any worldline without another incoming or outgoing graviton or worldline field. The graviton should either touch no worldline or connect to a worldline vertex at which one has another worldline field or graviton field. We will mark all active gravitons red in our diagrams. As an example we have a
\begin{align*}
        \begin{tikzpicture}[baseline={([yshift=0ex]current bounding box.south)},scale=1]
                \coordinate (inA) at (0.4,.7);
                \coordinate (outA) at (3.6,.7);
                \coordinate (inB) at (0.4,-.7);
                \coordinate (outB) at (3.6,-.7);
                \coordinate (xA) at (1,.7);
                \coordinate (xxA) at (1.825,.7) ;
                \coordinate (xxB) at (1.825,-.7) ;
                \coordinate (yA) at (1.60+0.4,.7);
                \coordinate (yyA) at (2.375,0) ;
                \coordinate (zA) at (3,.7);
                \coordinate (zzA) at (3.75,.7) ;
                \coordinate (xB) at (1,-.7);
                \coordinate (yB) at (1.6+0.4,-.7);
                \coordinate (zB) at (3,-.7);
                \coordinate (zzB) at (3.75,-.7);
                \coordinate (xM) at (1,0);
                \coordinate (yM) at (1.6+0.4,0);
                \coordinate (zM) at (3,-0);
                \draw [dotted] (inA) -- (outA);
                \draw [dotted] (inB) -- (outB)  ;
                \draw [draw=none] (xA) to[out=40,in=140] (zA);
                \draw [photon] (xA) -- (xB);
                \draw [photon] (zA) -- (zB);
                \draw [photon,red] (xM) -- (zA);
                \draw [fill] (xA) circle (.08);
                 \draw [fill] (xM) circle (.08);
                \draw [fill] (zA) circle (.08);
                \draw [fill] (xB) circle (.08);
                \draw [fill] (zB) circle (.08);
              \end{tikzpicture}\; .
\end{align*}
As one can see there is only a single graviton which does not connect to a vertex of the type
\begin{align*}
\begin{tikzpicture}[baseline={([yshift=-1ex]current bounding box.south)},scale=1]
        \coordinate (in) at (-1,0);
        \coordinate (out) at (1,0);
        \coordinate (x) at (0,0);
        \draw [dotted] (in) -- (x);
        \draw [dotted] (x) -- (out);
        \draw [photon] (x) -- (k);
        \draw [fill] (x) circle (.08);
        \end{tikzpicture}.
\end{align*}       
Furthermore we will often leave the $\i0$ prescriptions implicit in this thesis, as many features of the integrals are independent of them. Therefore, we will often drop the $\sigma$ indices and we will use for planar integrals the notation:
\begin{align}
        {{\cal P}\;^{\text{mPM-jSF}}}_{\{n\}}^{\{\sigma\}}=\text{P}^{m,j}(n). \label{eq:integralNotation}
\end{align}
\section{Integration by parts and Master integrals}
\label{sec:IBPs}
The integral families introduced in the previous section play a central role in the next step of the calculation. From here on we work in $d=4-2\epsilon$ dimensions and adopt the form~\refeq{eq:generalFamily4d} throughout. Casting all scalar integrals into this common form makes it possible to derive systematic algebraic relations between them and so reduce the potentially enormous set of integrals encountered in a given amplitude to a small basis. The most important such relations are the \emph{integration-by-parts} (IBP) identities~\cite{Chetyrkin:1981qh}. For PM integrals they read
\begin{align}
        \int_{\ell_1,...,\ell_n}\dfrac{\partial}{\partial \ell_j^\mu} q_{IBP}^\mu 
        \frac{
          \prod_{i=1}^L\dd^{(n_i-1)}(\ell_i\cdot v_{\alpha_i})
          }{
          \prod_{i=1}^{L}D_{i}^{n_{i+L}}
          \prod_{I<J}D_{IJ}^{n_{IJ}}
          }
      =0, \label{eq:feyn}
\end{align}
where $q_{\mathrm{IBP}}^\mu$ is a loop or external momentum, and we have used that boundary terms vanish in dimensional regularisation. This equation is also often referred to as an \emph{IBP vector}. 

Since every scalar product produced by the differentiation can be re-expressed in terms of the propagators and ISPs of the integral family, \refeq{eq:feyn} amounts to a linear relation among integrals of the same family. A small subtlety in our setup is the presence of delta functions in the numerator. They can be handled on the same footing as ordinary denominators by means of the \emph{reverse-unitarity} identity ~\cite{Anastasiou:2002yz,Anastasiou:2002qz,Anastasiou:2003yy,Anastasiou:2015yha}
\begin{align}
        \frac{\dd^{(n)}(\omega)}{(-1)^n\, n!} \;=\; \frac{i}{(\omega+\i0)^{n+1}} - \frac{i}{(\omega-\i0)^{n+1}},
\end{align}
which converts a delta-function derivative into the discontinuity of a linear propagator. The delta functions then participate in the IBP machinery exactly as the linear propagators do, with the convention that any resulting integral having $n<0$ for one of the delta-function entries vanishes identically.

In practice, IBP relations are often exploited through the Laporta algorithm~\cite{Laporta:2000dsw}, implemented in dedicated reduction packages such as \texttt{Kira}~\cite{Lange:2021edb,Lange:2025fba,Klappert:2020nbg,Maierhofer:2017gsa}, \texttt{FIRE}~\cite{Smirnov:2025prc,Smirnov:2023yhb}, \texttt{Blade}~\cite{Guan:2024byi} or \texttt{Reduze}~\cite{vonManteuffel:2012np}. Conceptually, one writes~\refeq{eq:feyn} with the propagator powers $\{n_i\}$ kept symbolic, then evaluates them on a finite grid of integer values; this generates a large linear system in the integrals of the family. Provided the system is sufficiently large, it can be shown that every integral $I_{\{n\}}$ of the family reduces to a finite number of \emph{master integrals} (MIs)~\cite{Smirnov:2010hn},
\begin{align}
        I_{\{n\}} \;=\; \sum_i c_i(\vec z, d)\, M_i= \; \sum_i |q|^{m_i} \tilde{c}_i(\gamma, d)\, M_i,
\end{align}
with rational coefficients $c_i(\vec z, d)$ that depend on the spacetime dimension $d$ and the kinematic variables $\vec z$. In our case $\vec z = (\gamma, |q|)$. 
This reduction often requires solving a large system of linear equations. The row reduction and determination of the coefficients $c_i(\vec z,d)$ often is done using \emph{finite field methods}~\cite{Peraro:2016wsq,vonManteuffel:2014ixa}. Instead of a symbolic solution, the system gets solved at certain kinematic points multiple times modulo a prime field. The fully analytic expression can then be reconstructed from these; see App.~\ref{App:FiniteField} for a more thorough discussion.

Since $|q|$ is the only scale carrying a mass dimension we can set it without loss of generality to $|q|=1$ and reconstruct $m_i$ always by mass dimensional analysis, when needed. Therefore it is enough to derive the coefficients $\tilde{c}_i(\gamma, d)$. 

A structural feature of the reduction deserves to be highlighted. Differentiating a propagator raises its power by one,
\begin{align}
        \frac{\partial}{\partial \ell_j^\mu}\frac{1}{D_i^{n_i}} \;\propto\; \frac{n_i}{D_i^{n_i+1}},
\end{align}
so the IBP relations applied to $I_{\{n\}}$ can never introduce a denominator absent from the original integrand. As a consequence, one may always choose the IBP system, and hence the master basis, in such a way that the master integrals appearing in the reduction of $I_{\{n\}}$ carry only the same denominators as $I_{\{n\}}$, or a subset of them. This naturally leads to organising the family into \emph{sectors}. We can define a sector $S$ of an integral $I_{\{n \}}$ as the set:
\begin{align}
        S \;=\; \{\,\Theta(n_i - 1/2)\,\},
\end{align}
with $i =1,..., \frac{L(L+1)}{2}+3 L $,
specifying which denominators are present. The subsectors of $S$ are obtained by removing one or more denominators, which graphically corresponds to pinching the associated propagators in the Feynman graph (cf.\ Fig.~\ref{fig:wlcut}). Once enough propagators have been pinched, the integrand becomes scaleless and the resulting sector vanishes identically in dimensional regularisation; we refer to such sectors as \emph{trivial}. On the other hand when computing our observables at a given loop order we will face integrals with a maximal number of propagators. These sectors will be referred to as the \emph{topsectors}.

As an illustration, Fig.~\ref{fig:wlcut} shows the most intricate topsector arising at 3PM--1SF together with its non-trivial subsectors. As discussed, one descends from the topsector to its subsectors by pinching propagators in all admissible ways:
\begin{figure}[h!]
        \centering
        \begin{tikzpicture}[baseline={([yshift=-1ex]current bounding box.south)},scale=1]
                \coordinate (inA) at (0.4,.7);
                \coordinate (outA) at (3.6,.7);
                \coordinate (inB) at (0.4,-.7);
                \coordinate (outB) at (3.6,-.7);
                \coordinate (xA) at (1,.7);
                \coordinate (xxA) at (1.825,.7) ;
                \coordinate (xxB) at (1.825,-.7) ;
                \coordinate (yA) at (1.60+0.4,.7);
                \coordinate (yyA) at (2.375,0) ;
                \coordinate (zA) at (3,.7);
                \coordinate (zzA) at (3.75,.7) ;
                \coordinate (xB) at (1,-.7);
                \coordinate (yB) at (1.6+0.4,-.7);
                \coordinate (zB) at (3,-.7);
                \coordinate (zzB) at (3.75,-.7);
                \coordinate (xM) at (1,0);
                \coordinate (yM) at (1.6+0.4,0);
                \coordinate (zM) at (3,-0);
                \draw [dotted] (inA) -- (outA);
                \draw [dotted] (inB) -- (outB) node [midway, below]{\vspace{1cm}$\swarrow \quad\searrow$} ;
                \draw [draw=none] (xA) to[out=40,in=140] (zA);
                \draw [photon] (xA) -- (xB);
                \draw [photon] (zA) -- (zB);
                \draw [photon,red] (xM) -- (zM);
              
                \draw [fill] (xA) circle (.08);
                 \draw [fill] (xM) circle (.08);
                \draw [fill] (zA) circle (.08);
                \draw [fill] (xB) circle (.08);
                 \draw [fill] (zM) circle (.08);
                \draw [fill] (zB) circle (.08);
              \end{tikzpicture}\\
              \begin{tikzpicture}[baseline={([yshift=-1ex]current bounding box.south)},scale=1]
                \coordinate (inA) at (0.4,.7);
                \coordinate (outA) at (3.6,.7);
                \coordinate (inB) at (0.4,-.7);
                \coordinate (outB) at (3.6,-.7);
                \coordinate (xA) at (1,.7);
                \coordinate (xxA) at (1.825,.7) ;
                \coordinate (xxB) at (1.825,-.7) ;
                \coordinate (yA) at (1.60+0.4,.7);
                \coordinate (yyA) at (2.375,0) ;
                \coordinate (zA) at (3,.7);
                \coordinate (zzA) at (3.75,.7) ;
                \coordinate (xB) at (1,-.7);
                \coordinate (yB) at (1.6+0.4,-.7);
                \coordinate (zB) at (3,-.7);
                \coordinate (zzB) at (3.75,-.7);
                \coordinate (xM) at (1,0);
                \coordinate (yM) at (1.6+0.4,0);
                \coordinate (zM) at (3,-0);
                \draw [dotted] (inA) -- (outA);
                \draw [dotted] (inB) -- (outB) node [midway, , below=0.5cm]{\white{$\swarrow \qquad \downarrow$}} ;
                \draw [draw=none] (xA) to[out=40,in=140] (zA);
                \draw [photon] (xA) -- (yM);
                \draw [photon] (xB) -- (yM);
                \draw [photon] (zA) -- (yM);
                \draw [photon] (zB) -- (yM);
              
                \draw [fill] (xA) circle (.08);
                 \draw [fill] (yM) circle (.08);
                \draw [fill] (zA) circle (.08);
                \draw [fill] (xB) circle (.08);
                \draw [fill] (zB) circle (.08);
              \end{tikzpicture} 
              \begin{tikzpicture}[baseline={([yshift=-1ex]current bounding box.south)},scale=1]
                \coordinate (inA) at (0.4,.7);
                \coordinate (outA) at (3.6,.7);
                \coordinate (inB) at (0.4,-.7);
                \coordinate (outB) at (3.6,-.7);
                \coordinate (xA) at (1,.7);
                \coordinate (xxA) at (1.825,.7) ;
                \coordinate (xxB) at (1.825,-.7) ;
                \coordinate (yA) at (1.60+0.4,.7);
                \coordinate (yyA) at (2.375,0) ;
                \coordinate (zA) at (3,.7);
                \coordinate (zzA) at (3.75,.7) ;
                \coordinate (xB) at (1,-.7);
                \coordinate (yB) at (1.6+0.4,-.7);
                \coordinate (zB) at (3,-.7);
                \coordinate (zzB) at (3.75,-.7);
                \coordinate (xM) at (1,0);
                \coordinate (yM) at (1.6+0.4,0);
                \coordinate (zM) at (3,-0);
                \draw [dotted] (inA) -- (outA);
                \draw [dotted] (inB) -- (outB) node [midway, below=0.5cm]{ \hspace{-1cm}$\swarrow \qquad \downarrow$} ;
                \draw [draw=none] (xA) to[out=40,in=140] (zA);
                \draw [photon] (xA) -- (xB);
                \draw [photon] (zA) -- (zB);
                \draw [photon,red] (xM) -- (zA);
              
                \draw [fill] (xA) circle (.08);
                 \draw [fill] (xM) circle (.08);
                \draw [fill] (zA) circle (.08);
                \draw [fill] (xB) circle (.08);
                \draw [fill] (zB) circle (.08);
              \end{tikzpicture}\\
              \begin{tikzpicture}[baseline={([yshift=-1ex]current bounding box.south)},scale=1]
                \coordinate (inA) at (0.4,.7);
                \coordinate (outA) at (3.6,.7);
                \coordinate (inB) at (0.4,-.7);
                \coordinate (outB) at (3.6,-.7);
                \coordinate (xA) at (1,.7);
                \coordinate (xxA) at (1.825,.7) ;
                \coordinate (xxB) at (1.825,-.7) ;
                \coordinate (yA) at (1.60+0.4,.7);
                \coordinate (yyA) at (2.375,0) ;
                \coordinate (zA) at (3,.7);
                \coordinate (zzA) at (3.75,.7) ;
                \coordinate (xB) at (1,-.7);
                \coordinate (yB) at (1.6+0.4,-.7);
                \coordinate (zB) at (3,-.7);
                \coordinate (zzB) at (3.75,-.7);
                \coordinate (xM) at (1,0);
                \coordinate (yM) at (1.6+0.4,0);
                \coordinate (zM) at (3,-0);
                \draw [dotted] (inA) -- (outA);
                \draw [dotted] (inB) -- (outB) node [midway, below]{} ;
                \draw [draw=none] (xA) to[out=40,in=140] (zA);
                \draw [photon] (xA) -- (xB);
                \draw [photon] (zA) -- (zB);
                \draw [photon,red] (xB) -- (zA);
              
                \draw [fill] (xA) circle (.08);
                \draw [fill] (zA) circle (.08);
                \draw [fill] (xB) circle (.08);
                \draw [fill] (zB) circle (.08);
              \end{tikzpicture}
              \begin{tikzpicture}[baseline={([yshift=-1ex]current bounding box.south)},scale=1]
                \coordinate (inA) at (0.4,.7);
                \coordinate (outA) at (3.4,.7);
                \coordinate (inB) at (0.4,-.7);
                \coordinate (outB) at (3.4,-.7);
                \coordinate (xA) at (0.9,.7);
                \coordinate (xB) at (0.9,-.7);
                \coordinate (yA) at (1.9,.7);
                \coordinate (yB) at (1.9,-.7);
                \coordinate (zA) at (2.9,.7);
                \coordinate (zB) at (2.9,-.7);
                \draw [dotted] (inA) -- (outA);
                \draw [dotted] (inB) -- (outB) node [midway, below]{} ;
                \draw [draw=none] (xA) to[out=40,in=140] (zA);
                \draw [photon] (xA) -- (xB);
           
                \draw [photon] (zA) -- (zB);
                \draw [photon,red][out=40,in=140] (xA) to[out=-40,in=-140]  (zA);
               
                \draw [fill] (xA) circle (.08);
           
                \draw [fill] (zA) circle (.08);
        
                \draw [fill] (zB) circle (.08);
           
                \draw [fill] (xB) circle (.08);
        
              \end{tikzpicture}
              \caption{The H-graph topsector appearing at 3PM and all its subsectors. }
        \label{fig:wlcut}
\end{figure}
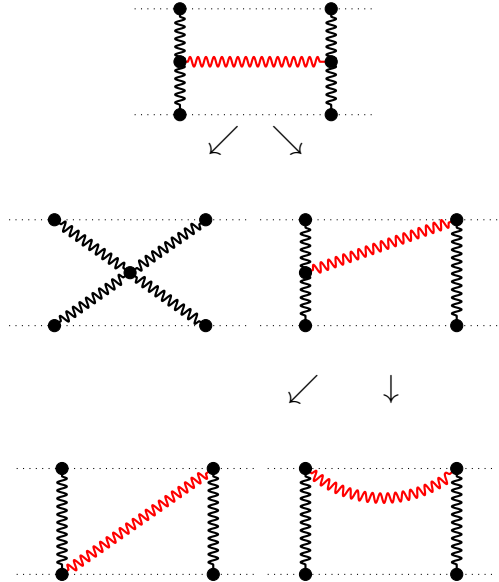

\section{B- and V-type integrals}
A useful observation is that the quantity
\begin{align}
        P \;=\; \sum_{i=1}^{L}(n_i - 1) \;+\; \sum_{i=L+1}^{2L}n_i \; \mod 2, \label{eq:parity}
\end{align}
is preserved under IBP reduction for integrals of the family defined in eq.~\refeq{eq:generalFamily4d}.

Integrals with an odd or even total number of powers of linear propagators (or delta functions) are independent under IBP reduction. This is again clear by mass dimensional analysis. The parity determines if the mass dimension of the integral has an odd or even power. The IBP reduction can only produce coefficients $\sim (q^2)^{n} \sim |q|^{2n}$, which are even in the mass dimension. The integral family splits up into two sub-families.

In the impulse computation, this means that the integrals contributing to the part of the impulse proportional to $b^\mu/q^\mu$ decouple completely from those proportional to $v_i^\mu$ in \refeq{eq:vecStructImp} and \refeq{eq:bvecStructImp}. 

We will refer to them therefore from now on as B-type (even) and V-type (odd) integrals.

\section{Symmetry relations}
A further class of relations, complementary to the IBP identities, allows us to reduce the number of master integrals. These are the so-called \emph{symmetry relations}, and they exploit the fact that within a given integral family there is typically a residual freedom in how the loop momenta are labelled. Consider the two loop example
\begin{align}
       I = \int_{\ell_1,\ell_2}
        \frac{
          \dd^{(n_1-1)}(\ell_1\cdot v_{1})\,\dd^{(n_2-1)}(\ell_2\cdot v_{2})
          }{
          (\ell_1-\ell_2)^2\, \ell_1^2\, (\ell_2-q)^2
          }\,.
\end{align}
A generic shift or relabelling of the loop momenta produces an integral that no longer belongs to our family. There is, however, a distinguished subset of shifts and relabellings that maps the integral back into the same family. For instance, the relabelling $\ell_1\leftrightarrow \ell_2$, $v_1\leftrightarrow v_2$, combined with Lorentz invariance, sends $I$ to
\begin{align}
        I = \int_{\ell_1,\ell_2}
         \frac{
           \dd^{(n_2-1)}(\ell_1\cdot v_{1})\,\dd^{(n_1-1)}(\ell_2\cdot v_{2})
           }{
           (\ell_1-\ell_2)^2\, \ell_2^2\, (\ell_1-q)^2
           }\,,
\end{align}
which is again manifestly an element of the family. 

We can also use Lorentz symmetry to send $v_1\leftrightarrow v_2$ to show that $n$SF and $\bar{n}$SF contributions are given in terms of the same integral families.

Each such mapping yields an additional linear relation between integrals. Identifying all of these relations is in general non-trivial, but they are systematically detected by standard IBP-reduction packages. 

A caveat specific to our setting concerns the $i0$ prescriptions, which we have suppressed in the example. With Feynman $i0$ prescriptions the graviton propagators are invariant under shifts and relabellings, but the situation is more subtle for the linear propagators and for retarded graviton propagators: shifts of the form $\ell\to-\ell$ can flip a prescription, e.g.
\begin{align}
        \frac{1}{\ell\cdot v_1 + \i0}\;\longrightarrow\;-\frac{1}{\ell\cdot v_1 - \i0}\,,\qquad
        \frac{1}{(\ell^0+\i0)^2-\vec{\ell}^2}\;\longrightarrow\;\frac{1}{(\ell^0-\i0)^2-\vec{\ell}^2}\,.
\end{align}
Some care is therefore required in deciding which symmetry relations are admissible. 
In practice, we will use no symmetry relations when working with retarded graviton propagators.
In the next section we examine one particular symmetry in detail, since it sharply illustrates the difference between the Feynman and retarded prescriptions and shows how the choice of retarded propagators breaks part of the symmetry available with Feynman propagators.

\section{Parity symmetry of integrals}

The symmetry operation we will be looking at in this chapter will be 
\begin{align}
        \ell_i\;\to\;-\ell_i\,\text{ and }q\;\to\;-q.
\end{align}
It is clear that the integral must be invariant under this transformation.
The way this symmetry acts on an individual integral depends sensitively on the choice of propagator prescription, and on its parity $P$, defined in \refeq{eq:parity}. Exploiting it leads to a substantial simplification of the calculation, both by setting whole classes of integrals to zero and by exposing others as products of lower-loop building blocks.

We treat the case of Feynman propagators first, which is relevant for the conservative part of the impulse, before turning to retarded propagators, which are required to obtain the full $\Delta p$.

\subsection{Feynman propagators}
\label{sec:feynmanparity}
We will assume for this chapter integrals of the form
\begin{align}
        {\cal P}^{\mathrm{F}}{}_{\{n\}}^{\{\sigma\}}
        =
        \int_{\ell_1,\dots,\ell_{\text{L}}}
        \frac{
          \prod_{i=1}^L\dd(\ell_i\cdot v_{\alpha_i})
          }{
          \prod_{i=1}^{L}D_{i}^{n_{i}}(\sigma_{i})
          \prod_{I<J}D_{IJ}^{n_{IJ}} 
          }\,,
\end{align}
where
\begin{align}
         D_{i}=\ell_i\cdot v_{{\bar \alpha}_i}+\sigma_i\, \i0\,,
\end{align}
and the graviton propagators $D_{IJ}$ are
\begin{align}
        D_{ij}&=(\ell_i-\ell_j)^2 +  \i0 \,,\quad   
        D_{qi}=(\ell_i-q)^2\,,\quad
        D_{0i}=\ell_i^2\,.
\end{align}
In general one can have derivatives of delta functions in the integral but one can always use IBP reductions to remove those. To simplify the discussion we therefore assume the integrals to have no derivatives acting on them.

With these assumptions $\ell_i\to -\ell_i$ and $q \to -q $ yields
\begin{align}
        &\ell_i\cdot v_{{\bar \alpha}_i}+\sigma_i\, \i0 \rightarrow -(\ell_i\cdot v_{{\bar \alpha}_i}-\sigma_i\, \i0)\nn \\
        &\Rightarrow {\cal P}^{\mathrm{F}}{}_{\{n\}}^{\{\sigma\}}
        =(-1)^{\sum_{i=1}^{L} n_i}\, {\cal P}^{\mathrm{F}}{}_{\{n\}}^{\{-\sigma\}}
       \label{equ:feynParity}
\end{align}
i.e.\ the operation flips the sign of every worldline $i0$ prescription and produces an overall sign $(-1)^{\sum_i n_i}$ controlled by the total power of the linear propagators. In practice this is the same as sending $v_i \to -v_i$, so that we will also refer to this symmetry as ``parity''. As noted in the previous section, integrals with an even and an odd number of worldline-propagator powers are not mixed by the IBP relations, so they organise themselves into two effectively independent families. Combined with~\refeq{equ:feynParity}, this motivates the terminology: integrals with an \emph{even} sum $\sum_i n_i$ are called \emph{parity-even}, and those with an \emph{odd} sum \emph{parity-odd}.

When all $n_i\le 0$ on the worldline propagators (so the $i0$ prescriptions play no role), the symmetry collapses to
\begin{align}
        {\cal P}^{\mathrm{F}}{}_{\{n\}}^{\{\}}=(-1)^{\sum_{i=1}^{L} n_i}\, {\cal P}^{\mathrm{F}}{}_{\{n\}}^{\{\}}\,,
\end{align}
so that ${\cal P}^{\mathrm{F}}{}_{\{n\}}^{\{\}} = 0$ whenever $\sum_i n_i$, and equivalently the parity of the integral, is odd. 

This single observation already eliminates a large class of integrals from the calculation. In particular, all integrals coupling to $v_i^\mu$ and which have no WL propagators in the denominator vanish identically.
\\
Furthermore every integral with $\sum_i n_i$ odd but $n_i\ge 1$ for at least one worldline propagator factorises into lower loop-order integrals. To see this, assume (without loss of generality, after IBP reduction) that the relevant master integrals satisfy $n_i = 1$ whenever the corresponding linear propagator appears in the denominator. The identity
\begin{align}
        \frac{1}{\ell_n\cdot v_m \pm \i0}\;=\;\mathrm{PV}\,\frac{1}{\ell_n\cdot v_m}\;\mp\; i\pi\,\delta(\ell_n\cdot v_m)\,,
\end{align}
where $\mathrm{PV}$ denotes the Cauchy principal value, decomposes each linear propagator into a real (principal-value) and an imaginary (delta-function) piece. The two pieces behave differently under $\ell_i\to-\ell_i$,
\begin{align}
       - \frac{1}{\ell_n\cdot v_m \mp \i0}\;=\;-\,\mathrm{PV}\,\frac{1}{\ell_n\cdot v_m}\;\mp\; i\pi\,\delta(\ell_n\cdot v_m)\,,
\end{align}
the principal-value part being odd under the parity and the delta-function part even. Combined with the parity argument above, this implies that every non-vanishing contribution to a parity-odd integral must come from a configuration in which at least one worldline propagator has been replaced by a delta function, that is, in which at least one worldline propagator has been put on-shell.

Concretely, every parity-odd integral can be rewritten as a sum of \emph{cut} integrals in which one or more linear propagators are replaced by delta functions. Each such cut splits the Feynman graph into two pieces and produces a convolution of two lower-loop integrals. After Fourier transform to impact-parameter space the convolution becomes an ordinary product, and the parity-odd integral is exposed as a product of lower-loop building blocks.

The simplest non-trivial case is the one-loop integral
\begin{align}
        I = \int_{\ell}\frac{\dd(\ell\cdot v_1)}{\ell^2\,(\ell-q)^2\,(\ell\cdot v_2 + \i0)}
        = \int_{\ell}\frac{\dd(\ell\cdot v_1)}{\ell^2\,(\ell-q)^2}\,\mathrm{PV}\,\frac{1}{\ell\cdot v_2}
          - 2\,i  \int_{\ell}\frac{\dd(\ell\cdot v_1)\,\dd(\ell\cdot v_2)}{\ell^2\,(\ell-q)^2}\,.
\end{align}
By the symmetry argument the principal-value piece vanishes,
\begin{align}
        \int_{\ell}\frac{\dd(\ell\cdot v_1)}{\ell^2\,(\ell-q)^2}\,\mathrm{PV}\,\frac{1}{\ell\cdot v_2} \;=\; 0\,,
\end{align}
so that
\begin{align}
        I = -2 \,i  \int_{\ell}\frac{\dd(\ell\cdot v_1)\,\dd(\ell\cdot v_2)}{\ell^2\,(\ell-q)^2}\,.
\end{align}
Integrating out the two delta functions leaves
\begin{align}
        I = -\dfrac{1}{\sqrt{\gamma^2-1}}\int_{\boldsymbol{\ell}_\perp}\frac{2 \, i}{\boldsymbol{\ell}_\perp^2\,(\boldsymbol{\ell}_\perp - \boldsymbol q)^2}\,,
\end{align}
where $\boldsymbol{\ell}_\perp$ denotes the component of $\ell$ orthogonal to $v_1$ and $v_2$ and $\int_{\boldsymbol{\ell}_\perp}=\int d^{d-2}\boldsymbol{\ell}_\perp/((2 \pi)^{d-2})$. Fourier-transforming to impact-parameter space,
\begin{align}
        \int_{\boldsymbol q} e^{- i \boldsymbol q \cdot \boldsymbol b}
        \,\int_{\boldsymbol{\ell}_\perp}\frac{1}{\boldsymbol{\ell}_\perp^2\,(\boldsymbol{\ell}_\perp - \boldsymbol q)^2}
         \;=\;
         \int_{\boldsymbol{q}} e^{- i \boldsymbol{q}\cdot \boldsymbol b}\,\int_{\boldsymbol{q} '}\frac{1}{\boldsymbol{q}'^2\,(\boldsymbol{q}' - \boldsymbol q)^2}\,,
\end{align}
and using the convolution theorem we get
\begin{align}
        \dfrac{1}{\gamma^2-1} \int_{\boldsymbol q} e^{-i \boldsymbol q\cdot \boldsymbol b}\int_{\boldsymbol{\ell}_\perp}\frac{1}{\boldsymbol{\ell}_\perp^2\,(\boldsymbol{\ell}_\perp - \boldsymbol q)^2}
        \;=\;\Bigl(\int_q e^{iq\cdot b}\,\frac{\dd(q\cdot v_1)\,\dd(q\cdot v_2)}{q^2}\Bigr)^2,
\end{align}
so that, after Fourier transform, this 2PM one-loop integral reduces to the square of a 1PM tree-level building block. The argument generalises: every integral containing a loop in which both worldlines are cut factorises, upon Fourier transform to impact-parameter space, into a product of lower-loop integrals. 

 Diagrammatically this can be written as:

        \begin{align}
        &\begin{tikzpicture}[baseline={([yshift=-1ex]current bounding box.center)},scale=1]
                \coordinate (inA) at (0.3,.7);
                \coordinate (outA) at (3.6,.7);
                \coordinate (inB) at (0.3,-.7);
                \coordinate (outB) at (3.6,-.7);
                \coordinate (xA) at (0.6,.7);
                \coordinate (yA) at (1.5,.7);
                \coordinate (zA) at (2.4,.7);
                \coordinate (wA) at (3.3,.7);
                \coordinate (xB) at (0.6,-.7);
                \coordinate (yB) at (1.5,-.7);
                \coordinate (zB) at (2.4,-.7);
                \coordinate (wB) at (3.3,-.7);
                \coordinate (xM) at (0.6,0);
                \coordinate (yM) at (1.5,0);
                \coordinate (zM) at (2.4,0);
                \coordinate (wM) at (3.3,0);
                \coordinate (xyMB) at (1.05,-.35);
                \coordinate (yzMB) at (1.95,-.35);
                \coordinate (zwMB) at (2.85,-.35);
                \coordinate (yzMA) at (1.95,.35);
                \draw [dotted] (inA) -- (outA);
                 \draw [dotted] (inB) -- (outB);
                \draw [photon] (yA) --  (yB);
                \draw [photon] (zA) --  (zB);
                \draw [photon] (zM) --  (wM);
                \draw [photon] (yM) --  (xM);
                \draw [fill] (zM) circle (.08);
                \draw [fill] (yM) circle (.08);
                \draw [] (xM) circle (.2);
                 \draw (xM) node[cross=4.6pt] {};
                 \draw [] (wM) circle (.2);
                 \draw (wM) node[cross=4.6pt] {};
                 \draw [fill] (yA) circle (.08);
                 \draw [fill] (yB) circle (.08);
                 \draw [fill] (zA) circle (.08);
                 \draw [fill] (zB) circle (.08);
                 \draw [white] (-.5,0) node[] {\LARGE{$\sim \int_{q'}$}};
                \draw [draw=none] (xB) to[out=-40,in=-140] (zB);
              \end{tikzpicture}
              \begin{tikzpicture}[baseline={([yshift=0ex]current bounding box.center)},scale=1]
                \coordinate (inA) at (0.3,.7);
                \coordinate (outA) at (5.2,.7);
                \coordinate (inB) at (0.3,-.7);
                \coordinate (outB) at (5.2,-.7);
                \coordinate (xA) at (0.6,.7);
                \coordinate (yA) at (1.5,.7);
                \coordinate (zA) at (4.0,.7);
                \coordinate (wA) at (4.9,.7);
                \coordinate (xB) at (0.6,-.7);
                \coordinate (yB) at (1.5,-.7);
                \coordinate (zB) at (4.0,-.7);
                \coordinate (wB) at (4.9,-.7);
                \coordinate (xM) at (0.6,0);
                \coordinate (yM) at (1.5,0);
                \coordinate (zM) at (4.0,0);
                \coordinate (wM) at (4.9,0);
                \coordinate (xyMB) at (1.05,-.35);
                \coordinate (yzMB) at (1.95,-.35);
                \coordinate (zwMB) at (2.85,-.35);
                \coordinate (yzMA) at (1.95,.35);
                \draw [dotted] (inA) -- (yA);
                \draw [dotted] (zA) -- (outA);
                \draw [dotted] (inB) -- (yB);
                \draw [dotted] (zB) -- (outB);
                \draw [photon] (yA) --  (yB);
                \draw [photon] (zA) --  (zB);
                \draw [photon] (zM) --  (wM);
                \draw [photon] (yM) --  (xM);
                \draw [fill] (zM) circle (.08);
                \draw [fill] (yM) circle (.08);
                \draw [] (xM) circle (.2);
                 \draw (xM) node[cross=4.6pt] {};
                 \draw [] (wM) circle (.2);
                 \draw (wM) node[cross=4.6pt] {};
                 \draw [fill] (yA) circle (.08);
                 \draw [fill] (yB) circle (.08);
                 \draw [fill] (zA) circle (.08);
                 \draw [fill] (zB) circle (.08);
                 \draw (-.5,0) node[] {\LARGE{$\sim \int_{q'}$}};
                 \draw (2.6,0) node[] {$(q-q') \;\times$};
                 \draw (5.7,0) node[] {$(q')$};
                \draw [draw=none] (xB) to[out=-40,in=-140] (zB);
              \end{tikzpicture} \\&
              \begin{tikzpicture}[baseline={([yshift=-1ex]current bounding box.center)},scale=1]
                \coordinate (inA) at (0.3,.7);
                \coordinate (outA) at (3.6,.7);
                \coordinate (inB) at (0.3,-.7);
                \coordinate (outB) at (3.6,-.7);
                \coordinate (xA) at (0.6,.7);
                \coordinate (yA) at (1.5,.7);
                \coordinate (zA) at (2.4,.7);
                \coordinate (wA) at (3.3,.7);
                \coordinate (xB) at (0.6,-.7);
                \coordinate (yB) at (1.5,-.7);
                \coordinate (zB) at (2.4,-.7);
                \coordinate (wB) at (3.3,-.7);
                \coordinate (xM) at (0.6,0);
                \coordinate (yM) at (1.5,0);
                \coordinate (zM) at (2.4,0);
                \coordinate (wM) at (3.3,0);
                \coordinate (xyMB) at (1.05,-.35);
                \coordinate (yzMB) at (1.95,-.35);
                \coordinate (zwMB) at (2.85,-.35);
                \coordinate (yzMA) at (1.95,.35);
                \draw [dotted] (inA) -- (outA);
                 \draw [dotted] (inB) -- (outB);
                \draw [photon] (yA) --  (yB);
                \draw [photon] (zA) --  (zB);
                \draw [photon] (zM) --  (wM);
                \draw [photon] (yM) --  (xM);
                \draw [fill] (zM) circle (.08);
                \draw [fill] (yM) circle (.08);
                \draw [] (xM) circle (.2);
                 \draw (xM) node[cross=4.6pt] {};
                 \draw [] (wM) circle (.2);
                 \draw (wM) node[cross=4.6pt] {};
                 \draw [fill] (yA) circle (.08);
                 \draw [fill] (yB) circle (.08);
                 \draw [fill] (zA) circle (.08);
                 \draw [fill] (zB) circle (.08);
                 \draw  (-1.3,0) node[] {\LARGE{$ \Rightarrow \int_{q} e^{i q\cdot b}$}};
                \draw [draw=none] (xB) to[out=-40,in=-140] (zB);
              \end{tikzpicture}
              \begin{tikzpicture}[baseline={([yshift=0ex]current bounding box.center)},scale=1]
                \coordinate (inA) at (0.3,.7);
                \coordinate (outA) at (6.1,.7);
                \coordinate (inB) at (0.3,-.7);
                \coordinate (outB) at (6.1,-.7);
                \coordinate (xA) at (0.6,.7);
                \coordinate (yA) at (1.5,.7);
                \coordinate (zA) at (4.9,.7);
                \coordinate (wA) at (5.8,.7);
                \coordinate (xB) at (0.6,-.7);
                \coordinate (yB) at (1.5,-.7);
                \coordinate (zB) at (4.9,-.7);
                \coordinate (wB) at (5.8,-.7);
                \coordinate (xM) at (0.6,0);
                \coordinate (yM) at (1.5,0);
                \coordinate (zM) at (4.9,0);
                \coordinate (wM) at (5.8,0);
                \coordinate (xyMB) at (1.05,-.35);
                \coordinate (yzMB) at (1.95,-.35);
                \coordinate (zwMB) at (2.85,-.35);
                \coordinate (yzMA) at (1.95,.35);
                \draw [dotted] (inA) -- (yA);
                \draw [dotted] (zA) -- (outA);
                \draw [dotted] (inB) -- (yB);
                \draw [dotted] (zB) -- (outB);
                \draw [photon] (yA) --  (yB);
                \draw [photon] (zA) --  (zB);
                \draw [photon] (zM) --  (wM);
                \draw [photon] (yM) --  (xM);
                \draw [fill] (zM) circle (.08);
                \draw [fill] (yM) circle (.08);
                \draw [] (xM) circle (.2);
                 \draw (xM) node[cross=4.6pt] {};
                 \draw [] (wM) circle (.2);
                 \draw (wM) node[cross=4.6pt] {};
                 \draw [fill] (yA) circle (.08);
                 \draw [fill] (yB) circle (.08);
                 \draw [fill] (zA) circle (.08);
                 \draw [fill] (zB) circle (.08);
                 \draw (-1.3,0) node[] {\LARGE{$\sim$}\Huge{$\bigg($}\LARGE{$\int_{q} e^{iq \cdot b}$}};
                 \draw (3.2,0) node[] {\Huge{$\bigg)\bigg($}\LARGE{$\int_{q} e^{iq \cdot b}$}};
                 \draw (6.5,0) node[] {\Huge{$\bigg)$}};
                \draw [draw=none] (xB) to[out=-40,in=-140] (zB);
              \end{tikzpicture} ,
                 \end{align}
where $\begin{tikzpicture} \draw (0,0) node[cross=4pt] {};
        \draw [] (0,0) circle (.2);
\end{tikzpicture}$ stands for the remainder of the diagram on the two sides of the cut.

This property has a clear physical interpretation. Feynman propagators describe conservative scattering, $\Delta p_1+\Delta p_2 = 0$, and the mass-shell condition $(p_i+\Delta p_i)^2 = p_i^2$ with $p_i = m_i v_i$ then fixes the projections of $\Delta p_1$ along $v_1$ and $v_2$ at a given PM order in terms of lower-order results,
\begin{align}
        \Delta p_1\cdot p_1 &= -\tfrac{1}{2}\,\Delta p_1^2\,,\\
        \Delta p_1\cdot p_2 &= \phantom{-}\tfrac{1}{2}\,\Delta p_1^2\,.
\end{align}
These relations factorise the $v_i^\mu$ components of $\Delta p_1$ into lower-loop pieces; the parity argument above is the integral-level statement of the same factorisation. The relations give no information on the $b^\mu$ component of $\Delta p_1$, since $b\cdot v_i = 0$ --- in line with the symmetry properties of the parity-even integrals, which control precisely that part of the impulse.

The same kind of argument applies to time-symmetric combinations of retarded and advanced propagators.
\subsection{Retarded propagators}
The picture changes considerably once retarded propagators are introduced. Recall that the retarded graviton propagator can be written as
\begin{align}
        \dfrac{1}{(l^0+\i0)^2-\vec{l}^2}=\dfrac{1}{({l^0}^2+l^0 \i0  )-\vec{l}^2}.
\end{align}
This form is not manifestly Lorentz invariant. A Lorentz-covariant rewriting is available however: for any timelike four-vector $v_i$ with $v_i^2 = 1$ and $v_i^0>0$,
\begin{align}
        \frac{1}{\ell^2 + \mathrm{sgn}(\ell\cdot v_i)\,\i0}\,,
\end{align}
since on the mass shell $\ell^2 = 0$ the sign of $\ell^0$ agrees with that of $\ell\cdot v_i$. We will work with this version of the retarded propagators from now on. In contrast to the Feynman case, however, the retarded graviton propagators are not invariant under the operation $\ell_i \rightarrow - \ell_i$, but the $i0$ prescription gets flipped.

Having a general L-loop planar integral family:
\begin{align}
        {\cal P}_{\{n\}}^{\{\sigma\}}
        =
        \int_{\ell_1,\dots,\ell_{\text{L}}}
        \frac{
          \prod_{i=1}^L\dd(\ell_i\cdot v_{\alpha_i})
          }{
          \prod_{i=1}^{L}D_{i}^{n_{i}}(\sigma_{i})
          \prod_{I<J}D_{IJ}^{n_{IJ}}
          }\,,
      \end{align}
      where
      \begin{align}
         D_{i}&=\ell_i\cdot v_{{\bar \alpha}_i}+\sigma_i \i0\,,
      \end{align}
      and the graviton propagators $D_{IJ}$ are
      \begin{align}\label{DIJs2}
      D_{ij}=(\ell_i-\ell_j)^2 + \sigma_{i,j}&\text{sign}((\ell_i-\ell_j)\cdot v_1) \i0 ,  
        D_{qi}=(\ell_i-q)^2
        \,,
        D_{0i}=\ell_i^2
        \,,
      \end{align}
      the inversion $\ell_i\to -\ell_i$ and $q \to -q $ yields
      \begin{align}
              {\cal P}_{\{n\}}^{\{\sigma\}}=(-1)^{\sum_{i=1}^L n_i}\, {\cal P}_{\{n\}}^{\{- \sigma\}}\,.
      \end{align}
The retarded propagators thus break the symmetry~\refeq{equ:feynParity} which was used above to set whole classes of parity-odd integrals to zero. Physically this is exactly as expected: retarded (or advanced) propagators correspond to the full physical answer, including the radiative sector, so that $E_{\rm rad}\neq 0$. 
\section{Partial fraction identities}
The last technique we employ to simplify the integrand is the use of \emph{partial-fraction identities}. We use them at 0SF and 1SF order to 'planarise' the integrand. Starting at 4PM order, we can get integrals that we cannot bring into the form \refeq{eq:generalFamily4d} by using shift and relabelling operations. Therefore we must employ so-called partial fraction identities such as
\begin{align}\label{eq:PF}
        \begin{tikzpicture}[baseline={([yshift=-1ex]current bounding box.center)},scale=.7]
          \coordinate (inA) at (0.2,.6);
          \coordinate (outA) at (3.8,.6);
          \coordinate (inB) at (0.2,-.6);
          \coordinate (outB) at (3.8,-.6);
          \coordinate (xA) at (.8,.6);
          \coordinate (xyA) at (1.5,.6);
          \coordinate (yA) at (2,.6);
          \coordinate (yzA) at (2.5,.6);
          \coordinate (zA) at (3.2,.6);
          \coordinate (xB) at (.8,-.6);
          \coordinate (yB) at (2,-.6);
          \coordinate (zB) at (3.2,-.6);
          \draw [fill] (xA) circle (.08);
          \draw [fill] (yA) circle (.08);
          \draw [fill] (zA) circle (.08);
          \draw [dotted] (inA) -- (outA);
          \draw [zParticle] (xA) -- (yA);
          \draw [zParticle] (zA) -- (yA);
          \draw [photon] (xA) -- (xB);
          \draw [photon] (yA) -- (yB);
          \draw [photon] (zA) -- (zB);
          \draw (xyA) node [above] {$\ell_1\!\cdot\!v_1\,\,\,$};
          \draw (yzA) node [above] {$\,\,\,\ell_2\!\cdot\!v_1$};
          \draw [draw=none] (xB) to[out=-40,in=-140] (zB);
        \end{tikzpicture}\!&=\!
      \begin{tikzpicture}[baseline={([yshift=-1ex]current bounding box.center)},scale=.7]
          \coordinate (inA) at (0.2,.6);
          \coordinate (outA) at (3.8,.6);
          \coordinate (inB) at (0.2,-.6);
          \coordinate (outB) at (3.8,-.6);
          \coordinate (xA) at (.8,.6);
          \coordinate (xyA) at (1.5,.6);
          \coordinate (yA) at (2,.6);
          \coordinate (yzA) at (2.5,.6);
          \coordinate (zA) at (3.2,.6);
          \coordinate (xB) at (.8,-.6);
          \coordinate (yB) at (2,-.6);
          \coordinate (zB) at (3.2,-.6);
          \coordinate (yzAB) at (2.6,0);
          \draw [fill] (xA) circle (.08);
          \draw [fill] (yA) circle (.08);
          \draw [fill] (zA) circle (.08);
          \draw [dotted] (inA) -- (outA);
          \draw [zParticle] (xA) -- (yA);
          \draw [zParticle] (yA) -- (zA);
          \draw [photon] (xA) -- (xB);
          \draw [photon] (yA) -- (zB);
          \filldraw[white] (yzAB) circle (6pt);
          \draw [photon] (zA) -- (yB);
          \draw (xyA) node [above] {$\ell_1\!\cdot\!v_1\,\,\,$};
          \draw (yzA) node [above] {$\,\,\,\,\,\ell_{12}\!\cdot\!v_1$};
          \draw [draw=none] (xB) to[out=-40,in=-140] (zB);
        \end{tikzpicture}\,+\,
      \begin{tikzpicture}[baseline={([yshift=-1ex]current bounding box.center)},scale=.7]
          \coordinate (inA) at (0.2,.6);
          \coordinate (outA) at (3.8,.6);
          \coordinate (inB) at (0.2,-.6);
          \coordinate (outB) at (3.8,-.6);
          \coordinate (xA) at (.8,.6);
          \coordinate (xyA) at (1.5,.6);
          \coordinate (yA) at (2,.6);
          \coordinate (yzA) at (2.5,.6);
          \coordinate (zA) at (3.2,.6);
          \coordinate (xB) at (.8,-.6);
          \coordinate (yB) at (2,-.6);
          \coordinate (zB) at (3.2,-.6);
          \coordinate (xyAB) at (1.4,0);
          \draw [fill] (xA) circle (.08);
          \draw [fill] (yA) circle (.08);
          \draw [fill] (zA) circle (.08);
          \draw [dotted] (inA) -- (outA);
          \draw [zParticle] (yA) -- (xA);
          \draw [zParticle] (zA) -- (yA);
          \draw [photon] (zA) -- (zB);
          \draw [photon] (yA) -- (xB);
          \filldraw[white] (xyAB) circle (6pt);
          \draw [photon] (xA) -- (yB);
          \draw (xyA) node [above] {$\ell_{12}\!\cdot\!v_1\,\,\,\,\,$};
          \draw (yzA) node [above] {$\,\,\,\ell_2\!\cdot\!v_1$};
          \draw [draw=none] (xB) to[out=-40,in=-140] (zB);
        \end{tikzpicture}\nn\\[-1em]
        \frac1{(\ell_1\cdot v_1)(\ell_2\cdot v_1)}&=
        \frac1{(\ell_1\cdot v_1)(\ell_{12}\cdot v_1)}+
        \frac1{(\ell_{12}\cdot v_1)(\ell_2\cdot v_1)}\,,
      \end{align}
with $\ell_{12}=\ell_1+\ell_2$.
Pictorially, such identities remove ``jumping'' worldline lines from the graphs, disentangling crossings between graviton and worldline edges. Their iterated application produces an integrand expressed entirely in terms of fully planar topologies.

One of the simplest examples is a three loop graph:
\begin{align}
      \begin{tikzpicture}[baseline={([yshift=-1ex]current bounding box.center)},scale=1]
                  \coordinate (inA) at (0.4,.7);
                \coordinate (outA) at (4.4,.7);
                \coordinate (inB) at (0.4,-.7);
                \coordinate (outB) at (4.4,-.7);
                \coordinate (xA) at (0.9,.7);
                \coordinate (xB) at (0.9,-.7);
                \coordinate (yA) at (1.9,.7);
                \coordinate (yB) at (1.9,-.7);
                \coordinate (zA) at (2.9,.7);
                \coordinate (zB) at (2.9,-.7);
                \coordinate (aA) at (3.9,.7);
                \coordinate (aB) at (3.9,-.7);
                \draw [dotted] (inA) -- (outA);
                \draw [dotted] (inB) -- (outB) node [midway, below]{} ;
                \draw [draw=none] (xA) to[out=40,in=140] (zA);
                \draw [photon] (xA) -- (xB);
                \draw [photon,red] (yA) -- (yB);
                \draw [photon] (zA) -- (zB);
                \draw [photon] (aA) -- (aB);
                \draw [zParticle] (xA) -- (yA);
                \draw [zParticle] (yB) to[out=-40,in=-140]  (aB);
                \draw [zParticle] (zA) -- (yA);
                \draw [fill] (xA) circle (.08);
                \draw [fill] (yA) circle (.08);
                \draw [fill] (zA) circle (.08);
                \draw [fill] (aA) circle (.08);
                \draw [fill] (aB) circle (.08);
                \draw [fill] (zB) circle (.08);
                \draw [fill] (yB) circle (.08);
                \draw [fill] (xB) circle (.08);
        \end{tikzpicture}\!&=\!
      \begin{tikzpicture}[baseline={([yshift=-1ex]current bounding box.center)},scale=1]
        \coordinate (inA) at (0.4,.7);
      \coordinate (outA) at (4.4,.7);
      \coordinate (inB) at (0.4,-.7);
      \coordinate (outB) at (4.4,-.7);
      \coordinate (xA) at (0.9,.7);
      \coordinate (xB) at (0.9,-.7);
      \coordinate (yA) at (1.9,.7);
      \coordinate (yB) at (1.9,-.7);
      \coordinate (zA) at (2.9,.7);
      \coordinate (zB) at (2.9,-.7);
      \coordinate (aA) at (3.9,.7);
      \coordinate (aB) at (3.9,-.7);
      \draw [dotted] (inA) -- (outA);
      \draw [dotted] (inB) -- (outB) node [midway, below]{} ;
      \draw [draw=none] (xA) to[out=40,in=140] (zA);
      \draw [photon] (xA) -- (xB);
      \draw [photon] (yA) -- (yB);
      \draw [photon,red] (zA) -- (zB);
      \draw [photon] (aA) -- (aB);
      \draw [zParticle] (xA) -- (yA);
      \draw [zParticle] (zB) to  (aB);
      \draw [zParticle] (yA) -- (zA);
      \draw [zParticle,white] (yB) to[out=-40,in=-140]  (aB);
      \draw [fill] (xA) circle (.08);
      \draw [fill] (yA) circle (.08);
      \draw [fill] (zA) circle (.08);
      \draw [fill] (aA) circle (.08);
      \draw [fill] (aB) circle (.08);
      \draw [fill] (zB) circle (.08);
      \draw [fill] (yB) circle (.08);
      \draw [fill] (xB) circle (.08);
\end{tikzpicture}\, +\,
\begin{tikzpicture}[baseline={([yshift=-1ex]current bounding box.center)},scale=1]
        \coordinate (inA) at (0.4,.7);
      \coordinate (outA) at (4.4,.7);
      \coordinate (inB) at (0.4,-.7);
      \coordinate (outB) at (4.4,-.7);
      \coordinate (xA) at (0.9,.7);
      \coordinate (xB) at (0.9,-.7);
      \coordinate (yA) at (1.9,.7);
      \coordinate (yB) at (1.9,-.7);
      \coordinate (zA) at (2.9,.7);
      \coordinate (zB) at (2.9,-.7);
      \coordinate (aA) at (3.9,.7);
      \coordinate (aB) at (3.9,-.7);
      \draw [dotted] (inA) -- (outA);
      \draw [dotted] (inB) -- (outB) node [midway, below]{} ;
      \draw [draw=none] (xA) to[out=40,in=140] (zA);
      \draw [photon] (xA) -- (xB);
      \draw [photon,red] (yA) -- (yB);
      \draw [photon] (zA) -- (zB);
      \draw [photon] (aA) -- (aB);
      \draw [zParticle] (zA) -- (yA);
      \draw [zParticle] (aA) to  (zA);
      \draw [zParticle] (yB) -- (xB);
      \draw [zParticle,white] (yB) to[out=-40,in=-140]  (aB);
      \draw [fill] (xA) circle (.08);
      \draw [fill] (yA) circle (.08);
      \draw [fill] (zA) circle (.08);
      \draw [fill] (aA) circle (.08);
      \draw [fill] (aB) circle (.08);
      \draw [fill] (zB) circle (.08);
      \draw [fill] (yB) circle (.08);
      \draw [fill] (xB) circle (.08);
\end{tikzpicture}
      \end{align}
               
The idea is always similar. As in the example above at 0SF one has either multiple ``jumping'' worldline propagators on a single worldline. Then one can use these partial fraction identities to remove them iteratively. At 1SF one can also have a jumping propagator on the second worldline, as in the example above. Then the strategy is always to shuffle the jumping worldline to the right or left side to disentangle the graphs. This makes it possible to fully disentangle the graphs at 0SF and 1SF. The structure gets more complicated at 2SF, so that this is no longer possible. 
\chapter{Structure of the reduced integrand}
\label{sec:struct}
All the discussed reduction steps can simplify the structure of the integrand massively. In particular, we can use partial fractions at $0$/$\bar{0}$SF and $1$/$\bar{1}$SF order to have at any loop order just a single integral family. Furthermore, we can use IBP reductions to reduce the number of relevant topsectors. 

In general our integrand will have a large number of topsectors. Many of those will, however, reduce to integrals with fewer graviton propagators, which lie in subsectors of other, irreducible topsectors that cannot be reduced further. We will discuss these sectors in detail and which topsectors reduce to lower sectors. The relevant topologies also depend on the theory, as the integrands in electrodynamics will only have a subset of the topologies appearing in Einstein gravity. 

\section{Reducible topsectors}
In this section, we discuss topsectors that can be fully reduced to lower sectors. We will discuss two types of sectors, called mushroom or triangle type. We also refer to ref.~\cite{Frellesvig:2024zph}, which discusses very similar reduction rules.
\subsection{Mushroom topsectors}
\label{sec:mushroom}
The first type of topsector that can be reduced to a subsector by IBP reduction is self-energy bubbles with possible internal loops:
\begin{align}
        &\begin{tikzpicture}[baseline={([yshift=-1ex]current bounding box.center)},scale=3]
                \coordinate (inA) at (0.2,0);
                \coordinate (outA) at (3.8,0);
                \coordinate (inB) at (0.2,0);
                \coordinate (outB) at (3.8,0);
                \coordinate (xA) at (.2,0);
                \coordinate (xyA) at (1.5,0);
                \coordinate (yA) at (2,0);
                \coordinate (yzA) at (2.5,0);
                \coordinate (zA) at (3.2,0);
                \coordinate (xB) at (.8,0);
                \coordinate (yB) at (2,0);
                \coordinate (yM) at (2,0);
                \coordinate (zB) at (3.2,0);
                \coordinate (xyAB) at (1.4,0);
                \draw [fill] (yA) circle (.03);
                \draw [fill]  (.6,.35) circle (.03);
                \draw [fill] (.9,.45) circle (.03);
                \draw [fill] (1.3,.45) circle (.03);
                \draw [fill] (1.6,.35) circle (.03);
                \draw [fill]  (.6,.0) circle (.03);
                \draw [fill] (.9,.0) circle (.03);
                \draw [fill] (1.3,.0) circle (.03);
                \draw [fill] (1.6,.0) circle (.03);
                \draw [dotted] (inA) --node [midway, below]{$\rightarrow \ell_1$}  (outA);
                 \draw [zUndirected] (zA) -- (yA);
                \draw [photon] (xA) [in=120,out=60] to   (yA);
                \draw [photon] (.6,0) to  (.6,.37);
                \draw [photon] (.9,0) --node [midway, right]{$\hspace{0.3cm}...$}  (.9,.47);
                \draw [photon] (1.3,0) to  (1.3,.47);
                \draw [photon] (1.6,0) to  (1.6,.37);
                \node at (.47,.15) {\mbox{\scriptsize$k_1$}};
                \node at (.47,.15) {\mbox{\Large$\circlearrowright$}};
                \node at (.75,.28) {\mbox{\scriptsize$k_2$}};
                \node at (.75,.15) {\mbox{\Large$\circlearrowright$}};
                \node at (1.75,.15) {\mbox{\scriptsize$k_{l}$}};
                \node at (1.75,.15) {\mbox{\Large$\circlearrowright$}};
                \node at (1.45,.28) {\mbox{\scriptsize$k_{l-1}$}};
                \node at (1.45,.15) {\mbox{\Large$\circlearrowright$}};
                \draw [] (yA) to (zA);
                \draw [] (xA) circle (.1);
                \draw (xA) node[cross=6pt] {};
                \draw [] (zA) circle (.1);
                \draw (zA) node[cross=6pt] {};
                \draw [draw=none] (xB) to[out=-40,in=-140] (zB);
              \end{tikzpicture}\nn
        \end{align}
        \vspace{-1.8cm}
        \begin{align}
              & = 
              \int_{\ell_1,...,\ell_n,k_1,...,k_l}\dfrac{\dd(\ell_1 \cdot v_1 )\dd((k_1-\ell_1) \cdot v_2)\prod_{i=2}^{l}\dd((k_1-k_i) \cdot v_2)}{(\ell_1 \cdot v_2) \prod_{i=1}^{l}(k_i)^2 \prod_{i=1}^{l-1}(k_i-k_{i+1})^2} N(\ell_1,...,\ell_n),
\end{align} 
where $k_1,...,k_l$ are the internal momenta of the self energy bubble and $\begin{tikzpicture} \draw (0,0) node[cross=4pt] {};
        \draw [] (0,0) circle (.2);
\end{tikzpicture}$ depicts the remaining structure of the graph $N(\ell_1,...,\ell_n)$, which is independent of $k_i$. These types of self-energy bubbles are often attached to a worldline propagator $1/(\ell_1 \cdot v_2)$, which can be removed by IBP reduction.
We show here a relation for the corner integral (the integral of a sector with all propagators raised to power 1) to keep the discussion simple but this can be done for arbitrary propagator powers. We can look only at the self-energy bubble integral and use dimensional analysis to fix its dependence on $\omega=\ell_1 \cdot v_2$:
\begin{align}
     & B(\omega)=  \int_{k_1,...,k_l}\dfrac{\dd(k_1 \cdot v_2-\omega)\prod_{i=2}^{l}\dd((k_1-k_i) \cdot v_2)}{ \prod_{i=1}^{l}(k_i)^2 \prod_{i=1}^{l-1}(k_i-k_{i+1})^2}=n_B \omega^{l( d -5)+2}\nn \\
     & \Rightarrow \dfrac{d}{d \omega} B(\omega) = \omega^{-1} (l( d -5)+2)B(\omega)\nn\\
     & \Rightarrow \dfrac{d}{d \omega} (B(\omega) \omega)= (l( d -5)+3)B(\omega)\nn\\
&=\int_{k_1,...,k_l}\omega\dfrac{ \dd'(k_1 \cdot v_2-\omega)\prod_{i=1}^{l}\dd((k_1-k_i) \cdot v_2)}{ \prod_{i=1}^{l}(k_i)^2 \prod_{i=1}^{l-1}(k_i-k_{i+1})^2}+B(\omega)\\
& \Rightarrow  B(\omega)= (l( d -5)+2)^{-1}\int_{k_1,...,k_l}\omega\dfrac{ \dd'(k_1 \cdot v_2-\omega)\prod_{i=1}^{l}\dd((k_1-k_i) \cdot v_2)}{ \prod_{i=1}^{l}(k_i)^2 \prod_{i=1}^{l-1}(k_i-k_{i+1})^2},
\end{align}
where $n_B$ is some numerical factor.
This relation removes the worldline from the integral. Diagrammatically this relation would have the form
\begin{align}
\begin{tikzpicture}[baseline={([yshift=0ex]current bounding box.center)},scale=1.5]
        \coordinate (inA) at (0.2,0);
        \coordinate (outA) at (3.8,0);
        \coordinate (inB) at (0.2,0);
        \coordinate (outB) at (3.8,0);
        \coordinate (xA) at (.2,0);
        \coordinate (xyA) at (1.5,0);
        \coordinate (yA) at (2,0);
        \coordinate (yzA) at (2.5,0);
        \coordinate (zA) at (3.2,0);
        \coordinate (xB) at (.8,0);
        \coordinate (yB) at (2,0);
        \coordinate (yM) at (2,0);
        \coordinate (zB) at (3.2,0);
        \coordinate (xyAB) at (1.4,0);
        \draw [fill] (yA) circle (.03);
        \draw [fill]  (.6,.35) circle (.03);
        \draw [fill] (.9,.45) circle (.03);
        \draw [fill] (1.3,.45) circle (.03);
        \draw [fill] (1.6,.35) circle (.03);
        \draw [fill]  (.6,.0) circle (.03);
        \draw [fill] (.9,.0) circle (.03);
        \draw [fill] (1.3,.0) circle (.03);
        \draw [fill] (1.6,.0) circle (.03);
        \draw [dotted] (inA) -- (outA);
         \draw [zUndirected] (zA) -- (yA);
        \draw [photon] (xA) [in=120,out=60] to   (yA);
        \draw [photon] (.6,0) to  (.6,.37);
        \draw [photon] (.9,0) --node [midway, right]{$\hspace{-0.05cm}...$}  (.9,.47);
        \draw [photon] (1.3,0) to  (1.3,.47);
        \draw [photon] (1.6,0) to  (1.6,.37);
        \draw [] (yA) to (zA);
        \draw [] (xA) circle (.1);
        \draw (xA) node[cross=3pt] {};
        \draw [] (zA) circle (.1);
        \draw (zA) node[cross=3pt] {};
        \draw [draw=none] (xB) to[out=-40,in=-140] (zB);
      \end{tikzpicture}
      =-(l( d -5)+2)^{-1}
      \begin{tikzpicture}[baseline={([yshift=-1ex]current bounding box.center)},scale=1.5]
        \coordinate (inA) at (0.2,0);
        \coordinate (outA) at (2,0);
        \coordinate (inB) at (0.2,0);
        \coordinate (outB) at (2,0);
        \coordinate (xA) at (.2,0);
        \coordinate (xyA) at (1.5,0);
        \coordinate (yA) at (2,0);
        \coordinate (yzA) at (2.5,0);
        \coordinate (zA) at (2,0);
        \coordinate (xB) at (.8,0);
        \coordinate (yB) at (2,0);
        \coordinate (yM) at (2,0);
        \coordinate (zB) at (2,0);
        \coordinate (xyAB) at (1.4,0);
        \draw [fill]  (.6,.35) circle (.03);
        \draw [fill] (.9,.45) circle (.03);
        \draw [fill] (1.3,.45) circle (.03);
        \draw [fill] (1.6,.35) circle (.03);
        \draw [fill]  (.6,.0) circle (.03);
        \draw [fill] (.9,.0) circle (.03);
        \draw [fill] (1.3,.0) circle (.03);
        \draw [fill] (1.6,.0) circle (.03);
        \draw [dotted] (inA) -- (outA);
         \draw [zUndirected] (zA) -- (yA);
        \draw [photon] (xA) [in=120,out=60] to   (yA);
        \draw [photon] (.6,0) to  (.6,.37);
        \draw [photon] (.9,0) --node [midway, right]{$\hspace{-0.05cm}...$}  (.9,.47);
        \draw [photon] (1.3,0) to  (1.3,.47);
        \draw [photon] (1.6,0) to  (1.6,.37);
        \draw [] (yA) to (zA);
        \draw [] (xA) circle (.1);
        \draw (xA) node[cross=3pt] {};
        \draw [] (zA) circle (.1);
        \draw (zA) node[cross=3pt] {};
        \draw [draw=none] (xB) to[out=-40,in=-140] (zB);
        \draw [fill,blue!70] (0.45,0) circle (.04);
      \end{tikzpicture}.
\end{align}
The same kind of reduction can be done for arbitrary powers of denominators. We have checked this explicitly up to 5PM.
This simplifies the calculation and also solves another issue that can appear. In the integrand we can find integrals containing terms like:
\begin{align}
        \dfrac{1}{(\ell_1 \cdot v_2+\i0)(\ell_1 \cdot v_2-\i0)}.
\end{align}
This $\i0$ can in principle lead to difficulties and it is not fully clear how to handle these terms. Since the IBP relation removes these terms, this gives us already a hint that the $\i0$ prescription on the WLs here does not matter. A more careful analysis of these integrals confirms this. The $\i0$ prescription only matters if the incoming energy $\ell_1\cdot v_2$ of the self-energy bubble is zero. In this case, however, the bubble integral itself becomes scaleless and is trivially zero.
\subsection{Triangle topsectors}
\label{sec:triangles}
The second kind of topsector, which reduces completely to lower sectors, is of the kind:
\begin{align}
        \begin{tikzpicture}[baseline={([yshift=-1ex]current bounding box.center)},scale=.8]
                \coordinate (inA) at (0.2,.6);
                \coordinate (outA) at (3.8,.6);
                \coordinate (inB) at (0.2,-.6);
                \coordinate (outB) at (3.8,-.6);
                \coordinate (xA) at (.8,.6);
                \coordinate (xyA) at (1.5,.6);
                \coordinate (yA) at (2,.6);
                \coordinate (yzA) at (2.5,.6);
                \coordinate (zA) at (3.2,.6);
                \coordinate (xB) at (.8,-.6);
                \coordinate (yB) at (2,-.6);
                \coordinate (yM) at (2,0);
                \coordinate (zB) at (3.2,-.6);
                \coordinate (xyAB) at (1.4,0);
                \draw [fill] (xA) circle (.08);
                \draw [fill] (yA) circle (.08);
                \draw [dotted] (inA) -- (outA);
                \draw [photon] (xA) -- (xyAB);
                \draw [photon] (yA) -- (xyAB);
                \draw [] (yB) circle (.2);
                \draw (yB) node[cross=4pt] {};
                \draw [fill] (xyAB) circle (.08);
                \draw [photon] (xyAB) -- (yB);
                \draw [draw=none] (xB) to[out=-40,in=-140] (zB);
              \end{tikzpicture}  = \int_{\ell_1,...,\ell_n} \dfrac{\dd(\ell_1 \cdot v_1 )\dd(\ell_2 \cdot v_1)}{(\ell_1-\ell_2)^2(\ell_1)^2 (\ell_2)^2 }N(\ell_2,...,\ell_n),
\end{align}
where $\begin{tikzpicture} \draw (0,0) node[cross=4pt] {};
        \draw [] (0,0) circle (.2);
\end{tikzpicture}$ denotes $N(\ell_2,...,\ell_n)$.
For these sectors we can always use IBP relations to remove the $\ell_2^2$ propagator. We show this for the single powers of the propagators but this can also be done for arbitrary powers:
\begin{align}
        &\int_{\ell_1,...,\ell_n} \dfrac{d }{d \ell_1^\mu}\ell_1^\mu\dfrac{\dd(\ell_1 \cdot v_1 )\dd(\ell_2 \cdot v_1)}{(\ell_1-\ell_2)^2(\ell_1)^2 (\ell_2)^2 }N(\ell_2,...,\ell_n)\nn \\
        =&(\eta_\mu^\mu-3)\int_{\ell_1,...,\ell_n} \dfrac{\dd(\ell_1 \cdot v_1 )\dd(\ell_2 \cdot v_1)}{(\ell_1-\ell_2)^2(\ell_1)^2 (\ell_2)^2 }N(\ell_2,...,\ell_n)\nn \\
        &\; \; +2 \int_{\ell_1,...,\ell_n} \dfrac{\dd(\ell_1 \cdot v_1 )\dd(\ell_2 \cdot v_1) \ell_1\cdot \ell_2 }{((\ell_1-\ell_2)^2)^2(\ell_1)^2 (\ell_2)^2 }N(\ell_2,...,\ell_n)\nn \\
        =&(d-4) \int_{\ell_1,...,\ell_n} \dfrac{\dd(\ell_1 \cdot v_1 )\dd(\ell_2 \cdot v_1)}{(\ell_1-\ell_2)^2(\ell_1)^2 (\ell_2)^2 }N(\ell_2,...,\ell_n)+\int_{\ell_1,...,\ell_n} \dfrac{\dd(\ell_1 \cdot v_1 )\dd(\ell_2 \cdot v_1)}{((\ell_1-\ell_2)^2)^2(\ell_1)^2  }N(\ell_2,...,\ell_n)=0,
\end{align}
where we dropped scaleless integrals.

Diagrammatically this relation translates to:
\begin{align}
        \begin{tikzpicture}[baseline={([yshift=-1ex]current bounding box.center)},scale=1]
                \coordinate (inA) at (0.2,.6);
                \coordinate (outA) at (3.8,.6);
                \coordinate (inB) at (0.2,-.6);
                \coordinate (outB) at (3.8,-.6);
                \coordinate (xA) at (.8,.6);
                \coordinate (xyA) at (1.5,.6);
                \coordinate (yA) at (2,.6);
                \coordinate (yzA) at (2.5,.6);
                \coordinate (zA) at (3.2,.6);
                \coordinate (xB) at (.8,-.6);
                \coordinate (yB) at (2,-.6);
                \coordinate (yM) at (2,0);
                \coordinate (zB) at (3.2,-.6);
                \coordinate (xyAB) at (1.4,0);
                \draw [fill] (xA) circle (.08);
                \draw [fill] (yA) circle (.08);
                \draw [dotted] (inA) -- (outA);
                \draw [photon] (xA) -- (xyAB);
                \draw [photon] (yA) -- (xyAB);
                \draw [] (yB) circle (.2);
                \draw (yB) node[cross=4pt] {};
                \draw [fill] (xyAB) circle (.08);
                \draw [photon] (xyAB) -- (yB);
                \draw [draw=none] (xB) to[out=-40,in=-140] (zB);
              \end{tikzpicture}  = \dfrac{1}{4-d} \begin{tikzpicture}[baseline={([yshift=-1ex]current bounding box.center)},scale=1]
                \coordinate (inA) at (0.2,.6);
                \coordinate (outA) at (3.8,.6);
                \coordinate (inB) at (0.2,-.6);
                \coordinate (outB) at (3.8,-.6);
                \coordinate (xA) at (.8,.6);
                \coordinate (xyA) at (1.5,.6);
                \coordinate (yA) at (2,.6);
                \coordinate (yzA) at (2.5,.6);
                \coordinate (zA) at (3.2,.6);
                \coordinate (xB) at (.8,-.6);
                \coordinate (yB) at (2,-.6);
                \coordinate (yM) at (2,0);
                \coordinate (zB) at (3.2,-.6);
                \coordinate (xyAB) at (1.4,0);
                \draw [fill] (xA) circle (.08);
                \draw [fill] (yA) circle (.08);
                \draw [dotted] (inA) -- (outA);
                \draw [photon] (xA) -- (xyAB);
                \draw [photon] (yA) -- (xyAB);
                 \draw [] (1.4,-0.15) circle (.2);
                \draw (1.4,-0.15) node[cross=4pt] {};
                \draw [draw=none] (xB) to[out=-40,in=-140] (zB);
                \draw [fill,blue!70] (1.10,.3) circle (.1);
              \end{tikzpicture} 
\end{align}
As we  can see we can always remove the $\ell_2^2$ propagator. This will reduce the number of irreducible topsectors crucially. Physically we can connect this type of reduction to a resumming of the Schwarzschild metric. In a self-force EFT~\cite{Cheung:2024byb,Kosmopoulos:2023bwc, Mougiakakos:2024nku,Mougiakakos:2024lif,Hoogeveen:2025tew} one can resum the Schwarzschild metric and one could get this type of vertex in a PM expansion. When working in a PM-EFT this is done by the IBP reduction. 
\section{Irreducible topsectors}
\subsection{0SF-order}
At leading self-force order all Feynman diagrams at $n$-PM order are characterised by a $m_1 m_2^n$ mass scaling. 
Every worldline vertex comes with a mass scaling of $m_i$.
Therefore, the 0SF integrand is given as a sum of diagrams that ``touch'' the first worldline only a single time. Multiple worldline vertices connected by worldline propagators count as a single ``touch'' due to their $\dfrac{1}{m_i}$ scaling:
\begin{align*}
        \begin{tikzpicture}[baseline={([yshift=-1ex]current bounding box.south)},scale=1.2]
                \coordinate (inA) at (0.4,.7);
                \coordinate (outA) at (3.6,.7);
                \coordinate (inB) at (0.4,-.7);
                \coordinate (outB) at (3.6,-.7);
                \coordinate (xA) at (1,.7);
                \coordinate (xxA) at (1.825,.7) ;
                \coordinate (xxM) at (1.525,.7) ;
                \coordinate (yyM) at (2.475,.7) ;
                \coordinate (xxB) at (1.525,-.7) ;
                \coordinate (yA) at (1.60+0.4,.7);
                \coordinate (yyA) at (2.375,0) ;
                \coordinate (yyB) at (2.475,-.7) ;
                \coordinate (zA) at (3,.7);
                \coordinate (zzA) at (3.75,.7) ;
                \coordinate (xB) at (1,-.7);
                \coordinate (yB) at (1.6+0.4,-.7);
                \coordinate (zB) at (3,-.7);
                \coordinate (zzB) at (3.75,-.7);
                \coordinate (xM) at (1,0);
                \coordinate (yM) at (1.6+0.4,0);
                \coordinate (zM) at (3,-0);
                \draw [dotted] (inA) -- (outA);
                \draw [dotted] (inB) -- (outB) node [midway, below]{} ;
                \draw [draw=none] (xA) to[out=40,in=140] (zA);
                \draw [photon] (xA) -- (xB);
                \draw [photon] (zA) -- (zB);
                \draw [zUndirected] (xA) -- (zA);
          \node[above] at (xA) {\tiny $\tiny{\sim m_1}$};
          \node[right] at (3.2,0) { $\tiny{\sim m_1}$};
          \node[above] at (zA) {\tiny $\tiny{\sim m_1}$};
          \node[below] at (2,.7) {\tiny $\tiny{\sim m_1^{-1}}$};
                \draw [fill] (xA) circle (.08);
                \draw [fill] (zA) circle (.08);
                \draw [fill] (xB) circle (.08);
                \draw [fill] (zB) circle (.08);
              \end{tikzpicture}.
\end{align*}
The structures in these diagrams can all be categorised into non-planarities coming from the worldline propagators and multiple graviton interactions in the bulk. 
The first can be removed by iterative application of partial fraction identities, so that all diagrams get planarised.
The graviton interactions are all of the ``triangle topsector'' structure discussed in \ref{sec:triangles} and will be automatically removed by integration by parts reduction.

The relevant master integrals will be ladder-type integrals and all graviton interactions are reduced to multiple graviton vertices on the worldline. At a given loop order there is only one relevant ladder topsector:
\begin{align*}
\begin{tikzpicture}[baseline={([yshift=-1ex]current bounding box.south)},scale=1.2]
        \coordinate (inA) at (0.4,.7);
        \coordinate (outA) at (3.6,.7);
        \coordinate (inB) at (0.4,-.7);
        \coordinate (outB) at (3.6,-.7);
        \coordinate (xA) at (1,.7);
        \coordinate (xxA) at (1.825,.7) ;
        \coordinate (xxM) at (1.525,.7) ;
        \coordinate (yyM) at (2.475,.7) ;
        \coordinate (xxB) at (1.525,-.7) ;
        \coordinate (yA) at (1.60+0.4,.7);
        \coordinate (yyA) at (2.375,0) ;
        \coordinate (yyB) at (2.475,-.7) ;
        \coordinate (zA) at (3,.7);
        \coordinate (zzA) at (3.75,.7) ;
        \coordinate (xB) at (1,-.7);
        \coordinate (yB) at (1.6+0.4,-.7);
        \coordinate (zB) at (3,-.7);
        \coordinate (zzB) at (3.75,-.7);
        \coordinate (xM) at (1,0);
        \coordinate (yM) at (1.6+0.4,0);
        \coordinate (zM) at (3,-0);
        \draw [dotted] (inA) -- (outA);
        \draw [dotted] (inB) -- (outB) node [midway, below]{} ;
        \draw [draw=none] (xA) to[out=40,in=140] (zA);
        \draw [photon] (xA) -- (xB);
        \draw [photon] (zA) -- (zB);
        \draw [zUndirected] (xA) -- (zA);
        \draw [photon] (xxM) -- (xxB);
        \draw [photon] (yyM) -- (yyB);
         \node[below=5pt] at (2, 0.7) { $.....$};
        \draw [fill] (xA) circle (.08);
        \draw [fill] (xxM) circle (.08);
        \draw [fill] (xxB) circle (.08);
        \draw [fill] (yyM) circle (.08);
        \draw [fill] (yyB) circle (.08);
        \draw [fill] (zA) circle (.08);
        \draw [fill] (xB) circle (.08);
        \draw [fill] (zB) circle (.08);
      \end{tikzpicture}       
\end{align*}
\subsection{1SF-order}
At subleading self-force order, things are more involved, but we can still write down all relevant topsectors. The relevant Feynman diagrams will have two points that touch the first worldline and their mass scaling will be $m_1^2 m_2^{n-1}$ at $n$-PM order. Technically, the integrand at 2PM also has 1SF terms in that sense, but we will neglect these since these terms are simply given by symmetry of 0SF terms. The first genuine 1SF contribution starts at 3PM.  
At this order, the most complicated topsector is the so-called H-graph, which contains two 3-graviton vertices. At 4PM, the most involved topsector will be the tennis-court graph containing three 3-graviton vertices. This pattern will continue.
At $n$PM, the most involved topsector is given as a graph containing $n-1$ 3-graviton vertices:
\begin{align*}
\begin{tikzpicture}[baseline={([yshift=-1ex]current bounding box.south)},scale=1.2]
        \coordinate (inA) at (0.4,.7);
        \coordinate (outA) at (3.6,.7);
        \coordinate (inB) at (0.4,-.7);
        \coordinate (outB) at (3.6,-.7);
        \coordinate (xA) at (1,.7);
        \coordinate (xxA) at (1.825,.7) ;
        \coordinate (xxM) at (1.525,0) ;
        \coordinate (yyM) at (2.475,0) ;
        \coordinate (xxB) at (1.525,-.7) ;
        \coordinate (yA) at (1.60+0.4,.7);
        \coordinate (yyA) at (2.375,0) ;
        \coordinate (yyB) at (2.475,-.7) ;
        \coordinate (zA) at (3,.7);
        \coordinate (zzA) at (3.75,.7) ;
        \coordinate (xB) at (1,-.7);
        \coordinate (yB) at (1.6+0.4,-.7);
        \coordinate (zB) at (3,-.7);
        \coordinate (zzB) at (3.75,-.7);
        \coordinate (xM) at (1,0);
        \coordinate (yM) at (1.6+0.4,0);
        \coordinate (zM) at (3,-0);
        \draw [dotted] (inA) -- (outA);
        \draw [dotted] (inB) -- (outB) node [midway, below]{} ;
        \draw [draw=none] (xA) to[out=40,in=140] (zA);
        \draw [photon] (xA) -- (xB);
        \draw [photon] (zA) -- (zB);
        \draw [photon,red] (xM) -- (zM);
        \draw [photon] (xxM) -- (xxB);
        \draw [photon] (yyM) -- (yyB);
         \node[below=5pt] at (2, 0) { $.....$};
        \draw [fill] (xA) circle (.08);
        \draw [fill] (xxM) circle (.08);
        \draw [fill] (xxB) circle (.08);
        \draw [fill] (yyM) circle (.08);
        \draw [fill] (yyB) circle (.08);
         \draw [fill] (xM) circle (.08);
        \draw [fill] (zA) circle (.08);
        \draw [fill] (xB) circle (.08);
         \draw [fill] (zM) circle (.08);
        \draw [fill] (zB) circle (.08);
      \end{tikzpicture}          
\end{align*}

In a WEFT~\cite{Kalin:2020mvi}, this topsector would correspond to new contributions to the effective action at this PM order. This topsector and its subsectors will also play the main role in the following chapters, since all special functions arising will be coming from subsectors of this topsector.

All other relevant topsectors can then iteratively be constructed by taking the topsectors of lower loop orders and connecting them via worldlines in all different possibilities without creating any triangles of type \ref{sec:triangles}. These identities from \ref{sec:triangles} and \ref{sec:mushroom} will then reduce all integrals from topsectors to a subsector of these.
\subsubsection{3PM}
We will look at 3PM as an example.
We can first draw the irreducible topsector containing no worldlines, which at 3PM is the H-graph:
\begin{align*}
  
\end{align*}

Now we can look at the topsectors of lower loop orders. To build a 3PM graph, we can connect a 2PM topsector with a 1PM topsector. There is just one topsector at each of these orders:
\begin{align*}
        %
\end{align*}
There is only one possible way to combine those that leads to the right mass scaling:
\begin{align*}
        %
\end{align*}
\subsubsection{4PM}

At 4PM, we have the tennis-court as a new topsector:
\begin{align*}
    %
\end{align*}

The other topsectors are given by connections of the 1PM topsector and the two 3PM topsectors:
\begin{align*}
        %
\end{align*}

\subsubsection{5PM}
At fifth Post-Minkowskian order, we have six irreducible topsectors:
\begin{align*}
        %
        \end{align*}

\subsection{2SF-order}
At second self-force order things get again more involved. We cannot remove all non-planarity coming from the worldlines using partial fraction identities anymore. Furthermore, we can get non-planarity from graviton propagators. We are not aware of an all order statement on how to find irreducible topsectors. We will therefore simply draw all irreducible topsectors at 5PM-2SF order.
There are 7 irreducible planar topsectors:
\begin{align*}
          %
        \end{align*}
        Furthermore there are 3 topsectors which are non-planar due to their worldline structure:
        \begin{align*}
          %
        \end{align*}
        Finally, there are three non-planar integrals due to their graviton structure. These will be connected to the non-linear memory effect~\cite{Christodoulou:1991cr,Wiseman:1991ss}:
        \begin{align*}
        %
        \end{align*}
\chapter{The method of differential equations at 3PM order}
\label{sec:methdiff}
Tensor and integration-by-parts reductions decrease the number of integrals which must be evaluated in order to compute observables. At 1SF order and beyond, however, the direct integration of the surviving master integrals is rarely feasible - already at two loops, and certainly at higher loop orders. We therefore require an efficient and systematic procedure for computing all of the master integrals which appear.

We begin with two general observations. By Lorentz invariance, our integrals can depend only on the two scalar invariants $q^2 = -|q|^2$ and $v_1\cdot v_2 = \gamma$, together with the dimensional regulator $\epsilon = 2-d/2$. Both $\gamma$ and $\epsilon$ are dimensionless, while $|q|$ alone carries a mass dimension. The dependence on $|q|$ is therefore trivial and is fixed by simple dimensional analysis. At two loops, for instance, dimensional analysis dictates 
\begin{align}
        {{\cal P}^{\text{3PM-1SF}}}_{\{n\}}^{\{\sigma\}}\propto |q|^{2 d - \sum_{i=1}^{4} n_i - \sum_{i=5}^{9} 2 n_i}.
\end{align}
We may therefore set $|q|=1$ throughout, restoring the $|q|$-dependence at the end of the calculation. Our integrals are then functions only of $\gamma$ and $\epsilon$.

The second important observation is that the physical observables, and the amplitudes which feed into them, are required only in the limit $\epsilon\to 0$. Dimensional regularisation is introduced because individual Feynman diagrams and individual master integrals can be divergent, while the final observable is finite once the limit is taken. We therefore need our integrals only in the form of a Laurent series in $\epsilon$, rather than as fully resolved functions of the regulator.

Our task thus reduces to determining, in a systematic manner, the dependence of the master integrals on the single parameter $\gamma$, given as a Laurent series in $\epsilon$. The contrast between 0SF and 1SF is here particularly sharp. As shown in~\refeq{eq:0sfgamma}, the $\gamma$-dependence of the 0SF integrals factorises completely, rendering this task trivial. At 1SF this is no longer the case, and the determination of the $\gamma$-dependence becomes the principal obstacle in the computation of the impulse.

\begin{figure}[h!]
        \centering
        \includegraphics[width=0.25\textwidth]{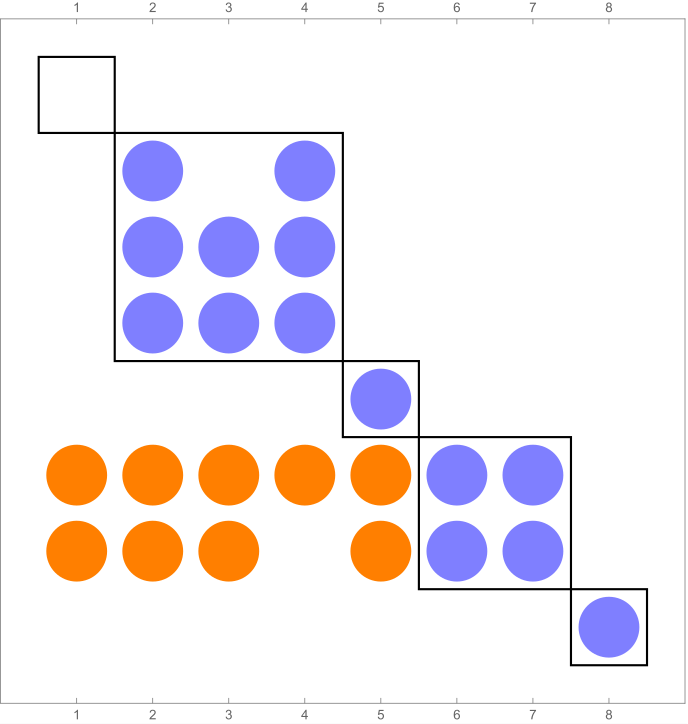}
        \caption{Block diagonal structure of the differential equation matrix at 3PM order shown in a heat plot. The diagonal block entries are marked in purple and the off-diagonal entries are marked in orange. Zero entries are left blank.}\label{fig:3pmB}
\end{figure}

The method we adopt is the so-called \emph{method of differential equations}~\cite{Kotikov:1990kg, Remiddi:1997ny, Gehrmann:1999as}. Rather than attempting a direct integration, we differentiate each master integral with respect to $\gamma$. Differentiating a master integral produces a sum of further integrals which obey two important properties, and which together render the method very efficient. First, the derivative of an integral always lies in the same integral family. Secondly, differentiation cannot introduce new denominators, so that all resulting integrals belong to the same topology or to a subtopology of the original master integral. Each derivative can therefore be IBP-reduced back onto master integrals of the same topology or its subtopologies. Collecting all relevant master integrals into a vector $\vec I$, we arrive at the matrix equation

\begin{align}
        \dfrac{d \vec{I}}{d \gamma}=M(\gamma,\epsilon) \vec{I}.
\end{align}
If the integrals in $\vec I$ are ordered from lower to higher topologies, the matrix $M(\gamma,\epsilon)$ assumes a block-triangular structure~\cite{Caron-Huot:2014lda}, as shown schematically in Fig.~\ref{fig:3pmB}. This structure follows directly from the fact that the derivative of an integral can only depend on integrals of the same or of a lower topology. The diagonal blocks constitute the homogeneous parts of the differential equations, with their sizes given by the number of master integrals in the corresponding sectors, while the off-diagonal blocks describe the coupling of each sector to its subsectors and form the inhomogeneous parts of the system.
This block structure already exhibits the first advantage of the method. Solving a system of $n$ coupled differential equations is generally infeasible once $n$ becomes large, but the block-triangular form ensures that the system can be solved bottom-up, sector by sector.

Even with as few as two master integrals, the homogeneous part of the differential equations can be already non-trivial to solve, and a direct approach to the full system would not make effective use of the Laurent expansion in~$\epsilon$.

Were our differential equations in a so-called \emph{$\epsilon$-form} or \emph{canonical form}~\cite{Henn:2013pwa}\footnote{We will use the terms $\epsilon$-form and canonical form mostly equivalently. The definition of what a canonical form is in the presence of more complicated functions may be up to a question of definition (see also~\cite{Duhr:2025lbz}). We will not discuss these subtleties in this thesis.},
\begin{align}
        \dfrac{d \vec{I}}{d \gamma}=\epsilon M(\gamma) \vec{I},
\end{align}
this task would be essentially trivial: each order in $\epsilon$ of $\vec I$ would depend only on lower orders, so that
\begin{align}
        \vec{I}^{(0)}&=\vec{g}_0,\\
        \vec{I}^{(n)}&=\int \d\gamma\, M(\gamma)\, \vec{I}^{(n-1)}+\vec{g}_n, \label{eq:iteratedsol}
\end{align}
where $\vec{I}=\sum_{i=0}^\infty \epsilon^{i-m} \vec{I}^{(i)}$ and $m$ denotes the leading order in $\epsilon$. Formally, the solution can be written as a path-ordered exponential,
\begin{align}
        \vec{I}=\mathcal{P}\exp\!\left[\int \d\gamma\, \epsilon\, M(\gamma)\right]\vec{j}, \label{eq:canSol}
\end{align}
with the boundary constants $\vec j$.
The structure~\refeq{eq:iteratedsol} naturally produces \emph{Chen iterated integrals}~\cite{ChenSymbol},
\begin{align}\label{eq:integralDef}
	{\cal I}(\varphi_1,\varphi_2,\ldots,\varphi_n;x)
	\coloneqq\!\!
	 \int_0^x\!\!
	 \d x'
	 \varphi_1(x'){\cal I}(\varphi_2,\ldots,\varphi_n;x'),
\end{align}
with the base case ${\cal I}(;x)=1$, where the kernels $\varphi_i$ are the linearly independent entries of the matrix $M(\gamma)$. 
A particularly important special case is that of \emph{multiple polylogarithms} (MPLs)~\cite{Kummer, Remiddi:1999ew, Goncharov:1995, Goncharov:1998kja, Vollinga:2004sn, Duhr:2011zq,Duhr:2019tlz}, for which the integration kernels take the simple form $\varphi_a(x)=1/(x-a)$. 
These functions also give rise to the notion of \emph{transcendental weight}~\cite{Arkani-Hamed:2010pyv,Kotikov:2010gf}.
The weight in the case of MPLs is defined as the number of iterated integrations. Furthermore, it is useful to assign $\epsilon$ a transcendental weight of $-1$. By this definition the solution in \refeq{eq:canSol} will give rise to integrals of uniform transcendental weight in an $\epsilon$ expansion, which means that all terms in the integral solution have the same weight.
\\
So in an ideal scenario, our differential equations would have a factorised $\epsilon$-dependence and would take the form
\begin{align}
        M(\gamma)=\sum_a \dfrac{m_a}{\gamma-a}, \label{eq:idealDE}
\end{align}
where the $m_a$ are constant matrices. A differential equation of this form would render the solution trivial, and would furthermore yield a result for the observable in which all cancellations among different functions are immediately manifest.

In general, of course, our differential equations will not take such a simple form, and the partial factorisation of $\sqrt{\gamma^2-1}$ in~\refeq{eq:1sfgamma} already makes it clear that square roots of this type must be expected in the DE system. Nevertheless, motivated by the manifest advantages of the form~\refeq{eq:idealDE}, we trade the problem of solving the differential equations for that of finding a basis of master integrals in which the differential equations take this form.

Two remarks are in order here. It is, a priori, neither clear nor proven that an $\epsilon$-factorised form can always be achieved. We are nevertheless aware of no example in which it has not been possible to do so. In what follows, we discuss several strategies for finding a basis which yields an $\epsilon$-factorised differential equation system at 3PM order. These strategies remain mostly applicable at higher loops, although they may become less practical there, and additional techniques may be required to handle the more involved integral structures.

\section{dlog-integrals}
Integrals which can be written as integrations of so-called \emph{$\d\!\log$-forms}~\cite{Arkani-Hamed:2014via} always lead to an $\epsilon$-factorised differential equation system, which itself can be solved in terms of iterated integrals over $\d\!\log$ one-forms~\cite{Henn:2014qga,Henn:2020lye, Wasser:2022kwg, Chen:2020uyk, Dlapa:2021qsl, Chen:2022lzr}. Given an integral
\begin{align}
        I=\int \Omega,
\end{align}
where $\Omega$ is an $n$-form, we shall call this integral a \emph{$\d\!\log$-integral} if $\Omega$ can be written as
\begin{align}
        \Omega= \d\!\log{\alpha_1} \wedge\cdots\wedge \d\!\log{\alpha_n},
\end{align}
or as a sum of such forms, where the $\alpha_i$ are algebraic functions.

It is, however, often difficult to ascertain directly in momentum space whether a given integral is of $\d\!\log$ form. It is therefore useful to re-parametrise the integral in order to facilitate the integrand analysis. The most common parametrisations are the Feynman and Baikov parametrisations. We employ the Baikov parametrisation throughout, although the underlying idea applies equally to other parametrisations.

\subsection{The standard and the loop-by-loop Baikov representation}
We do not derive the Baikov representation~\cite{Baikov:1996iu, Baikov:1996rk, Frellesvig:2017aai, Frellesvig:2024ymq} here, but recall its main idea, which is to switch from an integration over the loop momenta to one in which the propagators or ISPs themselves serve as the integration variables. Any $L$-loop Feynman integral $I$ can be written as
\begin{align}
        I=\dfrac{(-i)^L \pi^{(L-n)/2} \mathcal{E}^{(E-d+1)/2}}{\prod_{l=1}^{L} \Gamma((d+1-E-l)/2)}\int_\mathcal{C} d^n x \dfrac{x_{P+1}^{-\alpha_{P+1}}\cdots x_{n}^{-\alpha_{n}}}{x_1^{\alpha_1}\cdots x_P^{\alpha_P}} \mathcal{B}^{(d-E-L-1)/2} ,
\end{align}
where the $x_i$ are the propagators $D_i$ of our integral family raised to the power $\alpha_i$, $E$ is the number of independent external momenta (in our case $E=3$), and $n$ is the number of propagators in the integral family. Furthermore, $\mathcal{B}=\det(G(\ell_1,\ldots,\ell_{\text{L}},q,v_1,v_2))$ is the Baikov polynomial, while $\mathcal{E}=\det(G(q,v_1,v_2))$ is the external Baikov polynomial. $G$ is here a Gram matrix given by the scalar products of the set of momenta $\{ k \}$:
\begin{align}
        G(\{ k \})_{i,j}=k_i \cdot k_j.
\end{align}

The integration contour is given by the roots of the Baikov polynomial, but plays no role in our integrand analysis.

For multi-loop integrals it is often more convenient to work in the loop-by-loop Baikov representation, in which the same idea is applied to each loop separately. The loop-by-loop Baikov representation of an integral $I$ reads
\begin{align}
        I=\dfrac{(-i)^L \pi^{(L-\tilde{n})/2} }{\prod_{l=1}^{L} \Gamma((d-E_{\text{L}})/2)}\int_\mathcal{C} d^{\tilde{n}} x \dfrac{x_{P+1}^{-\alpha_{P+1}}\cdots x_{\tilde{n}}^{-\alpha_{\tilde{n}}}}{x_1^{\alpha_1}\cdots x_P^{\alpha_P}} \prod_{l=1}^L\!\Bigl(\mathcal{E}_{\text{l}}^{(E_{\text{l}}-d+1)/2} \mathcal{B}_{\text{l}}^{(d-E_{\text{l}}-2)/2}\Bigr) , \label{eq:LBLBaikov}
\end{align}
where $\mathcal{E}_{\text{l}}$ and $\mathcal{B}_{\text{l}}$ are the Baikov polynomials of an internal sub-loop, and $E_{\text{l}}$ is the number of independent external momenta of that sub-loop. The advantage of this parametrisation is that one does not need to integrate over all $n$ irreducible scalar products appearing in the integral family, but only a subset whose size $\tilde n$ depends on the integral itself and on the choice of loop momenta.

Both of these representations are also implemented in \texttt{Mathematica} packages like~\cite{Frellesvig:2024ymq}.
\subsection{Integrand analysis}
\label{sec:integrandan}
We will use the Baikov analysis now for our 3PM integrals. This discussion is completely independent of the choice of the $\i0$-prescription. We will use the notation introduced in~\refeq{eq:integralNotation} and use the loop-by-loop Baikov representation introduced in the previous subsection.
\subsection*{Sector: $1,1,0,0,0,1,1,1,1$}
In order to understand how one analyses whether an integral is of $\d\!\log$ form, we begin with a simple 3PM example, namely the integral
\begin{align}
        I_1=\int_{\ell_1,\ell_2} \dfrac{\dd(\ell_1 \cdot v_1) \dd(\ell_2 \cdot v_2)}{\ell_1^2 \ell_2^2 (\ell_1-q)^2(\ell_2-q)^2}\,.
\end{align}
The simplest analysis we can carry out in the Baikov parametrisation is one on the so-called \emph{maximal cut}, which in practice amounts to replacing all denominators by delta functions~\cite{Britto:2024mna}. By reverse unitarity, the integral still satisfies the same IBP relations, but all subsectors are set to zero. At the level of the differential equations, this means that one is examining only the diagonal block~\cite{Primo:2016ebd, Primo:2017ipr, Frellesvig:2017aai, Bosma:2017ens}. For the present example the maximal-cut analysis is, in fact, equivalent to the full one, since all subsectors of $I_1$ are scaleless and therefore trivially zero.

We employ the loop-by-loop Baikov parametrisation~\refeq{eq:LBLBaikov} choosing $\ell_2$ as the first sub-loop and $\ell_1$ as the second sub-loop, so that
\begin{align}
        &I_1 \propto  \int_\mathcal{C} d^{7} x \dfrac{1}{x_1\cdots x_6} \mathcal{B}_{1}^{-\epsilon } \mathcal{B}_{2}^{-\epsilon -\frac{1}{2}}
\end{align}    
where the Baikov polynomials are given by
\begin{align*}
        \mathcal{B}_{1}&=\det(G(\{ \ell_2, v_2,q \}))= 4 x_2^2-x_4^2-(x_6+1)^2+2 x_4 (x_6-1)\; ,\nn \\
        \mathcal{B}_{2}&=\det(G(\{ \ell_1,v_1, v_2,q \}))=  \gamma^2 \left(x_3^2-2 (x_5-1) x_3+(x_5+1)^2\right)\nn\\
        &-8 \gamma x_1 x_7+4 x_1^2-x_3^2-(x_5+1)^2+4 x_7^2+2 x_3 (x_5-1) ,
\end{align*}
and
 $x_{1/2}=\ell_{1/2}\cdot v_{1/2}$, $x_{3/4}=\ell_{1/2}^2$, $x_{5/6}=(\ell_{1/2}-q)^2$ and $x_7=\ell_{1}\cdot v_{2}$.

The two linear delta functions are treated as ordinary propagators, an operation which can always be performed and which is compensated by an appropriate change of integration contour, taking only the residues around $x_{1/2}=0$. Since we are working on the maximal cut, the integration contour plays no further role, and the maximal cut in Baikov representation is simply given by taking the residues around zero for all propagators. In our example, taking the residues for $x_1$ through $x_6$ around zero yields
\begin{align}
        I_1 \propto \int \d x_7  \dfrac{1}{\sqrt{\gamma^2+4 x_7^2-1}}\!\left(\dfrac{\gamma^2-1}{\gamma^2+4 x_7^2-1}\right)^{\!\epsilon}.
\end{align}
The integral now depends on a single Baikov variable, $x_7$. We set $\epsilon=0$ and continue our analysis in $d=4$. The reason why factors of the form $(f(\gamma,x_i))^\epsilon$ may always be neglected at this stage will become apparent shortly.

We are left with the simple integral
\begin{align}
        I_1|_{\epsilon \rightarrow 0} \propto \int \d x_7  \dfrac{1}{\sqrt{\gamma^2+4 x_7^2-1}} \propto \int \d\!\log{\alpha_1},
\end{align}
where $\alpha_1= (\sqrt{\gamma^2+4 x_7^2-1}+2 x_7)/(\sqrt{\gamma^2+4 x_7^2-1}-2 x_7)$. The integral is manifestly of $\d\!\log$ form at $\epsilon=0$.

Reinstating the previously dropped factor,
\begin{align}
        \left(\dfrac{\gamma^2-1}{\gamma^2+4 x_7^2-1}\right)^{\!\epsilon}=1+\epsilon\bigl(\log(\gamma^2-1)-\log(\gamma^2+4 x_7^2-1)\bigr)+\mathcal{O}(\epsilon^2),
\end{align}
we obtain
\begin{align}
        I_1 \propto \int \d\!\log{\alpha_1}\, \Bigl( 1+\epsilon\bigl(\log(\gamma^2-1)-\log(\gamma^2+4 x_7^2-1)\bigr)+\mathcal{O}(\epsilon^2) \Bigr).
\end{align}
The factor produces only logarithmic contributions to the integrand, so that at every order in $\epsilon$ one obtains higher-weight polylogarithmic functions. This is precisely the structure required for canonical differential equations.

In fact, computing the differential equation for $I_1$ explicitly yields a trivial $\epsilon$-factorised form,
\begin{align}
        \dfrac{d I_1}{d \gamma}= 0.
\end{align}
We see that $I_1$ is in fact a $\gamma$-independent constant. This is unsurprising once $I_1$ is expressed in the family~\refeq{eq:1sfgamma}: a non-trivial $\gamma$-dependence requires the propagator $(\ell_1-\ell_2)^2$ to be present.
\subsection*{Sector: $1,1,0,0,1,0,1,1,0$}

We turn to a second, less trivial example,
\begin{align}
        I_2=\int_{\ell_1,\ell_2} \dfrac{\dd(\ell_1 \cdot v_1) \dd(\ell_2 \cdot v_2)}{(\ell_1-\ell_2)^2 \ell_1^2 (\ell_2-q)^2}.
\end{align}
This integral belongs to a sector with three master integrals; in order to obtain an $\epsilon$-factorised differential equation on the maximal cut, we must therefore identify three $\d\!\log$-integrals in the sector.

We begin by analysing the corner integral $I_2$, again in the loop-by-loop Baikov parametrisation. We choose here $\ell_1$ as the first sub-loop and $\ell_2$ as the second one. 
\begin{align}
        &I_2 \propto  \int_\mathcal{C} d^{7} x \dfrac{1}{x_1\cdots x_5}\mathcal{E}_{1}^{-\frac{1}{2}+\epsilon } \mathcal{B}_{1}^{-\epsilon } \mathcal{B}_{2}^{-\epsilon -\frac{1}{2}}
\end{align}    
where the Baikov polynomials are given by
\begin{align*}
        \mathcal{B}_{1}&=\det(G(\{\ell_1, \ell_2, v_1 \}))\; ,\nn \\
        \mathcal{E}_{1}&=\det(G(\{ \ell_2, v_1 \}))\; ,\nn \\
        \mathcal{B}_{2}&=\det(G(\{ \ell_2,v_1, v_2,q \}))
\end{align*}
and
 $x_{1/2}=\ell_{1/2}\cdot v_{1/2}$, $x_{3}=(\ell_{1}-\ell_2)^2$, $x_{4}=\ell_{1}^2$, $x_5=(\ell_{2}-q)^2$, $x_6=\ell_2\cdot v_1$ and $x_7=\ell_2^2$.
As before, we take residues at zero for all propagators $x_1,...,x_5$ and we set $\epsilon=0$, so that
\begin{align}
        I_2|_{\epsilon \rightarrow 0} \propto \int \d x_6\, d x_7\, \frac{1}{\sqrt{x_7-x_6^2}\, \sqrt{\gamma^2 (x_7+1)^2+4 x_6^2-(x_7+1)^2}}.
\end{align}
The integrand contains two square roots, both quadratic in $x_6$ and linear or quadratic in $x_7$. Rescaling $u_1=x_6/(1+x_7)$ removes $x_7$ from one of the square roots and allows the $x_7$-integration to be carried out,
\begin{align}
        I_2|_{\epsilon \rightarrow 0} \propto \int \d u_1\,  d x_7\, \frac{1}{ \sqrt{\gamma^2+4 u_1^2-1}\, \sqrt{x_7-u_1^2 (x_7+1)^2}}\propto \int \d u_1\, \d\!\log{\alpha_1}\, \dfrac{1}{u_1 \sqrt{\gamma^2+4 u_1^2-1}},
\end{align}
where $\alpha_1=2 \sqrt{x_7-u_1^2 (x_7+1)^2}-i (2 u_1^2 (x_7+1)-1)/u_1$. The remaining integral over $u_1$ has a residue at $u_1=0$ and is also straightforward to evaluate,
\begin{align}
        I_2|_{\epsilon \rightarrow 0} \propto \int  \d\!\log{\alpha_1} \wedge \d\!\log{\alpha_2}\,\dfrac{1}{\sqrt{\gamma^2-1}},
\end{align}
where $\alpha_2=(\sqrt{1-\gamma^2}-i \sqrt{\gamma^2+4 u_1^2-1})/(\sqrt{1-\gamma^2}+i \sqrt{\gamma^2+4 u_1^2-1})$.

We see that $I_2$ is proportional to an integration of $\d\!\log{\alpha_1} \wedge \d\!\log{\alpha_2}/\sqrt{\gamma^2-1}$, which is not in pure $\d\!\log$-form. It is, however, immediate that normalising $I_2$ by a factor of $\sqrt{\gamma^2-1}$ produces a $\d\!\log$-integral. The factor $\sqrt{\gamma^2-1}$ is referred to as the \emph{leading singularity} of the integral. The first canonical integral of this sector is therefore $\sqrt{\gamma^2-1}\, I_2$.

We must now identify two further $\d\!\log$-integrals. To produce a second candidate, we either insert a scalar product or raise the power of one of the denominators. It is easily seen that a numerator insertion of $x_6$ would not yield a logarithmic singularity in the integrand. We therefore put a dot on a graviton propagator for our second candidate
\begin{align}
        I_3=\int_{\ell_1,\ell_2} \dfrac{\dd(\ell_1 \cdot v_1) \dd(\ell_2 \cdot v_2)}{((\ell_1-\ell_2)^2)^2\, \ell_1^2\, (\ell_2-q)^2}.
\end{align}
As before, we parametrise this integral as
\begin{align}
       I_3 \propto \int \d^{7} x \dfrac{1}{x_1 x_2 x_3^2 x_4 x_5} \prod_{l=1}^L\!\Bigl(\mathcal{E}_{\text{L}}^{(E_{\text{L}}-d+1)/2} \mathcal{B}_{\text{L}}^{(d-E_{\text{L}}-2)/2}\Bigr).
\end{align}
Residues in $x_1, x_2, x_4, x_5$ at zero are taken straightforwardly, but $x_3$ now carries a double pole. To reduce this to a simple pole we perform an integration by parts; in dimensional regularisation all boundary terms vanish, leaving
\begin{align}
        I_3 \propto \int \d^{7} x \dfrac{1}{x_1 x_2 x_3 x_4 x_5}\, \dfrac{\partial}{\partial x_3}\prod_{l=1}^L\!\Bigl(\mathcal{E}_{\text{L}}^{(E_{\text{L}}-d+1)/2} \mathcal{B}_{\text{L}}^{(d-E_{\text{L}}-2)/2}\Bigr).
\end{align}
The integrand now carries only simple poles and we obtain
 \begin{align}
        I_3|_{\epsilon \rightarrow 0} \propto \int \d x_6\, \d x_7\,  \frac{\epsilon}{2 \pi^3 x_7\, \sqrt{x_7-x_6^2}\, \sqrt{\gamma^2 (x_7+1)^2+4 x_6^2-(x_7+1)^2}}.
 \end{align}
The integrand is closely analogous to that of $I_2$, but the integration by parts has produced an explicit factor of $\epsilon$. The double pole in $x_3$ caused a drop of one in transcendental weight, which must therefore be compensated. Furthermore, the integrand acquires an additional residue at $x_7=0$. Performing the same change of variable as before, both integrations can be carried out and we find
\begin{align}
        I_3|_{\epsilon \rightarrow 0} \propto \int  \d\!\log{\alpha_3} \wedge \d\!\log{\alpha_2}\,\dfrac{\epsilon}{\sqrt{\gamma^2-1}},
\end{align}
where $\alpha_3=2 \sqrt{-x_7^2 u_1^2-2 x_7 u_1^2-u_1^2+x_7}/x_7-i (2 x_7 u_1^2+2 u_1^2-x_7)/(u_1 x_7)$. The leading singularity of $I_3$ is therefore the same as that of $I_2$, and our second canonical integral is $\epsilon^{-1}\sqrt{\gamma^2-1}\, I_3$.

The sector contains three master integrals, so we still need to identify a third canonical one. Unfortunately, finding a third linearly independent master integral in $\d\!\log$-form is rather difficult here, and we therefore adopt a different strategy. Rather than analysing the integrand directly, we use the differential equations themselves to construct a canonical basis.

\subsection{Using differential equations to find canonical integrals}
The structure of the differential equation matrix can itself be used to find canonical integrals~\cite{Lee:2014ioa, Gehrmann:2014bfa, Meyer:2017joq, Dlapa:2020cwj, Lee:2020zfb}. We start from a basis $\vec I$ which is not canonical, and seek a transformation $T$ so that $\vec J = T \vec I$ is canonical.

This is a very common strategy, especially for the off-diagonal blocks describing the corrections from subsector couplings, which we will discuss further below. For the diagonal blocks the same idea also applies, and dedicated algorithms such as \texttt{CANONICA}~\cite{Meyer:2017joq}, \texttt{Libra}~\cite{Lee:2020zfb} or \texttt{INITIAL}~\cite{Dlapa:2020cwj,Dlapa:2022wdu} are designed to find the corresponding transformation. We will not enter into the details of these algorithms here, but remark that they do not always succeed---particularly when the number of masters in a sector is high, or when the starting basis $\vec I$ is ``bad''.

We define a ``bad'' basis as one which requires a complicated transformation $T(\gamma,\epsilon)$. Even when no third $\d\!\log$-integral can be found, it is therefore worthwhile to look for a ``good'' third master integral, which only requires a minimal transformation $T(\gamma,\epsilon)$. When one or more canonical integrals are already at hand, a particularly practical choice is to use the derivative of one of these as an additional master integral. Canonical integrals are formally given by
\begin{align}
        \vec{J}=\mathcal{P}\exp\!\left[\int \d\gamma\, \epsilon\, M(\gamma)\right]\vec{j},
\end{align}
so that their derivatives read
\begin{align}
        \dfrac{d \vec{J}}{d \gamma}=\epsilon\, M(\gamma)\, \mathcal{P}\exp\!\left[\int \d\gamma\, \epsilon\, M(\gamma)\right]\vec{j}.
\end{align}
Even without explicit knowledge of $M(\gamma)$, we see that only a rescaling by $\epsilon$ together with a $\gamma$-dependent transformation is required to convert the derivative of a canonical integral into a canonical one.

In the case at hand we therefore choose $I_4=\epsilon^{-1}\, d(\sqrt{\gamma^2-1}\, I_2)/d\gamma$. The resulting differential equations take the form
\begin{align}
        \dfrac{d \vec{I}}{d \gamma}= \left(
                \begin{array}{ccc}
                 0 & 0 & 0  \\
                0 & 0 & 0 \\
                 -\frac{6 }{\left(\gamma ^2-1\right)^2} & 0 & - \frac{\gamma  }{\gamma ^2-1} \\
                \end{array}
                \right)\vec{I} +\epsilon \left(
                \begin{array}{ccc}
                 0 & 0 & 1  \\
                 -\frac{12 \gamma   }{\gamma ^2-1} & -\frac{2 \gamma   }{\gamma ^2-1} & 2  \\
                 -\frac{6 \left(2 \left(\gamma ^2-1\right)  \right)}{\left(\gamma ^2-1\right)^2} & -\frac{2  }{\gamma ^2-1} & \frac{6 \gamma }{\gamma ^2-1} \\
                \end{array}
                \right) \vec{I},
\end{align}
where $\vec{I}=(\sqrt{\gamma^2-1}\, I_2,\, \epsilon^{-1} \sqrt{\gamma^2-1}\, I_3,\, I_4)$. The first two rows are already proportional to $\epsilon$, and only the $\epsilon^0$ piece in the third row remains to be removed.

The first transformation we perform is
\begin{align}
        T_1=\left(
                \begin{array}{ccc}
                 1 & 0 & 0 \\
                 0 & 1 & 0 \\
                 0 & 0 & \sqrt{\gamma ^2-1} \\
                \end{array}
                \right),
\end{align}
which renders all diagonal entries proportional to $\epsilon$:
\begin{align}
        \dfrac{d (T_1 \vec{I})}{d \gamma}=\left(
                \begin{array}{ccc}
                 0 & 0 & \frac{\epsilon }{\sqrt{\gamma ^2-1}} \\
                 -\frac{12 \gamma  \epsilon }{\gamma ^2-1} & -\frac{2 \gamma  \epsilon }{\gamma ^2-1} & \frac{2 \epsilon }{\sqrt{\gamma ^2-1}} \\
                 -\frac{6 \left(2 \left(\gamma ^2-1\right) \epsilon +1\right)}{\left(\gamma ^2-1\right)^{3/2}} & -\frac{2 \epsilon }{\sqrt{\gamma ^2-1}} & \frac{6 \gamma  \epsilon }{\gamma ^2-1} \\
                \end{array}
                \right)T_1 \vec{I}.
\end{align}
Only one entry remains which is not in $\epsilon$-form. We correct the third integral by
\begin{align}
        T_2=\left(
                \begin{array}{ccc}
                 1 & 0 & 0 \\
                 0 & 1 & 0 \\
                 -\frac{6 \gamma }{\sqrt{\gamma ^2-1}} & 0 & 1 \\
                \end{array}
                \right),
\end{align}
after which the canonical basis $\vec J = T_2 T_1 \vec I$ admits the differential equations
\begin{align}
        \dfrac{d \vec{J}}{d \gamma}= \left(
\begin{array}{ccc}
 \frac{6 \gamma  \epsilon }{\gamma ^2-1} & 0 & \frac{\epsilon }{\sqrt{\gamma ^2-1}} \\
 0 & -\frac{2 \gamma  \epsilon }{\gamma ^2-1} & \frac{2 \epsilon }{\sqrt{\gamma ^2-1}} \\
 -\frac{12 \epsilon }{\sqrt{\gamma ^2-1}} & -\frac{2 \epsilon }{\sqrt{\gamma ^2-1}} & 0 \\
\end{array}
\right) \vec{J},
\end{align}
which are now in $\epsilon$-form.

Let us summarise the steps that have been taken. After choosing the derivative of the first integral as a third basis element, we exploited the fact that there exists an $\epsilon$-independent transformation to a canonical basis. We first rescaled the third integral by $\sqrt{\gamma^2-1}$ to bring its diagonal entry into $\epsilon$-form, and then used a shift of the third integral by the canonical ones to bring the remaining off-diagonal entries into $\epsilon$-form. Both transformations are easily found by an ansatz, demanding that the $\epsilon^0$ pieces vanish. Solving the resulting first-order differential equation then yields the required transformation. We see that, even though our starting basis was not canonical, the necessary transformations could be found by hand without recourse to dedicated algorithms.

As anticipated, the $\epsilon$-factorisation also occurs in this non-trivial example with three master integrals. One may verify that integrating any of the differential forms appearing in the matrix produces only logarithmic functions, although the matrix is not yet in the form~\refeq{eq:idealDE}. The square root $\sqrt{\gamma^2-1}$ which normalised our differential equations re-enters the differential-equation matrix, and in order to bring the system to the form~\refeq{eq:idealDE} we must change variables in such a way as to rationalise this square root. Doing so simplifies both the resulting function space and the manipulation of the differential equations. We therefore introduce the variable 
\begin{align}
        x=\gamma-\sqrt{\gamma^2-1}\;.
\end{align}
 In this new variable the differential equations read
\begin{align}
        \dfrac{d \vec{J}}{d x}=\epsilon \left(
                \begin{array}{ccc}
                 \frac{6  }{x-1}-\frac{6  }{x}+\frac{6  }{x+1} & 0 & -\frac{1 }{x} \\
                 0 & -\frac{2  }{x-1}+\frac{2  }{x}-\frac{2  }{x+1} & -\frac{2  }{x} \\
                 \frac{12  }{x} & \frac{2  }{x} & 0 \\
                \end{array}
                \right)\vec{J}.
\end{align}
It is a matter of taste whether one wants to work with the variable $x$ or $\gamma$. We will continue here using $\gamma$ for now. Once we want to solve the differential equations it is better to use $x$ to get MPLs as solutions of the differential equations.

We follow the same strategy in every sector. In all sectors a full set of $\d\!\log$-integrals can be identified, and we may write down a basis which is expected to admit an $\epsilon$-factorised matrix on the diagonal blocks. Our integrand analysis was so far valid only on the maximal cut, however, and therefore neglects the coupling to lower sectors, i.e.\ the off-diagonal entries. We choose the following basis for all eight master integrals,
\begin{align}
       I_1&=\text{P}^{3,1}(1,1,0,0,0,1,1,1,1)\, , \nn\\
       I_2&=\sqrt{\gamma^2-1}\text{P}^{3,1}(1,1,0,0,1,0,1,1,0)\, , \nn\\
       I_3&=\sqrt{\gamma^2-1}/\epsilon\, \text{P}^{3,1}(1,1,0,0,2,0,1,1,0)\, , \nn\\
       I_4&=-6 \gamma\, \text{P}^{3,1}(1,1,0,0,1,0,1,1,0)+\sqrt{\gamma^2-1} \dfrac{d}{d \gamma} \!\left( \sqrt{\gamma^2-1}/\epsilon \, \text{P}^{3,1}(1,1,0,0,1,0,1,1,0) \right) \, , \nn\\
       I_5&=\sqrt{\gamma^2-1}/\epsilon\, \text{P}^{3,1}(1,1,0,0,2,0,1,0,1)\, , \nn\\
       I_6&=\sqrt{\gamma^2-1}\, \text{P}^{3,1}(1,1,0,0,1,1,1,1,1)\, , \nn\\
       I_7&=\text{P}^{3,1}(1,1,-1,-1,1,1,1,1,1)\, , \nn\\
       I_8&=(\gamma^2-1)\text{P}^{3,1}(1,1,1,1,1,1,0,0,1),
\end{align}
which obey the differential equations
\begin{align}
        \dfrac{d \vec{I}}{d \gamma}=\left(
                \begin{array}{cccccccc}
                 0 & 0 & 0 & 0 & 0 & 0 & 0 & 0 \\
                 0 & \frac{6 \gamma  \epsilon }{\gamma ^2-1} & 0 & \frac{\epsilon }{\sqrt{\gamma ^2-1}} & 0 & 0 & 0 & 0 \\
                 0 & 0 & -\frac{2 \gamma  \epsilon }{\gamma ^2-1} & \frac{2 \epsilon }{\sqrt{\gamma ^2-1}} & 0 & 0 & 0 & 0 \\
                 0 & -\frac{12 \epsilon }{\sqrt{\gamma ^2-1}} & -\frac{2 \epsilon }{\sqrt{\gamma ^2-1}} & 0 & 0 & 0 & 0 & 0 \\
                 0 & 0 & 0 & 0 & -\frac{2 \gamma  \epsilon }{\gamma ^2-1} & 0 & 0 & 0 \\
                 \frac{4 \gamma  \epsilon }{\sqrt{\gamma ^2-1}} & 0 & -\frac{4 \gamma  \epsilon }{\gamma ^2-1} & -\frac{2 \epsilon }{\sqrt{\gamma ^2-1}} & \frac{4 \gamma  \epsilon }{\gamma ^2-1} & -\frac{2 \gamma  \epsilon }{\gamma ^2-1} & \frac{8 \epsilon }{\sqrt{\gamma ^2-1}} & 0 \\
                 \epsilon -\frac{1}{2} & -\frac{3 \epsilon }{\sqrt{\gamma ^2-1}} & -\frac{2 \epsilon }{\sqrt{\gamma ^2-1}} & 0 & \frac{2 \epsilon }{\sqrt{\gamma ^2-1}} & -\frac{\epsilon }{2 \sqrt{\gamma ^2-1}} & \frac{2 \gamma  \epsilon }{\gamma ^2-1} & 0 \\
                 0 & 0 & 0 & 0 & 0 & 0 & 0 & 0 \\
                \end{array}
                \right) \vec{I}. \label{equ:deAllMis}
\end{align}

Surprisingly, we find that almost the entire matrix is already in $\epsilon$-form. In general this does not need to be the case, although as a useful rule of thumb, integrals which are canonical on the maximal cut typically remain a ``good'' basis even for the couplings to the lower sectors. In our case, only one off-diagonal entry fails to be in $\epsilon$-form. Since this is a recurring problem, often amenable to a simple algorithmic treatment, we discuss it in more generality in the next subsection.

\subsection{The off-diagonal blocks}
\label{sec:off}

The off-diagonal blocks of the matrix constitute the inhomogeneous parts of the differential equations. In most cases they are easier to handle, in the sense that there is a clear strategy as long as the integrals are polylogarithmic. We illustrate the algorithm \cite{Lee:2014ioa,Meyer:2017joq,Lee:2020zfb} on a simple but representative example. We work here with the variable $x$ and assume that this rationalises all appearing roots. One can also generalise this approach to square roots. Implementations of this algorithm often demand a rationalisation of all square roots.\\
Suppose our system contains two diagonal blocks, both of which are already in canonical form, with singularities only at $x=a$ and $x=\infty$,
\begin{align}
        M=\begin{pmatrix}
                \dfrac{ \epsilon A}{x-a} & 0 \\
                C(x,\epsilon) & \dfrac{ \epsilon B}{x-a}
                \end{pmatrix},
\end{align}
where $A$ and $B$ are constant matrices and the matrix $C$ depends on $x$ as
\begin{align}
        C(x,\epsilon) = \sum_{i=1}^{n} \dfrac{C_i(\epsilon)}{(x-a)^i}+  \sum_{i=0}^{m} C_{-i}(\epsilon)\, x^i.
\end{align}
We know from~\refeq{eq:idealDE} that the differential equations should be proportional to $\epsilon$ and contain only $\d\!\log$-forms; in other words, every term other than the simple pole $\frac{1}{x-a}$ must be removed, i.e. all higher poles and all polynomial terms $x^i$ with $i \ge 0$.

Although the diagonal blocks here have poles only at $x=a$, the strategy applies equally well in the presence of further poles. A transformation of the form
\begin{align}
        T=\begin{pmatrix}
                \mathbf{1} & 0 \\
                D(x,\epsilon) & \mathbf{1}
                \end{pmatrix}
\end{align}
maps our matrix into
\begin{align}
        \tilde{M}=\begin{pmatrix}
                \dfrac{ \epsilon A}{x-a} & 0 \\
                D'(x,\epsilon)+C(x,\epsilon)+ \left(\dfrac{ \epsilon D(x,\epsilon) A}{x-a}-\dfrac{ \epsilon B D(x,\epsilon)}{x-a}\right) & \dfrac{ \epsilon B}{x-a}
                \end{pmatrix}.
\end{align}
Choosing the ansatz
\begin{align}
        D(x,\epsilon)=\dfrac{d(\epsilon)}{(x-a)^{n-1}},
\end{align}
with a matrix of rational functions $d(\epsilon)$,
the highest pole in $x$ can be removed by solving
\begin{align}
        -(n-1)\, d(\epsilon)+C_n(\epsilon)+\epsilon( d(\epsilon) A-  B d(\epsilon))\,=0.
\end{align}
This ansatz can be used iteratively to remove all higher poles at $x=a$. If we have higher poles, we can iteratively remove them one by one, and similarly all terms $x^i$ with $i\geq 0$ can be removed by the ansatz
\begin{align}
        D(x,\epsilon)=d(\epsilon)\, x^{i}.
\end{align}
The differential equations are thereby brought to a form containing only simple poles, so that every integration produces only logarithmic functions. 

In a larger matrix containing several sectors there are correspondingly more off-diagonal blocks,
\begin{align}
        M=\begin{pmatrix}
                \dfrac{ \epsilon D_1}{x-a} & 0 & 0 \\
                C_1(x,\epsilon) & \dfrac{ \epsilon D_2}{x-a} & 0\\
                C_3(x,\epsilon) & C_2(x,\epsilon) & \dfrac{ \epsilon D_3}{x-a}
                \end{pmatrix},
\end{align}
and one proceeds sector by sector, from top to bottom (from lower to higher sectors) and from right to left, transforming first $C_1$ and $C_2$ and last $C_3$.

To complete the canonical form, only an $x$-independent transformation $T_\epsilon(\epsilon)$ remains, with which to factorise $\epsilon$. Following ref.~\cite{Lee:2014ioa} and assuming a differential equation matrix of the form
\begin{align}
        M(x,\epsilon)=\sum_k \dfrac{m_k(\epsilon)}{x-x_k},
\end{align}
the transformation $T_\epsilon(\epsilon)$ can be found by solving
\begin{align}
        \dfrac{m_k(\epsilon)}{\epsilon}\, T(\epsilon,\mu)=T(\epsilon,\mu)\, \dfrac{m_k(\mu)}{\mu} \label{eq:epsT}
\end{align}
for all $k$. We can choose in principle any rational number for $\mu$ as long as $T_\epsilon(\epsilon)=T(\epsilon,\mu)$ is invertible. This algorithm is implemented for example in \texttt{CANONICA}~\cite{Meyer:2017joq} and \texttt{Libra}~\cite{Lee:2020zfb}.

The procedure described here is fully general, and exploits prior knowledge of which differential forms are expected to appear in the canonical equations. In more difficult cases such prior knowledge will, however, no longer be available. Returning to the system~\refeq{equ:deAllMis}, applying this general procedure by hand would be considerably more involved than necessary. A closer inspection of~\refeq{equ:deAllMis} reveals, however, that only a single entry is not proportional to $\epsilon$. Rather than removing all non-$\d\!\log$ kernels, we may therefore aim directly at removing entries order by order in $\epsilon$. Concretely,
\begin{align}
        M(\gamma,\epsilon)=\left(
                \begin{array}{cccccccc}
                 0 & 0 & 0 & 0 & 0 & 0 & 0 & 0 \\
                 0 & 0 & 0 & 0 & 0 & 0 & 0 & 0 \\
                 0 & 0 & 0 & 0 & 0 & 0 & 0 & 0 \\
                 0 & 0 & 0 & 0 & 0 & 0 & 0 & 0 \\
                 0 & 0 & 0 & 0 & 0 & 0 & 0 & 0 \\
                 0 & 0 & 0 & 0 & 0 & 0 & 0 & 0 \\
                 -\frac{1}{2} & 0 & 0 & 0 & 0 & 0 & 0 & 0 \\
                 0 & 0 & 0 & 0 & 0 & 0 & 0 & 0 \\
                \end{array}
                \right)+\mathcal{O}(\epsilon),
\end{align}
so that with the ansatz
\begin{align}
        T(\gamma)=\left(
                \begin{array}{cccccccc}
                 0 & 0 & 0 & 0 & 0 & 0 & 0 & 0 \\
                 0 & 0 & 0 & 0 & 0 & 0 & 0 & 0 \\
                 0 & 0 & 0 & 0 & 0 & 0 & 0 & 0 \\
                 0 & 0 & 0 & 0 & 0 & 0 & 0 & 0 \\
                 0 & 0 & 0 & 0 & 0 & 0 & 0 & 0 \\
                 0 & 0 & 0 & 0 & 0 & 0 & 0 & 0 \\
                f(\gamma) & 0 & 0 & 0 & 0 & 0 & 0 & 0 \\
                 0 & 0 & 0 & 0 & 0 & 0 & 0 & 0 \\
                \end{array}
                \right)
\end{align}
the choice $f(\gamma)=\gamma/2$ brings the differential equations into the form
\begin{align}
        \dfrac{d T(\gamma) \vec{I}}{d \gamma}=\epsilon \left(
                \begin{array}{cccccccc}
                 0 & 0 & 0 & 0 & 0 & 0 & 0 & 0 \\
                 0 & \frac{6 \gamma }{\gamma ^2-1} & 0 & \frac{1}{\sqrt{\gamma ^2-1}} & 0 & 0 & 0 & 0 \\
                 0 & 0 & -\frac{2 \gamma }{\gamma ^2-1} & \frac{2}{\sqrt{\gamma ^2-1}} & 0 & 0 & 0 & 0 \\
                 0 & -\frac{12}{\sqrt{\gamma ^2-1}} & -\frac{2}{\sqrt{\gamma ^2-1}} & 0 & 0 & 0 & 0 & 0 \\
                 0 & 0 & 0 & 0 & -\frac{2 \gamma }{\gamma ^2-1} & 0 & 0 & 0 \\
                 0 & 0 & -\frac{4 \gamma }{\gamma ^2-1} & -\frac{2}{\sqrt{\gamma ^2-1}} & \frac{4 \gamma }{\gamma ^2-1} & -\frac{2 \gamma }{\gamma ^2-1} & \frac{8}{\sqrt{\gamma ^2-1}} & 0 \\
                 \frac{1}{1-\gamma ^2} & -\frac{3}{\sqrt{\gamma ^2-1}} & -\frac{2}{\sqrt{\gamma ^2-1}} & 0 & \frac{2}{\sqrt{\gamma ^2-1}} & -\frac{1}{2 \sqrt{\gamma ^2-1}} & \frac{2 \gamma }{\gamma ^2-1} & 0 \\
                 0 & 0 & 0 & 0 & 0 & 0 & 0 & 0 \\
                \end{array}
                \right)T(\gamma) \vec{I}, \label{equ:epsDE}
\end{align}
which is both in $\d\!\log$-form and proportional to $\epsilon$. We emphasise that this simple ansatz worked only because of our ``good'' initial choice of basis; with a different basis the $\epsilon$-structure of the differential-equation matrix may be more involved, and the same simple strategy may no longer be available. The advantage of this approach is that no prior knowledge of the differential forms expected to appear in the matrix is required.

The full canonical basis $\vec{\mathcal J}$ is then
\begin{align}
        \mathcal{J}_1&=\text{P}^{3,1}(1,1,0,0,0,1,1,1,1)\, , \nn\\
        \mathcal{J}_2&=\sqrt{\gamma^2-1}\text{P}^{3,1}(1,1,0,0,1,0,1,1,0)\, , \nn\\
        \mathcal{J}_3&=\sqrt{\gamma^2-1}/\epsilon\, \text{P}^{3,1}(1,1,0,0,2,0,1,1,0)\, , \nn\\
        \mathcal{J}_4&=-6 \gamma\, \text{P}^{3,1}(1,1,0,0,1,0,1,1,0)+\sqrt{\gamma^2-1} \dfrac{d}{d \gamma} \!\left( \sqrt{\gamma^2-1}/\epsilon \, \text{P}^{3,1}(1,1,0,0,1,0,1,1,0) \right) \, , \nn\\
        \mathcal{J}_5&=\sqrt{\gamma^2-1}/\epsilon\, \text{P}^{3,1}(1,1,0,0,2,0,1,0,1)\, , \nn\\
        \mathcal{J}_6&=\sqrt{\gamma^2-1}\, \text{P}^{3,1}(1,1,0,0,1,1,1,1,1)\, , \nn\\
        \mathcal{J}_7&=\text{P}^{3,1}(1,1,-1,-1,1,1,1,1,1)+\tfrac{1}{2} \gamma\, \text{P}^{3,1}(1,1,0,0,0,1,1,1,1)\, , \nn\\
        \mathcal{J}_8&=(\gamma^2-1)\text{P}^{3,1}(1,1,1,1,1,1,0,0,1).
\end{align}

It is also useful to introduce a second, ``pre-canonical'' basis, which requires only a minimal transformation to the canonical one but contains no square roots. We use this basis for the IBP reduction, where square roots would otherwise complicate matters,
\begin{align}
        \mathcal{I}_1&=\text{P}^{3,1}(1,1,0,0,0,1,1,1,1)\, , \nn\\
        \mathcal{I}_2&=\text{P}^{3,1}(1,1,0,0,1,0,1,1,0)\, , \nn\\
        \mathcal{I}_3&=\text{P}^{3,1}(1,1,0,0,2,0,1,1,0)\, , \nn\\
        \mathcal{I}_4&=\sqrt{\gamma^2-1}\, \dfrac{d}{d \gamma}\! \left( \sqrt{\gamma^2-1}\, \text{P}^{3,1}(1,1,0,0,1,0,1,1,0) \right)=  \text{P}^{3,1}(2,1,-1,0,1,0,1,1,0)\, , \nn\\
        \mathcal{I}_5&=\text{P}^{3,1}(1,1,0,0,2,0,1,0,1)\, , \nn\\
        \mathcal{I}_6&=\text{P}^{3,1}(1,1,0,0,1,1,1,1,1)\, , \nn\\
        \mathcal{I}_7&=\text{P}^{3,1}(1,1,-1,-1,1,1,1,1,1)\, , \nn\\
        \mathcal{I}_8&=\text{P}^{3,1}(1,1,1,1,1,1,0,0,1),
\end{align}
so that $T_{\rm can}\, \vec{\mathcal{I}}=\vec{\mathcal{J}}$ with
\be
        T_{\rm can}=\left(
                \begin{array}{cccccccc}
                 1 & 0 & 0 & 0 & 0 & 0 & 0 & 0 \\
                 0 & \sqrt{\gamma ^2-1} & 0 & 0 & 0 & 0 & 0 & 0 \\
                 0 & 0 & \frac{\sqrt{\gamma ^2-1}}{\epsilon } & 0 & 0 & 0 & 0 & 0 \\
                 0 & -6 \gamma  & 0 & \frac{1}{\epsilon } & 0 & 0 & 0 & 0 \\
                 0 & 0 & 0 & 0 & \frac{\sqrt{\gamma ^2-1}}{\epsilon } & 0 & 0 & 0 \\
                 0 & 0 & 0 & 0 & 0 & \sqrt{\gamma ^2-1} & 0 & 0 \\
                 \frac{\gamma }{2} & 0 & 0 & 0 & 0 & 0 & 1 & 0 \\
                 0 & 0 & 0 & 0 & 0 & 0 & 0 & \gamma ^2-1 \\
                \end{array}
                \right).
\ee

We have derived a canonical basis for the B-type integrals here. As we can see, it is fully independent of any V-type integrals, and we may treat them separately. We could now derive, in the same way, a basis for the whole V-type sector. It follows the same principle but is much easier than for the B-type integrals as it only has $1\times 1$ sectors. We may refer to \cite{Jakobsen:2022psy} for the V-types.
\subsection{Function class and integration kernels}
The strategies described above work very well so long as our Feynman integrals are $\d \! \log$-integrals. Once more involved functions (elliptic and beyond) appear, the situation becomes considerably more delicate. We will discuss these more general cases in the next chapter. Here we want to indicate how one can determine whether such functions are present in a given block.

One option is again an integrand analysis, which we have presented in Sec.~\ref{sec:integrandan}. Alternatively, the function space can be read off from the differential equations themselves. Restricting attention to a diagonal block, in which the differential equations take the form
\begin{align}
        \dfrac{d (I_1,...,I_n)}{dx}= \hat{M}(x,\epsilon)(I_1,...,I_n),
\end{align}
the $n$ coupled first-order equations can be transformed into a single $n$-th order equation by the basis change $(I_1,\ldots,I_n)\to(I_1,I_1',\ldots,I_1^{(n-1)})$. This yields a Picard-Fuchs equation of order $n$ for a single integral,
\begin{align}
        \sum_{i=0}^n a_i(x,\epsilon)\, I_1^{(i)}=0.
\end{align}
At $\epsilon=0$, this equation determines the function space. If the $n$-th order operator factorises into a product of $n$ first-order operators, the integral is purely logarithmic; if it does not factorise, more involved transcendental functions probably arise.
\section{Boundary matching}
As already discussed we can write the solution to eq.~\refeq{equ:epsDE} formally as a path-ordered exponential:
\be
	\vec{\mathcal J}(x;\epsilon)= \mathcal{P} e^{\epsilon \int_1^x \mathrm dx M(x)} \vec{j} \, ,
\label{eq:path_exp}	
\ee
where the vector $\vec{j}$ is a set of boundary constants. We chose as the lower integration limit $x=1$, i.e.\ the small-velocity limit $v=0$ ($\gamma=1$), where $\gamma=(1-v^2)^{-1/2}$. We use $x$ as integration variable to rationalise all appearing square roots.  

This expression can be expanded order by order in $\epsilon$, giving rise to multiple polylogarithms (MPLs). The remaining task is then to fix the boundary values $\vec j$.

In principle, both the path-ordered exponential and the boundary values could be expanded in $\epsilon$, with each constant fixed by direct comparison against a slow-velocity expansion of the relevant integral. The differential equations themselves, however, contain useful information which both reduces the number of boundary constants that must be evaluated and allows the boundary constants to be treated as functions of $\epsilon$, without expansion. Both features simplify the matching considerably.

The general idea is therefore to solve the differential equations~\refeq{equ:epsDE} perturbatively in $1-x$, while retaining their full $\epsilon$-dependence. The solution should structurally have the form
\begin{align}
	\vec{\mathcal{J}}(x;\epsilon) = \sum_{k,m,n}  \vec{g}_{k,m,n}(\epsilon)(1-x)^{k+m\epsilon}\log^{n}(1-x) \,.
\end{align}
The crucial new feature is the appearance of terms of the form $(1-x)^{m\epsilon}$, which carry useful physical information; an expansion in $\epsilon$ at this stage would obscure them.

All of these scalings can in fact be derived from the $1-x$ pole $A$ in the differential equations,
\be
        M(x)=\dfrac{A}{1-x}+\mathcal{O}((1-x)^0). \label{eq:deExpanded}
\ee
Following Wasow's method~\cite{WASOW1966378,Bruser:2018jnc}, we write
\be
\vec{\mathcal{J} } = \hat S_0(x;\epsilon) \cdot \vec j + \mathcal O((1-x)).
\ee
Inserting this into~\refeq{equ:epsDE} yields
\begin{align}
        \dfrac{\mathrm d \hat{S}_0(x;\epsilon)}{\mathrm dx}=\epsilon\, \dfrac{ A}{1-x}\,  \hat{S}_0(x;\epsilon),
\end{align}
which is solved by the matrix exponential
\be
        S_0(x;\epsilon)=\exp\!\bigl(- \log(1-x)\, \epsilon\, A\bigr)=(1-x)^{- \epsilon A}.
\ee
This is also the slow-velocity limit of the path-ordered exponential,
\be
\mathcal{P}\, e^{\epsilon \int_1^x \mathrm dx\, M(x)} \;\xrightarrow{x\to 1}\; \mathcal{P}\, e^{\epsilon \int_1^x \mathrm dx\, A/(1-x)}=\exp\!\bigl(- \log(1-x)\, \epsilon\, A\bigr).
\ee
We see that all admissible exponents $m$ in the scalings $(1-x)^{m\epsilon}$ are determined by the eigenvalues of the matrix $A$. We find
\be
 S_0(x;\epsilon)=\left(
        \begin{array}{cccccccc}
         1 & 0 & 0 & 0 & 0 & 0 & 0 & 0 \\
         0 & \delta ^{6 \epsilon } & 0 & 0 & 0 & 0 & 0 & 0 \\
         0 & 0 & \delta ^{-2 \epsilon } & 0 & 0 & 0 & 0 & 0 \\
         0 & 0 & 0 & 1 & 0 & 0 & 0 & 0 \\
         0 & 0 & 0 & 0 & \delta ^{-2 \epsilon } & 0 & 0 & 0 \\
         0 & 0 & -4 \epsilon  \delta ^{-2 \epsilon } \log (\delta ) & 0 & 4 \epsilon  \delta ^{-2 \epsilon } \log (\delta ) & \delta ^{-2 \epsilon } & 0 & 0 \\
         \frac{1}{2} \left(1-\delta ^{2 \epsilon }\right) & 0 & 0 & 0 & 0 & 0 & \delta ^{2 \epsilon } & 0 \\
         0 & 0 & 0 & 0 & 0 & 0 & 0 & 1 \\
        \end{array}
        \right),
\ee
where $\delta=1-x$. The possible velocity scalings are $m\in\{0,-2,2,6\}$. These different scalings arise from different regions of the integrals, which we discuss in detail in the next section. For the present discussion it suffices to know that in the PN limit all scalings with $m>0$ are unphysical, and that the regions can be treated separately.

This already allows us to fix some of the boundary constants. Demanding that all scalings with $m>0$ vanish in the solution
\be
        S_0(x;\epsilon)\cdot \vec{j}=\!\left(j_1,\, j_2 \delta^{6 \epsilon },\, j_3 \delta^{-2 \epsilon },\, j_4,\, j_5 \delta^{-2 \epsilon },\, \delta^{-2 \epsilon }\!\left(4 (j_5-j_3) \epsilon  \log (\delta )+j_6\right),\, \!\left(j_7-\tfrac{j_1}{2}\right) \delta^{2 \epsilon }+\tfrac{j_1}{2},\, j_8\right), \label{eq:scalings}
\ee
we find $j_2=0$ and $j_7=j_1/2$.

We are then left with the boundaries of the two physical regions corresponding to $\delta^0$ and $\delta^{-2\epsilon}$, namely the potential and the radiative region. As mentioned above, these can be treated independently of one another: in each region all other scalings are demanded to vanish, and the contributions are summed at the end.

\subsection*{The potential region}

The potential region corresponds to the $\delta^0$ scalings. We use~\refeq{eq:scalings} once more, demanding additionally that all $\delta^{-2\epsilon}$ scalings vanish, and find the relations
\be
        j_3=0, \quad j_5=0, \quad j_6=0,
\ee
so that
\be
        S_0(x;\epsilon)\cdot \vec{j}=\!\left(j_1,\, 0,\, 0,\, j_4,\, 0,\, 0,\, \tfrac{j_1}{2},\, j_8\right), \label{eq:potentialboundaries}
\ee
leaving only three boundary constants to fix.

To compute these constants it is useful to expand the solution to higher orders in $\delta$. We make the ansatz
\be
        \vec{\mathcal{J}}=\!\left(\sum_{i=0}^n\sum_{k=0}^{i} P_{i,k}\, \delta^i \log^k\delta \right)\!S_0(x;\epsilon)\, \vec{j},
\ee
plug it into the differential equations~\refeq{equ:epsDE}, and expand in $\delta$,
\begin{align}
        \dfrac{d \vec{\mathcal{J}}}{d x}=& -\!\!\left(\sum_{i=0}^n\sum_{k=0}^{i} P_{i,k}\, \delta^{i-1} \!\left(i \log^k\delta+k \log^{k-1}\delta\right)\!\right)\!S_0(x;\epsilon)\, \vec{j} \nn \\
        &+\epsilon\! \left(\sum_{i=0}^n\sum_{k=0}^{i} P_{i,k}\, \delta^{i-1} \log^k\delta \right)\! A\, S_0(x;\epsilon)\, \vec{j} \nn \\
        =&\epsilon\! \left(\dfrac{A}{\delta}+\sum_{i=0}^n M_i \delta^i\right)\! \!\left(\sum_{i=0}^n\sum_{k=0}^{i} P_{i,k}\, \delta^i \log^k\delta \right)\!S_0(x;\epsilon)\, \vec{j},
\end{align}
where $M(x)=A/\delta+\sum_i M_i \delta^i$. Comparing both sides order by order in $\delta$ and $\log\delta$ produces a linear system, which is solved iteratively. Trivially $P_{0,0}=\mathds{1}$ and $P_{0,i}=0$ for $i>0$. Since the differential equations contain no $\log\delta$ terms in their expansion, we may further set $P_{i,k}=0$ for $k>0$.

In the present case it suffices to compute $\vec{\mathcal J}$ to order $\delta$,
\be
\vec{\mathcal{J}}=\!\left(j_1,\, -\tfrac{\delta\, j_4 \epsilon }{6 \epsilon -1},\, \tfrac{2 \delta\, j_4 \epsilon }{2 \epsilon +1},\, j_4,\, 0,\, \tfrac{8 \delta\, j_1 \epsilon^2}{8 \epsilon^2+6 \epsilon +1}-\tfrac{2 \delta\, j_4 (6 \epsilon^2+\epsilon )}{(2 \epsilon +1)^2}+\tfrac{4 \delta\, j_1 \epsilon }{4 \epsilon +1},\, \tfrac{1}{2} j_1 (\delta\, \epsilon +1)-\tfrac{1}{2} \delta\, j_1 \epsilon ,\, j_8\right).
\ee
The constants are then matched against the slow-velocity expansion (see Sec.~\ref{sec:regions}) of the pre-canonical integrals
\begin{align}
	\mathcal{I}_i(\delta;\epsilon) = \sum_{k,m}  f_{i,k,m}(\epsilon)\, \delta^{k+m\epsilon} =(T_{\rm can}^{-1}\, \vec{\mathcal{J}})_i,
\label{eq:expandedInt}
\end{align}
expanded on both sides, yielding
\be
        j_1=f_{1,0,0},\quad j_4=\tfrac{(1-6 \epsilon)\, f_{2,0,0}}{\epsilon},\quad j_8=f_{8,-2,0}.
\ee
The coefficients $f_{i,j,k}$ are determined in the next section.

\subsection*{The radiative region}

The second physical region is the so-called radiative region, associated with the velocity scaling $\delta^{-2\epsilon}$. As before, we demand all other scalings to vanish in~\refeq{eq:scalings}. In addition, since the expansion of our pre-canonical integrals contains no $\log\delta$ terms, we further require these to vanish. We find
\be
        j_1=0,\, j_2=0,\, j_4=0,\, j_5=j_3,\, j_7=0,\, j_8=0,
\ee
and the two remaining coefficients are matched as
\be
        j_3=\tfrac{f_{5,-1,-2}}{\epsilon},\quad j_6=f_{6,-1,-2}.
\ee

\section{Boundary integrals and the method of regions}
\label{sec:regions}
The previous section showed that we need to determine certain boundary constants by computing the slow-velocity expansion of some integrals in the pre-canonical basis. Specifically, we need this expansion in the limit $\delta\to 0$, i.e., $v\to 0$, $x\to 1$, $\gamma\to 1$, while keeping the $\epsilon$-dependence exact, in order to extract the coefficients of the $\delta^{n+m\epsilon}$ scalings.

To this end, we employ the so-called \emph{method of regions}~\cite{Beneke:1997zp}, which provides a general framework for asymptotic expansions of Feynman integrals. We expand our integral by expanding its integrand in a chosen limit. The difficulty here is that we do not know a priori how the loop momenta over which we integrate scale with respect to the expansion parameter. The method of regions resolves this by identifying multiple integration regions, each with a distinct loop-momentum scaling, the full expansion being given by the sum over all non-scaleless regions.

Determining all relevant regions for a generic Feynman integral can be a highly non-trivial task. There exist systematic methods which find the relevant regions, although it is in general not guaranteed that all regions are recovered by such methods~\cite{Pak:2010pt,Jantzen:2012mw,Ma:2025emu}. For Post-Minkowskian integrals the regions are, fortunately, comparatively simple. Two scalings are relevant,
\begin{equation}\label{PRregions}
        \ell^\text{P}=(\ell^0,\Bell)\sim(v,1)\,,  \qquad
         \ell^\text{R}=(\ell^0,\Bell)\sim(v,v) \, .
\end{equation}
It is convenient to use the form~\refeq{eq:1sfgamma} of the 3PM-1SF family, in which the propagators read
\begin{align}
        \hat{D}_{1/2}=\boldsymbol{\ell}_{1/2}\cdot \boldsymbol{e}, \hat{D}_{3}=(\boldsymbol{\ell}_{1}- \boldsymbol{\ell}_{2})^2-2 (\gamma-1)(\boldsymbol{\ell}_{1}\cdot \boldsymbol{e})(\boldsymbol{\ell}_{2}\cdot \boldsymbol{e}),\;
        \hat{D}_{4/5}=(\boldsymbol{\ell}_{1/2}- \boldsymbol{q})^2,\; \hat{D}_{6/7}=(\boldsymbol{\ell}_{1/2})^2.
\end{align}
Two different regions arise from these scalings at this order, which we will discuss in the following.
\subsection*{Potential region}
The simplest region is the so-called \emph{potential region}, in which all gravitons are taken to have potential scaling. This corresponds to a ``naive'' expansion of the integrand. In this representation of the family it is immediate that, as $\gamma\to 1$, the propagators expand at leading order to
        \begin{align}
                \hat{D}_{1/2}\to\boldsymbol{\ell}_{1/2}\cdot \boldsymbol{e},\;\; \hat{D}_{3}\to(\boldsymbol{\ell}_{1}- \boldsymbol{\ell}_{2})^2,\;\;
                \hat{D}_{4/5}\to(\boldsymbol{\ell}_{1/2}- \boldsymbol{q})^2,\;\; \hat{D}_{6/7}\to(\boldsymbol{\ell}_{1/2})^2.
        \end{align}
The expanded family is precisely the 0SF family~\refeq{eq:0sfgamma}. The 0SF integrals therefore play the role of the slow-velocity boundaries of the 1SF integrals. For a general integral one obtains
\begin{align}
       { \mathcal{P}^{\text{3PM-1SF}}}^{ \{ \sigma_1,\sigma_2,\sigma_3 \}}_{\{1,1, n_1,n_2,n_3,n_4,n_5,n_6,n_7\}}\to { \mathcal{P}^{\text{3PM-0SF}}}^{\{\sigma_1,\sigma_2 \}}_{\{1,1,n_1,n_2,n_3,n_4,n_5,n_6,n_7\}} +\mathcal{O}(v^{-n_1-n_2+1}).
\end{align}

The coefficients which remain to be determined are
\be
        j_1=f_{1,0,0},\quad j_4=\tfrac{(1-6 \epsilon)\, f_{2,0,0}}{\epsilon},\quad j_8=f_{8,-2,0},
\ee
and the potential expansion yields
\begin{align}
        f_{1,0,0}&=\text{P}^{3,0}(1,1,0,0,0,1,1,1,1)\big|_{\gamma=\sqrt{2}}, \\
        f_{2,0,0}&=\text{P}^{3,0}(1,1,0,0,1,0,1,1,0)\big|_{\gamma=\sqrt{2}}, \\
        f_{8,-2,0}&=\text{P}^{3,0}(1,1,1,1,1,1,0,0,1)\big|_{\gamma=\sqrt{2}},
\end{align}
where the $\gamma$-dependence $\sim \sqrt{\gamma^2-1}^n$ has been removed by setting $\gamma=\sqrt 2$.

The expansion can also be understood diagrammatically. It is obtained from the 1SF diagrams simply by pushing all worldlines to one side, pinching all dotted lines and attaching a graviton at the bottom carrying the outgoing momentum $q$, e.g.
\begin{figure*}[h!]\hfill
        \begin{subfigure}{ 0.2 \textwidth}
                \begin{tikzpicture}[baseline={([yshift=-1ex]current bounding box.south)},scale=1.2]
                        \coordinate (inA) at (0.4,.7);
                        \coordinate (outA) at (3.4,.7);
                        \coordinate (inB) at (0.4,-.7);
                        \coordinate (outB) at (3.4,-.7);
                        \coordinate (xA) at (0.9,.7);
                        \coordinate (xB) at (0.9,-.7);
                        \coordinate (xM) at (0.9,0);
                        \coordinate (xAM) at (0.9,.35);
                        \coordinate (xBM) at (0.9,-.35);
                        \coordinate (yA) at (1.9,.7);
                        \coordinate (yB) at (1.9,-.7);
                        \coordinate (yC) at (1.9,-1.2);
                        \coordinate (xC) at (0.9,-1.2);
                        \coordinate (zC) at (2.9,-1.2);
                        \coordinate (yM) at (1.9,0);
                        \coordinate (yAM) at (1.9,.35);
                        \coordinate (yBM) at (1.9,-.35);
                        \coordinate (zA) at (2.9,.7);
                        \coordinate (zB) at (2.9,-.7);
                        \coordinate (zM) at (2.9,0);
                        \coordinate (zAM) at (2.9,.35);
                        \coordinate (zBM) at (2.9,-.35);
                        \draw [dotted] (inA) -- (outA);
                        \draw [dotted] (inB) -- (outB);
                        \draw [photon] (xA) -- (xB);
                        \draw [photon] (yA) -- (yB);
                        \draw [photon] (zA) --node [midway, right]{$\qquad\rightarrow$}  (zB);
                        \draw [zUndirected] (xA) -- (yA);
                        \draw [zUndirected] (yB) -- (zB);
                        \draw [photon,white] (yB) -- (yC);
                        \foreach \v in {yA,yB,xA,xB,zA,zB}
                        \draw[fill] (\v) circle (.08);
                      \end{tikzpicture}
        \end{subfigure}\hfill
        \begin{subfigure}{ 0.2 \textwidth}
                \begin{tikzpicture}[baseline={([yshift=-1ex]current bounding box.south)},scale=1.2]
                        \coordinate (inA) at (0.4,.7);
                        \coordinate (outA) at (3.4,.7);
                        \coordinate (inB) at (0.4,-.7);
                        \coordinate (outB) at (3.4,-.7);
                        \coordinate (xA) at (0.9,.7);
                        \coordinate (xB) at (0.9,-.7);
                        \coordinate (xM) at (0.9,0);
                        \coordinate (xAM) at (0.9,.35);
                        \coordinate (xBM) at (0.9,-.35);
                        \coordinate (yA) at (1.9,.7);
                        \coordinate (yB) at (1.9,-.7);
                        \coordinate (yC) at (1.9,-1.2);
                        \coordinate (xC) at (0.9,-1.2);
                        \coordinate (zC) at (2.9,-1.2);
                        \coordinate (yM) at (1.9,0);
                        \coordinate (yAM) at (1.9,.35);
                        \coordinate (yBM) at (1.9,-.35);
                        \coordinate (zA) at (2.9,.7);
                        \coordinate (zB) at (2.9,-.7);
                        \coordinate (zM) at (2.9,0);
                        \coordinate (zAM) at (2.9,.35);
                        \coordinate (zBM) at (2.9,-.35);
                        \draw [dotted] (inA) -- (outA);
                        \draw [dotted] (inB) -- (outB);
                        \draw [photon] (xA) -- (xB);
                        \draw [photon] (yA) -- (yB);
                        \draw [photon] (zA) --node [midway, right]{$\qquad \rightarrow$} (zB);
                        \draw [photon,white] (yB) -- (yC);
                        \draw [zUndirected] (xA) -- (yA);
                        \draw [zUndirected] (yA) -- (zA);
                        \foreach \v in {yA,yB,xA,xB,zA,zB}
                        \draw[fill] (\v) circle (.08);
                      \end{tikzpicture}
        \end{subfigure}\hfill
        \begin{subfigure}{ 0.2 \textwidth}
                \begin{tikzpicture}[baseline={([yshift=-1ex]current bounding box.south)},scale=1.2]
                        \coordinate (inA) at (0.4,.7);
                        \coordinate (outA) at (3.4,.7);
                        \coordinate (inB) at (0.4,-.7);
                        \coordinate (outB) at (3.4,-.7);
                        \coordinate (xA) at (0.9,.7);
                        \coordinate (xB) at (0.9,-.7);
                        \coordinate (xM) at (0.9,0);
                        \coordinate (xAM) at (0.9,.35);
                        \coordinate (xBM) at (0.9,-.35);
                        \coordinate (yA) at (1.9,.7);
                        \coordinate (yB) at (1.9,-.7);
                        \coordinate (yC) at (1.9,-1.2);
                        \coordinate (xC) at (0.9,-1.2);
                        \coordinate (zC) at (2.9,-1.2);
                        \coordinate (yM) at (1.9,0);
                        \coordinate (yAM) at (1.9,.35);
                        \coordinate (yBM) at (1.9,-.35);
                        \coordinate (zA) at (2.9,.7);
                        \coordinate (zB) at (2.9,-.7);
                        \coordinate (zM) at (2.9,0);
                        \coordinate (zAM) at (2.9,.35);
                        \coordinate (zBM) at (2.9,-.35);
                        \draw [dotted] (inA) --node [midway, above]{$\rightarrow \vec{e}$}  (outA);
                        \draw [photon] (yB) --node [midway, right]{$\downarrow \vec{q}$}  (yC);
                        \draw [photon] (xA) to[out=-45,in=180] (yB);
                        \draw [photon] (yA) -- (yB);
                        \draw [zUndirected] (xA) -- (yA);
                        \draw [zUndirected] (yA) -- (zA);
                        \draw [photon] (zA) to[out=-90-45,in=0] (yB);
                        \foreach \v in {yA,yB,xA,yC,zA}
                        \draw[fill] (\v) circle (.08);
                      \end{tikzpicture}
        \end{subfigure}.\hfill $\qquad $
\end{figure*}

\begin{figure}[h!]\hfill
        \begin{subfigure}{ 0.2 \textwidth}
                \begin{tikzpicture}[baseline={([yshift=-1ex]current bounding box.south)},scale=1.2]
                        \coordinate (inA) at (0.4,.7);
                        \coordinate (outA) at (3.4,.7);
                        \coordinate (inB) at (0.4,-.7);
                        \coordinate (outB) at (3.4,-.7);
                        \coordinate (xA) at (0.9,.7);
                        \coordinate (xB) at (0.9,-.7);
                        \coordinate (xM) at (0.9,0);
                        \coordinate (xAM) at (0.9,.35);
                        \coordinate (xBM) at (0.9,-.35);
                        \coordinate (yA) at (1.9,.7);
                        \coordinate (yB) at (1.9,-.7);
                        \coordinate (yC) at (1.9,-1.2);
                        \coordinate (xC) at (0.9,-1.2);
                        \coordinate (zC) at (2.9,-1.2);
                        \coordinate (yM) at (1.9,0);
                        \coordinate (yAM) at (1.9,.35);
                        \coordinate (yBM) at (1.9,-.35);
                        \coordinate (zA) at (2.9,.7);
                        \coordinate (zB) at (2.9,-.7);
                        \coordinate (zM) at (2.9,0);
                        \coordinate (zAM) at (2.9,.35);
                        \coordinate (zBM) at (2.9,-.35);
                        \draw [dotted] (inA) --node [midway, above]{$\rightarrow \vec{e}$}  (outA);
                        \draw [photon] (yB) --node [midway, right]{$\downarrow \vec{q}$} (yC);
                        \draw[photon] ($(yA)!0.5!(yM)$) circle [radius=0.35,rotate=90];
                        \draw[photon] ($(yB)!0.5!(yM)$) circle [radius=0.35,rotate=90];
                        \foreach \v in {yA,yB,yM,yC}
                        \draw[fill] (\v) circle (.08);
                      \end{tikzpicture}\caption{$\text{P}^{3,0}(1,1,0,0,0,1,1,1,1)$}
        \end{subfigure}\hfill
        \begin{subfigure}{ 0.2 \textwidth}
                \begin{tikzpicture}[baseline={([yshift=-1ex]current bounding box.south)},scale=1.2]
                        \coordinate (inA) at (0.4,.7);
                        \coordinate (outA) at (3.4,.7);
                        \coordinate (inB) at (0.4,-.7);
                        \coordinate (outB) at (3.4,-.7);
                        \coordinate (xA) at (0.9,.7);
                        \coordinate (xB) at (0.9,-.7);
                        \coordinate (xM) at (0.9,0);
                        \coordinate (xAM) at (0.9,.35);
                        \coordinate (xBM) at (0.9,-.35);
                        \coordinate (yA) at (1.9,.7);
                        \coordinate (yB) at (1.9,-.7);
                        \coordinate (yC) at (1.9,-1.2);
                        \coordinate (xC) at (0.9,-1.2);
                        \coordinate (zC) at (2.9,-1.2);
                        \coordinate (yM) at (1.9,0);
                        \coordinate (yAM) at (1.9,.35);
                        \coordinate (yBM) at (1.9,-.35);
                        \coordinate (zA) at (2.9,.7);
                        \coordinate (zB) at (2.9,-.7);
                        \coordinate (zM) at (2.9,0);
                        \coordinate (zAM) at (2.9,.35);
                        \coordinate (zBM) at (2.9,-.35);
                        \draw [dotted] (inA) --node [midway, above]{$\rightarrow \vec{e}$}  (outA);
                        \draw [photon] (yB) --node [midway, right]{$\downarrow \vec{q}$} (yC);
                        \draw[photon] ($(yA)!0.5!(yB)$) circle [radius=0.7,rotate=90];
                        \draw [photon] (yB) -- (yA);
                        \foreach \v in {yA,yB,yC}
                        \draw[fill] (\v) circle (.08);
                      \end{tikzpicture}\caption{$\text{P}^{3,0}(1,1,0,0,1,0,1,1,0)$}
        \end{subfigure}
        \hfill
        \begin{subfigure}{ 0.2 \textwidth}
                \begin{tikzpicture}[baseline={([yshift=-1ex]current bounding box.south)},scale=1.2]
                        \coordinate (inA) at (0.4,.7);
                        \coordinate (outA) at (3.4,.7);
                        \coordinate (inB) at (0.4,-.7);
                        \coordinate (outB) at (3.4,-.7);
                        \coordinate (xA) at (0.9,.7);
                        \coordinate (xB) at (0.9,-.7);
                        \coordinate (xM) at (0.9,0);
                        \coordinate (xAM) at (0.9,.35);
                        \coordinate (xBM) at (0.9,-.35);
                        \coordinate (yA) at (1.9,.7);
                        \coordinate (yB) at (1.9,-.7);
                        \coordinate (yC) at (1.9,-1.2);
                        \coordinate (xC) at (0.9,-1.2);
                        \coordinate (zC) at (2.9,-1.2);
                        \coordinate (yM) at (1.9,0);
                        \coordinate (yAM) at (1.9,.35);
                        \coordinate (yBM) at (1.9,-.35);
                        \coordinate (zA) at (2.9,.7);
                        \coordinate (zB) at (2.9,-.7);
                        \coordinate (zM) at (2.9,0);
                        \coordinate (zAM) at (2.9,.35);
                        \coordinate (zBM) at (2.9,-.35);
                        \draw [dotted] (inA) --node [midway, above]{$\rightarrow \vec{e}$}  (outA);
                        \draw [photon] (yB) --node [midway, right]{$\downarrow \vec{q}$} (yC);
                        \draw [photon] (xA) to[out=-45,in=180] (yB);
                        \draw [photon] (yA) -- (yB);
                        \draw [zUndirected] (xA) -- (yA);
                        \draw [zUndirected] (yA) -- (zA);
                        \draw [photon] (zA) to[out=-90-45,in=0] (yB);
                        \foreach \v in {yA,yB,xA,yC,zA}
                        \draw[fill] (\v) circle (.08);
                      \end{tikzpicture}\caption{$\text{P}^{3,0}(1,1,1,1,1,1,0,0,1)$}
        \end{subfigure}\hfill $\qquad $ \caption{The potential boundary integrals required for the calculation of the impulse at 3PM order.}\label{fig:pBound3PM}
\end{figure}
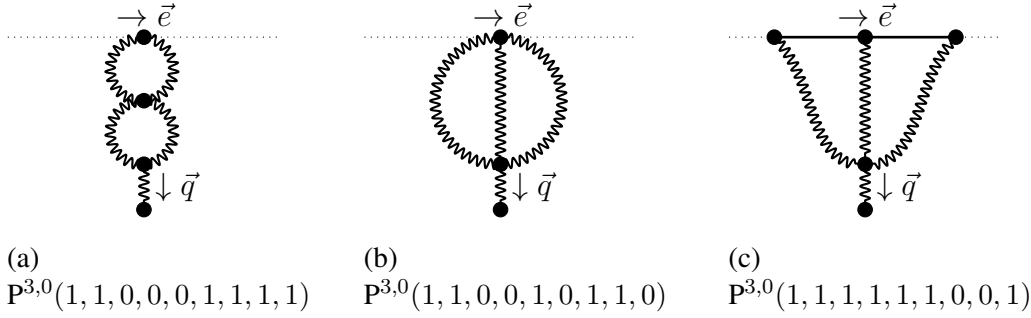

In this notation, the three needed boundary integrals are given by the diagrams depicted in Fig.~\ref{fig:pBound3PM}.

\subsection*{Radiative region}

Having dealt with the potential region, we turn to the possible radiative scalings. It is immediate from the form of the propagators that the rescaling $\boldsymbol{\ell}_{1/2}\sim v$ would yield a factorised, scaleless integral in the slow-velocity expansion, and we may therefore disregard such regions. The form of $\hat D_3$ is, however, special. Choosing the rescaling $\boldsymbol k = (\boldsymbol\ell_1-\boldsymbol\ell_2)\to v\boldsymbol k$ and expanding in $v$ does not yield a scaleless integral, but rather
\begin{align}
        \hat{D}_{1/2} \to \boldsymbol{\ell}_{1}\cdot \boldsymbol{e},\;\;  \hat{D}_{3}\to v^2\bigl(\boldsymbol{k}^2- (\boldsymbol{\ell}_{1}\cdot \boldsymbol{e})(\boldsymbol{\ell}_{1}\cdot \boldsymbol{e})\bigr),\;\;
        \hat{D}_{4/5} \to (\boldsymbol{\ell}_{1}- \boldsymbol{q})^2,\;\; \hat{D}_{6/7} \to (\boldsymbol{\ell}_{1})^2.
\end{align}
The integration measure furthermore contributes a factor of $v^{3-2\epsilon}$, precisely the $v^{-2\epsilon}$ scaling found earlier in the solution of the differential-equation matrix.

Transforming back to the original family, we find
\begin{align}
        { \mathcal{P}^{\text{3PM-1SF}}}^{\{ \sigma_1,\sigma_2,\sigma_3 \}}_{\{1,1,n_3,\ldots,n_9 \}}\to { \mathcal{P}^{\text{3PM-1SF}}}^{ \{ 0,0,\sigma_3 \}}_{1,1,0,n_3+n_4,n_5,n_6+n_7,0,n_8+n_9,0} +\mathcal{O}(v^{3-n_3-n_4-2 n_5- 2 \epsilon+1}).
 \end{align}
The expansion thus maps back into the same 3PM-1SF family, and always into the same topology. Diagrammatically, the active propagator is pushed up to form a mushroom; all worldlines are similarly pushed up and attached to the mushroom, and the worldlines themselves can be removed by a subsequent IBP reduction:
\begin{figure*}[h!]$\qquad $
        \begin{subfigure}{ 0.25 \textwidth}
                \begin{tikzpicture}[baseline={([yshift=-1ex]current bounding box.south)},scale=1.2]
                        \coordinate (inA) at (0.4,.7);
                        \coordinate (outA) at (3.4,.7);
                        \coordinate (inB) at (0.4,-.7);
                        \coordinate (outB) at (3.4,-.7);
                        \coordinate (xA) at (0.9,.7);
                        \coordinate (xB) at (0.9,-.7);
                        \coordinate (xM) at (0.9,0);
                        \coordinate (xAM) at (0.9,.35);
                        \coordinate (xBM) at (0.9,-.35);
                        \coordinate (yA) at (1.9,.7);
                        \coordinate (yB) at (1.9,-.7);
                        \coordinate (yC) at (1.9,-1.2);
                        \coordinate (xC) at (0.9,-1.2);
                        \coordinate (zC) at (2.9,-1.2);
                        \coordinate (yM) at (1.9,0);
                        \coordinate (yAM) at (1.9,.35);
                        \coordinate (yBM) at (1.9,-.35);
                        \coordinate (zA) at (2.9,.7);
                        \coordinate (zB) at (2.9,-.7);
                        \coordinate (zM) at (2.9,0);
                        \coordinate (zAM) at (2.9,.35);
                        \coordinate (zBM) at (2.9,-.35);
                        \draw [dotted] (inA) -- (outA);
                        \draw [dotted] (inB) -- (outB);
                        \draw [photon] (xA) -- (xB);
                        \draw [photon,red] (yA) -- (yB);
                        \draw [photon] (zA) --node [midway, right]{$\qquad \quad \rightarrow$}  (zB);
                        \draw [zUndirected] (xA) -- (yA);
                        \draw [zUndirected] (yB) -- (zB);
                        \foreach \v in {yA,yB,xA,xB,zA,zB}
                        \draw[fill] (\v) circle (.08);
                      \end{tikzpicture}
        \end{subfigure}\hfill
        \begin{subfigure}{ 0.25 \textwidth}
                \begin{tikzpicture}[baseline={([yshift=-1ex]current bounding box.south)},scale=1.2]
                        \coordinate (inA) at (0.3,.7);
                         \coordinate (outA) at (3.6,.7);
                         \coordinate (inB) at (0.3,-.7);
                         \coordinate (outB) at (3.6,-.7);
                         \coordinate (xA) at (0.6,.7);
                         \coordinate (yA) at (1.5,.7);
                         \coordinate (zA) at (2.4,.7);
                         \coordinate (wA) at (3.3,.7);
                         \coordinate (xB) at (0.6,-.7);
                         \coordinate (yB) at (1.5,-.7);
                         \coordinate (zB) at (2.4,-.7);
                         \coordinate (wB) at (3.3,-.7);
                         \coordinate (xM) at (0.6,0);
                         \coordinate (yM) at (1.5,0);
                         \coordinate (zM) at (2.4,0);
                         \coordinate (wM) at (3.3,0);
                         \coordinate (xyMB) at (1.05,-.35);
                         \coordinate (yzMB) at (1.95,-.35);
                         \coordinate (zwMB) at (2.85,-.35);
                         \coordinate (yzMA) at (1.95,.35);
                         \draw [dotted] (inA) -- (outA);
                         \draw [dotted] (inB) -- (outB);
                         \draw [photon] (xA) -- (xB) ;
                         \draw [zUndirected] (xA) -- (yA) ;
                         \draw [zUndirected] (zA) -- (wA) ;
                         \draw [photon] (xA) -- (xB) ;
                         \draw [photon] (wB) to node [midway, right]{$\qquad\rightarrow$}  (wA) ;
                         \draw [photon,red][out=40,in=140] (yA) to[out=40,in=140] (zA);
                [out=40,in=140] (xA) to[out=40,in=140]
                         \draw [fill] (yA) circle (.08);
                         \draw [fill] (zA) circle (.08);
                         \draw [fill] (xB) circle (.08);
                         \draw [fill] (xA) circle (.08);
                         \draw [fill] (wA) circle (.08);
                         \draw [fill] (wB) circle (.08);
                       \end{tikzpicture}
        \end{subfigure}
        \hfill
        \begin{subfigure}{ 0.25 \textwidth}
                \begin{tikzpicture}[baseline={([yshift=-1ex]current bounding box.south)},scale=1.2]
                        \coordinate (inA) at (0.4,.7);
                        \coordinate (outA) at (3.4,.7);
                        \coordinate (inB) at (0.4,-.7);
                        \coordinate (outB) at (3.4,-.7);
                        \coordinate (xA) at (0.9,.7);
                        \coordinate (xB) at (0.9,-.7);
                        \coordinate (xM) at (0.9,0);
                        \coordinate (xAM) at (0.9,.35);
                        \coordinate (xBM) at (0.9,-.35);
                        \coordinate (yA) at (1.9,.7);
                        \coordinate (yB) at (1.9,-.7);
                        \coordinate (yC) at (1.9,-1.2);
                        \coordinate (xC) at (0.9,-1.2);
                        \coordinate (zC) at (2.9,-1.2);
                        \coordinate (yM) at (1.9,0);
                        \coordinate (yAM) at (1.9,.35);
                        \coordinate (yBM) at (1.9,-.35);
                        \coordinate (zA) at (2.9,.7);
                        \coordinate (zB) at (2.9,-.7);
                        \coordinate (zM) at (2.9,0);
                        \coordinate (zAM) at (2.9,.35);
                        \coordinate (zBM) at (2.9,-.35);
                        \draw [dotted] (inA) -- (outA);
                        \draw [dotted] (inB) -- (outB);
                        \draw [photon] (xA) -- (xB);
                        \draw [photon,red][out=40,in=140] (xA) to[out=40,in=140] (zA);
                        \draw [photon] (zA) -- (zB);
                        \foreach \v in {xA,xB,zA,zB}
                        \draw[fill] (\v) circle (.08);
                      \end{tikzpicture}
        \end{subfigure}.
        $\qquad \qquad\qquad\qquad\qquad\qquad \qquad\qquad\qquad\qquad\qquad \qquad\qquad\qquad\qquad$
\end{figure*}

For the remaining coefficients we find
 \begin{align}
 f_{5,-1,-2}&=\text{P}^{3,1}(1,1,0,0,2,1,0,1,0)\big|_{\gamma=\sqrt{2}}, \\
 f_{6,-1,-2}&=0.
 \end{align}
One boundary constant is therefore trivially zero, leaving only a single boundary integral to be evaluated.

We leave the computation of all boundary constants implicit; notice that the computation of the boundary integrals is the first time that we have to explicitly use the $\i0$ prescription of the graviton or worldline propagators. The computation can be done in many cases via direct integration in Schwinger parametrisation. Another possibility is to use numerical programs like \texttt{AMFlow}~\cite{Liu:2022chg} or \texttt{AmpRed}~\cite{Chen:2019mqc,Chen:2019fzm,Chen:2020wsh,Chen:2024xwt} and apply PSLQ to the numerical values to reconstruct the analytic result.
\chapter{Geometries in Feynman integrals}
\label{sec:geom}
In the previous chapter we reviewed the standard machinery for evaluating multi-loop Feynman integrals. We now turn to the integrals giving rise to an enlarged function space. The integrals we have encountered so far have all been expressible as combinations of rational and polylogarithmic functions. It is well known, however, that loop integrals can in general involve substantially richer functions. In particle physics such functions typically arise as soon as the propagators carry internal masses; the most familiar example is probably the two-loop electron self-energy, which already evaluates to elliptic functions. We shall see that this elliptic case is just the simplest representative of a larger class of integrals which evaluate to period integrals of so-called Calabi--Yau (CY) varieties. Especially for one parameter CY cases and elliptic cases there has been a lot of progress in recent years on how to deal with these integrals~\cite{Adams:2016xah, Adams:2018bsn,Vanhove:2014wqa, Pogel:2022ken, Frellesvig:2021hkr, Dlapa:2022wdu, Frellesvig:2023iwr, Gorges:2023zgv, Pogel:2022yat,e-collaboration:2025frv,Bree:2025tug,Maggio:2025jel,Duhr:2025ppd,Duhr:2022dxb,Duhr:2024uid, Becchetti:2025rrz,Doran:2023yzu, Becchetti:2025oyb}. These CY-type integrals will play a central role at fourth and fifth Post-Minkowskian order.

To understand how such geometries enter Feynman integrals, we first take a step back from gravity and study a simple, particle-physics-flavoured example, before returning to the PM setup and showing how the same geometric picture emerges there. Our presentation in this chapter closely follows ref.~\cite{Duhr:2025lbz}.

\section{An elliptic example}
\label{main}
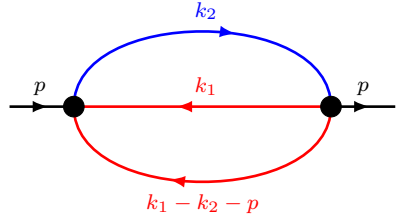
\begin{figure}[!h]
\centering
\begin{tikzpicture}[baseline={([yshift=-1ex]current bounding box.south)},scale=1.7]
        \coordinate (inA) at (0.4,.0);
        \coordinate (outA) at (3.4,.0);
        \coordinate (xA) at (0.9,.0);
        \coordinate (yA) at (1.9,.0);
        \coordinate (zA) at (2.9,.0);
        \draw [zParticle] (inA) --node [midway, above,font=\scriptsize, text width=.2 cm]{$p$} (xA);
        \draw [zParticle,red] (zA) to [in=-90,out=-90]node [midway, below,font=\scriptsize]{$k_1-k_2-p$} (xA);
        \draw [zParticle,blue] (xA) to [in=90,out=90]node [midway, above,font=\scriptsize, text width=.2 cm]{$k_2$} (zA);
        \draw [zParticle,red] (zA) --node [midway, above,font=\scriptsize, text width=.2 cm]{$k_1$} (xA);
        \draw [zParticle] (zA) --node [midway, above,font=\scriptsize]{$p$} (outA);
        \foreach \v in {xA,zA}
        \draw[fill] (\v) circle (.08);
      \end{tikzpicture}
\caption{The two loop sunrise graph with two different masses \red{$m$} and \blue{$M$}. }
\label{fig:sunrise}
\end{figure}

Probably the simplest Feynman integral exhibiting a non-trivial underlying geometry is the so-called sunrise graph, shown in Fig.~\ref{fig:sunrise}. It is part of a broader class of ``banana'' families, which is maybe the most studied class of Feynman diagrams~\cite{Gorges:2023zgv,Duhr:2024bzt, Forner:2024ojj,Sabry, Broadhurst:1987ei, Bauberger:1994by, Bauberger:1994hx, Laporta:2004rb, Bloch:2013tra, Adams:2013nia, Remiddi:2013joa, Adams:2014vja, Adams:2015gva, Adams:2015ydq, Remiddi:2016gno, Adams:2016xah, Remiddi:2017har, Hidding:2017jkk,Vanhove:2026fth,Bloch:2016izu,Pogel:2022ken,Pogel:2022yos,Pogel:2022yat,Pogel:2022vat,Klemm:2019dbm,Bonisch:2020qmm,Broedel:2019kmn}. Our purpose here is purely to understand the method: we use the sunrise to illustrate how an integral becomes attached to a geometry, and to show in practice how the method of refs.~\cite{Gorges:2023zgv,Duhr:2025lbz} works. The same method will then be applied to the more involved integrals that arise in the PM calculation. 

Concretely, we consider the sunrise integral family
\begin{equation}
\label{def:sunrise}
I_{\nu_1,\dots,\nu_{5}}(\uz;d) = \int \left(\prod_{j=1}^2\frac{\mathrm d^dk_j}{i\pi^{d/2}}\right)\frac{(k_1 \cdot p)^{-\nu_{4}}(k_2 \cdot p)^{-\nu_{5}}}{(k_1^2-m^2)^{\nu_1}(k_2^2-M^2)^{\nu_2}((k_1-k_2-p)^2-m^2)^{\nu_3}}\, ,
\end{equation}
which depends on two dimensionless variables $\uz = \{m^2/s,\,M^2/s\}$, with $s=p^2$ the only dimensionful scale.

We will see that for $M^2 = 0$ this integral is polylogarithmic, just like the PM integrals encountered in the previous chapter, while for $M^2 \neq 0$ it gives rise to elliptic functions. The strategy used in the elliptic case is closely parallel to the one followed earlier in section~\ref{sec:integrandan}: an integrand analysis brings the differential equations into a form in which the rotation to a canonical basis is straightforward. The integrand analysis can in fact be extended in such a way that a canonical form can be found also for substantially more involved geometries~\cite{Forner:2026vby}; here, however, we shall not pursue this and instead choose a pre-canonical basis from which the canonical form is reached by a fixed sequence of rotations.

For simplicity we work on the maximal cut and use Baikov representation. In contrast to the PM integrals we work in $d = 2 - 2\epsilon$ dimensions. For the corner integral $I_1 = I_{1,1,1,0,0}(\uz)$ we find
\begin{align}
{\rm LS}\left( I_1 \right) &\,\propto \oint \frac{\mathrm dx_5}{\sqrt{M^2+s+2 x_5} \sqrt{M^2 s-x_5^2}
   \sqrt{4 m^2-M^2-s-2 x_5}}= \oint \frac{\mathrm dx_5}{\sqrt{P_4(x_5)}}\,,
   \label{eq:cutsunrise}
\end{align}
with $x_5=k_2 \cdot p$,
\begin{align}
P_4(x_5) &= (M^2+s+2 x_5)(M^2 s-x_5^2)(4 m^2-M^2-s-2 x_5)\,, \\ 
&= 4(x_5-a_1)(x_5-a_2)(x_5-a_3)(x_5-a_4)\,, \label{eq:P4}
\end{align}
${\rm LS}$ denotes the leading singularity
and 
\begin{align}
    a_1 = - \sqrt{M^2 s}\,, \quad a_2 = - \frac{M^2+s}{2}
    \,, \quad a_3 = \frac{4m^2-M^2-s}{2}\,, \quad a_4 = + \sqrt{M^2 s}\,.
\label{eq:rootsquar}
\end{align}
We see here the first connection to an elliptic curve. The polynomial in the denominator of the integral defines an elliptic curve as
\begin{align}
\mathcal E:\quad y^2 = P_4(x_5) =  (M^2+s+2 x_5)(M^2 s-x_5^2)(4 m^2-M^2-s-2 x_5)   \,, \label{eq:Esunset}
\end{align}
to which we may therefore associate the sunrise integral. Inspecting the roots~\refeq{eq:rootsquar}, we see that for $M=0$ one has $a_1=a_4=0$, so that the curve $\mathcal E$ degenerates. We start our discussion in this degenerate, polylogarithmic limit. We will neglect in all steps contributions from lower sectors in this chapter and always work on the maximal cut. 
\subsubsection*{$M=0$: the polylogarithmic case}
\begin{figure}[h!]
        \centering
        \includegraphics[width=0.8\textwidth]{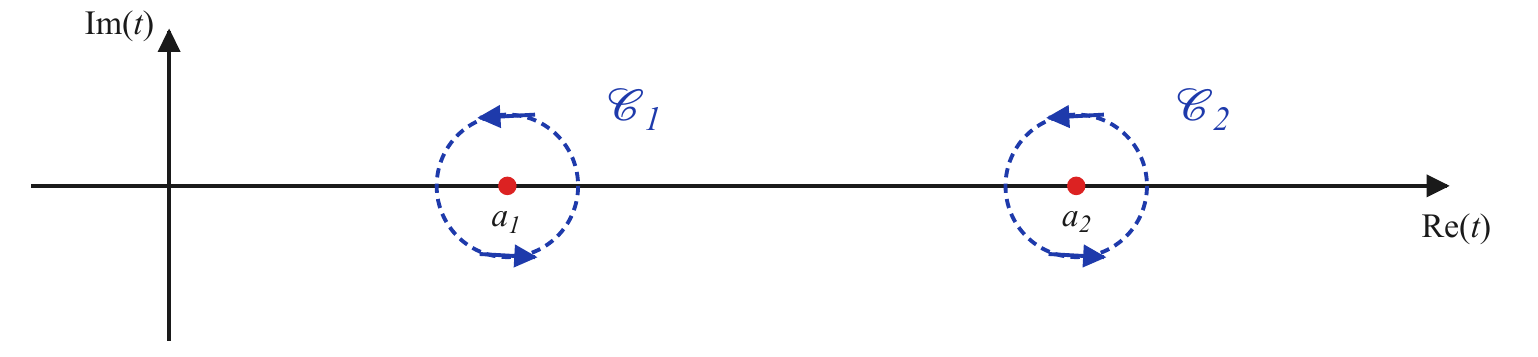}
         \caption{The possible integration contours for the logarithmic sunrise.}
        \label{fig:contoursLog}
    \end{figure}

After IBP reduction, the sector $M=0$ contains two master integrals. An integrand analysis on the maximal cut of the corner integral, obtained from~\refeq{eq:cutsunrise} by setting $M=0$, gives
\begin{equation}
        {\rm LS}\left( I_1 \right)\Big|_{M^2=0} \propto \oint \frac{\mathrm dx_5}{x_5\, \sqrt{s+2 x_5} 
           \sqrt{4 m^2-s-2 x_5}}\,.
           \label{eq:cutsunrise2}
\end{equation}
One of the three square-root factors of $P_4$ has degenerated, and the other two can be rationalised by the substitution $x_5 = \tfrac{m^2 (t+1)^2-s\,t}{2 t}$,

\begin{align}
        {\rm LS}\left( I_1 \right)\Big|_{M^2=0} \propto \oint \frac{d t}{m^2 (t+1)^2-s t}\; .
\end{align}
Rather than attempting to bring this expression directly into $\mathrm d\!\log$ form, we examine the available integration contours. Two independent contours $\mathcal C_1$ and $\mathcal C_2$ may now be chosen as small loops encircling the two poles in the $t$-variable, located  $t = \frac{\pm\sqrt{s(s-4m^2)} - 2m^2 + s}{2 m^2}=a_{1/2}$ as shown in Fig.~\ref{fig:contoursLog}
By the global residue theorem~\cite{GriffithsHarris}, the integrals along $\mathcal C_1$ and $\mathcal C_2$ produce the same leading singularity,
\begin{equation}
        \begin{aligned}
          \oint_{\mathcal C_1}\frac{d t}{m^2 (t+1)^2-s t}&\propto
           \oint_{\mathcal C_2}\frac{d t}{m^2 (t+1)^2-s t}
           \propto\frac{1}{\sqrt{s(s-4m^2)}}\,.
        \end{aligned}
        \end{equation}
We therefore have a leading unit singularity (after multiplying with $\sqrt{s(s-4m^2)}$), so the integral can be normalised into $\mathrm d\!\log$ form.

The same procedure applied to a second integral with the numerator insertion $x_5$ yields the very simple result 
\begin{equation}
        {\rm LS}\left( I_2 \right)\Big|_{M^2=0} \propto \oint \frac{\mathrm dx_5\,}{ \sqrt{s+2 x_5} 
           \sqrt{4 m^2-s-2 x_5}}\propto \oint \dfrac{dt}{t}\propto 1\,.
           \label{eq:cutsunrise2a}
\end{equation}
Its integrand has poles only at $t = 0$ and $t = \infty$; integrating around either yields a leading singularity proportional to unity, so $I_2$ requires no further normalisation.

These observations admit a clean geometric reading. On the one hand we may associate the integral to the geometry~\refeq{eq:Esunset}, which in the limit $M\to 0$ degenerates to two copies of the Riemann sphere joined along a branch cut. This is topologically equivalent to a sphere. On the other hand, after rationalisation of the surviving square root, we end up integrating in projective space against an integrand on the Riemann sphere, punctured at the poles of the integrand. This is precisely the geometric picture associated to $\d\!\log$ integrals.

The corresponding canonical basis is
\begin{equation}
    J_1 = \sqrt{s(s-4m^2)} I_1 \,, \qquad J_2 = I_2\,,
\end{equation}
satisfying the differential equation in $s$
\begin{align} \label{eq:caneqmassless}
   \partial_s\! \begin{pmatrix}
       J_1 \\  J_2
   \end{pmatrix}  =
\epsilon   
   \left(
\begin{array}{cc}
 \frac{2 (s-m^2)}{s (s-4 m^2)} & \frac{3}{\sqrt{s(s-4m^2)}} \\
 -\frac{1}{2 \sqrt{s(s-4m^2)}} & -\frac{1}{2 s} \\
\end{array}
\right)\begin{pmatrix}
        J_1 \\  J_2
   \end{pmatrix} \,.
\end{align} 
\subsubsection*{$M \neq 0$: the elliptic case}
For generic $M\neq 0$, the sector contains three master integrals, one more than in the polylogarithmic case. We first track what happens to the two masters chosen above, and then identify the additional one.

\begin{figure}[h!]
        \centering
        \includegraphics[width=0.8\textwidth]{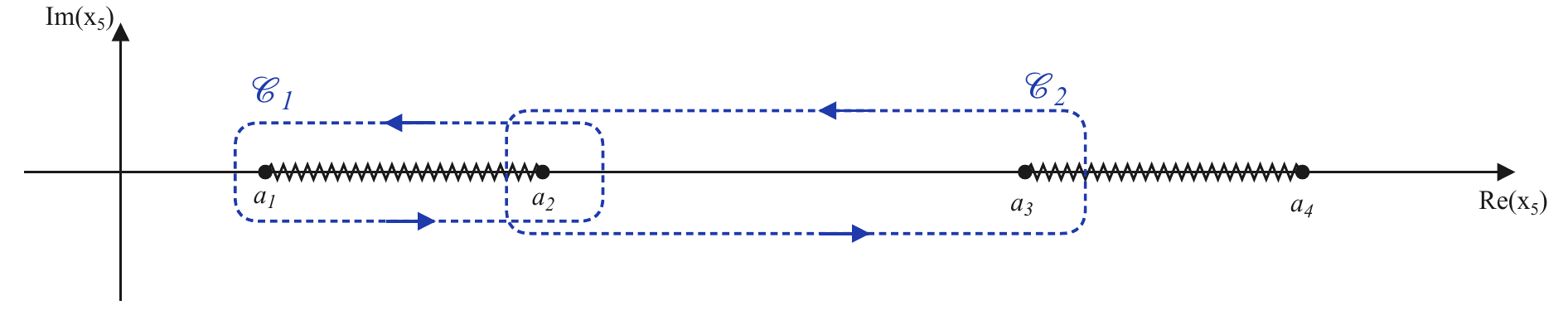}
        \caption{The possible integration contours for the two mass sunrise.}
        \label{fig:contoursEll}
    \end{figure}

For $I_1$, equation~\refeq{eq:cutsunrise} gives
\begin{align}
        {\rm LS}\left( I_1 \right) &\,\propto \oint \frac{\mathrm dx_5}{\sqrt{M^2+s+2 x_5} \sqrt{M^2 s-x_5^2}
   \sqrt{4 m^2-M^2-s-2 x_5}}
        \end{align}
Two independent integration contours $\mathcal C_{1/2}$ are now available. There is a contour encircling $a_1$ and $a_2$, and a second one encircling $a_2$ and $a_3$, as shown in Fig.~\ref{fig:contoursEll}. 
Unlike in the degenerate case, the integrals along these two contours are linearly independent, and define two periods
\begin{equation}
        \pi_0 = \oint_{\mathcal{C}_1}\frac{\mathrm d x_5}{\sqrt{P_4(x_5)}}\,, \qquad
        \pi_1 = \oint_{\mathcal{C}_2} \frac{\mathrm d x_5}{\sqrt{P_4(x_5)}}\,.  \label{eq:persun}
        \end{equation}
 They can be evaluated in terms of the complete elliptic integral of the first kind,
        
\begin{align}
\K(\lambda) = \int_0^1\frac{\rd x}{\sqrt{(1-x^2)(1-\lambda \, x^2)}}\,.
\end{align}
Geometrically the picture is now more complex. Due to the two branch cuts we may need two copies of the Riemann sphere with two branch cuts. Glueing both together along them gives rise to a torus topology on which our integration is defined.

\begin{figure}[h!]
        \centering
        \includegraphics[width=0.4\textwidth]{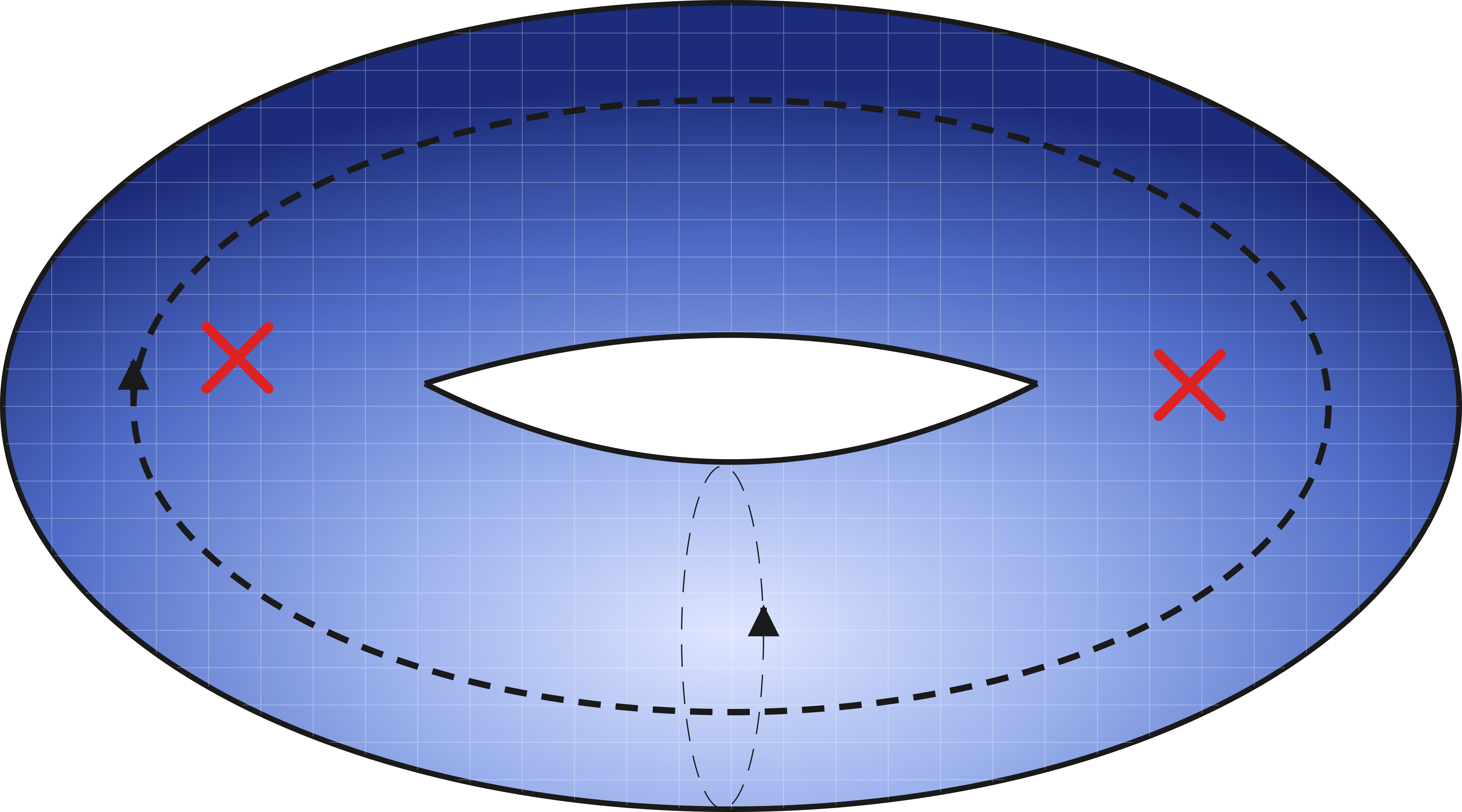}
         \caption{The punctured torus as geometry associated to $I_2$.}
        \label{fig:torusPunct}
    \end{figure}

For the second integral $I_2$ we find 
\begin{align}
        {\rm LS}\left( I_2 \right) &\,\propto \oint \frac{\mathrm dx_5 \; x_5}{\sqrt{M^2+s+2 x_5} \sqrt{M^2 s-x_5^2}
   \sqrt{4 m^2-M^2-s-2 x_5}}
        \end{align}
In addition to the four branch points, this integrand now develops a pole at $x_5 = \infty$. Making the pole manifest by the change of variables $x_5 = 1/t$,
\begin{align}
        {\rm LS}\left( I_2 \right) &\propto \oint \frac{ \mathrm d t}{t\, \sqrt{4(1-t a_1)(1-ta_2)(1-ta_3)(1-ta_4)}} 
        \nonumber \\
        &\sim \oint \mathrm dt \left[ \frac{ 1}{2t } - \frac{s+M^2-2m^2}{4} + \mathcal{O}(t)\right] \,,
        \label{eq:cutI2}
        \end{align}
we see that the integral admits a third independent contour $\mathcal{C}_3$ encircling the pole at $x_5 = \infty$. The integration along $\mathcal C_3$ once again gives a leading singularity proportional to unity, while the integrations along $\mathcal C_{1/2}$ are now naturally interpreted as elliptic integrals of the third kind. The third-kind elliptic integrals are characterised by first-order inhomogeneous differential equations whose inhomogeneities involve the first- and second-kind integrals. The geometry associated to $I_2$ would be a torus with two punctures; shown in Fig.~\ref{fig:torusPunct}.

We have therefore identified, in this sector associated to an elliptic curve, three master integrals, two of which are naturally associated to a differential of the first kind ($I_1$) and a differential of the third kind ($I_2$), respectively. A third master integral, naturally associated to a differential of the second kind, is still missing. Rather than constructing it directly, we exploit the fact that the derivative of a first-kind form is, modulo $\epsilon$-corrections, a combination of first-kind and second-kind forms. We have already seen that derivatives of canonical integrals require only a multiplicative rescaling by $\epsilon$ to compensate for the weight drop, plus a constant rotation in $\epsilon$. Our basis is therefore taken to be
\begin{align}
        \widetilde{I}_1 = I_1 =  I_{1,1,1,0,0}(\underline z)\,, \qquad \widetilde{I}_2 = \partial_{z}\widetilde{I}_1\,,\qquad
        \widetilde{I}_3 =I_2= I_{1,1,1,0,-1}(\underline z)\,.
        \end{align}
The choice of dimensionless variable in the partial derivative is in principle arbitrary; here we take $z=m^2/s$.

The differential equations for this basis read
\begin{align}\label{eq:DEeps3x3a}
        \partial_z\!\begin{pmatrix} \widetilde{I_1} \\ \widetilde{I_2} \\ \widetilde{I_3} \end{pmatrix}
        = \left[ A(\underline z) + \epsilon \, B(\underline z) + \epsilon^2 \, C(\underline z) \right]
        \!\begin{pmatrix} \widetilde{I_1} \\ \widetilde{I_2} \\ \widetilde{I_3} \end{pmatrix} \, , 
    \end{align}
    with
    \begin{equation}
    \label{eq:DEeps3x3}
        A(\underline z) = \begin{pmatrix}
        0 & 1 & 0 \\
        a_{21}(\underline z) & a_{22}(\underline z) & 0 \\
        a_{31}(\underline z) & a_{32}(\underline z) & 0
        \end{pmatrix}  , \
        B(\underline z) =  \begin{pmatrix}
        0 & 0 & 0 \\
        b_{21}(\underline z) & b_{22}(\underline z) & b_{23}(\underline z)  \\
        b_{31}(\underline z) & 0 & b_{33}(\underline z)
        \end{pmatrix} , \
        C(\underline z) = \begin{pmatrix}
        0 & 0 & 0 \\
        c_{21}(\underline z) & 0 & c_{23}(\underline z)  \\
        0 & 0 & 0
        \end{pmatrix} \, ,
    \end{equation}
    and entries 
    
    \begin{alignat}{3}
        a_{21}(\underline z) =& \ \frac{2 (1-6 m^2+M^2)}{m^2 
        R} \, , 
        \quad && 
        a_{22}(\underline z) = -\frac{48 m^4-16 m^2
       (M^2+1)+(M^2-1)^2 }{m^2 R} \, , 
       \nonumber \\
        a_{31}(\underline z) =& \ 1 \, ,
        \quad && 
        a_{32}(\underline z) = \ \frac{1}{2} (4 m^2-M^2-1) \, , 
        \nonumber \\
        b_{21}(\underline z) =& \ \frac{6 M^2-56 m^2 + 6}{m^2 R} \, ,
        \quad &&
        b_{22}(\underline z) = \ -\frac{80 m^4-24 m^2
       (M^2+1)+(M^2-1)^2}{m^2 R} \, , \nonumber \\
        b_{23}(\underline z) =& \ -\frac{12}{m^2 R} \, , 
        \quad &&
        b_{31}(\underline z) = \ 2 \, , 
        \qquad  
        b_{33}(\underline z) = \ 0 \, , \nonumber \\
        c_{21}(\underline z) =& \ \frac{4 (1-16 m^2+M^2)}{m^2 R} \, ,
        \quad &&
        c_{23}(\underline z) = \ -\frac{24}{m^2 R} \, .
        \label{eq:sunriseentriesexplicit}
    \end{alignat}
    where $R = 16 m^4-8 m^2 \left(M^2+1\right)+\left(M^2-1\right)^2$ and $s=1$. 

These differential equations are clearly not in canonical form, but they already exhibit several useful structural features. The $\epsilon^0$ part of the DE, the matrix $A(\underline z)$, is the most important part. We see that the third column is completely zero. This is exactly what one expects from the integrand analysis on the maximal cut: first- and second-kind forms are independent of the third-kind form, while the third-kind integrals are themselves defined by inhomogeneous first-order equations sourced by the first- and second-kind integrals. The vanishing of the diagonal entry in the third column further reflects the unit residue at infinity carried by $\widetilde{I}_3$.

This block structure is exactly what is needed to apply the algorithm of ref.~\cite{Gorges:2023zgv} to the upper $2\times 2$ block of the $\epsilon^0$ part. From this block we read off the second-order Picard-Fuchs operator
\begin{equation}
        \left[ \partial_z^2 -a_{22}(\underline z) \partial_z  -a_{21}(\underline z) \right]\pi_i=\mathcal{L}_2\pi_i = 0 \, , \qquad i=0,1 \, .\label{eq:sunrise_PF}
    \end{equation}
This second-order differential operator has a two-dimensional space of solutions, spanned precisely by the two periods $\pi_0,\pi_1$ defined by the contours~\refeq{eq:persun}. The solutions of the operator are thus the periods of the elliptic curve, and the operator itself is connected to the geometry via the periods.

\subsubsection*{The splitting of the Wronskian}

We want to bring the DE of \refeq{eq:DEeps3x3a} into canonical form. The idea is to split our solutions into a semi-simple part and a unipotent part. The unipotent part fulfils a unipotent differential equation and is the part of our solutions that will satisfy canonical differential equations. Therefore, a rotation with the inverse of the semi-simple part should remove the unwanted parts of our basis. 

It is useful to work locally around singular points of our differential operator to do this split. We will use the Frobenius method to expand the solutions of our periods at a point $z_0$ in a generalised power series. The solutions of an operator connected to a $n$-dimensional variety have at most divergencies up to $\log^n (z-z_0)$. The singular points at which this maximal power appears in the solutions are called points of \emph{Maximal Unipotent Monodromy} (MUM). In our case $m^2=0$ is such a point, so that the so-called \emph{Frobenius basis} is given by:
\begin{align}
        \label{eq:ForbeniusElliptic}
        \varpi_0(z) = 1 + \sum_{j=1}^\infty c_j z^j \textrm{~~~and~~~}  \varpi_{1}(z) = \varpi_0(z) \log(z) + \sum_{j=1}^\infty d_j z^j \,,
       \end{align}
with $\mathcal{L}_2\varpi_i=0 $ and $c_j$ and $d_j$ constants.

The corresponding Wronskian and its first-order matrix differential equation are
\begin{align} \label{eq:Wrdef}
        W = \begin{pmatrix} \varpi_0 & \varpi_{1} \\ \partial_z \varpi_0 & \partial_z \varpi_{1}\end{pmatrix}\,, \qquad 
        \mbox{with} \qquad
        \partial_z W = \begin{pmatrix}
        0 & 1  \\
        a_{21}(\underline z) & a_{22}(\underline z)
        \end{pmatrix}  W\,.
        \end{align}
Following ref.~\cite{Broedel:2018qkq}, we assign to the period ratio 
\begin{align}\label{eq:tau_def}
        \tau(z) = \frac{\varpi_{1}(z)}{\varpi_0(z)} = \log(z) + \ord(z)\,
        \end{align}
a transcendental weight of one. Near the MUM-point $\tau$ behaves as a logarithm, and one finds
\begin{align}
        \partial_z \tau = \dfrac{\partial_z \varpi_{1} \varpi_0-\partial_z \varpi_0 \varpi_{1}}{\varpi_0^2}=\dfrac{\Delta}{\varpi_0^2},
\end{align}
where 
\begin{align}
        \Delta \;=\; \det W \;=\; \frac{9 \left(M^2-1\right)^2 \left(M^4+4 M^2+1\right)^2}{4 m^2 \left(4 M^4+13 M^2+4\right)^2 R}\,.
\end{align}
In contrast to the polylogarithmic case, this is a strictly local statement, valid only in a neighbourhood of the MUM-point.

We use here the bilinear relation satisfied by the Wronskian
\begin{align}
        W \Sigma W^T=Z(\underline z),
\end{align}
where
\begin{align}
        \Sigma=\begin{pmatrix}
                 0 & 1 \\
                 -1 & 0
        \end{pmatrix}\,, \qquad Z = \Delta\,\Sigma\,.
\end{align}
We now use this relation to factorise the Wronskian into the announced \emph{semi-simple} and \emph{unipotent} parts, following refs.~\cite{Broedel:2018qkq, Gorges:2023zgv}:
\begin{equation}
        \label{eq:sesiunipotsplitting}
          \underbrace{  \begin{pmatrix}
                \varpi_0 & \varpi_1  \\
                \partial_z \varpi_0 & \partial_z\varpi_1
            \end{pmatrix}}_{W} =   
              \underbrace{  \begin{pmatrix}
                \varpi_0 & 0 \\
                \partial_z \varpi_0 & \frac{\Delta}{ \varpi_0}
            \end{pmatrix}}_{W^\text{SS}}
            \underbrace{\begin{pmatrix}
                1 & \tau \\
                0 & 1
            \end{pmatrix} }_{W^\text{u}} \,.
        \end{equation}
The unipotent factor satisfies, as the name suggests, a unipotent (purely off-diagonal) differential equation,
        \begin{equation}
            \partial_z W^\text{u} = \begin{pmatrix}
                0 & \partial_z\tau \\
                0 & 0
            \end{pmatrix} W^\text{u} = 
            \begin{pmatrix}
                0 & \frac{\Delta}{\varpi_0^2} \\
                0 & 0
            \end{pmatrix} W^\text{u}\,.
            \label{eq:diffeqWu}
        \end{equation}
We now rotate by the inverse of the semi-simple factor $W^{\text{SS}}$, with an additional relative rescaling by $\epsilon$ on the second master to compensate for the weight drop induced by taking a derivative (cf.\ the discussion in the previous chapter). The rotation reads
\begin{equation}
        \label{eq:DWss_ell3}
            \begin{pmatrix}
            \widetilde{J}_1 \\ \widetilde{J}_2 \\ \widetilde{J}_3\end{pmatrix} = 
            \begin{pmatrix}
                \epsilon & 0& 0 \\
               0 & 1& 0 \\
                0 & 0& \epsilon \\
            \end{pmatrix}
            \begin{pmatrix}
                \big(W^{\textrm{ss}}\big)^{-1} & \begin{matrix} 0 \\ 0 \end{matrix} \\
                \begin{matrix} 0 \ \ & \ \ 0 \end{matrix} & 1
            \end{pmatrix}
            \begin{pmatrix}
            \widetilde{I}_1 \\ \widetilde{I}_2 \\ \widetilde{I}_3 \end{pmatrix}\,.
        \end{equation}
After this rotation the differential equations take the form 
\begin{align}\label{eq:deqJell3}
        \partial_z  \begin{pmatrix}
            \widetilde{J}_1 \\ \widetilde{J}_2 \\ \widetilde{J}_3
        \end{pmatrix} &= 
        \left[
        \widetilde{A}(\underline{z})+\eps\,\widetilde{B}(\underline{z})
        \right]
        \begin{pmatrix}
            \widetilde{J}_1 \\ \widetilde{J}_2 \\ \widetilde{J}_3
        \end{pmatrix}\,,
    \end{align}
    with
\begin{align}
\widetilde{A}(\underline{z}) &= \begin{pmatrix}
        0 &  0 & 0 \\
        \frac{\varpi_0^2}{\Delta} b_{21}(\underline z) + \frac{\varpi_0 \, \partial_z \varpi_0}{\Delta} b_{22}(\underline z) & 0 & \frac{\varpi_0}{\Delta} b_{23}(\underline z) \\
        \varpi_0 \, a_{31}(\underline z) + \partial_z \varpi_0 \, a_{32}(\underline z) & 0 & 0
    \end{pmatrix} \, , \\
    \widetilde{B}(\underline{z}) &=  \begin{pmatrix}
        0 &  \frac{\Delta}{\varpi_0^2} & 0 \\
        \frac{\varpi_0^2}{\Delta} c_{21}(\underline z) & b_{22}(\underline z) & \frac{\varpi_0}{\Delta} c_{23}(\underline z) \\
        \varpi_0 \, b_{31}(\underline z) & \frac{\Delta}{\varpi_0} a_{32}(\underline z) & b_{33}(\underline z)
    \end{pmatrix}\,. \label{eq:Btilde_3x3}
\end{align}
We do not yet have an $\epsilon$-form, but we have brought the DE into a form in which the $\epsilon^0$ part $\widetilde{A}$ is nilpotent and can get removed by a rotation with the following ansatz
\begin{equation}
        \label{eq:G_rot_ell3}
       \begin{pmatrix}
       J_1 \\ J_2 \\ J_3 \end{pmatrix} = G(\underline{z})\begin{pmatrix}
       \widetilde{J}_1 \\ \widetilde{J}_2 \\ \widetilde{J}_3 \end{pmatrix}\,,\qquad G(\underline{z}) = \begin{pmatrix} 1&0&0\\ g_1(\underline{z}) &1 & g_2(\underline{z}) \\ g_3(\underline{z}) &0&1\end{pmatrix}\,.
       \end{equation}
Imposing that the rotated system be $\epsilon$-factorised and solving for the $g_i$ yields three first-order differential equations,
\begin{align}
        \partial_z g_1(\underline{z}) &= - \frac{\varpi_0^2}{\Delta} b_{21}(\underline z) - \frac{\varpi_0 \, \partial_z \varpi_0}{\Delta} b_{22}(\underline z) - g_2(\underline{z}) \left(\varpi_0 a_{31}(\underline z) +\partial_z \varpi_0 \, a_{32}(\underline z) \right) \, , \\
        \partial_z g_2(\underline{z}) &= - \frac{\varpi_0}{\Delta} b_{23}(\underline z) \, , \\
        \partial_z g_3(\underline{z}) &= - \varpi_0 a_{31}(\underline z) -\partial_z \varpi_0 \, a_{32}(\underline z) \, . \label{eq:g3diffeq}
    \end{align}
The first of these can be solved in closed form in terms of $\varpi_0$ and its derivatives. The other two cannot; their solutions $g_2,g_3$ require us to introduce one new transcendental function  characterised by the first-order ODEs above. This new function is closely related to the elliptic integrals of the third kind.

After this final rotation the system is in $\epsilon$-factorised form,
\begin{align}\label{eq:deqJell_final_3}
        \partial_z  \begin{pmatrix}
            J_1 \\ J_2 \\ J_3
        \end{pmatrix} &= 
    \eps\,\mathcal{B}(\underline{z})
        \begin{pmatrix}
            J_1 \\ J_2 \\ J_3
        \end{pmatrix}\, ,
    \end{align}
    with
    \begin{align} \label{eq:deqB_final_3}
    \mathcal{B}(\underline{z}) = G(\underline{z})\widetilde{B}(\underline{z})G(\underline{z})^{-1}\,.
    \end{align}
We have therefore constructed an $\epsilon$-factorised, canonical basis for a sector associated to an elliptic curve. Note an important feature of the construction: the local Frobenius expansion of the periods was used only as a conceptual handle on the analytic structure (in particular, to identify the semi-simple/unipotent split), while the rotations themselves were defined entirely through relations following from the Picard-Fuchs equation~\refeq{eq:sunrise_PF}. Any other choice of period basis satisfying the same relations could therefore have been used. The procedure as a whole consists of the following five steps, which we shall apply unchanged to the more involved geometries appearing later:
\begin{enumerate}
        \item Perform an integrand analysis on the maximal cut to identify differential forms of the first and third kind, choosing them so that the third-kind integrals decouple from the rest at $\epsilon = 0$.
        \item Use the derivative of the first-kind integral as the second-kind integral.
        \item Split the period Wronskian into a semi-simple and a unipotent part.
        \item Rotate by the inverse of the semi-simple part and adjust the $\epsilon$-weights.
        \item Remove the residual lower-order-in-$\epsilon$ terms by an ansatz.
\end{enumerate}

\section{Review of CY varieties and their periods}
\label{eq:CY_ops}
The broader class of geometries relevant for the PM integrals at fourth and fifth order are \emph{Calabi--Yau} (CY) varieties. Elliptic curves are the one-dimensional members of this family. CY varieties are studied in physics primarily in the context of string theory and carry a rich mathematical structure that we cannot cover in any depth here (see also ref.~\cite{Bonisch:2021yfw,Vanhove:2014wqa,Klemm:2018lvx,Bourjaily:2018yfy}). We restrict ourselves to a brief summary of those features needed in the rest of this thesis to construct canonical forms.

A CY $n$-fold is an $n$-dimensional K\"ahler variety $X$ with vanishing first Chern class. Its metric obeys Einstein's vacuum equations in $2n$ spacetime dimensions. 
The middle-dimensional cohomology has a pure Hodge structure of weight $n$, which means that we can decompose the cohomology as
\begin{align}
H^n(X,\mathbb{C}) = \bigoplus_{p+q=n}H^{p,q}(X)\,,
\end{align}
where $H^{p,q}(X)$ is generated by $(p,q)$-forms. The associated Hodge filtration is
\begin{align}\label{eq:CY_Hodge}
F^p = \bigoplus_{k=p}^{n} H^{k,n-k}(X)\,,\qquad 0\subset F^n\subseteq F^{n-1}\subseteq\ldots\subseteq F^0 = H^n(X,\mathbb{C})\,.
\end{align}
The Hodge numbers are $h^{p,q} = \dim H^{p,q}(X)$ and $b_n = \sum_{p+q=n}h^{p,q}$. 

The vanishing of the first Chern class implies $h^{n,0} = \dim H^{n,0}(X) = 1$, so that $X$ admits a unique (up to scale) holomorphic top form $\Omega$. Choosing a basis $\mathcal C_1,\dots,\mathcal C_{b_n}$ of the middle homology $H_n(X,\mathbb Z)$ ---  a choice of independent integration contours --- gives a vector of periods
\begin{align}
\pi = \big(\pi_0,\ldots,\pi_{b_n-1}\big) = \Big(\int_{\mathcal{C}_1}\Omega,\ldots,\int_{\mathcal{C}_{b_n}}\Omega\Big)\,,
\end{align}
generalising the two periods $\pi_0,\pi_1$ encountered in the elliptic case.

For our purposes it suffices to consider a particular family of CY $n$-folds whose Hodge numbers in the middle row of the Hodge diamond all equal one (for $n=2$ we are only interested in the transcendental substructure),
\begin{align}\label{eq:Hodge_numbers_CY_ops}
h^{n,0} = h^{n-1,1} = \ldots = h^{1,n-1}=h^{0,n} = 1\,, \qquad b_n=n+1\,.
\end{align}
The periods of such a family are annihilated by an $(n+1)$-th order Picard-Fuchs operator,
\begin{align}
\label{eq:CYop}
 \mathcal L_0^{(n+1)}\, \pi_i = 0\,, \quad \text{for } i=0,1,\hdots,n \,,
 \end{align}
 with
\begin{equation}
    \mathcal L_0^{(n+1)} = \partial_z^{n+1} + p_n(z) \partial_z^n + \hdots + p_0(z) \,,
\label{eq:pfcy}
\end{equation}
where the $p_i(z)$ are rational functions and the operator is Fuchsian. We assume throughout that the Picard-Fuchs operator possesses a MUM-point, which without loss of generality we place at $z=0$. Operators of this kind are known as \emph{CY operators} in the mathematics literature; their properties have been studied in considerable detail (see e.g.~refs.~\cite{Almkvist2, BognerCY, BognerThesis, CYoperators, Almkvist:2023yja}) and they will be the central technical object underlying the canonicalisation procedure in the following chapters.
They must satisfy a number of further conditions~\cite{BognerThesis}. One of the most important is that the operator be \emph{essentially self-adjoint}~\cite{BognerThesis}. The adjoint of $\mathcal L_0^{(n+1)}$ is~\cite{Bonisch:2021yfw, Almkvist:2023yja}
\begin{equation}
        \mathcal L_0^{*(n+1)} = \sum_{i=0}^{n+1} \left( -\partial_z  \right)^i p_i(z) \, ,
    \label{eq:adjop}
\end{equation}
and $\mathcal L_0^{(n+1)}$ is said to be essentially self-adjoint if there exists an algebraic function $A(z)$ with
\begin{equation}
        \mathcal L_0^{(n+1)}\, A(z) = A(z)\, \mathcal L_0^{*(n+1)}\,.
    \label{eq:essselfadj}
\end{equation}
In direct analogy with the elliptic case, the Frobenius basis of $\mathcal L_0^{(n+1)}$ around the MUM-point $z=0$ takes the form
\begin{equation}
    \varpi_i = z^r\sum_{j=0}^i\frac{1}{j!}\log^j(z) \, S_{i-j}(z)  \, ,
\label{eq:frob}
\end{equation}
for $i \in \{0,\dots,n\}$, where $S_i(z) = \sum_{j=0}^\infty \sigma_{i,j}\, z^j$ are local holomorphic series and $r$ the so-called local exponent. The maximal logarithmic singularity of any local solution is now $\log^n(z)$.

We can write down the Wronskian in this basis
\begin{equation}\label{eq:CY_Wronskian}
    W_n(z) = \begin{pmatrix}
                            \varpi_0 & \varpi_1 & \hdots & \varpi_{n} \\
                            \partial_z\varpi_0 & \partial_z\varpi_1 & \hdots & \partial_z\varpi_{n} \\ 
                            \vdots & \vdots & \vdots & \vdots \\
                            \partial_z^{n}\varpi_0 & \partial_z^{n}\varpi_1 & \hdots & \partial_z^{n}\varpi_{n}
                       \end{pmatrix} \, .
\end{equation}
We have seen earlier that we need to find quadratic relations between the periods and their derivatives to perform the semi-simple split
\begin{align}
        W_n(z)=W^{\text{SS}}_n(z) \; W^{\text{U}}_n(z).
\end{align}
We demand again that the unipotent part $W^{\text{U}}_n(z)$ should satisfy a unipotent differential equation.

For higher dimensional CY varieties at a MUM-point we choose the intersection matrix 
\begin{align}
\Sigma_n = (-1)^n \Sigma_n^T=\begin{pmatrix}0 & \ldots & 0&1\\0 & \ldots & -1&0 \\ \vdots&\iddots&\vdots&\\(-1)^n & \ldots & 0&0\end{pmatrix} \, .
\end{align}
Griffiths transversality~\cite{GriffithsHarris} then implies the bilinear period relation~\cite{Bonisch:2021yfw}
\begin{align}\label{eq:qual_Rel_wronskian}
        W_n(z)\Sigma_nW_n(z)^T = Z_n(z).
        \end{align}
with 
\begin{align}
        {Z_{n}(z)}_{i,n-i+2} = (-1)^{i+1}C_n(z)\,
        \label{eq:defnptcoupling}
        \end{align}
        and ${Z_{n}(z)}_{i,j}=0$ if $n+1-i \ge j$.

Here $C_n(z)$ is the so-called \emph{$n$-point coupling}, determined directly by the Picard-Fuchs operator~\refeq{eq:pfcy} via
        \begin{align}
            \frac{\partial_z C_n(z)}{C_n(z)} = -\frac{2}{n+1}\, p_n(z) \,.
        \end{align}
The cases that will appear in the PM calculation are $n=2$ and $n=3$. For completeness we record the explicit form of $W^{\text{SS}}_n$ for $n=1,2,3$. Their explicit derivation can be found in App.~\ref{semisimple}:
\begin{align}
        \label{eq:SeSiExplicit}
             W^\text{SS}_1(z) &= \begin{pmatrix}
                \varpi_0 & 0 \\
                \varpi_0' & \frac{C_1(z)}{\varpi_0}
            \end{pmatrix} \, , \\
         \nonumber    W^\text{SS}_2(z) &= \begin{pmatrix}
                \varpi_0 & 0 & 0 \\
                \varpi_0' & {\scriptstyle\sqrt{C_2(z)}} & 0 \\
                \frac{\varpi_0'^2}{2\varpi_0}+\frac{2p_2(z)^2+3 p_2'(z)-9p_1(z)}{18}\varpi_0 - \frac{p_2(z)}{3}\varpi_0' \ \ & \left(\frac{\varpi_0'}{\varpi_0} - \frac{p_2(z)}{3} \right){\scriptstyle \sqrt{C_2(z)}} \ \ & \frac{C_2(z)}{\varpi_0}
            \end{pmatrix} \, , \\
        \nonumber     W^\text{SS}_3(z) &= \begin{pmatrix}
                \varpi_0 & 0 & 0 & 0 \\
                \varpi_0' & \frac{\varpi_0}{z\alpha_1} & 0 & 0 \\
                \varpi_0'' & \frac{2\varpi_0'}{z\alpha_1}-\left( \frac{1}{z^2\alpha_1} + \frac{\alpha_1'}{z\alpha_1^2} \right)\varpi_0 & \frac{zC_3(z)\alpha_1}{\varpi_0} & 0 \\
                \varpi_0''' & \frac{2\varpi_0'^2}{z\varpi_0\alpha_1} 
                - \frac{\varpi_0''}{z\alpha_1} 
                + u_1(z)\varpi_0-u_2(z)\varpi_0' \ & 
                \left(\frac{z\varpi_0'\alpha_1}{\varpi_0^2}
                -\frac{zp_3(z)\alpha_1}{2\varpi_0}\right) {\scriptstyle C_3(z)}\ & \frac{{\scriptstyle C_3(z)}}{\varpi_0}
            \end{pmatrix} \, , \label{eq:semisimpleCY}
            \end{align} 
        with
        \begin{align}
                \alpha_1(z)&\,=  \frac{\varpi_0^2}{z \left(\varpi_0 \varpi_1'-\varpi_1 \varpi_0'\right)} \\
             u_1(z) &\,= \frac{p_3(z)\alpha_1'}{2z\alpha_1^2} + \frac{2p_3(z)+zp_3(z)^2+2zp_3'(z)-4zp_2(z)}{4z^2\alpha_1} \, , \\
             u_2(z) &\,= \frac{\alpha_1'}{z\alpha_1^2} + \frac{1+zp_3(z)}{z^2\alpha_1} \, .
            \end{align}
        Additionally, we get from the bilinear relations~\refeq{eq:qual_Rel_wronskian} additional relations for $n=2$ and $n=3$, respectively,
        \begin{align}
         \nonumber  \varpi_0'' &\,= \frac{\varpi_0'^2}{2\varpi_0}+\frac{2p_2(z)^2+3 p_2'(z)-9p_1(z)}{18}\varpi_0 - \frac{p_2(z)}{3}\varpi_0' \qquad &&\text{for} \ n=2 \, , \\
         \label{eq:quadexplicitly}   \alpha_1'' &\,= (zp_3(z)-2)\frac{\varpi_0'\alpha_1}{z\varpi_0}-2\frac{\varpi_0'\alpha_1'}{\varpi_0}+4\frac{\varpi_0''\alpha_1}{\varpi_0}-2\frac{\varpi_0'^2\alpha_1}{\varpi_0^2}+2\frac{\alpha_1'^2}{\alpha_1^2}  && \\
                &\quad -\frac{2zp_3(z)+z^2p_3(z)^2+2z^2p_3'(z)-4z^2p_2(z)-8}{4z^2}\alpha_1 - \frac{zp_3(z)-4}{2z}\alpha_1' \qquad &&\text{for} \ n=3\nonumber \, .
        \end{align}
With these relations we are equipped to construct a canonical basis for all integrals appearing in the PM calculation up to 5PM order. In the following chapters we will see how the methods of this chapter are applied to the computation of the impulse.

\chapter{Integrals at 4PM order}
\label{sec:4pm}
We will now look at the integrals contributing to the 4PM order of the impulse. 
This order was first calculated in~\cite{Bern:2021dqo,Bern:2021yeh,Dlapa:2021npj,Dlapa:2021vgp,Dlapa:2022lmu}. The result was confirmed in~\cite{Damgaard:2023ttc} and extended by spin effects in~\cite{Jakobsen:2023hig,Jakobsen:2023ndj}.

We will apply the methods discussed in the previous chapters to the integrals required for the impulse at fourth Post-Minkowskian order. The good news is that for most integrals the procedure is essentially identical to the two loop case.  The new features are confined to a few specific subtleties, which we discuss in detail. 
\section{The 4PM-1SF integral family}
At 4PM order the impulse receives contributions from the 0SF and 1SF sectors. We do not discuss the probe limit here. The integrals appearing in the 1SF family can be brought into a planar form, with the result that they all belong to the family
\begin{subequations}
        \begin{align}
          {\mathcal{P}^{\text{4PM-1SF}}}^{(\sigma_1,\sigma_2,\ldots,\sigma_5)}_{n_1,n_2,\ldots,n_{15}}
          :=
          \int_{\ell_1,\ell_2,\ell_3}
          \frac{
            \dd^{(n_1-1)}(\ell_1\cdot v_{1})
            \dd^{(n_2-1)}(\ell_2\cdot v_{1})
            \dd^{(n_3-1)}(\ell_3\cdot v_{2})
          }{
            D_1^{n_4}
            D_2^{n_5}
            ...
            D_{12}^{n_{15}}
          }
        \end{align}
        with
        \begin{align}
          D_{1}
          &=
          \ell_{1}\cdot v_{2} +\sigma_{1}\i0, \;
          \nn
          D_{2}
          =
          \ell_{2}\cdot v_{2} +\sigma_{2}\i0, \;
          \nn
          D_{3}
          =
          \ell_{3}\cdot v_{1} +\sigma_{3}\i0, \;
          \nn
          \\
          D_{4}
          &=
          (\ell_1-\ell_3)^2+\sigma_4 \sign{ (\ell_1-\ell_3)\cdot v_1 }\; \i0, \;
          \nn
          D_{5}
          =
          (\ell_2-\ell_3)^2+\sigma_5  \sign{ (\ell_2-\ell_3)\cdot v_1} \; \i0, \;
          \nn
          \\
          D_{6}
          &=
          (\ell_1-\ell_2)^2, \;
          \nn
          \\
          D_{7}
          &=\ell_1^2, \;
          D_{8}
          =\ell_2^2, \;
          D_{9}
          =\ell_3^2, \;
          D_{10}
          =(\ell_1+q)^2, \;
          D_{11}
          =(\ell_2+q)^2, \;
          D_{12}
          =(\ell_3+q)^2
          \nn
        \end{align}
      \end{subequations}
      As at 3PM, we may go to a frame in which the delta functions can be integrated out, after which it is manifest that $D_4$ and $D_5$ are both active propagators:
\begin{align}
        {{\cal P}^{\text{4PM-1SF}}}_{\{1,1,1,n\}}^{\{\sigma\}}
                =\dfrac{(-1)^{\sum_{i=4,i\neq6}^{15} n_i}}{{\sqrt{\gamma^2-1}}^{n_4+n_5+n_6}}  \int_{\boldsymbol{\ell_1}\boldsymbol{\ell_2}\boldsymbol{\ell_3}} \dfrac{1}{\prod_{i=1}^{12} \hat{D}_i^{n_{i+3}}}, 
        \end{align}
      where the 12 propagators are given as
      \begin{align}
        \hat{D}_{1/2/3}=\boldsymbol{\ell}_{1/2/3}\cdot \boldsymbol{e}, \hat{D}_{4/5}=(\boldsymbol{\ell}_{1/2}- \boldsymbol{\ell}_{3})^2-2 (\gamma-1)\boldsymbol{\ell}_{1/2}\cdot \boldsymbol{e}\boldsymbol{\ell}_{3}\cdot \boldsymbol{e},\nn \\
         \hat{D}_{6}=(\boldsymbol{\ell}_{1}- \boldsymbol{\ell}_{2})^2,
        \hat{D}_{7/8/9}=(\boldsymbol{\ell}_{1/2/3}- \boldsymbol{q})^2,\hat{D}_{10/11/12}=(\boldsymbol{\ell}_{1/2/3})^2
\end{align}
For the actual computations we will however use the integral family in its original form, with the delta functions kept explicit.
\section{Integral reduction and symmetries}
The IBP reduction at three loops can be carried out with any of the publicly available IBP packages, such as \texttt{Kira} or \texttt{FIRE}. The reduction takes somewhat longer than at 3PM, but with finite-field methods it remains very fast. There is, however, one key subtlety relative to 3PM: at three loops the integrals can carry up to two active gravitons, and these propagators are sensitive to their $i0$ prescriptions. As a consequence, one finds integrals with identical propagator powers but different $i0$ prescriptions, e.g.
\begin{align*}
            \begin{tikzpicture}[baseline={([yshift=-1ex]current bounding box.south)},scale=1.2]
              \coordinate (inA) at (0.4,.7);
              \coordinate (outA) at (3.6,.7);
              \coordinate (inB) at (0.4,-.7);
              \coordinate (outB) at (3.6,-.7);
              \coordinate (xA) at (1,.7);
              \coordinate (xxA) at (1.825,.7) ;
              \coordinate (xxB) at (1.825,-.7) ;
              \coordinate (yA) at (1.60+0.4,.7);
              \coordinate (yyA) at (2.375,0) ;
              \coordinate (zA) at (3,.7);
              \coordinate (zzA) at (3.75,.7) ;
              \coordinate (xB) at (1,-.7);
              \coordinate (yB) at (1.6+0.4,-.7);
              \coordinate (zB) at (3,-.7);
              \coordinate (zzB) at (3.75,-.7);
              \coordinate (xM) at (1,0);
              \coordinate (yM) at (1.6+0.4,0);
              \coordinate (zM) at (3,-0);
              \draw [dotted] (inA) -- (outA);
              \draw [dotted] (inB) -- (outB) node [midway, below]{} ;
              \draw [draw=none] (xA) to[out=40,in=140] (zA);
              \draw [photon] (xA) -- (xB);
              \draw [photon] (zA) -- (zB);
              \draw [photon,red] (xM) --node [midway, above]{$\rightarrow$} (yM);
                \draw [photon,red] (yM) --node [midway, above]{$\leftarrow$} (zM);
              \draw [photon] (yM) -- (yB);
        \draw [fill] (yB) circle (.08);
        \draw [fill] (yM) circle (.08);
              \draw [fill] (xA) circle (.08);
               \draw [fill] (xM) circle (.08);
              \draw [fill] (zA) circle (.08);
              \draw [fill] (xB) circle (.08);
               \draw [fill] (zM) circle (.08);
              \draw [fill] (zB) circle (.08);
            \end{tikzpicture}
            \begin{tikzpicture}[baseline={([yshift=-1ex]current bounding box.south)},scale=1.2]
                \coordinate (inA) at (0.4,.7);
                \coordinate (outA) at (3.6,.7);
                \coordinate (inB) at (0.4,-.7);
                \coordinate (outB) at (3.6,-.7);
                \coordinate (xA) at (1,.7);
                \coordinate (xxA) at (1.825,.7) ;
                \coordinate (xxB) at (1.825,-.7) ;
                \coordinate (yA) at (1.60+0.4,.7);
                \coordinate (yyA) at (2.375,0) ;
                \coordinate (zA) at (3,.7);
                \coordinate (zzA) at (3.75,.7) ;
                \coordinate (xB) at (1,-.7);
                \coordinate (yB) at (1.6+0.4,-.7);
                \coordinate (zB) at (3,-.7);
                \coordinate (zzB) at (3.75,-.7);
                \coordinate (xM) at (1,0);
                \coordinate (yM) at (1.6+0.4,0);
                \coordinate (zM) at (3,-0);
                \draw [dotted] (inA) -- (outA);
                \draw [dotted] (inB) -- (outB) node [midway, below]{} ;
                \draw [draw=none] (xA) to[out=40,in=140] (zA);
                \draw [photon] (xA) -- (xB);
                \draw [photon] (zA) -- (zB);
                \draw [photon,red] (xM) --node [midway, above]{$\rightarrow$} (yM);
                \draw [photon,red] (yM) --node [midway, above]{$\rightarrow$} (zM);
                \draw [photon] (yM) -- (yB);
          \draw [fill] (yB) circle (.08);
          \draw [fill] (yM) circle (.08);
                \draw [fill] (xA) circle (.08);
                 \draw [fill] (xM) circle (.08);
                \draw [fill] (zA) circle (.08);
                \draw [fill] (xB) circle (.08);
                 \draw [fill] (zM) circle (.08);
                \draw [fill] (zB) circle (.08);
              \end{tikzpicture}   .
        \end{align*}
IBP software would usually detect a symmetry mirroring the tennis-court diagram along the vertical axis. This symmetry is present only when both active propagators point inwards (left diagram); on the right-hand diagram, where both propagators point in the same direction, it is broken.

In principle one could treat the two cases separately, but this would be cumbersome. A more practical alternative is the following: since both integrals satisfy the same IBP relations and differ only in their symmetry properties, one can simply IBP-reduce them simultaneously while disabling all symmetry relations. This produces a larger set of master integrals and, naively, larger diagonal blocks in the differential equation. We shall see below, however, that the additional masters can be disentangled from those that would have been obtained had symmetries been used: the result is only a few extra diagonal blocks in the DE system, with no increase in the complexity of the homogeneous part of the equations.
\section{The symmetry split of the differential equations/ how to use symmetries when you are not allowed to use symmetries}
In this section we will discuss how to efficiently handle the extra MIs in our differential equations, which we encounter if we disable symmetry relations in our IBP reduction. We examine one of the subsectors of the tennis-court topology to illustrate the idea:
\begin{align*}
   
\end{align*}
When the symmetry relations in our IBP reduction are turned on, this sector contains four master integrals for the B-type integrals, which we may take as
\begin{align*}
%
    
\end{align*}
This is not a particularly good basis as far as canonicalisation is concerned, but it makes the treatment of the symmetries especially simple.

When the IBP symmetries are turned off, this same sector contains six master integrals, which we may take as
\begin{align*}
%
    
\end{align*}
The differential equations in this basis $\vec{I}$ are, unsurprisingly, a coupled system of six equations, with the schematic structure 
\begin{align}
    \dfrac{d \vec I}{d x}  =  \left(
    \begin{array}{ccccccc}
     \times  & \times  &\times  &\times  &\times  &\times      \\
     \times  & \times  &\times  &\times  &\times  &\times      \\ 
     \times  & \times  &\times  &\times  &\times  &\times      \\ 
     \times  & \times  &\times  &\times  &\times  &\times      \\ 
     \times  & \times  &\times  &\times  &\times  &\times      \\ 
     \times  & \times  &\times  &\times  &\times  &\times      \\ 
    \end{array}
    \right) \vec{I}\, ,
\end{align}
where $\times$ denotes a (generically) non-zero entry and $0$ would denote a zero entry. As one sees, no zero entries are present in this form.

The two additional masters are precisely the symmetric counterparts of master integrals 2 and 3. We exploit this by performing a basis transformation into a basis whose elements are either symmetric or anti-symmetric under the relevant symmetry operation. Master integrals 1 and 4 are already symmetric; for the remaining integrals we form the linear combinations
\begin{align*}
      %
 
\end{align*}
We have thus constructed two additional symmetric and two anti-symmetric master integrals. Once the symmetric and anti-symmetric masters are grouped together, the differential equation matrix factorises:

\begin{align}
                \left(
            \begin{array}{ccccccc}
             \times  & \times  &\times  &\times  &\times  &\times      \\
             \times  & \times  &\times  &\times  &\times  &\times      \\ 
             \times  & \times  &\times  &\times  &\times  &\times      \\ 
             \times  & \times  &\times  &\times  &\times  &\times      \\ 
             \times  & \times  &\times  &\times  &\times  &\times      \\ 
             \times  & \times  &\times  &\times  &\times  &\times      \\ 
            \end{array}
            \right) \, 
\rightarrow
        \left(
    \begin{array}{ccccccc}
     \times  & \times  &\times  &\times  &0  &0      \\
     \times  & \times  &\times  &\times  &0  &0      \\ 
     \times  & \times  &\times  &\times  &0  &0      \\ 
     \times  & \times  &\times  &\times  &0  &0      \\ 
     0  & 0  &0  &0  &\times  &\times      \\ 
     0  & 0  &0  &0  &\times  &\times      \\ 
    \end{array}
    \right) \, ,
\end{align}
so that the $6\times 6$ block decomposes into a $4\times 4$ and a $2\times 2$ block. The construction itself is completely general. At three loops it is not strictly necessary to perform this split when solving the diagonal blocks. A $6\times 6$ block can still be handled by standard algorithms, and the number of additional masters is comparatively small. But the procedure becomes essential at higher PM orders. Additionally the same observation can also be used to streamline the IBP reduction itself. Note that we have not made any assumptions on the $\i0$ prescriptions of any gravitons or used any symmetry relations, but we have just performed a basis rotation.

To summarise the discussion: if we had -- when using Feynman propagators -- a symmetry relation in our IBP reduction, represented by the operator $\hat S$ with $\hat S^2=1$, then we would want to choose -- when using retarded propagators -- master integrals $I_i$ in a way so that they are eigenstates of the operator with eigenvalues $\pm 1$ and $\hat S I_i =\pm I_i$.

\section{A sector related to a K3 surface at 4PM-1SF order}
\label{sec:K34pm}
\begin{figure}[h]
        \centering
    \begin{tikzpicture}[baseline={([yshift=-1ex]current bounding box.south)},scale=1.2]
      \coordinate (inA) at (0.4,.7);
      \coordinate (outA) at (3.6,.7);
      \coordinate (inB) at (0.4,-.7);
      \coordinate (outB) at (3.6,-.7);
      \coordinate (xA) at (1,.7);
      \coordinate (xxA) at (1.825,.7) ;
      \coordinate (xxB) at (1.825,-.7) ;
      \coordinate (yA) at (1.60+0.4,.7);
      \coordinate (yyA) at (2.375,0) ;
      \coordinate (zA) at (3,.7);
      \coordinate (zzA) at (3.75,.7) ;
      \coordinate (xB) at (1,-.7);
      \coordinate (yB) at (1.6+0.4,-.7);
      \coordinate (zB) at (3,-.7);
      \coordinate (zzB) at (3.75,-.7);
      \coordinate (xM) at (1,0);
      \coordinate (yM) at (1.6+0.4,0);
      \coordinate (zM) at (3,-0);
      \draw [dotted] (inA) -- (outA);
      \draw [dotted] (inB) -- (outB) node [midway, below]{} ;
      \draw [draw=none] (xA) to[out=40,in=140] (zA);
      \draw [photon] (xA) -- (xB);
      \draw [photon] (zM) -- (yB);
      \draw [photon] (zA) -- (zB);
      \draw [photon,red] (xA) -- (zM);
      \draw [fill] (zM) circle (.08);
    
      \draw [fill] (xA) circle (.08);
      \draw [fill] (zA) circle (.08);
      
      \draw [fill] (xB) circle (.08);
      \draw [fill] (yB) circle (.08);
      \draw [fill] (zB) circle (.08);
    \end{tikzpicture}
        \caption{The sector at 4PM-1SF order, which is connected to a K3 surface.}
        \label{fig:K31sfGW}
    \end{figure}
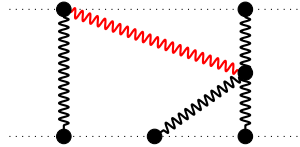

We saw in Chapter~\ref{sec:geom} that Feynman integrals in particle physics do not necessarily evaluate to rational or polylogarithmic functions. The same phenomenon occurs in Post-Minkowskian Feynman integrals starting at three loops. The relevant 4PM--1SF sector is shown in Fig.~\ref{fig:K31sfGW}.

We begin by computing the leading singularity of the corner integral via a loop-by-loop Baikov analysis, which yields
\begin{align}
        {\rm{LS}}{(\text{P}^{4,1}(1,1,1,0,0,0,1,0,1,1,0,0,0,1,1))}=x \int \dfrac{\mathrm dt_1 \mathrm dt_2}{\sqrt{P_{4,4}(t_1,t_2)}} \,,
    \label{eq:K31SFLS}
    \end{align}
    where
    \begin{align}
        P_{4,4}(t_1,t_2) &= t_1 \left(t_1-t_2^2\right) \left(t_1^2 x^4-2 t_1^2 x^2+t_1^2+2 t_1 x^4-4 t_1 x^2+2 t_1+16 t_2^2 x^2+x^4-2 x^2+1\right) \, . 
    \end{align}
As in the elliptic sunrise example, we obtain an integral over the inverse square root of a polynomial, which we cannot rationalise. The new feature is that the polynomial is now quartic in two integration variables rather than one, and the resulting geometry is no longer immediately readable off the Baikov representation. The polynomial nevertheless points strongly toward a Calabi--Yau twofold/a K3 surface. Rather than analysing the Baikov polynomial directly, we proceed via the Picard-Fuchs operator of the integrand.

The simplest way to extract the Picard-Fuchs operator is from the differential equations of the integral, evaluated on the maximal cut in a derivative basis. Normalising the corner integral by $x$, we choose the master integrals
\begin{align}
        I_1 =& \frac{1  }{x}\text{P}^{4,1}(1,1,1,0,0,0,1,0,1,1,0,0,0,1,1) \, , \quad I_2 =   \partial_x I_1 \, , \quad I_3 = \partial_x^2 I_1 \, 
    \end{align}
so that 
\begin{align}
        \partial_x I=M(x,\epsilon) I \,.
    \end{align}
    where $\vec{I}=(I_1,I_2,I_3)$ and the differential equation matrix is
            \begin{align}
             M(x;\epsilon)=  \left(
                \begin{array}{ccc}
                 0 & 1 & 0 \\
                 0 & 0 & 1 \\
                 \frac{-8 x^2 (\epsilon -1)^2+136 \epsilon ^2+4}{x \left(x^2-1\right)^2} & \frac{x^4 \left(-4 \epsilon ^2+20 \epsilon -19\right)+8 x^2 \left(17 \epsilon ^2+2\right)-(2 \epsilon +1)^2}{x^2 \left(x^2-1\right)^2} & \frac{x^2 (4 \epsilon -9)+4 \epsilon +3}{x \left(x^2-1\right)} \\
                \end{array}
                \right) \, . \label{eq:K31sf}
            \end{align}
We can read-off the Picard-Fuchs operator by setting $\epsilon=0$:
\begin{equation}
        \begin{aligned}
            \mathcal{L}^{(3)}_{\text{K3,1}} &=-4 x^2 \left(2 x^2-1\right)-4  x^2 \left(3 x^2-2\right) \theta -6  (x-1) x^2 (x+1)\theta ^2-(x-1)^2 (x+1)^2 \theta ^3 \, ,
        \end{aligned}
        \label{eq:k31sfop}
        \end{equation}
where $\theta=x \partial_x$.

This operator satisfies all the conditions of a Calabi--Yau operator: it is Fuchsian, possesses MUM-points at $x=0$, $x=\pm 1$ and $x=\infty$, is essentially self-adjoint in the sense of~\refeq{eq:essselfadj} with $A(x) = 4/(x^2(1-x^2)^2)$, and its 2-point coupling reads $C_2(x) = 1/(x^2(1-x^2)^2)$. Moreover, $\mathcal{L}^{(3)}_{\text{K3,1}}$ is the symmetric square of a second-order operator,
\begin{align}
        \mathcal{L}^{(2)}_{L} &=  \theta_z ^2-z(\theta_z+\frac{1}{2})^2 \,, \label{eq:Legendre}
\end{align}
where $z=x^2$ and $\theta_z=z \partial_z$,
i.e.\ $\mathcal{L}^{(3)}_{\text{K3,1}} = \mathrm{Sym}^2 \mathcal{L}^{(2)}_{L}$. Concretely, the solutions of $\mathcal{L}^{(3)}_{\text{K3,1}}$ are the squares of those of $\mathcal{L}^{(2)}_{L}$. In our case, $\mathcal{L}^{(2)}_{\text{L}}$ is the Picard-Fuchs operator of the elliptic Legendre curve
\begin{align}
        Y^2= X(X-1)(X-x^2),
\end{align}
and the geometry associated to the third-order operator is therefore a Calabi--Yau twofold or K3, given explicitly by~\cite{Klemm:2024wtd}
\begin{align}
        Y^2=X(X-1)(X-x^2)\,Z(Z-1)(Z-1/x^2).
\end{align}
We can write the solutions of $\mathcal{L}^{(3)}_{\text{K3,1}}$ in an expansion around the MUM-point $x=0$:
\begin{align}
        \varpi_0(x)&=1+\frac{x^2}{2}+\frac{11 x^4}{32} +\frac{17 x^6}{64}+\mathcal{O}(x^8) \, ,\\
        \varpi_1(x)&=\varpi_0(x) \log(x)+S_1 = \varpi_0(x) \log(x) + \frac{x^2}{4}+\frac{29 x^4}{128} +\frac{151 x^6}{768}+\mathcal{O}(x^8) \, , \nonumber\\
        \varpi_2(x)&=\dfrac{1}{2} \varpi_0(x) \log^2(x) +S_1 \log(x)+ \frac{x^4}{32}+\frac{21 x^6}{512}+\mathcal{O}(x^8) \, . \nonumber
    \label{eq:frobK31sf}
    \end{align}
Equivalently, exploiting the symmetric-square structure, the holomorphic period can be written in closed form as
\begin{align}
        \varpi_0(x)&= \dfrac{4 K^2(x^2)}{\pi^2}.
\end{align}
Having identified the geometry, we can now canonicalise the differential equations along the lines of Section~\ref{main}. The Wronskian is
\begin{align}
        W=\left(
                \begin{array}{ccc}
                 \varpi_0 & \varpi_1 & \varpi_2 \\
                 \varpi_0' & \varpi_1'  & \varpi_2'  \\
                 \varpi_0'' & \varpi_1'' & \varpi_2'' \\
                \end{array}
                \right)
\end{align}
and the relevant intersection matrix is
\begin{align}
        \Sigma=\left(
                \begin{array}{ccc}
                 0 &0 & 1 \\
                 0& -1  &0 \\
                1 & 0& 0 \\
                \end{array}
                \right) .
\end{align}
The matrix $Z$ can be obtained either from the general formula~\refeq{eq:defnptcoupling} or from the local series expansions of the Wronskian 
order by order in $x$. We find
\begin{align}
        Z=\left(
                \begin{array}{ccc}
                 0 & 0 & \frac{1}{x^2 \left(x^2-1\right)^2} \\
                 0 & -\frac{1}{x^2 \left(x^2-1\right)^2} & \frac{3 x^2-1}{x^3 \left(x^2-1\right)^3} \\
                 \frac{1}{x^2 \left(x^2-1\right)^2} & \frac{3 x^2-1}{x^3 \left(x^2-1\right)^3} & \frac{-13 x^4+10 x^2-1}{x^4 \left(x^2-1\right)^4} \\
                \end{array}
                \right).
\end{align}
This expression for $Z$ in turn yields bilinear Griffiths-type relations among the periods. In particular,
\begin{align}
        \varpi_0''=\frac{\left(1-3 x^2\right) \varpi_0'}{x \left(x^2-1\right)}-\frac{2 \varpi_0}{x^2-1}+\frac{\varpi_0'^2}{2 \varpi_0}. \label{eq:gf4pm}
\end{align}
With~\refeq{eq:semisimpleCY}, or equivalently with the Griffiths relations, we can now perform the semi-simple/\allowbreak unipotent split of the Wronskian. The semi-simple factor reads
\begin{align}
        {W}^\text{SS}=\left(
    \begin{array}{ccc}
     \varpi_0 & 0 & 0 \\
     \varpi_0' & \sqrt{C_2} & 0 \\
     \varpi_0'' & \sqrt{C_2}\frac{\varpi_0'}{\varpi_0} + \partial_x \sqrt{C_2} & \frac{C_2}{\varpi_0} \\
    \end{array}
    \right)  \quad\text{with}\quad   C_2 = \frac{1}{x^2 \left(x^2-1\right)^2} \, .
    \label{eq:Grav_K3_W}
\end{align}
Rotating by the inverse of the semi-simple part and adjusting the weights with
            \begin{align}
                D(\epsilon) = \text{diag}(\epsilon^2,\epsilon,1)
                \label{eq:Grav_K3_D}
            \end{align}
brings the system close to canonical form. The terms of order $\epsilon^{-1}$ are then removed by the rotation
            \begin{align}
                T_1=\left(
                        \begin{array}{ccc}
                         1 & 0 & 0 \\
                         0 & 1 & 0 \\
                         -\frac{4 x \left(x^2-1\right)\varpi_0 \left(\left(x^2+1\right) \varpi_0'+2 x\varpi_0\right)}{3 \epsilon } & 0 & 1 \\
                        \end{array}
                        \right),
            \end{align}
and the residual $\epsilon^0$ terms by
            \begin{align}
                T_2=\left(
                        \begin{array}{ccc}
                         1 & 0 & 0 \\
                         -\frac{4}{3}  \left(x^2+1\right)\varpi_0 & 1 & 0 \\
                         \frac{2}{9} \left(5 x^4-314 x^2+5\right)\varpi_0^2 & -\frac{8}{3}  \left(x^2+1\right)\varpi_0 & 1 \\
                        \end{array}
                        \right).
            \end{align}
The canonical basis $\vec{J}=T_2 T_1 D(\epsilon) {W^{\text{SS}}}^{-1}\vec{I}$ then satisfies the differential equation
            \begin{align}
           \dfrac{d \vec{J}}{d x} =   \epsilon \left(
\begin{array}{ccc}
 \frac{4 \left(x^2+1\right)}{3 x \left(x^2-1\right)} & \frac{1}{x \left(x^2-1\right)\varpi_0} & 0 \\
 \frac{2 \left(x^4+110 x^2+1\right)\varpi_0}{3 x \left(x^2-1\right)} & \frac{4 \left(x^2+1\right)}{3 x \left(x^2-1\right)} & \frac{1}{x \left(x^2-1\right)\varpi_0} \\
 -\frac{16 \left(x^6-321 x^4-321 x^2+1\right)\varpi_0^2}{27 x \left(x^2-1\right)} & \frac{2 \left(x^4+110 x^2+1\right)\varpi_0}{3 x \left(x^2-1\right)} & \frac{4 \left(x^2+1\right)}{3 x \left(x^2-1\right)} \\
\end{array}
\right)\vec{J}.
            \end{align}

\subsubsection*{Continuation of the period}
We have now derived a canonical basis on the maximal cut for this sector. In the rotations to this basis the only relation we used is~\refeq{eq:gf4pm}, so in principle any function obeying it is an admissible choice of period. The solution $\varpi_0(x)$ clearly satisfies this relation and admits an expansion around $x=0$ free of logarithmic terms. Around the other singular points of the Picard-Fuchs operator, however, logarithms appear, e.g.
\begin{align}
        \varpi_0=\dfrac{1}{2} (\log{8}-\log{(1-x)})+\mathcal{O}(1-x).
\end{align}
This is problematic: the differential equations contain terms $\sim 1/\varpi_0$, which then have no well-defined expansion around $x=1$; yet $x=1$ is precisely the point at which we wish to fix our boundary values, and there a well-defined expansion is essential.

To fix the boundary values in the PN limit we therefore need a holomorphic solution of the Picard-Fuchs operator that also obeys the Griffiths relation~\refeq{eq:gf4pm}. Since $x=1$ is likewise a MUM-point of the operator, this solution is unique:
\begin{align}
        \hat{ \varpi}_0(\delta)&=1+\delta+\frac{7 \delta^2}{8} +\mathcal{O}(\delta^3) \, ,\\
        \hat{ \varpi}_1(\delta)&= \hat \varpi_0(\delta) \log(\delta)+\hat{S}_1 = \hat \varpi_0(\delta) \log(\delta) + \frac{\delta}{2}+\frac{11 \delta^2}{16}+\mathcal{O}(\delta^3) \, , \nonumber\\
        \hat{ \varpi}_2(\delta)&=\dfrac{1}{2}  \hat \varpi_0(\delta) \log^2(\delta) +\hat{S}_1 \log(\delta)+ \frac{\delta^2}{8}+\mathcal{O}(\delta^3) \, , \nonumber
    \end{align}
    where $\delta=1-x$. An explicit check confirms that $\hat{ \varpi}_0(\delta)$ obeys the Griffiths relation~\refeq{eq:gf4pm}, so it is indeed a valid choice.

This restricts us, however, to evaluating the integrals only within the radius of convergence of the series. Here this radius is $r_c=1$, set by the next singular point at $x=0$; no single solution can be expanded simultaneously around both $x=1$ and $x=0$. To continue the solution to a further singular point we must choose a holomorphic period there and fix the boundary data around it --- either by matching to the solution at a neighbouring point or by recomputing the boundaries directly from an expansion of the integrals at the new point. 

In our case we choose
\begin{align}
        \hat{ \varpi}_0(\delta)=\varpi_{\text{K3,1}}(x)=\dfrac{4 K^2(1-x^2)}{\pi^2} \; .
\end{align}

\section{An algorithm to canonicalise the off-diagonals including the K3 period}
\label{sec:K3off}
The transformation of the K3 sector introduces into the differential equations new integration kernels involving $\varpi_0$, $\varpi_0^2$ and $1/\varpi_0$. Unfortunately, these functions then propagate into all off-diagonal blocks below and left of the K3 diagonal block that have not yet been canonicalised; in those blocks one also generically encounters terms containing $\varpi_0'$.

How to canonicalise such terms is not a priori obvious. In the polylogarithmic case we saw two strategies. The conceptually simpler one is a careful integrand analysis on the maximal cut, structured so that the $\epsilon$-expansion becomes such that removing all $\mathcal{O}(\epsilon^n)$ pieces with $n<1$ produces the canonical form directly. This approach applies equally well in the present situation, but at the cost of substantial manual integrand analysis. The second strategy in the $\mathrm d\!\log$ case relied on knowing the integration kernels expected in the canonical form: whenever the result is a $\mathrm d\!\log$ form, one knows the differential forms a priori, since their integration must yield purely logarithmic terms. Once K3 periods $\varpi_0$ enter the kernels, this analysis is more involved and the admissible kernels are not obvious. Two general principles do, however, remain available: in a canonical form the differential equations should have at most simple poles in any local expansion, and they should not contain any explicit $\varpi_0'$~\cite{Duhr:2025lbz}. Building on these two requirements we construct an algorithm that generalises the one of Section~\ref{sec:off}. Once all unwanted kernels have been removed, the same final $x$-independent rotation as before brings the system into $\epsilon$-factorised form. For this algorithm we express our K3 periods as a square of the elliptic functions of first kind $K^2(1-x^2)$.

The differential equations have singular points at $x\in\{ -1,0,1,i,-i,\infty \}$. Around any $x_0\in\{-1,0,1,i,-i\}$ $K^2(1-x^2)$ admits an expansion of the form $\sim c_1 + c_2 \log (x-x_0)+\mathcal{O}(x-x_0)$ with $c_i$ constant. At $\infty$, with $x=1/t$, one has $K^2(1-1/t^2)=t^2(c_1 + c_2 \log t+\mathcal{O}(t))$; expanding the differential forms around $x=\infty$ one must additionally include the Jacobian $-1/t^2$. From this we conclude that, after a partial-fraction decomposition, the canonical differential equations can only contain integration kernels of the form
\begin{align}
        k_i=&\Big\{ \dfrac{1}{x-x_0},\;\dfrac{K^2(1-x^2)}{x-x_0},\;\dfrac{1}{(x-x_0) K^2(1-x^2)},\;  K^2(1-x^2)\, x^{n_1},\;  K^4(1-x^2)\, x^{n_2}\Big\}, \label{equ:kern}
\end{align}
with $n_1\in\{0,1\}$ and $n_2\in\{0,1,2,3\}$. We note for completeness that the kernel $1/(x\,K^2(1-x^2))$ does not have a well-defined expansion around $x=0,\infty$, so a different period choice would be required to expand around that point; this is irrelevant for the present analysis

The unwanted kernels can now be removed in two steps. We assume the differential equation has the structure
\begin{align}
        M=\left(\begin{array}{ll}
             \epsilon D_1(x,K^2(1-x^2)) & 0 \\
            A(x,\epsilon, K(1-x^2), E(1-x^2)) &  \epsilon D_2(x)\\
      \end{array}\right)
      \end{align}
 where $D_1$ is the (already canonicalised) K3 diagonal block, $E(x)$ the complete elliptic integral of second kind, $D_2$ is the diagonal block of a higher sector coupling to the K3 block, and $A$ has the form
\begin{align}
        A=\sum_{ij} \dfrac{m_{i j}(K(1-x^2),E(1-x^2),\epsilon)}{(x-x_i)^j}+\sum_i m_{\infty,i}(K(1-x^2),E(1-x^2),\epsilon)x^i,
    \end{align}
 with $m_{ij}$ allowed to contain terms of the form $K^2,\, K\cdot E,\, K^3 E,\, K^2 E^2,\, E^2,\, E/K^3,\, 1/K^2$. All transformations we consider are of the lower-triangular form
\begin{align}
        T=\left(\begin{array}{ll}
             \mathds{1} & 0 \\
            t &  \mathds{1},\\
      \end{array}\right)
      \end{align}
under which the differential-equation matrix becomes
\begin{align}
        M'=\left(\begin{array}{ll}
             \epsilon D_1 & 0 \\
            A+\dfrac{dt}{dx}+\epsilon (D_2 t-t D_1) &  \epsilon D_2\\
      \end{array}\right)
      \end{align}
\subsubsection*{Step 1: Removing poles $1/(x-x_0)^j$}
Since both $D_1$ and $D_2$ have only simple poles, we make the ansatz
\begin{align}
        t=\dfrac{1}{(x-x_i)^{j-1}}\, a(K(1-x^2),E(1-x^2),\epsilon)
\end{align}
for $a$ of the form
\begin{align}
        a= \sum_i b_i(\epsilon)\, c_i(K(1-x^2),E(1-x^2)),
\end{align}
where $c_i\in\bigl\{ K^2,\, K\cdot E,\, E^2,\, E/K,\, 1/K^2,\, 1\bigr\}$. After applying the transformation, performing a partial-fraction decomposition and extracting the coefficient of $1/(x-x_i)^j$, we read off the coefficient of every admissible $c_i$ and demand that each one vanish. This produces a linear system in the $b_i(\epsilon)$, whose solution determines $t$ and removes the $1/(x-x_i)^j$ pieces. Iterating over all $x_i$ from the highest pole order downwards eliminates all higher poles at every $x_i\neq\infty$.

\subsubsection*{Step 2: Removing poles at $\infty$ and removing $\varpi_0'$}
The remaining higher singularities at $\infty$, together with all derivative-of-period contributions, are removed by a second ansatz
\begin{align}
        t=\sum_i a_i(K,E)\, b_i(x),
\end{align}
where the $b_i(x)$ are polynomials of sufficiently high degree. Demanding that all terms not of the form~\refeq{equ:kern} vanish then fixes the coefficients.
  
 We have implemented this algorithm using the Mathematica packages \texttt{FiniteFlow}~\cite{Peraro:2019svx} and  \texttt{MultivariateApart}~\cite{Heller:2021qkz}.
The last transformation independent of $x$ can be found in the same way as in \refeq{eq:epsT}. This allows us again to fully automatise the canonicalisation of all off-diagonal elements of the differential equation matrix. 
\section{Canonicalisation of the full DE matrix}
We now have all ingredients necessary to canonicalise the full differential-equation matrix. Inside the diagonal blocks, the K3 sector is handled by the procedure of Section~\ref{sec:K34pm}, while the polylogarithmic blocks can be treated either by integrand analysis on the maximal cut (as at 3PM) or by the algorithm \texttt{CANONICA}. Thanks to the symmetry split of the previous section, the diagonal blocks have a maximum size of $4 \times 4$, well within the range that \texttt{CANONICA} can handle very rapidly. The off-diagonals are then removed by our own algorithm, so that the entire procedure is automated up to the determination of the K3-sector rotation.

\begin{figure}[h!]
        \centering
        \begin{subfigure}{0.49\textwidth}
        \includegraphics[width=\textwidth]{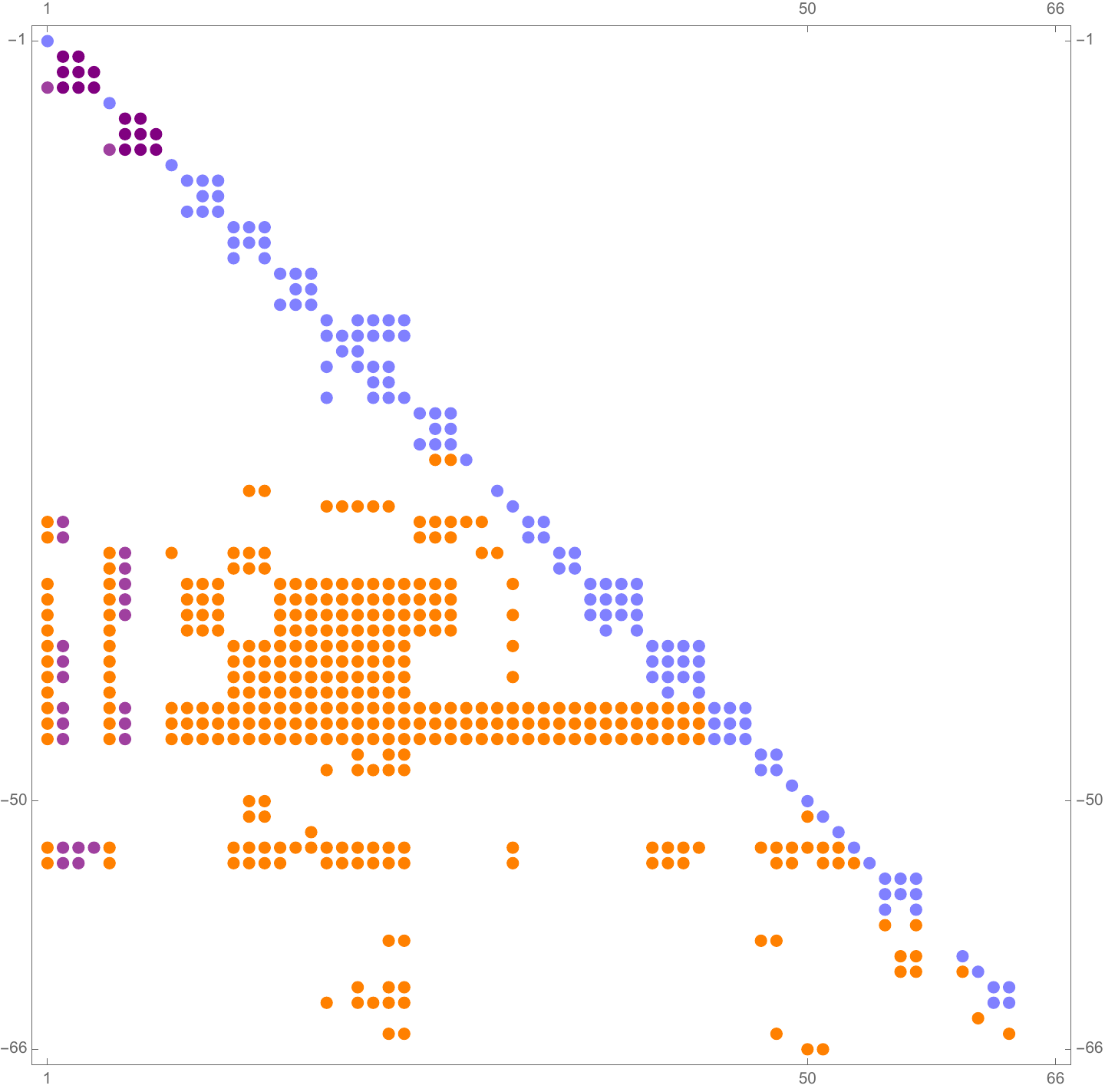}
        \caption{B-type integrals}
\end{subfigure}
        \hfill
        \begin{subfigure}{0.49\textwidth}
        \includegraphics[width=\textwidth]{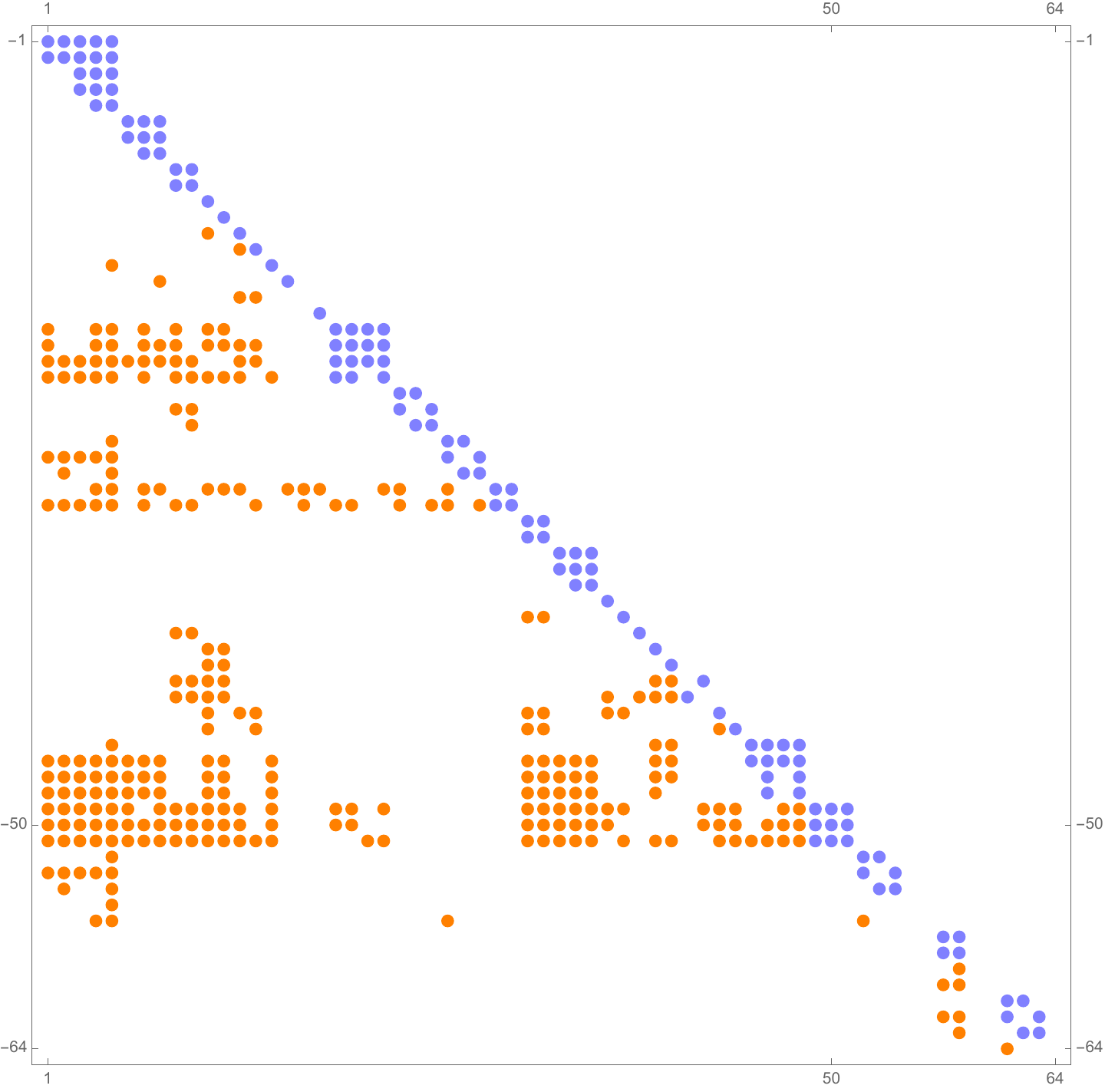}
        \caption{V-type integrals}
\end{subfigure}
\caption{Heat plot of the canonical differential equations of the b(a) and v(b) type integrals at 4PM order. Blue dots represent logarithmic diagonal blocks, orange dots represent logarithmic off-diagonal entries. Entries with a K3 period are marked by purple dots.}
\label{plot:heat4pm}
\end{figure}

The full canonical differential equations for the B- and V-type integrals are too large to reproduce here, but Fig.~\ref{plot:heat4pm} shows a heat plot of the matrix in which the colour of each dot encodes the type of kernel that appears at that position.

After this step, only the boundary conditions remain to be fixed in order to obtain the full 4PM result.

\section{Regions at 4PM}
At three loops the integral family contains two active gravitons rather than one as at two loops, so it is not surprising that we encounter additional regions. The momentum scalings themselves are unchanged from before:
\begin{equation}
        \ell^\text{P}=(\ell^0,\Bell)\sim(v,1)\,,  \qquad
         \ell^\text{R}=(\ell^0,\Bell)\sim(v,v) \, .
\end{equation}
The integral family $\mathcal{P}^{\text{4PM-1SF}}$ has two active propagators, $D_4$ and $D_5$. As at 3PM, these are constructed as differences of two loop momenta which are constrained against different velocities by the delta functions, and either propagator may be either potential or radiative.

The four naive combinations of scalings $PP,RP,PR,RR$ thus suggest four regions in which to expand. Note however that $D_4$ and $D_5$ are exchanged under the relabelling $\ell_1\leftrightarrow \ell_2$, so the $RP$ and $PR$ regions are related by symmetry and produce the same set of boundary integrals. We therefore treat them as a single region, while keeping in mind that two distinct expansions must actually be performed and their contributions added. 

\subsubsection*{Potential PP}
\begin{figure}[h!]
        \centering
        \begin{tikzpicture}[baseline={([yshift=-1ex]current bounding box.south)},scale=1.2]
                \coordinate (inA) at (0.4,.7);
                \coordinate (outA) at (3.4,.7);
                \coordinate (inB) at (0.4,-.7);
                \coordinate (outB) at (3.4,-.7);
                \coordinate (xA) at (0.9,.7);
                \coordinate (xB) at (0.9,-.7);
                \coordinate (xM) at (1.15,0);
                \coordinate (xAM) at (0.9,.35);
                \coordinate (xBM) at (0.9,-.35);
                \coordinate (yA) at (1.9,.7);
                \coordinate (yB) at (1.9,-.7);
                \coordinate (yC) at (1.9,-1.2);
                \coordinate (xC) at (0.9,-1.2);
                \coordinate (zC) at (2.9,-1.2);
                \coordinate (yM) at (1.9,0);
                \coordinate (yAM) at (1.9,.35);
                \coordinate (yBM) at (1.9,-.35);
                \coordinate (zA) at (2.9,.7);
                \coordinate (zB) at (2.9,-.7);
                \coordinate (zM) at (2.65,0);
                \coordinate (zAM) at (2.9,.35);
                \coordinate (zBM) at (2.9,-.35);
                \draw [dotted] (inA) --node [midway, above]{$\rightarrow \vec{e}$}  (outA);
                \draw [photon] (yB) --node [midway, right]{$\downarrow \vec{q}$} (yC);
                \draw[photon] ($(yA)!0.5!(yB)$) circle [radius=0.7,rotate=90];
                \draw [photon] (yB) -- (yA);
                \draw [photon] (yM) -- (zM);
              
                \foreach \v in {yA,yB,yC,yM,zM}
                \draw[fill] (\v) circle (.08);
        
              \end{tikzpicture}
              \begin{tikzpicture}[baseline={([yshift=-1ex]current bounding box.south)},scale=1.2]
                \coordinate (inA) at (0.4,.7);
                \coordinate (outA) at (3.4,.7);
                \coordinate (inB) at (0.4,-.7);
                \coordinate (outB) at (3.4,-.7);
                \coordinate (xA) at (0.9,.7);
                \coordinate (xB) at (0.9,-.7);
                \coordinate (xM) at (0.9,0);
                \coordinate (xAM) at (0.9,.35);
                \coordinate (xBM) at (0.9,-.35);
                \coordinate (yA) at (1.9,.7);
                \coordinate (yB) at (1.9,-.7);
                \coordinate (yC) at (1.9,-1.2);
                \coordinate (xC) at (0.9,-1.2);
                \coordinate (zC) at (2.9,-1.2);
                \coordinate (yM) at (1.9,0);
                \coordinate (yAM) at (1.9,.35);
                \coordinate (yBM) at (1.9,-.35);
                \coordinate (zA) at (2.9,.7);
                \coordinate (zB) at (2.9,-.7);
                \coordinate (zM) at (2.9,0);
                \coordinate (zAM) at (2.9,.35);
                \coordinate (zBM) at (2.9,-.35);
                \draw [dotted] (inA) --node [midway, above]{$\rightarrow \vec{e}$}  (outA);
                \draw [photon] (yB) --node [midway, right]{$\downarrow \vec{q}$} (yC);
                \draw[photon] ($(yA)!0.5!(yM)$) circle [radius=0.35,rotate=90];
                \draw[photon] ($(yB)!0.5!(yM)$) circle [radius=0.35,rotate=90];
                \draw [photon] (yB) -- (yM);
                \foreach \v in {yA,yB,yM,yC}
                \draw[fill] (\v) circle (.08);
        
              \end{tikzpicture}
              \begin{tikzpicture}[baseline={([yshift=-1ex]current bounding box.south)},scale=1.2]
                \coordinate (inA) at (0.4,.7);
                \coordinate (outA) at (3.4,.7);
                \coordinate (inB) at (0.4,-.7);
                \coordinate (outB) at (3.4,-.7);
                \coordinate (xA) at (0.9,.7);
                \coordinate (xB) at (0.9,-.7);
                \coordinate (xM) at (0.9,0);
                \coordinate (xAM) at (0.9,.35);
                \coordinate (xBM) at (0.9,-.35);
                \coordinate (yA) at (1.9,.7);
                \coordinate (yB) at (1.9,-.7);
                \coordinate (yC) at (1.9,-1.2);
                \coordinate (xC) at (0.9,-1.2);
                \coordinate (zC) at (2.9,-1.2);
                \coordinate (yM) at (1.9,0);
                \coordinate (yAM) at (1.9,.35);
                \coordinate (yBM) at (1.9,-.35);
                \coordinate (zA) at (2.9,.7);
                \coordinate (zB) at (2.9,-.7);
                \coordinate (zM) at (2.9,0);
                \coordinate (zAM) at (2.9,.35);
                \coordinate (zBM) at (2.9,-.35);
                \draw [dotted] (inA) --node [midway, above]{$\rightarrow \vec{e}$}  (outA);
                \draw [photon] (yB) --node [midway, right]{$\downarrow \vec{q}$} (yC);
                \draw [photon] (yA) to[out=-45,in=45] (yB);
                \draw [photon] (xA) to[out=-45,in=180] (yB);
                \draw [photon] (yA) to[out=-0,in=0,distance=1cm] (yB);
                \draw [zUndirected] (xA) -- (yA);
                \draw [photon] (yB) -- (yA);
                \draw [photon] (yB) -- (yA);
                \foreach \v in {yA,yB,xA,yC}
                \draw[fill] (\v) circle (.08);
        
              \end{tikzpicture}
              \begin{tikzpicture}[baseline={([yshift=-1ex]current bounding box.south)},scale=1.2]
                \coordinate (inA) at (0.4,.7);
                \coordinate (outA) at (3.4,.7);
                \coordinate (inB) at (0.4,-.7);
                \coordinate (outB) at (3.4,-.7);
                \coordinate (xA) at (0.9,.7);
                \coordinate (xB) at (0.9,-.7);
                \coordinate (xM) at (0.9,0);
                \coordinate (xAM) at (0.9,.35);
                \coordinate (xBM) at (0.9,-.35);
                \coordinate (yA) at (1.9,.7);
                \coordinate (yB) at (1.9,-.7);
                \coordinate (yC) at (1.9,-1.2);
                \coordinate (xC) at (0.9,-1.2);
                \coordinate (zC) at (2.9,-1.2);
                \coordinate (yM) at (1.9,0);
                \coordinate (yAM) at (1.9,.35);
                \coordinate (yBM) at (1.9,-.35);
                \coordinate (zA) at (2.9,.7);
                \coordinate (zB) at (2.9,-.7);
                \coordinate (zM) at (2.65,0);
                \coordinate (zAM) at (2.9,.35);
                \coordinate (zBM) at (2.9,-.35);
                \draw [dotted] (inA) --node [midway, above]{$\rightarrow \vec{e}$}  (outA);
                \draw [photon] (yB) --node [midway, right]{$\downarrow \vec{q}$} (yC);
                \draw [photon] (yM) -- (zM);
                \draw [photon] (xA) to[out=-45,in=180] (yB);
                \draw [photon] (yA) to[out=-0,in=0,distance=1cm] (yB);
                \draw [zUndirected] (xA) -- (yA);
                \draw [photon] (yB) -- (yA);
                \draw [photon] (yB) -- (yA);
                \foreach \v in {yA,yB,xA,yC,zM,yM}
                \draw[fill] (\v) circle (.08);
        
              \end{tikzpicture}
              \begin{tikzpicture}[baseline={([yshift=-1ex]current bounding box.south)},scale=1.2]
                \coordinate (inA) at (0.4,.7);
                \coordinate (outA) at (3.4,.7);
                \coordinate (inB) at (0.4,-.7);
                \coordinate (outB) at (3.4,-.7);
                \coordinate (xA) at (0.9,.7);
                \coordinate (xB) at (0.9,-.7);
                \coordinate (xM) at (0.9,0);
                \coordinate (xAM) at (0.9,.35);
                \coordinate (xBM) at (0.9,-.35);
                \coordinate (yA) at (1.9,.7);
                \coordinate (zyA) at (2.3,.7);
                \coordinate (xyA) at (1.5,.7);
                \coordinate (yB) at (1.9,-.7);
                \coordinate (yC) at (1.9,-1.2);
                \coordinate (xC) at (0.9,-1.2);
                \coordinate (zC) at (2.9,-1.2);
                \coordinate (yM) at (1.9,0);
                \coordinate (yAM) at (1.9,.35);
                \coordinate (yBM) at (1.9,-.35);
                \coordinate (zA) at (2.9,.7);
                \coordinate (zB) at (2.9,-.7);
                \coordinate (zM) at (2.9,0);
                \coordinate (zAM) at (2.9,.35);
                \coordinate (zBM) at (2.9,-.35);
                \draw [dotted] (inA) --node [midway, above]{$\rightarrow \vec{e}$}  (outA);
                \draw [photon] (yB) --node [midway, right]{$\downarrow \vec{q}$}  (yC);
                \draw [photon] (xyA) to (yB);
                \draw [photon] (xyA) to[out=-45-90,in=45+90] (yB);
                \draw [photon] (zyA) to[out=-45,in=45] (yB);
                \draw [zUndirected] (xyA) -- (zyA);

                \draw [photon] (zyA) to (yB);

                \foreach \v in {yB,yC,xyA,zyA}
                \draw[fill] (\v) circle (.08);
        
              \end{tikzpicture}
              \begin{tikzpicture}[baseline={([yshift=-1ex]current bounding box.south)},scale=1.2]
                \coordinate (inA) at (0.4,.7);
                \coordinate (outA) at (3.4,.7);
                \coordinate (inB) at (0.4,-.7);
                \coordinate (outB) at (3.4,-.7);
                \coordinate (xA) at (0.9,.7);
                \coordinate (xB) at (0.9,-.7);
                \coordinate (xM) at (0.9,0);
                \coordinate (xAM) at (0.9,.35);
                \coordinate (xBM) at (0.9,-.35);
                \coordinate (yA) at (1.9,.7);
                \coordinate (zyA) at (2.3,.7);
                \coordinate (xyA) at (1.5,.7);
                \coordinate (yB) at (1.9,-.7);
                \coordinate (yC) at (1.9,-1.2);
                \coordinate (xC) at (0.9,-1.2);
                \coordinate (zC) at (2.9,-1.2);
                \coordinate (yM) at (1.9,0);
                \coordinate (yAM) at (1.9,.35);
                \coordinate (yBM) at (1.9,-.35);
                \coordinate (zA) at (2.9,.7);
                \coordinate (zB) at (2.9,-.7);
                \coordinate (zM) at (2.9,0);
                \coordinate (zAM) at (2.9,.35);
                \coordinate (zBM) at (2.9,-.35);
                \draw [dotted] (inA) --node [midway, above]{$\rightarrow \vec{e}$}  (outA);
                \draw [photon] (yB) --node [midway, right]{$\downarrow \vec{q}$}  (yC);
                \draw [photon] (xA) to[out=-45,in=180] (yB);
                \draw [photon] (xyA) to (yB);
                
                \draw [zUndirected] (xA) -- (yA);
                \draw [zUndirected] (yA) -- (zA);
                \draw [photon] (zA) to[out=-90-45,in=0] (yB);
                \draw [photon] (zyA) to (yB);

                \foreach \v in {yB,xA,yC,zA,xyA,zyA}
                \draw[fill] (\v) circle (.08);
        
              \end{tikzpicture}
              \begin{tikzpicture}[baseline={([yshift=-1ex]current bounding box.south)},scale=1.2]
                \coordinate (inA) at (0.4,.7);
                \coordinate (outA) at (3.4,.7);
                \coordinate (inB) at (0.4,-.7);
                \coordinate (outB) at (3.4,-.7);
                \coordinate (xA) at (0.9,.7);
                \coordinate (xB) at (0.9,-.7);
                \coordinate (xM) at (0.9,0);
                \coordinate (xAM) at (0.9,.35);
                \coordinate (xBM) at (0.9,-.35);
                \coordinate (yA) at (1.9,.7);
                \coordinate (zyA) at (2.3,.7);
                \coordinate (xyA) at (1.5,.7);
                \coordinate (yB) at (1.9,-.7);
                \coordinate (yC) at (1.9,-1.2);
                \coordinate (xC) at (0.9,-1.2);
                \coordinate (zC) at (2.9,-1.2);
                \coordinate (yM) at (1.9,0);
                \coordinate (yAM) at (1.9,.35);
                \coordinate (yBM) at (1.9,-.35);
                \coordinate (zA) at (2.9,.7);
                \coordinate (zB) at (2.9,-.7);
                \coordinate (zM) at (2.9,0);
                \coordinate (zAM) at (2.9,.35);
                \coordinate (zBM) at (2.9,-.35);
                \draw [dotted] (inA) --node [midway, above]{$\rightarrow \vec{e}$}  (outA);
                \draw [photon] (yB) --node [midway, right]{$\downarrow \vec{q}$}  (yC);
                \draw [photon] (xyA) to[out=-45-90,in=180] (yB);
                \draw [photon] (xyA) to (yB);
                
                \draw [zUndirected] (xyA) -- (yA);
                \draw [zUndirected] (yA) -- (zA);
                \draw [photon] (zA) to[out=-90-45,in=0] (yB);
                \draw [photon] (zyA) to (yB);

                \foreach \v in {yB,yC,zA,xyA,zyA}
                \draw[fill] (\v) circle (.08);
        
              \end{tikzpicture}
              \begin{tikzpicture}[baseline={([yshift=-1ex]current bounding box.south)},scale=1.2]
                \coordinate (inA) at (0.4,.7);
                \coordinate (outA) at (3.4,.7);
                \coordinate (inB) at (0.4,-.7);
                \coordinate (outB) at (3.4,-.7);
                \coordinate (xA) at (0.9,.7);
                \coordinate (xB) at (0.9,-.7);
                \coordinate (xM) at (0.9,0);
                \coordinate (xAM) at (0.9,.35);
                \coordinate (xBM) at (0.9,-.35);
                \coordinate (yA) at (1.9,.7);
                \coordinate (zyA) at (2.3,.7);
                \coordinate (xyA) at (1.5,.7);
                \coordinate (yB) at (1.9,-.7);
                \coordinate (yC) at (1.9,-1.2);
                \coordinate (xC) at (0.9,-1.2);
                \coordinate (zC) at (2.9,-1.2);
                \coordinate (yM) at (1.9,0);
                \coordinate (yAM) at (1.9,.35);
                \coordinate (yBM) at (1.9,-.35);
                \coordinate (zA) at (2.9,.7);
                \coordinate (zB) at (2.9,-.7);
                \coordinate (zM) at (2.9,0);
                \coordinate (zAM) at (2.9,.35);
                \coordinate (zBM) at (2.9,-.35);
                \draw [dotted] (inA) --node [midway, above]{$\rightarrow \vec{e}$}  (outA);
                \draw [photon] (yB) --node [midway, right]{$\downarrow \vec{q}$}  (yC);
                \draw [photon] (yA) to[out=-90-45,in=90+45] (yB);
                \draw [photon] (yA) to[out=-45,in=45] (yB);
                \draw [photon] (xA) to[out=-45,in=180] (yB);
                \draw [zUndirected] (xA) -- (yA);
                \draw [zUndirected] (yA) -- (zA);
                \draw [photon] (zA) to[out=-90-45,in=0] (yB);
                \draw [photon] (zA) to[out=-90-45,in=0] (yB);

                \foreach \v in {yB,xA,yC,zA,yA}
                \draw[fill] (\v) circle (.08);
        
              \end{tikzpicture}
              \caption{The eight potential boundary integrals at 4PM order.}
              \label{fig:4pmPot}
\end{figure}
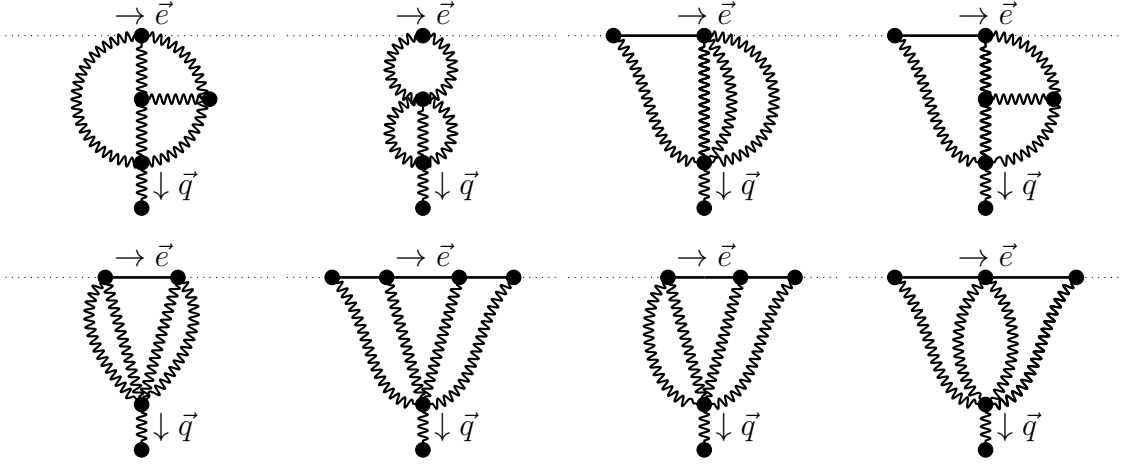

In the potential region one simply expands the integrand in $v$. The leading behaviour is
\begin{align}
        &{\mathcal{P}^{\text{4PM-1SF}}}^{\{\sigma_1, \sigma_2,\sigma_3,\sigma_4,\sigma_5 \}}_{1,1,1,a_1,...,a_{12}} = {\mathcal{P}^{\text{4PM-0SF}}}^{\{\sigma_1, \sigma_2,\sigma_3\}}_{1,1,1,a_1,...,a_{12}}+\mathcal{O}(v^{-a_1-a_2-a_3+1}).
\end{align}
The resulting integrals belong again to the 0SF integral family (see App.~\ref{App:0SFFam}). After IBP reduction we are left with eight boundary integrals, depicted in Fig.~\ref{fig:4pmPot}; for some of them several distinct combinations of worldline $i0$ prescriptions are possible. 
\subsubsection*{Radiation reaction region PR$+$RP}
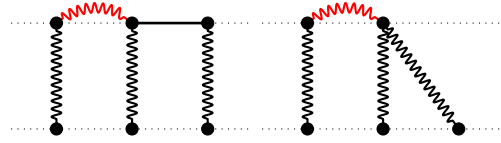
\begin{figure}[h!]\centering
        \begin{tikzpicture}[baseline={([yshift=-1ex]current bounding box.south)},scale=1]
          \coordinate (inA) at (0.4,.7);
          \coordinate (outA) at (3.6,.7);
          \coordinate (inB) at (0.4,-.7);
          \coordinate (outB) at (3.6,-.7);
          \coordinate (xA) at (1,.7);
          \coordinate (xxA) at (1.825,.7) ;
          \coordinate (xxB) at (1.825,-.7) ;
          \coordinate (yA) at (1.60+0.4,.7);
          \coordinate (yyA) at (2.375,0) ;
          \coordinate (zA) at (3,.7);
          \coordinate (zzA) at (3.75,.7) ;
          \coordinate (xB) at (1,-.7);
          \coordinate (yB) at (1.6+0.4,-.7);
          \coordinate (zB) at (3,-.7);
          \coordinate (zzB) at (3.75,-.7);
          \coordinate (xM) at (1,0);
          \coordinate (yM) at (1.6+0.4,0);
          \coordinate (zM) at (3,-0);
          \draw [dotted] (inA) -- (outA);
          \draw [dotted] (inB) -- (outB) node [midway, below]{} ;
          \draw [draw=none] (xA) to[out=40,in=140] (zA);
          \draw [photon] (xA) -- (xB);
          \draw [photon] (zA) -- (zB);
          \draw [zUndirected] (zA) -- (yA);
          \draw [photon,red][out=40,in=140] (xA) to[out=40,in=140] (yA);
          [out=40,in=140] (xA) to[out=40,in=140]
          \draw [photon] (yA) -- (yB);
        
    \draw [fill] (yB) circle (.08);
    \draw [fill] (yA) circle (.08);
          \draw [fill] (xA) circle (.08);
          \draw [fill] (xB) circle (.08);
          \draw [fill] (zA) circle (.08);
          \draw [fill] (zB) circle (.08);
        \end{tikzpicture}
        \begin{tikzpicture}[baseline={([yshift=-1ex]current bounding box.south)},scale=1]
                \coordinate (inA) at (0.4,.7);
                \coordinate (outA) at (3.6,.7);
                \coordinate (inB) at (0.4,-.7);
                \coordinate (outB) at (3.6,-.7);
                \coordinate (xA) at (1,.7);
                \coordinate (xxA) at (1.825,.7) ;
                \coordinate (xxB) at (1.825,-.7) ;
                \coordinate (yA) at (1.60+0.4,.7);
                \coordinate (yyA) at (2.375,0) ;
                \coordinate (zA) at (3,.7);
                \coordinate (zzA) at (3.75,.7) ;
                \coordinate (xB) at (1,-.7);
                \coordinate (yB) at (1.6+0.4,-.7);
                \coordinate (zB) at (3,-.7);
                \coordinate (zzB) at (3.75,-.7);
                \coordinate (xM) at (1,0);
                \coordinate (yM) at (1.6+0.4,0);
                \coordinate (zM) at (3,-0);
                \draw [dotted] (inA) -- (outA);
                \draw [dotted] (inB) -- (outB) node [midway, below]{} ;
                \draw [draw=none] (xA) to[out=40,in=140] (zA);
                \draw [photon] (xA) -- (xB);
                \draw [photon] (yA) -- (zB);
                \draw [photon,red][out=40,in=140] (xA) to[out=40,in=140] (yA);
                [out=40,in=140] (xA) to[out=40,in=140]
                \draw [photon] (yA) -- (yB);
              
          \draw [fill] (yB) circle (.08);
          \draw [fill] (yA) circle (.08);
                \draw [fill] (xA) circle (.08);
                \draw [fill] (xB) circle (.08);
                \draw [fill] (zB) circle (.08);
              \end{tikzpicture}
              \caption{The two relevant topologies for the $RP$ and $PR$ region. In both sectors we find one master integral for both B- and V-type.}
              \label{fig:PR4pm}
\end{figure}

The radiation-reaction region is obtained as before, by rescaling one of the two active propagators by $v$ and then expanding. The expansion in the $RP$ region reads
\begin{align}
       & {\mathcal{P}^{\text{4PM-1SF}}}^{\{\sigma_1, \sigma_2,\sigma_3,\sigma_4,\sigma_5 \}}_{1,1,1,a_1,...,a_{12}} = \nn \\
        &(-1)^{a_1+a_2}
        {\mathcal{P}^{\text{4PM-1SF}}}^{\{0, -\sigma_2,0,-\sigma_4,0 \}}_{1,1,1,a_1+a_3,a_2,0,a_4,0,a_5+a_6,a_7+a_9,a_8,0,a_{10}+a_{12},a_{11},0} +\mathcal{O}(v^{d-a_1-a_2-a_3-2 a_4}). \label{eq:4pmPRexp}
\end{align}
The $PR$ region follows from the same expansion after relabelling. As at 3PM, this region scales as $v^d \sim v^{-2\epsilon}$. We find a total of two master integrals for the B-type and two for the V-type, distributed across the topologies of Fig.~\ref{fig:PR4pm}; these are again mushroom integrals, here with one additional loop attached. The $i0$ prescription on the first worldline does not matter, but we must keep track of $\sigma_2$ in~\refeq{eq:4pmPRexp}.

\subsubsection*{Tail RR}
\begin{figure}\centering
        \begin{tikzpicture}[baseline={([yshift=-1ex]current bounding box.south)},scale=1]
                \coordinate (inA) at (0.4,.7);
                \coordinate (outA) at (3.6,.7);
                \coordinate (inB) at (0.4,-.7);
                \coordinate (outB) at (3.6,-.7);
                \coordinate (xA) at (1,.7);
                \coordinate (xxA) at (1.825,.7) ;
                \coordinate (xxB) at (1.825,-.7) ;
                \coordinate (yA) at (1.60+0.4,.7);
                \coordinate (yyA) at (2.375,0) ;
                \coordinate (zA) at (3,.7);
                \coordinate (zzA) at (3.75,.7) ;
                \coordinate (xB) at (1,-.7);
                \coordinate (yB) at (1.6+0.4,-.7);
                \coordinate (zB) at (3,-.7);
                \coordinate (zzB) at (3.75,-.7);
                \coordinate (xM) at (1,0);
                \coordinate (yM) at (1.6+0.4,1.1);
                \coordinate (zM) at (3,-0);
                \draw [dotted] (inA) -- (outA);
                \draw [dotted] (inB) -- (outB) node [midway, below]{} ;
                \draw [draw=none] (xA) to[out=40,in=140] (zA);
                \draw [photon] (zA)  --node [midway, right]{$\rightarrow$} (zB);
                \draw [photon] (xA) -- (xB);
                \draw [photon] (yA) -- (yM);
                \draw [photon,red][out=40,in=140] (xA) to[out=40,in=140] (zA);
       
          \draw [fill] (xA) circle (.08);
          \draw [fill] (yA) circle (.08);
          \draw [fill] (yM) circle (.08);
                \draw [fill] (zA) circle (.08);
                \draw [fill] (xB) circle (.08);
                \draw [fill] (zB) circle (.08);
              \end{tikzpicture}  
              \begin{tikzpicture}[baseline={([yshift=-1ex]current bounding box.south)},scale=1]
                \coordinate (inA) at (0.4,.7);
                \coordinate (outA) at (3.6,.7);
                \coordinate (inB) at (0.4,-.7);
                \coordinate (outB) at (3.6,-.7);
                \coordinate (xA) at (1,.7);
                \coordinate (xxA) at (1.825,.7) ;
                \coordinate (xxB) at (1.825,-.7) ;
                \coordinate (yA) at (1.60+0.4,.7);
                \coordinate (yyA) at (2.375,0) ;
                \coordinate (zA) at (3,.7);
                \coordinate (zzA) at (3.75,.7) ;
                \coordinate (xB) at (1,-.7);
                \coordinate (yB) at (1.6+0.4,-.7);
                \coordinate (zB) at (3,-.7);
                \coordinate (zzB) at (3.75,-.7);
                \coordinate (xM) at (1,0);
                \coordinate (yM) at (1.6+0.4,1.1);
                \coordinate (zM) at (3,-0);
                \draw [dotted] (inA) -- (outA);
                \draw [dotted] (inB) -- (outB) node [midway, below]{} ;
                \draw [draw=none] (xA) to[out=40,in=140] (zA);
                \draw [photon] (zA) -- (zB);
                \draw [photon] (xA) -- (xB);
                
                \draw [photon,red][out=40,in=140] (xA) to[out=40,in=140] (yA);
                \draw [photon,red][out=40,in=140] (yA) to[out=40,in=140] (zA);
       
          \draw [fill] (xA) circle (.08);
          \draw [fill] (yA) circle (.08);
                \draw [fill] (zA) circle (.08);
                \draw [fill] (xB) circle (.08);
                \draw [fill] (zB) circle (.08);
              \end{tikzpicture}
              \caption{The RR region has two MIs, which are both in the sector depicted on the RHS. We get for B- and V-type one MI each. Tail-type graphs like those on the LHS get IBP-reduced to the double-mushroom topology.} \label{fig:4pmRRBound}
\end{figure}
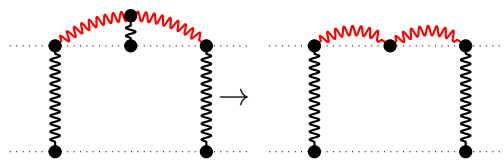

The expansion in the new tail region is obtained by rescaling \emph{both} active propagators $D_4$ and $D_5$ by $v$:
\begin{align}
        {\mathcal{P}^{\text{4PM-1SF}}}^{\{\sigma_1, \sigma_2,\sigma_3,\sigma_4,\sigma_5 \}}_{1,1,1,a_1,...,a_{12}} =& \nonumber \\
        &(-1)^{a_1+a_2} {\mathcal{P}^{\text{4PM-1SF}}}^{\{0, 0,0 ,\sigma_4,\sigma_5\}}_{\tiny{1,1,1,0,0,a_1+a_2+a_3,a_4,a_5,a_6,0,0,a_7+a_8+a_9,0,0,a_{10}+a_{11}+a_{12}}} \nonumber \\ &
        +\mathcal{O}(v^{2d-a_1-a_2-a_3-2 (a_4+a_5+a_6)-1}).
\end{align}
This expansion has an associated velocity scaling of $v^{-4\epsilon}$. Note that $D_6$ now also scales as $\sim v^2$ even though its graviton remains potential. The name of the region reflects the type of boundary integrals that appear: a radiative graviton interacts with a potential graviton, as shown on the LHS in Fig.~\ref{fig:4pmRRBound}. After IBP reduction all these integrals reduce to double mushrooms. We find for the B- and V-type families a single master each, in the topology shown on the RHS in Fig.~\ref{fig:4pmRRBound}.

\chapter{Integrals at 5PM-1SF order}
\label{sec:5pm1sf}
We will now look at the fifth Post-Minkowskian order and its integrals. This is the first order at which second self-force order effects play a role. As before, the different self-force order contributions are separate computations and we will first look at the 1SF integrals here. They were first computed fully in~\cite{Driesse:2024xad,Driesse:2024feo}, work to which the present author contributed. The B-type integrals were also computed for the 5PM scattering angle in supergravity~\cite{Bern:2024adl} and they were later also used for the 1SF part of the amplitudes computation~\cite{Bern:2025wyd}.
 
\section{The 5PM-1SF integral family}
At 5PM-1SF every integral can be brought into the family
        \begin{align}
          {\mathcal{P}^{\text{5PM-1SF}}}^{(\sigma_1,\sigma_2,\ldots,\sigma_7)}_{n_1,n_2,\ldots,n_{22}}
          :=
          \int_{\ell_1,\ell_2,\ell_3,\ell_4}
          \frac{
            \dd^{(n_1-1)}(\ell_1\cdot v_{1})
            \dd^{(n_2-1)}(\ell_2\cdot v_{1})
            \dd^{(n_3-1)}(\ell_3\cdot v_{1})
            \dd^{(n_4-1)}(\ell_4\cdot v_{2})
          }{
            D_1^{n_5}
            D_2^{n_6}
            ...
            D_{18}^{n_{22}}
          }
        \end{align}
        with
        \begin{align}
                & D_{1}
          =
          \ell_{1}\cdot v_{2} + \sigma_{1}\i0,\;
          \nn
          D_{2}
          =
          \ell_{2}\cdot v_{2} +\sigma_{2}\i0, \;
          \nn
          D_{3}
          =
          \ell_{3}\cdot v_{2} +\sigma_{3}\i0, \;
          \nn
          D_{4}
          =
          \ell_{4}\cdot v_{1} + \sigma_{4}\i0, \;
          \nn
          \\
          &D_{5/6/7}
          =
          (\ell_{1/2/3}-\ell_4)^2+\sigma_{5/6/7} \;\sign{(\ell_{1/2/3}-\ell_4)\cdot v_1} \; \i0 , \;
          \nn
          \\
          &D_{8/9}
          =
          (\ell_1-\ell_{2/3})^2, \;
          D_{10}
          =
          (\ell_2-\ell_{3})^2, \;
          \nn
          \\
          & D_{11/12/13/14}
          =\ell_{1/2/3/4}^2, \;
          D_{15/16/17/18}
          =(\ell_{1/2/3/4}-q)^2 \;.
          \nn
        \end{align}
The integrals have a total of 22 indices and a maximum of 13 propagators with 9 ISPs.

We pass to a frame to reveal $D_{5}$, $D_{6}$, $D_{7}$ to be the relevant active propagators of this family. The family reads in this form
      \begin{align}
        {{\cal P}^{\text{5PM-1SF}}}_{\{1,1,1,1,n\}}^{\{\sigma\}}
                =\dfrac{(-1)^{\sum_{i=5,i\neq 8}^{22} n_i}}{{\sqrt{\gamma^2-1}}^{n_5+n_6+n_7+n_8}}  \int_{\boldsymbol{\ell_1}\boldsymbol{\ell_2}\boldsymbol{\ell_3}\boldsymbol{\ell_4}} \dfrac{1}{\prod_{i=1}^{18} \hat{D}_i^{n_{i+4}}}, 
        \end{align}
      where the 18 propagators are given as
      \begin{align}
       & \hat{D}_{1/2/3/4}=\boldsymbol{\ell}_{1/2/3/4}\cdot \boldsymbol{e}, \hat{D}_{5/6/7}=(\boldsymbol{\ell}_{1/2/3}- \boldsymbol{\ell}_{4})^2-2 (\gamma-1)(\boldsymbol{\ell}_{1/2/3}\cdot \boldsymbol{e})\;(\boldsymbol{\ell}_{4}\cdot \boldsymbol{e}),\nn \\
       &  \hat{D}_{8/9}=(\boldsymbol{\ell}_{1}- \boldsymbol{\ell}_{2/3})^2,\hat{D}_{10}=(\boldsymbol{\ell}_{2}- \boldsymbol{\ell}_{3})^2,
        \hat{D}_{11/12/13/14}=(\boldsymbol{\ell}_{1/2/3/4})^2,\hat{D}_{15/16/17/18}=(\boldsymbol{\ell}_{1/2/3/4}- \boldsymbol{q})^2.
\end{align}
For the actual computations we will use again the form with explicit delta functions.
\section{The 4-loop IBP reduction}

The principal idea of IBP reduction was discussed in Section~\ref{sec:IBPs}. Normally there is little reason to worry about it, since dedicated programs are very fast and may be used essentially as black boxes. The Laporta algorithm itself is conceptually simple---one writes down a large set of equations and solves the resulting linear system---so that the only apparent complications are those of an efficient implementation in a user-friendly program.

At fifth Post-Minkowskian order, however, we are faced with four loop integrals carrying 22 indices, and the rank of the integrand grows steadily with each loop order. Whereas at 3PM the maximum numerator power is five, at 5PM we encounter integrals of rank up to nine. A naive attempt to feed such integrals into a standard IBP program rapidly runs into a serious obstacle: the number of equations becomes so large that the program runs out of memory and crashes. The step that previously required no thought thus becomes one of the principal bottlenecks of the calculation.

Of course, this is not the first calculation in which the IBP reduction itself becomes a technical challenge. Analogous problems arise routinely in collider physics, and a number of strategies for shrinking the linear system have been developed in that context. Popular approaches include the use of Syzygy equations~\cite{Kosower:2018obg, Smith:2025xes,Smirnov:2025prc,Gluza:2010ws, Schabinger:2011dz, Ita:2015tya, Larsen:2015ped, Abreu:2017xsl, Abreu:2017hqn, Bohm:2018bdy, Bendle:2019csk, Wu:2023upw, Wu:2025aeg, Page:2025gso}, the construction of block-triangular systems~\cite{Liu:2018dmc, Guan:2019bcx, Guan:2024byi} and machine learning~\cite{Zeng:2025xbh,vonHippel:2025okr,Berman:2026jgq,Song:2025pwy}. 

We adopt a different strategy. As a starting point, we employ the improved equation-selector algorithm implemented in \texttt{Kira} 3.0, see also ref.~\cite{Lange:2025fba} for details. The Laporta algorithm seeds far more equations than are strictly necessary, since a much smaller subset would already suffice to reduce all integrals to a minimal set of master integrals. The improved selector does not single out the optimal subset, but it produces a system that is already orders of magnitude smaller, and which scales well to four loops.

Even this is not, however, sufficient on its own. To see what further improvements are needed, consider the example topology shown below.
\begin{figure}[h!]
        \centering
        \begin{tikzpicture}[baseline={([yshift=-1ex]current bounding box.south)},scale=1.2]
                \coordinate (inA) at (0.3,.7);
                 \coordinate (outA) at (3.6,.7);
                 \coordinate (inB) at (0.3,-.7);
                 \coordinate (outB) at (3.6,-.7);
                 \coordinate (xA) at (0.6,.7);
                 \coordinate (yA) at (1.5,.7);
                 \coordinate (zA) at (2.4,.7);
                 \coordinate (wA) at (3.3,.7);
                 \coordinate (xB) at (0.6,-.7);
                 \coordinate (yB) at (1.5,-.7);
                 \coordinate (zB) at (2.4,-.7);
                 \coordinate (wB) at (3.3,-.7);
                 \coordinate (xM) at (0.6,0);
                 \coordinate (yM) at (1.5,0);
                 \coordinate (zM) at (2.4,0);
                 \coordinate (wM) at (3.3,0);
                 \coordinate (xyMB) at (1.05,-.35);
                 \coordinate (yzMB) at (1.95,-.35);
                 \coordinate (zwMB) at (2.85,-.35);
                 \coordinate (yzMA) at (1.95,.35);
                 \draw [dotted] (inA) -- (outA);
                 \draw [dotted] (inB) -- (outB);
                 \draw [photon] (xA) -- (xB) ;
                 \draw [photon] (wA) -- (wB);
                 \draw [photon,red] (xM) -- (yM);
                 \draw [photon,red] (yM) -- (zM);
                 \draw [photon,red] (zM) -- (wM);
                 \draw [photon] (yB) -- (yM);
                 \draw [photon] (zB) -- (zM);
                 \draw [fill] (xA) circle (.08);
                 \draw [fill] (wA) circle (.08);
                 \draw [fill] (xB) circle (.08);
                 \draw [fill] (yB) circle (.08);
                 \draw [fill] (zB) circle (.08);
                 \draw [fill] (wB) circle (.08);  
                 \draw [fill] (xM) circle (.08);
                 \draw [fill] (yM) circle (.08);
                 \draw [fill] (zM) circle (.08);  
                 \draw [fill] (wM) circle (.08);
        
               \end{tikzpicture}   
\end{figure}

In our index notation this topology corresponds to 
\begin{align}
        \text{P}^{5,1}(1,1,1,1,0,0,0,0,1,1,1,1,0,1,1,0,0,1,0,0,1,1).
\end{align}
To reduce an integral of rank nine, for example 
\begin{align}
        \text{P}^{5,1}(1,1,1,1,-9,0,0,0,1,1,1,1,0,1,1,0,0,1,0,0,1,1),
\end{align}
 the standard procedure is to seed every integral with up to nine numerators in this topsector and in all of its subsectors. The topsector alone then contributes $\sim 9^9 \sim 10^8$ seed integrals, while in the lowest non-trivial sector---which contains only nine propagators---the count rises to $\sim 13^9 \sim 10^{10}$. Generating all of these equations simultaneously in order to apply the equation selector exhausts the available memory long before the seeding is complete.

It is, in fact, unnecessary to seed so many integrals, particularly in the lower sectors. To reduce an integral of rank nine with 13 propagators it suffices to lower the maximum rank of the seeds by one for each propagator that is removed. Concretely, the topsector is seeded with up to nine numerators, the subsectors with 12 propagators only up to eight, and so on down to the lowest non-trivial subsectors with nine propagators, which require only up to five numerators. The lowest sector then needs $\sim 13^5 \sim 10^6$ rather than $\sim 13^9 \sim 10^{10}$ seed integrals, a saving of four orders of magnitude in that sector and of roughly two orders of magnitude overall. This trick of cutting seeds is generally applicable to Feynman integrals, and is also implemented in the \texttt{Kira} 3.0 algorithm.

Unfortunately these improvements would still not be efficient enough to reduce the 5PM-1SF integrand. We have to make more integral family specific optimisations. To understand these integrals better we look at topologies like these, which have one or more worldlines:
\begin{align*}
        \begin{tikzpicture}[baseline={([yshift=-1ex]current bounding box.south)},scale=1.2]
                \coordinate (inA) at (0.3,.7);
                 \coordinate (outA) at (3.6,.7);
                 \coordinate (inB) at (0.3,-.7);
                 \coordinate (outB) at (3.6,-.7);
                 \coordinate (1M) at (0.6,0);
                 \coordinate (2M) at (1.3,0);
                 \coordinate (3M) at (2.0,0);
                 \coordinate (4M) at (2.7,0);
                 \coordinate (5M) at (3.4,0);
                 \coordinate (xA) at (0.6,.7);
                 \coordinate (yA) at (1.5,.7);
                 \coordinate (zA) at (2.4,.7);
                 \coordinate (wA) at (3.3,.7);
                 \coordinate (xB) at (0.6,-.7);
                 \coordinate (yB) at (1.5,-.7);
                 \coordinate (zB) at (2.4,-.7);
                 \coordinate (wB) at (3.3,-.7);
                 \coordinate (xM) at (0.6,0);
                 \coordinate (yM) at (1.5,0);
                 \coordinate (zM) at (2.4,0);
                 \coordinate (wM) at (3.3,0);
                 \coordinate (xyMB) at (1.05,-.35);
                 \coordinate (yzMB) at (1.95,-.35);
                 \coordinate (zwM) at (2.85,0);
                 \coordinate (yMA) at (1.5,.35);
                 \draw [dotted] (inA) -- (outA);
                 \draw [dotted] (inB) -- (outB);
                 \draw [photon] (xA) -- (xB);
                 \draw [photon] (wA) -- (wB);
                 \draw [photon,red] (xM) -- (zM);
                 \draw [photon] (yB) -- (yM);
                 \draw [photon] (zB) -- (zM);
                 \draw [photon,red] (zM) -- (zA);
                 \draw [zUndirected] (zA) -- (wA);
                 \draw [fill] (xA) circle (.08);
                 \draw [fill] (zA) circle (.08);
                 \draw [fill] (wA) circle (.08);
                 \draw [fill] (xB) circle (.08);
                 \draw [fill] (yB) circle (.08);
                 \draw [fill] (zB) circle (.08);
                 \draw [fill] (wB) circle (.08);  
                 \draw [fill] (xM) circle (.08);
                 \draw [fill] (yM) circle (.08);
                 \draw [fill] (zM) circle (.08);  
               \end{tikzpicture}
               \begin{tikzpicture}[baseline={([yshift=-1ex]current bounding box.south)},scale=1.2]
                \coordinate (inA) at (0.3,.7);
                 \coordinate (outA) at (3.6,.7);
                 \coordinate (inB) at (0.3,-.7);
                 \coordinate (outB) at (3.6,-.7);
                 \coordinate (xA) at (0.6,.7);
                 \coordinate (yA) at (1.5,.7);
                 \coordinate (zA) at (2.4,.7);
                 \coordinate (wA) at (3.3,.7);
                 \coordinate (xB) at (0.6,-.7);
                 \coordinate (yB) at (1.5,-.7);
                 \coordinate (zB) at (2.4,-.7);
                 \coordinate (wB) at (3.3,-.7);
                 \coordinate (xM) at (0.6,0);
                 \coordinate (yM) at (1.5,0);
                 \coordinate (zM) at (2.4,0);
                 \coordinate (wM) at (3.3,0);
                 \coordinate (xyM) at (1.05,0);
                 \coordinate (zwM) at (2.85,0);
                 \coordinate (yzMA) at (1.95,.35);
                 \coordinate (yzMB) at (1.95,-.35);
                 \draw [dotted] (inA) -- (outA);
                 \draw [dotted] (inB) -- (outB);
                 \draw [photon] (xA) -- (xB);
                 \draw [photon] (wA) -- (wB);
                 \draw [photon,red] (yM) -- (zM);
                 \draw [photon,red] (yM) -- (yA);
                 \draw [photon] (yB) -- (yM);
                 \draw [photon] (zB) -- (zM);
                 \draw [photon,red] (zM) -- (zA);
                 \draw [zUndirected] (zA) -- (wA);
                 \draw [zUndirected] (xA) -- (yA);
                 \draw [fill] (xA) circle (.08);
                 \draw [fill] (yA) circle (.08);
                 \draw [fill] (zA) circle (.08);
                 \draw [fill] (wA) circle (.08);
                 \draw [fill] (xB) circle (.08);
                 \draw [fill] (yB) circle (.08);
                 \draw [fill] (zB) circle (.08);
                 \draw [fill] (wB) circle (.08);  
                 \draw [fill] (yM) circle (.08);
                 \draw [fill] (zM) circle (.08); 
               \end{tikzpicture}
\end{align*}

Such topologies arise naturally from the linear worldline propagators. Our Feynman rules dictate that these propagators always appear squared, so one generically expects one dot per worldline. Although such topologies are in principle easier to reduce---they involve fewer master integrals than the all-graviton ones---the dots make their IBP reduction more expensive. Mushroom-type topologies of the form
\begin{align*}
        \begin{tikzpicture}[baseline={([yshift=-1ex]current bounding box.south)},scale=1]
                \coordinate (inA) at (0.4,.7);
                \coordinate (outA) at (4.4,.7);
                \coordinate (inB) at (0.4,-.7);
                \coordinate (outB) at (4.4,-.7);
                \coordinate (xA) at (0.9,.7);
                \coordinate (xB) at (0.9,-.7);
                \coordinate (yA) at (1.9,.7);
                \coordinate (yB) at (1.9,-.7);
                \coordinate (yyA) at (2.9,.7);
                \coordinate (yyB) at (2.9,-.7);
                \coordinate (zA) at (3.9,.7);
                \coordinate (zB) at (3.9,-.7);
                \draw [dotted] (inA) -- (outA);
                \draw [dotted] (inB) -- (outB) node [midway, below]{} ;
                \draw [draw=none] (xA) to[out=40,in=140] (zA);
                \draw [photon] (xA) -- (xB);
                \draw [photon,red][out=40,in=140] (yB) to[out=40,in=140] (yyB);
                \draw [photon] (zA) -- (zB)node [midway, right]{$\qquad =$};
                \draw [zUndirected] (xB) -- (yB);
                \draw [zUndirected] (zB) -- (yyB);
                \draw [fill] (xA) circle (.08);
                \draw [fill,blue!70] (1.4,-.7) circle (.09);
                \draw [fill,blue!70] (3.4,-.7) circle (.09);
                \draw [fill] (yB) circle (.08);
                \draw [fill] (zA) circle (.08);
                \draw [fill] (zB) circle (.08);
                \draw [fill] (yyB) circle (.08);
                \draw [fill] (xB) circle (.08);
              \end{tikzpicture}
              \begin{tikzpicture}[baseline={([yshift=-1ex]current bounding box.south)},scale=1]
                \coordinate (inA) at (0.4,.7);
                \coordinate (outA) at (4.4,.7);
                \coordinate (inB) at (0.4,-.7);
                \coordinate (outB) at (4.4,-.7);
                \coordinate (xA) at (0.9,.7);
                \coordinate (xB) at (0.9,-.7);
                \coordinate (yA) at (1.9,.7);
                \coordinate (yB) at (1.9,-.7);
                \coordinate (yyA) at (2.9,.7);
                \coordinate (yyB) at (2.9,-.7);
                \coordinate (zA) at (3.9,.7);
                \coordinate (zB) at (3.9,-.7);
                \draw [dotted] (inA) -- (outA);
                \draw [dotted] (inB) -- (outB) node [midway, below]{} ;
                \draw [draw=none] (xA) to[out=40,in=140] (zA);
                \draw [photon] (xA) -- (xB);
                \draw [photon,red][out=40,in=140] (yB) to[out=40,in=140] (yyB);
                \draw [photon] (yyA) -- (yyB)node [midway, right]{};
                \draw [zUndirected] (xB) -- (yB);    
                \draw [fill] (xA) circle (.08);
                \draw [fill,blue!70] (1.4,-.7) circle (.09);
                \draw [fill,blue!70] (1.65,-.7) circle (.09);
                \draw [fill,blue!70] (1.15,-.7) circle (.09);
                \draw [fill] (yB) circle (.08);
                \draw [fill] (yyB) circle (.08);
                \draw [fill] (yyA) circle (.08);
                \draw [fill] (xB) circle (.08);
              \end{tikzpicture}
\end{align*}
are particularly problematic. Both worldlines carry the same energy, so that one ends up with three dots on the worldline; in the extreme case of a triple mushroom graph as many as seven dots can appear.

We are thus forced to reduce integrals carrying both dots and scalar products simultaneously. A useful structural observation comes to our aid. The more dots an integral carries, the fewer propagators it has, since each pair of fused propagators is replaced by a single propagator with two extra dots. Moreover, dots can only sit on the worldline propagators, and the highest powers of scalar products arise only from the linear ones. Since the mass dimension of a linear propagator is half that of a graviton propagator, nearly twice as many scalar products can be built from them.

So even though linear propagators are much easier structurally for the IBP reduction, they cause problems because of the structure of the integrand. Two observations follow. First, in view of their reduced mass dimension, the linear propagators must behave differently in the IBP vectors, since the raising of a graviton propagator can only be compensated by the lowering of two linear ones. Secondly, $\gamma$ can enter an IBP vector only through the identities generated by $v_i \cdot \partial/\partial \ell_i$, which always raise or lower linear propagators.

This suggests that linear and graviton propagators should be seeded differently and independently of one another. Doing so leads to a remarkable observation. When reducing nine linear scalar products in the all-graviton topology shown above, one must indeed seed up to nine scalar products on the four worldline propagators, but it suffices to seed only up to two on the gravitons. The total seeding effort then drops from $\sim 9^9 \sim 10^8$ to $\sim 4^9 \cdot 2^5 \sim 10^6$ integrals, a further two orders of magnitude. Similarly, when seeding dots, it is enough to place them on the worldline propagators alone. We do not have a first-principles derivation of why this is sufficient at the level of IBP vectors, and for the moment the statement remains an empirical observation; nevertheless it works, and we exploit it. Combined with the cutting of numerator powers in the lower sectors, this trick yields a reduction of several orders of magnitude in the total number of equations.

These ingredients together provide a working setup for the full IBP reduction of the 5PM-1SF integrand. The reduction itself still demands considerable computational resources. In order to exploit the observations above and to make full use of parallelisation, we split the integrand into many independent reduction jobs. The first split is between B- and V-type integrals, which are decoupled under IBP reduction. A further split is performed by topsector, with integrals lying in lower sectors pushed into the job files of the simplest topsector that contains them; then according to whether the integral carries a high number of scalar products on the linear or on the graviton propagators; and finally according to the number of dots. The resulting jobs are run in parallel on a computer cluster, and the full reduction consumes roughly $300\mathrm{k}$ CPU-core hours. Similar improvements and strategies were also developed independently for the 5PM calculation using the amplitudes approach~\cite{Bern:2024adl,Bern:2025wyd,Bern:2025zno,Smirnov:2025prc}.

After the IBP reduction the remaining obstacle is again computing the MIs via the method of differential equations.
\section{The K3 surface revisited at 5PM order}
It is hardly surprising that at four loops we again encounter integrals giving rise to periods of non-trivial geometries. Specifically, we find eight sectors---some of them related to one another by symmetry---which all yield the same K3 period that we already met at 4PM-1SF. Four of these sectors carry three master integrals each, and their canonicalisation proceeds exactly as at 4PM: one identifies the integral associated to the differential of the first kind, passes to the full derivative basis, performs the semi-simple/unipotent split of the Wronskian, and finishes off with the necessary corrections by hand.

The remaining four sectors each carry four master integrals. We saw something analogous in Chapter~\ref{main} for the two loop sunrise graph, where one had to choose an integral associated to an elliptic integral of the third kind. For that simple example the third-kind integral could be identified by direct integrand analysis. In the present case, however, the situation is more delicate. At four loops, and with a more involved underlying geometry, integrand analysis is no longer always feasible in practice, and we therefore proceed as follows. We first identify those integrals which map onto the elliptic differential of the first kind. These are typically easy to spot---they tend to be the corner integrals, or to require only a single numerator insertion of $\ell_4\cdot v_1$---and we use them, together with their first and second derivatives, as three of the four master integrals. The fourth master is then chosen by the following criterion. After IBP reduction, the K3 block, in the appropriate normalisation, should decouple from the fourth integral at $\epsilon = 0$. In practice we simply test candidate integrals until we find one which satisfies this condition.

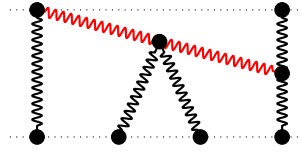
\begin{figure}[h!]
        \centering
        \begin{tikzpicture}[baseline={([yshift=-1ex]current bounding box.south)},scale=1.2]
                \coordinate (inA) at (0.3,.7);
                 \coordinate (outA) at (3.6,.7);
                 \coordinate (inB) at (0.3,-.7);
                 \coordinate (outB) at (3.6,-.7);
                 \coordinate (xA) at (0.6,.7);
                 \coordinate (yA) at (1.5,.7);
                 \coordinate (zA) at (2.4,.7);
                 \coordinate (wA) at (3.3,.7);
                 \coordinate (xB) at (0.6,-.7);
                 \coordinate (yB) at (1.5,-.7);
                 \coordinate (zB) at (2.4,-.7);
                 \coordinate (wB) at (3.3,-.7);
                 \coordinate (xM) at (0.6,0);
                 \coordinate (yM) at (1.95,0.35);
                 \coordinate (zM) at (2.4,0);
                 \coordinate (wM) at (3.3,0);
                 \coordinate (xyMB) at (1.05,-.35);
                 \coordinate (yzMB) at (1.95,-.35);
                 \coordinate (zwMB) at (2.85,-.35);
                 \coordinate (yzMA) at (1.95,.35);
                 \draw [dotted] (inA) -- (outA);
                 \draw [dotted] (inB) -- (outB);
                 \draw [photon] (xA) -- (xB) ;
                 \draw [photon] (wA) -- (wB);
                 \draw [photon,red] (xA) -- (wM);
                 \draw [photon] (yB) -- (yM);
                 \draw [photon] (zB) -- (yM);
                 \draw [fill] (xA) circle (.08);
                 \draw [fill] (wA) circle (.08);
                 \draw [fill] (xB) circle (.08);
                 \draw [fill] (yB) circle (.08);
                 \draw [fill] (zB) circle (.08);
                 \draw [fill] (wB) circle (.08);  
                 \draw [fill] (yM) circle (.08);
                 \draw [fill] (wM) circle (.08);
        
               \end{tikzpicture}   
               \caption{A sector at 5PM-1SF giving rise to the K3 surface encountered at 4PM}\label{fig:535pm1sfsec}
\end{figure}
We illustrate the procedure on one such sector (see Fig.~\ref{fig:535pm1sfsec}), choosing as basis
\begin{align}
       & I_1=\text{P}^{5,1}(1, 1, 1, 1, 0, 0, 0, -1, 1, 0, 1, 1, 0, 1, 1, 0, 0, 0, 0, 0, 1, 1)
        , \quad I_2 =   \partial_x I_1 \, , \quad I_3 = \partial_x^2 I_1 \,, \quad \nn\\
        &I_4 =\text{P}^{5,1}(1, 1, 2, 1, 0, 0, -1, -1, 1, 0, 1, 1, 0, 1, 1, 0, 0, 0, 0, 0, 1, 1) \, 
\end{align}
where $\vec I = (I_1,I_2,I_3,I_4)$. In this basis the differential equations read
\begin{align}
        M(x;\epsilon)=    \left(
\begin{array}{cccc}
 0 & 1 & 0 & 0 \\
 0 & 0 & 1 & 0 \\
 g(x) & \frac{x^4 \left(-4 \epsilon ^2+20 \epsilon -19\right)+8 x^2 \left(33 \epsilon ^2+2\right)-(2 \epsilon +1)^2}{x^2 \left(x^2-1\right)^2} & \frac{x^2 (4 \epsilon -9)+4 \epsilon +3}{x \left(x^2-1\right)} & -\frac{480 \epsilon }{x \left(x^2-1\right)^3} \\
 \frac{8 x \epsilon }{x^2-1} & 0 & 0 & 0 \\
\end{array}
\right)
\end{align}
where
\begin{align}
     g(x)=   -\frac{4 \left(2 x^4 (\epsilon -1)^2+x^2 \left(16 \epsilon ^3-68 \epsilon ^2+4 \epsilon -3\right)+16 \epsilon ^3+66 \epsilon ^2+1\right)}{x \left(x^2-1\right)^3}.
\end{align}
The K3 block clearly decouples from the fourth integral. In fact, in this case the fourth integral decouples completely from the K3 block in turn,
\begin{align}
      M(x;0)=  \left(
\begin{array}{cccc}
 0 & 1 & 0 & 0 \\
 0 & 0 & 1 & 0 \\
 \frac{4-8 x^2}{x \left(x^2-1\right)^2} & \frac{-19 x^4+16 x^2-1}{x^2 \left(x^2-1\right)^2} & -\frac{3-9 x^2}{x-x^3} & 0 \\
 0 & 0 & 0 & 0 \\
\end{array}
\right)
\end{align}
This pattern is shared by all four sectors which carry four master integrals, and considerably simplifies the canonicalisation. We rotate the K3 block by the inverse of the semi-simple factor~\refeq{eq:Grav_K3_W}, then by $D(\epsilon)=\text{diag}(\epsilon^2,\epsilon,1,\epsilon^2)$, and finally by
\begin{align}
      T=  \left(
\begin{array}{cccc}
 1 & 0 & 0 & 0 \\
 -\frac{4}{3}  \left(x^2+1\right) \varpi_0 & 1 & 0 & 0 \\
 \frac{2}{9} \left(5 x^4-602 x^2+5\right) \varpi_0^2+\frac{-\frac{8}{3}  x^2 \left(x^2-1\right) \varpi_0^2-\frac{4}{3} x \left(x^4-1\right) \varpi_0 \varpi_0'}{\epsilon } & -\frac{8}{3}  \left(x^2+1\right) \varpi_0 & 1 & 0 \\
 0 & 0 & 0 & 1 \\
\end{array}
\right)
\end{align}
to find the canonical form of this sector:
\begin{align}
        \dfrac{d \vec{J}}{d x} =   \epsilon  \left(
\begin{array}{cccc}
 \frac{4 \left(x^2+1\right)  }{3 x \left(x^2-1\right)} & \frac{ 1}{x \left(x^2-1\right) \varpi_0} & 0 & 0 \\
 \frac{2 \left(x^4+206 x^2+1\right)   \varpi_0}{3 x \left(x^2-1\right)} & \frac{4 \left(x^2+1\right)  }{3 x \left(x^2-1\right)} & \frac{1 }{x \left(x^2-1\right) \varpi_0} & 0 \\
 -\frac{16 \left(x^6-501 x^4-501 x^2+1\right)   \varpi_0^2}{27 x \left(x^2-1\right)} & \frac{2 \left(x^4+206 x^2+1\right)   \varpi_0}{3 x \left(x^2-1\right)} & \frac{4 \left(x^2+1\right)  }{3 x \left(x^2-1\right)} & -\frac{480 x   \varpi_0}{x^2-1} \\
 \frac{8 x   \varpi_0}{x^2-1} & 0 & 0 & 0 \\
\end{array}
\right)\vec{J},
\end{align}
where $\vec J = T\,D\,(W^{\rm SS})^{-1}\vec I$. Notably, although no integrand analysis was performed, $I_4$ required no correction whatsoever---the mere requirement that the K3 block and the fourth integral decouple at $\epsilon = 0$ was sufficient to land on a canonical integral.
\section{A sector related to a CY3 fold at 5PM-1SF}
The K3 surface is not the only non-trivial geometry encountered at 5PM-1SF, which was discussed in~\cite{Klemm:2024wtd}. We also find a sector containing the integral
\begin{align*}
        \begin{tikzpicture}[baseline={([yshift=-1ex]current bounding box.south)},scale=1.2]
                \coordinate (inA) at (0.4,.7);
                \coordinate (outA) at (4.35,.7);
                \coordinate (inB) at (0.4,-.7);
                \coordinate (outB) at (4.35,-.7);
                \coordinate (xA) at (1,.7);
                \coordinate (xxA) at (1.825,.7) ;
                \coordinate (xxB) at (1.825,-.7) ;
                \coordinate (yA) at (1.60,.7);
                \coordinate (yyA) at (2.375,0) ;
                \coordinate (zA) at (2.925,.7);
                \coordinate (zzA) at (3.75,.7) ;
                \coordinate (xB) at (1-0.2,-.7);
                \coordinate (yB) at (1.5,-.7);
                \coordinate (zB) at (2.925,-.7);
                \coordinate (zzB) at (3.75+0.2,-.7);
                \coordinate (xM) at (1,0);
                \coordinate (yM) at (1.5,0);
                \coordinate (zM) at (3,-0);
                \draw [dotted] (inA) -- (outA);
                \draw [dotted] (inB) -- (outB) node [midway, below] {};
                \draw [draw=none] (xA) to[out=40,in=140] (zA);
                \draw [photon] (xxA) -- (xB);
                \draw [photon] (yyA) -- (zB);
                \draw [photon] (zA) -- (zzB);
                \draw [photon] (yyA) -- (xxB);
                \draw [photon,red] (xxA) -- (yyA);
                 \draw [photon,red] (zA) -- (yyA);
                \draw [fill] (xxB) circle (.08);
                \draw [fill] (yyA) circle (.08);
                \draw [fill] (xxA) circle (.08);
                \draw [fill] (zA) circle (.08);
                \draw [fill] (zB) circle (.08);
                \draw [fill] (xB) circle (.08);
                \draw [fill] (zzB) circle (.08);
                \draw [fill,blue!70] (2.375,0.7) circle (.05);
              \end{tikzpicture},        
\end{align*}
which sits inside a $6\times 6$ sector. Using the symmetry split of the previous chapter, this sector decomposes into a $4\times 4$ and a $2\times 2$ block; the integral shown above belongs to the $4\times 4$ block, while the additional two integrals can safely be ignored. In the derivative basis
\begin{align}
        I_1 =& \frac{1  }{x}\text{P}^{5,1}(1,1,1,2,0,0,0,0,1, 0 ,1 ,1, 0, 1, 1 ,0,0 ,0, 0, 0 ,1, 0 ) \, , \quad \nn\\
         I_2 =&   \partial_x I_1 \, , \quad I_3 = \partial_x^2 I_1 \,, \quad I_4 = \partial_x^3 I_1 \, ,
    \end{align}
    the differential equations
    \begin{align}
        \partial_x I=M(x,\epsilon) I \,,
    \end{align}
    take, at $\epsilon = 0$, the form 
    \begin{align}
     M(x;0)=  \left(
        \begin{array}{cccc}
         0 & 1 & 0 & 0 \\
         0 & 0 & 1 & 0 \\
         0 & 0 & 0 & 1 \\
         -\frac{1}{x^4} & \frac{15 x^4+1}{x^3-x^7} & \frac{1-25 x^4}{x^2 \left(x^4-1\right)} & -\frac{2-10 x^4}{x-x^5} \\
        \end{array}
        \right) \, , \label{eq:cy3de}
    \end{align}
    from which a Picard-Fuchs operator can be read off.
    The full $\epsilon$-dependence is too involved to be displayed here; we instead show the highest $\epsilon$-order in each entry of the matrix,

   \begin{align}
        \left(
    \begin{array}{ccccccc}
     -  & 0  & - &-  \\
     -  & -  & 0 &- \\
     -  & -  & - &0 \\
    4  & 3  & 2&1  \\
    \end{array}
    \right) \, ,
    \label{eq:cy31sfeps}
    \end{align}
    where `$-$' marks vanishing entries.
    The corresponding Picard-Fuchs operator of $I_1$ at $\epsilon = 0$ reads
\begin{equation}
            \mathcal{L}^{(4)}_{\text{CY3,1}} = 
            \bigg[ \Big(\theta -1\Big)^4 - x^4\Big(\theta+1\Big)^4\bigg] \, .
        \end{equation}
        In contrast with 4PM, where the third-order operator we found turned out to be the symmetric square of a second-order operator, we encounter here a non-factorisable Picard-Fuchs operator of fourth order. A naive search for solutions of $\mathcal{L}^{(4)}_{\text{CY3,1}} f(x) = 0$ produces, among others, the hypergeometric function $f(x) = x \,_{4}F_3(\tfrac{1}{2},\tfrac{1}{2},\tfrac{1}{2},\tfrac{1}{2};1,1,1;x^4)$. It is natural to ask whether this operator can again be associated with a definite geometry; the order of the operator suggests a Calabi--Yau threefold, and indeed $\mathcal{L}^{(4)}_{\text{CY3,1}}$ is associated with a smooth one-parameter family of CY threefolds, parametrised by $x^4 = \psi^{-8}$ and defined as the resolution of four symmetric quadrics, as was shown in \cite{Klemm:2024wtd,Bonisch:2022mgw},
        \be
        x_j^2+y_j^2-2 \psi x_{j+1} y_{j+1}=0,\ j\in \mathbb{Z}/4 \mathbb{Z}
        \label{eq:CY1}
        \ee
        in the homogeneous coordinates $x_j, y_j$, $j=0,\ldots,3$, of $\mathbb{P}^7$. This hypergeometric CY-threefold motive previously appeared in studies of mirror symmetry~\cite{MR1201748}. It is one of the simplest known CY threefolds, with only four periods, so that all of them are directly accessible through our operator. Moreover, just as at 4PM the K3 operator was the symmetric square of the second-order operator associated with the Legendre curve, so does our $\mathcal{L}^{(4)}_{\text{CY3,1}}$ at 5PM-1SF relate to the Legendre operator---this time via the Hadamard product,
        \begin{align}
                \mathcal{\tilde{L}}^{(4)}_{\text{CY3,1}} = \mathcal{L}^{(2)}_{\text{L}} * \mathcal{L}^{(2)}_{\text{L}}\,,
        \end{align}
        that is, the solution of $\mathcal{\tilde{L}}^{(4)}_{\text{CY3,1}}$ is given by the Hadamard product of the solutions of the two second-order factors. We have to normalise our period with $x^{-1}$ to get $\mathcal{\tilde{L}}^{(4)}_{\text{CY3,1}}$ and use $z=x^4$ for $\mathcal{L}^{(2)}_{\text{L}}$ in eq.~\refeq{eq:Legendre} to see this relation. The Hadamard product itself is defined on power series by~\cite{Almkvist:2023yja}
        \begin{align}
                \sum_i a_i x^i \;*\; \sum_i b_i x^i= \sum_i a_i b_i x^i\;.
        \end{align}
In particular we find here
\begin{align}
        \,_{4}F_3(\tfrac{1}{2},\tfrac{1}{2},\tfrac{1}{2},\tfrac{1}{2};1,1,1;z)=\frac{2 K(z)}{\pi}*\frac{2 K(z)}{\pi}.
\end{align}
With the geometry now identified, we proceed to the canonicalisation of the differential equations around the MUM-point $x = 0$. The Frobenius basis of the periods is given by
        \begin{align}
                \varpi_0(x)&=x+\frac{x^5}{16}+\frac{81 x^9}{4096} +\frac{625 x^{13}}{65536}+\mathcal{O}(x^{14}) \, ,\\
                \varpi_1(x)&=\varpi_0(x) \log(x)+S_1 = \varpi_0(x) \log(x) + \frac{x^5}{16}+\frac{189 x^9}{8192} +\frac{4625 x^{13}}{393216}+\mathcal{O}(x^{14}) \, , \nonumber\\
                \varpi_2(x)&=\dfrac{1}{2} \varpi_0(x) \log^2(x) +S_1 \log(x)+S_2\nn \\
                &=
                \dfrac{1}{2} \varpi_0(x) \log^2(x) +S_1 \log(x) + \frac{x^5}{128}+\frac{729 x^9}{131072}+\frac{63925 x^{13}}{18874368}+\mathcal{O}(x^{14}), \nonumber\\
                \varpi_3(x)&=\dfrac{1}{6} \varpi_0(x) \log^3(x) +\dfrac{1}{2}S_1 \log^2(x)+S_2 \log(x)-\frac{x^5}{256}-\frac{537 x^9}{524288}-\frac{80735 x^{13}}{226492416}+\mathcal{O}(x^{14}) \, . \nonumber
            \label{eq:frobCY31sf}
            \end{align}
The Wronskian of the four periods can now be split into a unipotent and a semi-simple factor, as in~\refeq{eq:W_Wss_Wu},  
          
\begin{align}
        W^U=\left(
                \begin{array}{cccc}
                 1 & \frac{\varpi_1(x)}{\varpi_0(x)} & \frac{\varpi_2(x)}{\varpi_0(x)} & \frac{\varpi_3(x)}{\varpi_0(x)} \\
                 0 & 1 & \frac{\varpi_0(x) \varpi_2'(x)-\varpi_2(x) \varpi_0'(x)}{\varpi_0(x) \varpi_1'(x)-\varpi_1(x) \varpi_0'(x)} & \frac{\varpi_0(x) \varpi_3'(x)-\varpi_3(x) \varpi_0'(x)}{\varpi_0(x) \varpi_1'(x)-\varpi_1(x) \varpi_0'(x)} \\
                 0 & 0 & 1 & \frac{\varpi_1(x)}{\varpi_0(x)} \\
                 0 & 0 & 0 & 1 \\
                \end{array}
                \right)
\end{align}

\begin{align}\tiny
       W^{\text{SS}}= \left(
\begin{array}{cccc}
 \varpi_0 & 0 & 0 & 0 \\
 \varpi_0' & \frac{\varpi_0}{x \alpha } & 0 & 0 \\
 \varpi_0'' &-\frac{x \varpi_0 \alpha '+\alpha  \left(\varpi_0-2 x \varpi_0'\right)}{x^2 \alpha ^2} & \frac{\alpha }{(1-x) (x+1) \left(x^2+1\right) \varpi_0} & 0 \\
 \varpi_0''' & g(x) & -\frac{\alpha  \left(\left(1-5 x^4\right) \varpi_0-x \left(1-x^4\right) \varpi_0'\right)}{(1-x)^2 x (x+1)^2 \left(x^2+1\right)^2 \varpi_0^2} & \frac{1}{(1-x) x (x+1) \left(x^2+1\right) \varpi_0} \\
\end{array}
\right)
\end{align}
where 
\begin{align}
        g(x)&=\frac{2 x^2 \varpi_0 \alpha '^2-x \alpha  \left(3 x \alpha ' \varpi_0'+\varpi_0 \left(x \alpha ''-2 \alpha '\right)\right)+\alpha ^2 \left(3 x \left(x \varpi_0''-\varpi_0'\right)+2 \varpi_0\right)}{x^3 \alpha ^3}\\
        \alpha(x)&=  \frac{\varpi_0^2}{x \left(\varpi_0 \varpi_1'-\varpi_1 \varpi_0'\right)}
\end{align}
The semi-simple factor is structurally richer than at 4PM: it involves not one but two independent periods, the second of which we have rewritten in terms of the Yukawa coupling $\alpha$. The Griffiths relation for this second period reads
            \begin{align}
                \varpi_1'''=& \frac{\varpi_1\left(\left(5 x^4-1\right) \varpi_0'+x \left(x \left(x^4-1\right) \varpi_0'''+\left(5 x^4-1\right) \varpi_0''\right)\right)}{x^2 \left(x^4-1\right) \varpi_0}\nn\\
                &
                +\frac{-\left(5 x^4-1\right) \varpi_0 \left(x \varpi_1''+\varpi_1'\right)-\left(\left(x^4-1\right) x^2 \left(\varpi_0''\varpi_1'-\varpi_0' \varpi_1''\right)\right)}{x^2 \left(x^4-1\right) \varpi_0}\; ,
            \end{align}
            which can be recast as
            \begin{align}
                \alpha ''= \frac{2 \alpha}{x^2}-\frac{\left(3 x^4+1\right) \alpha '}{(x-1) x (x+1) \left(x^2+1\right)}+\frac{\frac{8 x^3 \alpha \varpi_0'}{(x-1) (x+1) \left(x^2+1\right)}-2 \alpha ' \varpi_0'+4 \alpha \varpi_0''}{\varpi_0}+\frac{2 \alpha '^2}{\alpha}-\frac{2 \alpha \varpi_0'^2}{\varpi_0^2}
            \end{align}
The Picard-Fuchs equation itself further yields
\begin{align}
        \varpi_0''''=-\frac{\varpi_0}{x^4}-\frac{\left(5 x^2-1\right) \left(5 x^2+1\right) \varpi_0''}{(x^2-1) x^2 \left(x^2+1\right)}-\frac{2 \left(5 x^4-1\right) \varpi_0'''}{(x^2-1) x  \left(x^2+1\right)}-\frac{\left(15 x^4+1\right) \varpi_0'}{(x^2-1) x^3  \left(x^2+1\right)}
\end{align}
We are now in possession of all the ingredients required to canonicalise the full matrix. We first rotate by the inverse of the semi-simple factor and rescale according to
            \begin{align}
                D(\epsilon) = \text{diag}(\epsilon^3,\epsilon^2,\epsilon,1) \, .
                \label{eq:Grav_cy3_D}
            \end{align}
After rescaling we will try to remove terms order by order in $\epsilon$, starting at $\mathcal{O}(\epsilon^{-1})$, by an ansatz containing the functions $\varpi_0$ and $\alpha$, similar to the K3 case. We find that it is not possible to remove all such terms by this strategy. We have to introduce four new functions $G_1,...,G_4$ to rotate to a canonical basis. The two needed rotations are          
            \begin{align}
                T_{G,1}=\left(
                        \begin{array}{cccc}
                         1 & 0 & 0 & 0 \\
                         0 & 1 & 0 & 0 \\
                         \frac{G_1(x)}{\epsilon } & 0 & 1 & 0 \\
                         \frac{G_2(x)}{\epsilon } & \frac{G_1(x)}{\epsilon } & 0 & 1 \\
                        \end{array}
                        \right)
            \end{align}
            and
            \begin{align}
                T_{G,2}=\left(
                        \begin{array}{cccc}
                         1 & 0 & 0 & 0 \\
                         G_3(x) & 1 & 0 & 0 \\
                         G_4(x) & 0 & 1 & 0 \\
                         -G_3(x) G_4(x) & -G_4(x) & -G_3(x) & 1 \\
                        \end{array}
                        \right),
            \end{align}
            where the functions $G_i$ obey first-order differential equations. 
            \begin{align}
                &G_1'= -\frac{96 x \left(x^4+1\right) \varpi_0^2}{(x-1)^2 (x+1)^2 \left(x^2+1\right)^2 \alpha}\\
                & G_2'= \varpi_0^2 \left(\frac{ 2 \left(5 x^8+28 x^6+262 x^4+28 x^2+5\right) \alpha'^2}{-3 (x-1) x (x+1) \left(x^2+1\right) \alpha^2}+\frac{16 \left(7 x^8+136 x^6+42 x^4+136 x^2+7\right) \alpha'}{3 (x-1)^2 (x+1)^2 \left(x^2+1\right)^2 \alpha}\right.\nn\\
                &\left. -\frac{16 \left(7 x^{12}+314 x^{10}+329 x^8+1340 x^6+329 x^4+314 x^2+7\right)}{3 (x-1)^3 x (x+1)^3 \left(x^2+1\right)^3}\right)-\frac{x G_1^2 \alpha^2}{(x-1) (x+1) \left(x^2+1\right) \varpi_0^2}\\
                &G_3'= -\frac{x G_1 \alpha^2}{(x-1) (x+1) \left(x^2+1\right) \varpi_0^2}\\
                & G_4'= \frac{G_2}{x \alpha}+\varpi_0^2 \Big(\frac{4 \left(5 x^8+28 x^6+262 x^4+28 x^2+5\right) \alpha'}{3 (x-1) x^2 (x+1) \left(x^2+1\right) \alpha^2}\nn\\
                &\quad \quad \; -\frac{16 \left(7 x^8+136 x^6+42 x^4+136 x^2+7\right)}{3 (x-1)^2 x (x+1)^2 \left(x^2+1\right)^2 \alpha}\Big)
            \end{align}
We see the canonicalisation requires the introduction of additional new functions similar to \refeq{eq:g3diffeq}, where we had to introduce one new function. In this example we could associate this new function to elliptic integrals of the third kind with an extra residue in the differential form. For the CY threefold the interpretation is rather more subtle. With only four masters at our disposal, none of them can be associated with a higher-dimensional analogue of the third-kind elliptic integral. We leave a deeper interpretation of these functions to future work; for our purposes it is sufficient to note that they are distinct from third-kind integrals, and that they can easily be evaluated by series expansions.

After these rotations the canonical basis $\vec{J}$ satisfies the differential equation

\begin{align}
    \dfrac{d \vec{J}}{dx}=\epsilon   \tiny \left(
        \begin{array}{cccc}
         \frac{2 \left(x^4-4 x^2+1\right)}{x \left(x^4-1\right)}-\frac{G_3}{x \alpha} & \frac{1}{x \alpha} & 0 & 0 \\
         k_3(x) & \frac{G_3}{x \alpha}+\frac{2 \left(x^4-4 x^2+1\right)}{x \left(x^4-1\right)} & -\frac{x \alpha^2}{\left(x^4-1\right) \varpi_0^2} & 0 \\
         k_2(x) & \frac{2 G_4}{x \alpha}-\frac{8 \left(5 x^8+28 x^6+262 x^4+28 x^2+5\right) \varpi_0^2}{3 x^3 \left(x^4-1\right) \alpha^2} & \frac{G_3}{x \alpha}+\frac{2 \left(x^4-4 x^2+1\right)}{x \left(x^4-1\right)} & \frac{1}{x \alpha} \\
         k_1(x) & k_2(x) & k_3(x) & \frac{2 \left(x^4-4 x^2+1\right)}{x \left(x^4-1\right)}-\frac{G_3}{x \alpha} \\
        \end{array}
        \right)\vec{J},
\end{align}
where
\begin{align}
     k_1(x)=&   G_3^2 \left(\frac{2 G_4}{x\alpha}-\frac{8 \left(5 x^8+28 x^6+262 x^4+28 x^2+5\right) \varpi_0^2}{3 x^3 \left(x^4-1\right)\alpha^2}\right) \nn \\
    & +\frac{96 \left(7 x^8+100 x^6+42 x^4+100 x^2+7\right) G_3 \varpi_0^2}{x \left(x^4-1\right)^2\alpha}-\frac{x G_4^2\alpha^2}{\left(x^4-1\right) \varpi_0^2} \nn \\
    &-\frac{4 \left(5 x^8+28 x^6+262 x^4+28 x^2+5\right) G_4\alpha}{3 x \left(x^4-1\right)^2}\nn \\
    &-\frac{4 \left(-119 x^{16}+280 x^{14}+62300 x^{12}+51240 x^{10}+165302 x^8\right) \varpi_0^2}{9 x^3 \left(x^4-1\right)^3} \nn\\
    &-\frac{4 \left(51240 x^6+62300 x^4+280 x^2-119\right) \varpi_0^2}{9 x^3 \left(x^4-1\right)^3} \\
    k_2(x)=&G_3 \left(\frac{8 \left(5 x^8+28 x^6+262 x^4+28 x^2+5\right) \varpi_0^2}{3 x^3 \left(x^4-1\right) \alpha^2}-\frac{2 G_4}{x \alpha}\right)\nn \\
    &-\frac{48 \left(7 x^8+100 x^6+42 x^4+100 x^2+7\right) \varpi_0^2}{x \left(x^4-1\right)^2 \alpha}\\
    k_3(x)=&-\frac{G_3^2}{x \alpha}+\frac{x G_4 \alpha^2}{\left(x^4-1\right) \varpi_0^2}+\frac{2 \left(5 x^8+28 x^6+262 x^4+28 x^2+5\right) \alpha}{3 x \left(x^4-1\right)^2}
\end{align}
We will use the notation $\varpi_0=\varpi_{\text{CY3,1}}$ and $\alpha_1=\alpha_{\text{CY3,1}}$ from now on for this period. 
\section{Canonicalisation of the full DE}
We have now discussed the two non-trivial geometries which appear in the differential equations at 5PM-1SF---the K3 surface, which arises in both the B-type and V-type integrals, and the new CY threefold, which arises only for the V-type. In addition, both systems contain numerous polylogarithmic diagonal blocks, of size up to $7\times 7$. Since both the number of integrals ($\sim 500$) and the number of blocks grow substantially at four loops, we employ the algorithm \texttt{CANONICA} to canonicalise the polylogarithmic diagonal blocks. The off-diagonal entries are then handled by the algorithm of Section~\ref{sec:K3off} for the B-type integrals, and similarly for the V-type up to those sectors which couple to the CY3. Since that algorithm does not extend to sectors involving the CY3 period, a different strategy is required there, to which we now turn.

\subsection{Couplings to the CY3}

Our treatment of the off-diagonals so far has relied on knowing in advance the integration kernels admissible in the canonical form, and on algorithmically removing all others. For the CY3, it is not a priori clear which integration kernels are admissible. We saw in Section~\ref{sec:off} that an alternative is to start from a ``good basis'' and remove residual terms order by order in $\epsilon$, in the same way as for the K3 and CY3 diagonal blocks. The challenge then lies in identifying a good basis from which the remaining transformations can be constructed without much effort.

One classical route to such a basis is integrand analysis in the Baikov representation. As already mentioned this becomes increasingly tedious at higher loop order, both because the analysis itself becomes harder and because of the sheer number of integrals. We adopt a different method, building on the ideas of \cite{Dlapa:2023hsl}. The strategy may be summarised as the use of IBPs, together with the already-canonicalised diagonal blocks, to automate the integrand analysis.

Once a canonical basis $\mathfrak{J}_k$ on the maximal cut is known, IBPs can be used to construct further integrals $I_{\rm can}$ which are themselves canonical on the maximal cut, since any such integral can be expressed as
\begin{align}
	I_{can}  = \sum_k c_k\,  \mathfrak J_k \, ,
\label{eq:maxcutrel}
\end{align}
where the $c_k$ are constants and all integrals are taken on the maximal cut. A mild generalisation of this is to admit candidate integrals on the maximal cut of the form
\begin{align}
	I_i^\text{candidate}  = f(\epsilon)g(x) \sum_k c_k\,  \mathfrak J_k \, , \label{eq:candCond}
\end{align}
which are canonical up to an overall normalisation. Here $g(x)$ plays the role of a leading singularity in the Baikov analysis.

Such candidates are expected to require only minimal corrections in order to be promoted to a canonical basis. For some sectors which are polylogarithmic on their diagonal block, we find that these candidates are still not quite a ``good basis'', and one is forced to relax the maximal cut to include further subsectors in the analysis. Demanding that Eq.~\refeq{eq:maxcutrel} hold across these subsectors yields additional conditions from the off-diagonal couplings. 

On the other hand one can have the opposite problem that one does not find enough candidates fulfilling the condition~\ref{eq:candCond}. Then one might relax the condition slightly, allowing some of the $c_k$ to depend on $x$, and one might still find suitable candidates.

The procedure is rendered efficient by the use of IBPs, which makes it possible to determine all transformations for the coupling to the CY3 by hand. Once a good basis is at hand, residual terms can be removed order by order in $\epsilon$. In analogy with the canonicalisation of the CY3 diagonal block, this requires the introduction of 11 new functions $H_1(x),\ldots,H_{11}(x)$, again defined by first-order differential equations whose explicit form is given in App.~\ref{app:cy31G}. Eight of these new functions arise from the mixings between the K3 and CY3 sectors, and their first-order ODEs involve the periods of both geometries.

\begin{figure}[h!]
        \centering
        \begin{subfigure}{0.49\textwidth}
        \includegraphics[width=\textwidth]{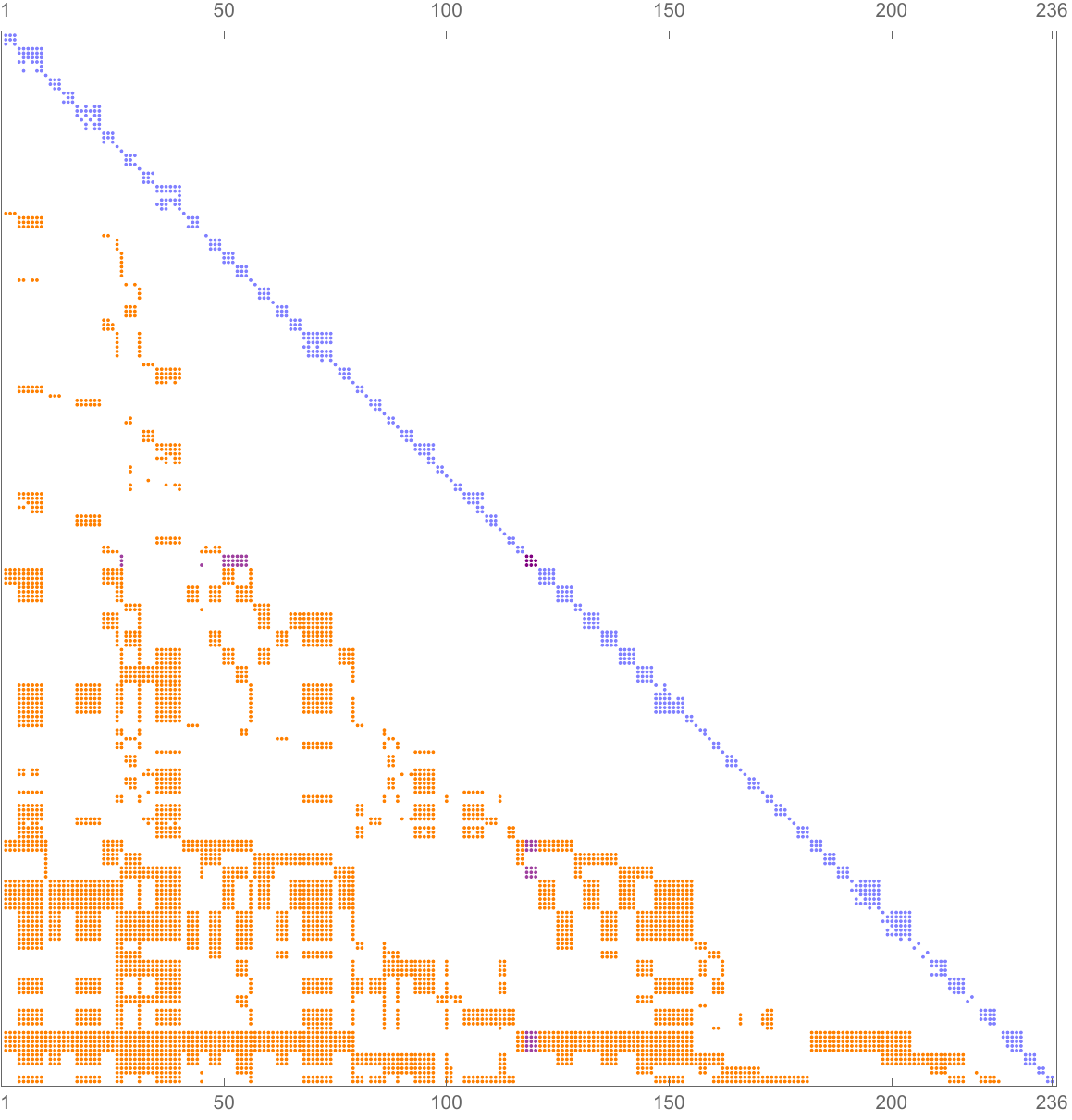}
        \caption{B-type integrals}
\end{subfigure}
        \hfill
        \begin{subfigure}{0.49\textwidth}
        \includegraphics[width=\textwidth]{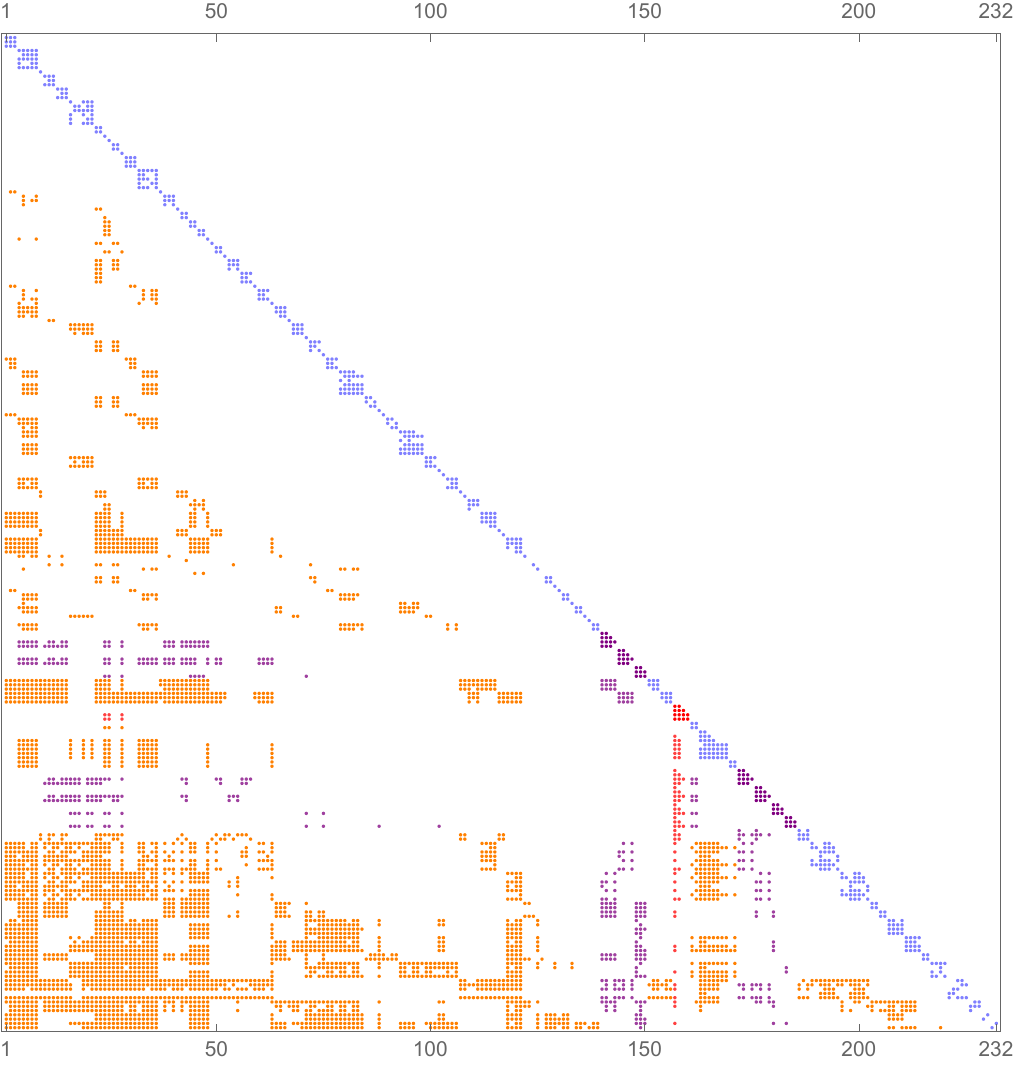}
        \caption{V-type integrals}
\end{subfigure}
\caption{In the heat plot of the differential equation matrices we mark logarithmic diagonal blocks blue and logarithmic off-diagonal sectors orange. The K3 sectors are marked purple and the CY3 sector is marked red.}\label{fig:heatPlot5pm1sf}
\end{figure}

This concludes the canonicalisation of the full differential-equation system. The heat plots of the fully canonicalised matrices for the B- and V-type integrals are shown in Fig.~\ref{fig:heatPlot5pm1sf}.

\section{Boundary fixing and continuation of the period}
We have now reached a canonical form for both the B-type and V-type integrals, with canonical bases $\vec{J}_{B/V}$ satisfying
\begin{align}
        \dfrac{d \vec{J}_{B/V}}{dx}=\epsilon \hat{A}_{B/V} \vec{J}_{B/V}.
\end{align}
The remaining step is the fixing of the boundary constants. The subtlety is that the K3 and CY3 sectors have been solved as expansions around the MUM-point $x=0$, while the boundary constants are most naturally fixed in the slow-velocity limit $x\to 1$. Holomorphic solutions of the periods at $x=1$ must therefore be chosen in order to expand the differential equations there, just as was done at 4PM.

At 4PM, and likewise for the K3 at 5PM-1SF, this choice is unique: $x=1$ is itself a MUM-point of the relevant operator, and only one holomorphic solution exists. For the CY3, however, the choice is no longer unique, and the final result depends on which combination of solutions is selected. Since the CY3 enters only $\hat{A}_V$, the discussion concerns the V-type integrals alone; for the B-type, the boundary fixing proceeds exactly as at lower orders.

We recall that the canonical integrals are to be expanded in $\delta=1-x$, with the $\epsilon$-dependence kept exact,
\begin{align}
	\vec J_V(\delta;\epsilon) = \sum_{k,m,n}  \vec{f}_{k,m,n}(\epsilon)(\delta)^{k+m\epsilon}\log^{n}(\delta) \,.
\label{eq:wasow}
\end{align}
At leading order this takes the form $\vec{J}_V = \hat S_0(x;\epsilon)\,\vec{j} + \mathcal{O}((1-x))$, where $\hat S_0(x;\epsilon)$ collects all the $\delta^{m\epsilon}$ factors and is determined by
\be
	-\dfrac{\mathrm d \hat{S}_0(\delta;\epsilon)}{\mathrm d\delta}=\epsilon \sum_{j=0}^2 \hat{A}_{V_{-1,j}}\dfrac{ \log^j{(\delta)}}{\delta}\,  \hat{S}_0(\delta;\epsilon) \, ,
\label{dgl2}
\ee
with $\hat{A}_{V_{i,j}}$ denoting the expansion coefficients of the differential-equation matrix,
\be
	\hat{A}_V
  =
   \sum_{j=0}^2 \hat{A}_{V_{-1,j}} \frac{\log^j(\delta)}{\delta}
   +
   \sum_{i,j\ge 0}
   \hat{A}_{V_{i,j}} \delta^i \log^j(\delta) \, .
\label{dglexp}
\ee
A new feature appears at 5PM order: the CY3 sector and its couplings can in general generate terms of order $\log(\delta)$, though whether they actually do so depends on the choice of period. At $x=1$ the CY3 has no MUM-point, so a linear combination of three of its solutions can be chosen which is free of logarithms. In addition, a second period can be selected for the definition of $\alpha$. In ref.~\cite{Driesse:2024feo} the choice was
\begin{equation}
        \begin{aligned}
                \varpi_{\text{CY3,1}} 	&= 1-(1-x)+\dfrac{(1-x)^3}{3}+\dfrac{(1-x)^4}{3}+\ldots \, ,\\
                \alpha_{\text{CY3,1}}	&= 1-2\cdot (1-x)+2\cdot(1-x)^2-\dfrac{(1-x)^3}{3}+\dfrac{(1-x)^4}{2}+\ldots \, .
\end{aligned}
\end{equation}
This generates logarithmic terms in the expansion of the differential equations, complicating the boundary fixing. While not wrong per se, it leads to a mixing of transcendental weights---once the matrix contains terms of order $\log\delta$, it already contains terms of weight one. This is undesirable, and is responsible for the four-fold iterated integrals reported in ref.~\cite{Driesse:2024feo}.

The complication can in fact be avoided by a particular choice of the periods, accompanied by a particular choice of integration constants in the first-order ODEs for the $G_i$ (which are defined only up to additive constants). The required choice of periods reads
\begin{equation}
        \begin{aligned}
                \varpi_{\text{CY3,1}} 	&= 1+\dfrac{(1-x)^2}{2}+\dfrac{(1-x)^3}{2}+\dfrac{(1-x)^4}{3}+\ldots \, ,\\
                \alpha_{\text{CY3,1}}	&= 1+2\cdot (1-x)^2+2\cdot(1-x)^3+\dfrac{41(1-x)^4}{12}+\ldots \, , \label{eq:periodchoice}
\end{aligned}
\end{equation}
with the additional property that $\varpi_0(v)=\varpi_0(-v)$ and $\alpha_1(v)=\alpha_1(-v)$. With this choice one finds $\hat{A}_{V_{i,j}}=0$ for $j>0$, and the boundary fixing then proceeds as at lower orders.
\section{Regions at 5PM-1SF}
The integral family at 5PM-1SF possesses three active gravitons. The momentum scalings remain unchanged from before,
\begin{equation}
        \ell^\text{P}=(\ell^0,\Bell)\sim(v,1)\,,  \qquad
         \ell^\text{R}=(\ell^0,\Bell)\sim(v,v) \, .
\end{equation}
There are eight relevant regions, $PPP$, $PRP$, $PPR$, $RPP$, $RRP$,$RPR$, $PRR$ and $RRR$, which we organise according to the number of radiative gravitons they contain.

\subsection*{Potential PPP}
\begin{figure}[h!]
        \centering
        \begin{tikzpicture}[baseline={([yshift=-1ex]current bounding box.south)},scale=1.2]
                \coordinate (inA) at (0.4,.7);
                \coordinate (outA) at (3.4,.7);
                \coordinate (inB) at (0.4,-.7);
                \coordinate (outB) at (3.4,-.7);
                \coordinate (xA) at (0.9,.7);
                \coordinate (xB) at (0.9,-.7);
                \coordinate (xM) at (1.15,0);
                \coordinate (xAM) at (0.9,.35);
                \coordinate (xBM) at (0.9,-.35);
                \coordinate (yA) at (1.9,.7);
                \coordinate (yB) at (1.9,-.7);
                \coordinate (yC) at (1.9,-1.2);
                \coordinate (xC) at (0.9,-1.2);
                \coordinate (zC) at (2.9,-1.2);
                \coordinate (yM) at (1.9,0);
                \coordinate (xyM) at (1.6,0);
                \coordinate (yzM) at (2.2,0);
                \coordinate (yAM) at (1.9,.35);
                \coordinate (yBM) at (1.9,-.35);
                \coordinate (zA) at (2.9,.7);
                \coordinate (zB) at (2.9,-.7);
                \coordinate (zM) at (2.65,0);
                \coordinate (zAM) at (2.9,.35);
                \coordinate (zBM) at (2.9,-.35);
                \draw [dotted] (inA) --node [midway, above]{$\rightarrow \vec{e}$}  (outA);
                \draw [photon] (yB) --node [midway, right]{$\downarrow \vec{q}$} (yC);
                \draw[photon] ($(yA)!0.5!(yB)$) circle [radius=0.7,rotate=90];
                \draw [photon] (xyM) -- (yB);
                \draw [photon] (yzM) -- (yB);
                \draw [photon] (xM) -- (zM);
              
                \foreach \v in {yA,yB,yC,xyM,yzM,zM,xM}
                \draw[fill] (\v) circle (.08);
        
              \end{tikzpicture}
              \begin{tikzpicture}[baseline={([yshift=-1ex]current bounding box.south)},scale=1.2]
                \coordinate (inA) at (0.4,.7);
                \coordinate (outA) at (3.4,.7);
                \coordinate (inB) at (0.4,-.7);
                \coordinate (outB) at (3.4,-.7);
                \coordinate (xA) at (0.9,.7);
                \coordinate (xB) at (0.9,-.7);
                \coordinate (xM) at (0.9,0);
                \coordinate (xAM) at (0.9,.35);
                \coordinate (xBM) at (0.9,-.35);
                \coordinate (yA) at (1.9,.7);
                \coordinate (yB) at (1.9,-.7);
                \coordinate (yC) at (1.9,-1.2);
                \coordinate (xC) at (0.9,-1.2);
                \coordinate (zC) at (2.9,-1.2);
                \coordinate (yM) at (1.9,0);
                \coordinate (yzM) at (2.225,0);
                \coordinate (yAM) at (1.9,.35);
                \coordinate (yBM) at (1.9,-.35);
                \coordinate (zA) at (2.9,.7);
                \coordinate (zB) at (2.9,-.7);
                \coordinate (zM) at (2.65,0);
                \coordinate (zAM) at (2.9,.35);
                \coordinate (zBM) at (2.9,-.35);
                \draw [dotted] (inA) --node [midway, above]{$\rightarrow \vec{e}$}  (outA);
                \draw [photon] (yB) --node [midway, right]{$\downarrow \vec{q}$} (yC);
                \draw [photon] (yM) -- (zM);
                \draw [photon] (xA) to[out=-45,in=180] (yB);
                \draw [photon] (yA) to[out=-0,in=0,distance=1cm] (yB);
                \draw [photon] (yzM) to[out=-90,in=45] (yB);
                \draw [zUndirected] (xA) -- (yA);
                \draw [photon] (yB) -- (yA);
                \draw [photon] (yB) -- (yA);
                \foreach \v in {yA,yB,xA,yC,zM,yM,yzM}
                \draw[fill] (\v) circle (.08);
        
              \end{tikzpicture}
              \begin{tikzpicture}[baseline={([yshift=-1ex]current bounding box.south)},scale=1.2]
                \coordinate (inA) at (0.4,.7);
                \coordinate (outA) at (3.4,.7);
                \coordinate (inB) at (0.4,-.7);
                \coordinate (outB) at (3.4,-.7);
                \coordinate (xA) at (0.9,.7);
                \coordinate (xB) at (0.9,-.7);
                \coordinate (xM) at (0.9,0);
                \coordinate (xAM) at (0.9,.35);
                \coordinate (xBM) at (0.9,-.35);
                \coordinate (yA) at (1.9,.7);
                \coordinate (zyA) at (2.3,.7);
                \coordinate (xyA) at (1.5,.7);
                \coordinate (yB) at (1.9,-.7);
                \coordinate (yC) at (1.9,-1.2);
                \coordinate (xC) at (0.9,-1.2);
                \coordinate (zC) at (2.9,-1.2);
                \coordinate (yM) at (1.9,0);
                \coordinate (yAM) at (1.9,.35);
                \coordinate (yBM) at (1.9,-.35);
                \coordinate (zA) at (2.9,.7);
                \coordinate (zB) at (2.9,-.7);
                \coordinate (zM) at (2.9,0);
                \coordinate (yzM) at (2.2,0);
                \coordinate (xyM) at (1.6,0);
                \coordinate (zAM) at (2.9,.35);
                \coordinate (zBM) at (2.9,-.35);
                \draw [dotted] (inA) --node [midway, above]{$\rightarrow \vec{e}$}  (outA);
                \draw [photon] (yB) --node [midway, right]{$\downarrow \vec{q}$}  (yC);
                \draw [photon] (yA) to[out=-90-45,in=90+45] (yB);
                \draw [photon] (yA) to[out=-45,in=45] (yB);
                \draw [photon] (xyM) to (yzM);
                \draw [photon] (xA) to[out=-45,in=180] (yB);
                \draw [zUndirected] (xA) -- (yA);
                \draw [zUndirected] (yA) -- (zA);
                \draw [photon] (zA) to[out=-90-45,in=0] (yB);
                \draw [photon] (zA) to[out=-90-45,in=0] (yB);

                \foreach \v in {yB,xA,yC,zA,yA,xyM,yzM}
                \draw[fill] (\v) circle (.08);
        
              \end{tikzpicture}
              \begin{tikzpicture}[baseline={([yshift=-1ex]current bounding box.south)},scale=1.2]
                \coordinate (inA) at (0.4,.7);
                \coordinate (outA) at (3.4,.7);
                \coordinate (inB) at (0.4,-.7);
                \coordinate (outB) at (3.4,-.7);
                \coordinate (xA) at (0.9,.7);
                \coordinate (xB) at (0.9,-.7);
                \coordinate (xM) at (0.9,0);
                \coordinate (xAM) at (0.9,.35);
                \coordinate (xBM) at (0.9,-.35);
                \coordinate (yA) at (1.9,.7);
                \coordinate (yB) at (1.9,-.7);
                \coordinate (yC) at (1.9,-1.2);
                \coordinate (xC) at (0.9,-1.2);
                \coordinate (zC) at (2.9,-1.2);
                \coordinate (yM) at (1.9,0);
                \coordinate (xyA) at (1.4,0.7);
                \coordinate (yAM) at (1.9,.35);
                \coordinate (yBM) at (1.9,-.35);
                \coordinate (zA) at (2.9,.7);
                \coordinate (zB) at (2.9,-.7);
                \coordinate (zM) at (2.65,0);
                \coordinate (zAM) at (2.9,.35);
                \coordinate (zBM) at (2.9,-.35);
                \draw [dotted] (inA) --node [midway, above]{$\rightarrow \vec{e}$}  (outA);
                \draw [photon] (yB) --node [midway, right]{$\downarrow \vec{q}$} (yC);
                \draw [photon] (yM) -- (zM);
                \draw [photon] (xA) to[out=-90,in=180] (yB);
                \draw [photon] (xyA) to[out=-90,in=120] (yB);
                \draw [photon] (yA) to[out=-0,in=0,distance=1cm] (yB);
                \draw [zUndirected] (xA) -- (yA);
                \draw [photon] (yB) -- (yA);
                \draw [photon] (yB) -- (yA);
                \foreach \v in {yA,yB,xA,yC,zM,yM,xyA}
                \draw[fill] (\v) circle (.08);
        
              \end{tikzpicture}

              \begin{tikzpicture}[baseline={([yshift=-1ex]current bounding box.south)},scale=1.2]
                \coordinate (inA) at (0.4,.7);
                \coordinate (outA) at (3.4,.7);
                \coordinate (inB) at (0.4,-.7);
                \coordinate (outB) at (3.4,-.7);
                \coordinate (xA) at (0.9,.7);
                \coordinate (xB) at (0.9,-.7);
                \coordinate (xM) at (0.9,0);
                \coordinate (xAM) at (0.9,.35);
                \coordinate (xBM) at (0.9,-.35);
                \coordinate (yA) at (1.9,.7);
                \coordinate (zyA) at (2.4,.7);
                \coordinate (xyA) at (1.4,.7);
                \coordinate (yB) at (1.9,-.7);
                \coordinate (yC) at (1.9,-1.2);
                \coordinate (xC) at (0.9,-1.2);
                \coordinate (zC) at (2.9,-1.2);
                \coordinate (yM) at (1.9,0);
                \coordinate (yAM) at (1.9,.35);
                \coordinate (yBM) at (1.9,-.35);
                \coordinate (zA) at (2.9,.7);
                \coordinate (zB) at (2.9,-.7);
                \coordinate (zM) at (2.9,0);
                \coordinate (zAM) at (2.9,.35);
                \coordinate (zBM) at (2.9,-.35);
                \draw [dotted] (inA) --node [midway, above]{$\rightarrow \vec{e}$}  (outA);
                \draw [photon] (yB) --node [midway, right]{$\downarrow \vec{q}$}  (yC);
                \draw [photon] (xA) to[out=-45,in=180] (yB);
                \draw [photon] (xyA) to (yB);
                
                \draw [zUndirected] (xA) -- (yA);
                \draw [zUndirected] (yA) -- (zA);
                \draw [photon] (zA) to[out=-90-45,in=0] (yB);
                \draw [photon] (zyA) to (yB);
                \draw [photon] (yA) to (yB);
                \foreach \v in {yB,xA,yC,zA,xyA,zyA,yA}
                \draw[fill] (\v) circle (.08);
        
              \end{tikzpicture}
              \caption{The five relevant topsectors of the potential boundary integrals for 5PM-1SF.}
              \label{fig:5pm1sfpot}
\end{figure}
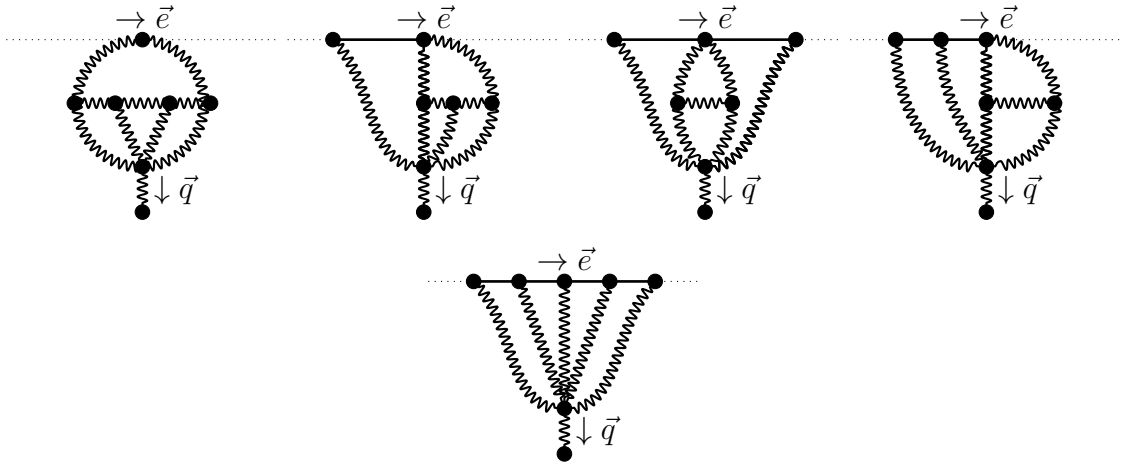

The potential region is in direct analogy with the lower-loop case,
\begin{align}
        &{\mathcal{P}^{\text{5PM-1SF}}}^{\{\sigma_1, \sigma_2,\sigma_3,\sigma_4,\sigma_5,\sigma_6,\sigma_7 \}}_{1,1,1,1,a_1,...,a_{18}} = {\mathcal{P}^{\text{5PM-0SF}}}^{\{\sigma_1, \sigma_2,\sigma_3,\sigma_4 \}}_{1,1,1,1,a_1,...,a_{18}}+\mathcal{O}(v^{-a_1-a_2-a_3-a_4+1}).
  \end{align}
The resulting integrals once again belong to the 0SF integral family and all belong to the topologies or subtopologies pictured in Fig.~\ref{fig:5pm1sfpot}.

\subsection*{Radiation reaction region PRP+PPR+RPP}
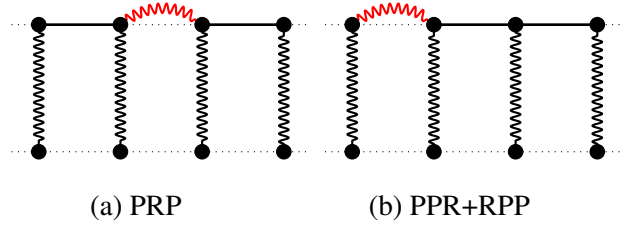
\begin{figure}[h!]\centering
        \begin{subfigure}{ 0.2 \textwidth}
                \begin{tikzpicture}[baseline={([yshift=-1ex]current bounding box.south)},scale=1.2]
                        \coordinate (inA) at (0.3,.7);
                         \coordinate (outA) at (3.6,.7);
                         \coordinate (inB) at (0.3,-.7);
                         \coordinate (outB) at (3.6,-.7);
                         \coordinate (xA) at (0.6,.7);
                         \coordinate (yA) at (1.5,.7);
                         \coordinate (zA) at (2.4,.7);
                         \coordinate (wA) at (3.3,.7);
                         \coordinate (xB) at (0.6,-.7);
                         \coordinate (yB) at (1.5,-.7);
                         \coordinate (zB) at (2.4,-.7);
                         \coordinate (wB) at (3.3,-.7);
                         \coordinate (xM) at (0.6,0);
                         \coordinate (yM) at (1.5,0);
                         \coordinate (zM) at (2.4,0);
                         \coordinate (wM) at (3.3,0);
                         \coordinate (xyMB) at (1.05,-.35);
                         \coordinate (yzMB) at (1.95,-.35);
                         \coordinate (zwMB) at (2.85,-.35);
                         \coordinate (yzMA) at (1.95,.35);
                         \draw [dotted] (inA) -- (outA);
                         \draw [dotted] (inB) -- (outB);
                         \draw [photon] (yA) -- (yB) ;
                         \draw [photon] (xA) -- (xB) ;
                         \draw [photon] (zA) -- (zB) ;
                         \draw [photon] (wB) -- (wA) ;
                         \draw [zUndirected] (zA) -- (wA) ;
                         \draw [zUndirected] (yA) -- (xA) ;
                         \draw [photon,red][out=40,in=140] (yA) to[out=40,in=140] (zA);
                [out=40,in=140] (xA) to[out=40,in=140]
                         \draw [fill] (yA) circle (.08);
                         \draw [fill] (wA) circle (.08);
                         \draw [fill] (xA) circle (.08);
                         \draw [fill] (zA) circle (.08);
                         \draw [fill] (xB) circle (.08);
                         \draw [fill] (yB) circle (.08);
                         \draw [fill] (zB) circle (.08);
                         \draw [fill] (wB) circle (.08);  
                        
                       \end{tikzpicture}\caption{PRP}
        \end{subfigure}\qquad
        \begin{subfigure}{ 0.2 \textwidth}
                \begin{tikzpicture}[baseline={([yshift=-1ex]current bounding box.south)},scale=1.2]
                        \coordinate (inA) at (0.3,.7);
                         \coordinate (outA) at (3.6,.7);
                         \coordinate (inB) at (0.3,-.7);
                         \coordinate (outB) at (3.6,-.7);
                         \coordinate (xA) at (0.6,.7);
                         \coordinate (yA) at (1.5,.7);
                         \coordinate (zA) at (2.4,.7);
                         \coordinate (wA) at (3.3,.7);
                         \coordinate (xB) at (0.6,-.7);
                         \coordinate (yB) at (1.5,-.7);
                         \coordinate (zB) at (2.4,-.7);
                         \coordinate (wB) at (3.3,-.7);
                         \coordinate (xM) at (0.6,0);
                         \coordinate (yM) at (1.5,0);
                         \coordinate (zM) at (2.4,0);
                         \coordinate (wM) at (3.3,0);
                         \coordinate (xyMB) at (1.05,-.35);
                         \coordinate (yzMB) at (1.95,-.35);
                         \coordinate (zwMB) at (2.85,-.35);
                         \coordinate (yzMA) at (1.95,.35);
                         \draw [dotted] (inA) -- (outA);
                         \draw [dotted] (inB) -- (outB);
                         \draw [photon] (xA) -- (xB) ;
                         \draw [photon] (yB) -- (yA) ;
                         \draw [photon] (zA) -- (zB) ;
                         \draw [photon] (wB) -- (wA) ;
                         \draw [zUndirected] (wA) -- (yA) ;
                         \draw [photon,red][out=40,in=140] (xA) to[out=40,in=140] (yA);
                [out=40,in=140] (xA) to[out=40,in=140]
                         \draw [fill] (yA) circle (.08);
                         \draw [fill] (wA) circle (.08);
                         \draw [fill] (zA) circle (.08);
                         \draw [fill] (xA) circle (.08);
                         \draw [fill] (xB) circle (.08);
                         \draw [fill] (yB) circle (.08);
                         \draw [fill] (zB) circle (.08);
                         \draw [fill] (wB) circle (.08);  
                        
                       \end{tikzpicture}\caption{PPR+RPP}
        \end{subfigure}
        \caption{The relevant topsectors of the boundaries in the radiation reaction regions at 5PM-1SF.}\label{fig:RadRe51}
\end{figure}

All regions containing a single radiative graviton can be rewritten within the same 5PM-1SF family. The expansion in the $RPP$ region reads
\begin{align}
       & {\mathcal{P}^{\text{5PM-1SF}}}^{\{\sigma_1, \sigma_2,\sigma_3,\sigma_4,\sigma_5,\sigma_6,\sigma_7 \}}_{1,1,1,1,a_1,...,a_{18}} = \nn \\
        &(-1)^{a_1+a_2+a_3}
        {\mathcal{P}^{\text{5PM-1SF}}}^{\{0, -\sigma_2,-\sigma_3,-\sigma_4,-\sigma_5,0,0 \}}_{1,1,1,1,0,a_2,a_3,a_1+a_4,a_5,a_6+a_8,a_7+a_9,0,0,a_{10},0,a_{12},a_{13},a_{11}+a_{14},0,a_{16},a_{17},a_{15}+a_{18}} \nn\\
        &+\mathcal{O}(v^{d-a_1-a_2-a_3-a_4-2 a_5}).
  \end{align}
The $PRP$ and $PPR$ regions follow from the same expansion upon relabelling. All three regions scale as $v^d \sim v^{-2\epsilon}$. The boundary integrals belong to the (sub-)sectors of the topologies depicted in Fig.~\ref{fig:RadRe51}, and are again mushroom integrals---this time with two additional loops attached.

\subsection*{Tail RRP+PRR+RPR}
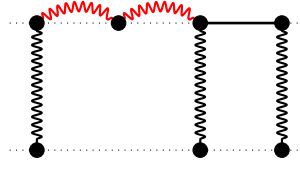
\begin{figure}[h!]\centering
        \begin{tikzpicture}[baseline={([yshift=-1ex]current bounding box.south)},scale=1.2]
                \coordinate (inA) at (0.3,.7);
                         \coordinate (outA) at (3.6,.7);
                         \coordinate (inB) at (0.3,-.7);
                         \coordinate (outB) at (3.6,-.7);
                         \coordinate (xA) at (0.6,.7);
                         \coordinate (yA) at (1.5,.7);
                         \coordinate (zA) at (2.4,.7);
                         \coordinate (wA) at (3.3,.7);
                         \coordinate (xB) at (0.6,-.7);
                         \coordinate (yB) at (1.5,-.7);
                         \coordinate (zB) at (2.4,-.7);
                         \coordinate (wB) at (3.3,-.7);
                         \coordinate (xM) at (0.6,0);
                         \coordinate (yM) at (1.5,1.05);
                         \coordinate (zM) at (2.4,0);
                         \coordinate (wM) at (3.3,0);
                         \coordinate (xyMB) at (1.05,-.35);
                         \coordinate (yzMB) at (1.95,-.35);
                         \coordinate (zwMB) at (2.85,-.35);
                         \coordinate (yzMA) at (1.95,.35);
                \draw [dotted] (inA) -- (outA);
                \draw [dotted] (inB) -- (outB) node [midway, below]{} ;
                \draw [draw=none] (xA) to[out=40,in=140] (zA);
                    \draw [zUndirected] (zA) -- (wA);
                \draw [photon] (zA) -- (zB);
                \draw [photon] (wA) -- (wB);
                \draw [photon] (xA) -- (xB);
                \draw [photon,red][out=40,in=140] (xA) to[out=40,in=140] (yA);
                \draw [photon,red][out=40,in=140] (yA) to[out=40,in=140] (zA);
       
          \draw [fill] (xA) circle (.08);
          \draw [fill] (yA) circle (.08);
          \draw [fill] (wB) circle (.08);
                \draw [fill] (zA) circle (.08);
                \draw [fill] (wA) circle (.08);
                \draw [fill] (xB) circle (.08);
                \draw [fill] (zB) circle (.08);
              \end{tikzpicture}
              \caption{The tail region boundary topology.}\label{fig:51tail}
\end{figure}

Three regions containing two radiative gravitons give rise to tail-type integrals. The boundary integrals here belong to a 2SF family, rather than to the 1SF family. The $RRP$ expansion is given by
\begin{align}
        &{\mathcal{P}^{\text{5PM-1SF}}}^{\{\sigma_1, \sigma_2,\sigma_3,\sigma_4,\sigma_5,\sigma_6,\sigma_7 \}}_{1,1,1,1,a_1,...,a_{18}} = \nonumber \\
        &(-1)^{a_1+a_2} {\mathcal{P}^{\text{5PM-2SF}}}^{\{0, 0,-\sigma_3,0,0,0,\sigma_5,\sigma_6\}}_{\tiny{1,1,1,1,0,0,a_3,a_1+a_2+a_4,
        0,0,a_5,a_6,a_8,a_7+a_9+a_{10},
        0,0,a_{13},a_{11}+a_{12}+a_{14},0,0,a_{17},a_{15}+a_{16}+a_{18}}}\nn\\
      &  +\mathcal{O}(v^{2d-a_1-a_2-a_3-a_4-2 (a_5+a_6+a_8)-1}).
\end{align}
This expansion carries the velocity scaling $v^{-4\epsilon}$ and we use the 5PM-2SF family, which is defined in the next chapter in \refeq{integralSchematic}. The other two expansions $PRR$ and $RPR$ can be done by relabelling and the same expansion. Although the integrals carry up to three radiative gravitons, only two of the three possible combinations actually contribute; one of the three expansions, depending on the parametrisation of the integral, gives rise to scaleless integrals and does not give any contribution. After IBP reduction, the resulting tail integrals can all be expressed in terms of the topology shown in Fig.~\ref{fig:51tail}.
\subsubsection*{Tail-of-tail RRR}
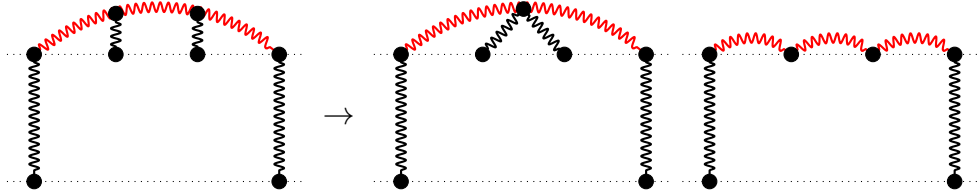
\begin{figure}[h!]\centering
        \begin{tikzpicture}[baseline={([yshift=-1ex]current bounding box.south)},scale=1.2]
                \coordinate (inA) at (0.3,.7);
                         \coordinate (outA) at (3.6,.7);
                         \coordinate (inB) at (0.3,-.7);
                         \coordinate (outB) at (3.6,-.7);
                         \coordinate (xA) at (0.6,.7);
                         \coordinate (yA) at (1.5,.7);
                         \coordinate (zA) at (2.4,.7);
                         \coordinate (wA) at (3.3,.7);
                         \coordinate (xB) at (0.6,-.7);
                         \coordinate (yB) at (1.5,-.7);
                         \coordinate (zB) at (2.4,-.7);
                         \coordinate (wB) at (3.3,-.7);
                         \coordinate (xM) at (0.6,0);
                         \coordinate (yM) at (1.5,1.15);
                         \coordinate (zM) at (2.4,1.15);
                         \coordinate (wM) at (3.3,0);
                         \coordinate (xyMB) at (1.05,-.35);
                         \coordinate (yzMB) at (1.95,-.35);
                         \coordinate (zwMB) at (2.85,-.35);
                         \coordinate (yzMA) at (1.95,.35);
                \draw [dotted] (inA) -- (outA);
                \draw [dotted] (inB) -- (outB)  ;
                \draw [draw=none] (xA) to[out=40,in=140] (zA);
                \draw [photon] (wA) -- node [midway, right]{$ \quad \rightarrow$}(wB);
          
                \draw [photon] (xA) -- (xB);
                \draw [photon] (yA) -- (yM);
                \draw [photon] (zA) -- (zM);
                \draw [photon,red][out=40,in=140] (xA) to[out=40,in=140] (wA);
       
          \draw [fill] (xA) circle (.08);
          \draw [fill] (wA) circle (.08);
          \draw [fill] (yM) circle (.08);
          \draw [fill] (zM) circle (.08);
          \draw [fill] (wB) circle (.08);
                \draw [fill] (zA) circle (.08);
                \draw [fill] (xB) circle (.08);
                \draw [fill] (yA) circle (.08);
              \end{tikzpicture}
              \begin{tikzpicture}[baseline={([yshift=-1ex]current bounding box.south)},scale=1.2]
                \coordinate (inA) at (0.3,.7);
                         \coordinate (outA) at (3.6,.7);
                         \coordinate (inB) at (0.3,-.7);
                         \coordinate (outB) at (3.6,-.7);
                         \coordinate (xA) at (0.6,.7);
                         \coordinate (yA) at (1.5,.7);
                         \coordinate (zA) at (2.4,.7);
                         \coordinate (wA) at (3.3,.7);
                         \coordinate (xB) at (0.6,-.7);
                         \coordinate (yB) at (1.5,-.7);
                         \coordinate (zB) at (2.4,-.7);
                         \coordinate (wB) at (3.3,-.7);
                         \coordinate (xM) at (0.6,0);
                         \coordinate (yM) at (1.95,1.2);
                         \coordinate (zM) at (1.95,1.15);
                         \coordinate (wM) at (3.3,0);
                         \coordinate (xyMB) at (1.05,-.35);
                         \coordinate (yzMB) at (1.95,-.35);
                         \coordinate (zwMB) at (2.85,-.35);
                         \coordinate (yzMA) at (1.95,.35);
                \draw [dotted] (inA) -- (outA);
                \draw [dotted] (inB) -- (outB)  ;
                \draw [draw=none] (xA) to[out=40,in=140] (zA);
                \draw [photon] (wA) -- (wB);
          
                \draw [photon] (xA) -- (xB);
                \draw [photon] (yA) -- (yM);
                \draw [photon] (zA) -- (yM);
                \draw [photon,red][out=40,in=140] (xA) to[out=40,in=140] (wA);
       
          \draw [fill] (xA) circle (.08);
          \draw [fill] (wA) circle (.08);
          \draw [fill] (yM) circle (.08);
          \draw [fill] (wB) circle (.08);
                \draw [fill] (zA) circle (.08);
                \draw [fill] (xB) circle (.08);
                \draw [fill] (yA) circle (.08);
              \end{tikzpicture}
              \begin{tikzpicture}[baseline={([yshift=-1ex]current bounding box.south)},scale=1.2]
                \coordinate (inA) at (0.3,.7);
                         \coordinate (outA) at (3.6,.7);
                         \coordinate (inB) at (0.3,-.7);
                         \coordinate (outB) at (3.6,-.7);
                         \coordinate (xA) at (0.6,.7);
                         \coordinate (yA) at (1.5,.7);
                         \coordinate (zA) at (2.4,.7);
                         \coordinate (wA) at (3.3,.7);
                         \coordinate (xB) at (0.6,-.7);
                         \coordinate (yB) at (1.5,-.7);
                         \coordinate (zB) at (2.4,-.7);
                         \coordinate (wB) at (3.3,-.7);
                         \coordinate (xM) at (0.6,0);
                         \coordinate (yM) at (1.5,1.15);
                         \coordinate (zM) at (2.4,1.15);
                         \coordinate (wM) at (3.3,0);
                         \coordinate (xyMB) at (1.05,-.35);
                         \coordinate (yzMB) at (1.95,-.35);
                         \coordinate (zwMB) at (2.85,-.35);
                         \coordinate (yzMA) at (1.95,.35);
                \draw [dotted] (inA) -- (outA);
                \draw [dotted] (inB) -- (outB) ;
                \draw [draw=none] (xA) to[out=40,in=140] (zA);
                \draw [photon] (wA) -- (wB);
          
                \draw [photon] (xA) -- (xB);
                \draw [photon,red][out=40,in=140] (xA) to[out=40,in=140] (yA);
                \draw [photon,red][out=40,in=140] (yA) to[out=40,in=140] (zA);
                \draw [photon,red][out=40,in=140] (zA) to[out=40,in=140] (wA);
       
          \draw [fill] (xA) circle (.08);
          \draw [fill] (wA) circle (.08);
          \draw [fill] (wB) circle (.08);
                \draw [fill] (zA) circle (.08);
                \draw [fill] (xB) circle (.08);
                \draw [fill] (yA) circle (.08);
              \end{tikzpicture}
              \caption{The topologies relevant for the $RRR$ region. The integral on the LHS with three radiative gravitons interacting with two potential gravitons is reducible under IBP reduction to the topologies on the right hand side.}
              \label{fig:tailtail}
\end{figure}

The genuinely new region at 5PM-1SF is the so-called tail-of-tail region, with three radiative gravitons. The $RRR$ expansion takes the form
\begin{align}
        &{\mathcal{P}^{\text{5PM-1SF}}}^{\{\sigma_1, \sigma_2,\sigma_3,\sigma_4,\sigma_5,\sigma_6,\sigma_7 \}}_{1,1,1,1,a_1,...,a_{18}} = \nonumber \\
        &(-1)^{a_1+a_2+a_3} {\mathcal{P}^{\text{5PM-1SF}}}^{\{0, 0,0,0,\sigma_5,\sigma_6,\sigma_7\}}_{\tiny{1,1,1,1,0,0,0,a_{1}+a_2+a_3+a_4,a_5,a_6,a_7,a_8,a_9,a_{10},0,0,0,a_{11}+a_{12}+a_{13}+a_{14},0,0,0,a_{15}+a_{16}+a_{17}+a_{18}}}\nn\\
       &+\mathcal{O}(v^{3d-a_1-a_2-a_3-a_4-2 (a_5 +a_6+a_7+a_8+a_9+a_{10})-2}).
\end{align}
This expansion carries the velocity scaling $v^{-6\epsilon}$. The boundary integrals once again belong to the 5PM-1SF integral family, and take the form of a radiative-graviton mushroom interacting with two potential gravitons. The topsector, shown on the left of Fig.~\ref{fig:tailtail}, contains three radiative gravitons. Under IBP reduction it reduces to master integrals; one MI contains only two radiative gravitons, shown on the right hand side of Fig.~\ref{fig:tailtail}. In total we find four master integrals, one each for the B- and V-type in each topology. We remark that even integrals carrying only two radiative gravitons can develop this $RRR$ region.
 
\vspace{0cm}\newpage
                
\chapter{Integrals at 5PM-2SF order}
\label{sec:5pm2sf}
The last category of integrals we will discuss in this thesis are integrals contributing at 5PM-2SF order. 
These integrals were first analytically computed in~\cite{Driesse:2026qiz}, work to which the present author contributed. The potential contributions were computed in~\cite{Bern:2025zno,Bern:2025wyd} in the form of a series expansion in the small velocity limit.

The 2SF integrals will be a little different from the 1SF integrals for two reasons. We will discuss only the integrals contributing to the conservative scattering angle. So we will only discuss the B-type integrals here and we will allow for the usage of symmetries in our IBP relations. Furthermore the integrals will not be planar anymore. We will see that we will have four different integral families from which only one will be planar. Partial fraction identities are not enough to planarise the integrand at 2SF.

\section{Integral families at second self-force order}
The four families all have the same general form:
\begin{align}\label{integralSchematic}
        {\cal F}_{\{n\}}^{\{\sigma\}}
        =
        \int_{\ell_1\ell_2\ell_3\ell_4}\!\!\!\!\!\!\!\!
        \frac{
          \prod_{i=1}^{2}\dd^{(n_i\!-\!1)}(\ell_i\cdot v_1)
          \prod_{j=3}^{4}\dd^{(n_j\!-\!1)}(\ell_j\cdot v_2)
          }{
          \prod_{i=1}^4 D_{i}^{F,n_{i+4}}(\sigma_{i})
          \prod_{i=1}^{14} D_{i+4}^{F,n_{i+8}}
          }\,,
      \end{align}

The first three integral families $P^{5,2},\text{NX1}, \text{NX2}$ share the same graviton propagator in the same structure as known from lower orders:
\begin{align}
       & D^F_{5/6}=(\ell_{1/2}-\ell_3)^2
        \,,&
        D^F_{7/8}&=(\ell_{1/2}-\ell_4)^2
        \,,&
        D^F_{9/10}&=(\ell_{1/3}-\ell_{2/4})^2
        \,,& \nn\\
       & D^F_{10+i}=(\ell_i-q)^2
        \,,&
        D^F_{14+i}&=\ell_i^2
        \,.
\end{align}
with $i \in [1,2,3,4]$ and $F \in [\text{P}^{5,2},\text{NX1},\text{NX2}]$. They differ, however, in the WL propagators:
\begin{align}
        & D^F_{1}=\ell_1 \cdot v_2, \qquad \nn \\
        &  D^{\text{P}}_{2}=\ell_2\cdot v_2, \qquad 
          D^{\text{NX1}/\text{NX2}}_{2}=(\ell_1 -\ell_2)\cdot v_2, \qquad \nn\\
        &   D^{F}_{3}=\ell_3 \cdot v_1, \qquad \nn\\
        &   D^{\text{P}/\text{NX1}}_{4}=\ell_4\cdot v_1, \qquad 
          D^{\text{NX2}}_{4}=(\ell_3 -\ell_4)\cdot v_1, \qquad 
\end{align}

In addition to these three families we have the non-planar $\text{NP}$ family, which also has a different graviton structure. Its propagators are defined as:
\begin{align}
        \begin{aligned}
          D^{\text{NP}}_1&=\ell_1\cdot v_2\,, &
          D^{\text{NP}}_2&=(\ell_1-\ell_2)\cdot v_2\,,\\
          D^{\text{NP}}_3&=\ell_3\cdot v_1\,, &
          D^{\text{NP}}_4&=\ell_4\cdot v_1\,.
        \end{aligned}\nn
        \end{align}
        \begin{align}
        \begin{aligned}
          D^{\text{NP}}_5&=(\ell_1-\ell_2-\ell_3+\ell_4)^2\,,	&& &&\\
          D^{\text{NP}}_6&=(\ell_1-\ell_3)^2\,, \qquad\qquad
          &&D^{\text{NP}}_7=(\ell_2-\ell_4)^2\,,				&& &&\\
          D^{\text{NP}}_8&=(\ell_1-\ell_2)^2\,, \qquad\qquad
          &&D^{\text{NP}}_9=(\ell_3-\ell_4)^2\,,				&&\\
          D^{\text{NP}}_{10}&=\ell_1^2\,, \quad
          D^{\text{NP}}_{11}=\ell_2^2\,, \quad
          &&D^{\text{NP}}_{12}=\ell_3^2\,, \quad
          &&D^{\text{NP}}_{13}=\ell_4^2\,,\\
          D^{\text{NP}}_{14}&=(\ell_1+q)^2\,, \qquad\qquad
          &&D^{\text{NP}}_{15}=(\ell_3+q)^2\,,				&&\\
          D^{\text{NP}}_{16}&=\ell_2\cdot\ell_3\,, \qquad\qquad
          &&D^{\text{NP}}_{17}=\ell_2\cdot q\,, \qquad
          &&D^{\text{NP}}_{18}=\ell_4\cdot q\,.
          \label{DNPs}
        \end{aligned}
        \end{align}
In particular the non planar NP family contributes to the increased complexity of the 2SF computation.

\section{The hidden $S_3$ symmetry of the non-planar family}
We have already seen at 4PM that an understanding of the underlying symmetries can considerably simplify both the canonicalisation of the differential equations and the IBP reduction itself. At 5PM-2SF this becomes even more crucial. Since we compute the conservative part of the 2SF scattering angle, we are free to exploit the full set of symmetry relations. Though the detection of symmetries is a standard feature of every modern IBP package, the situation here is somewhat different. The relevant integrals are non-planar topologies of the form
\begin{align*}
        \begin{tikzpicture}[baseline={([yshift=-1ex]current bounding box.south)},scale=1.5]
                \coordinate (inA)  at (0.6,0.6);
                \coordinate (outA) at (2.625,0.6);
                \coordinate (inB)  at (0.6,-0.6);
                \coordinate (outB) at (2.625,-0.6);
                \draw[dotted] (inA) -- (outA);
                \draw[dotted] (inB) -- (outB);
                \coordinate (aA) at (0.6,0.6);
                \coordinate (bA) at (1.85+0.1,0.6);
                \coordinate (cA) at (1.95,0.6);
                \coordinate (dA) at (2.625,0.6);
                \coordinate (eA) at (3.3,0.6);
                \coordinate (aM) at (0.6,0);
                \coordinate (bM) at (1.275,0);
                \coordinate (cM) at (1.95,0);
                \coordinate (dM) at (2.625,0);
                \coordinate (eM) at (3.3,0);
                \coordinate (aB) at (0.6,-0.6);
                \coordinate (bB) at (1.275,-0.6);
                \coordinate (cB) at (1.95,-0.6);
                \coordinate (dB) at (2.625,-0.6);
                \coordinate (eB) at (3.3,-0.6);
                \coordinate (bL) at (1.85+0.1,-0.35);
                \draw[photon,red] (aM) -- (dM);
                \draw[fill, white] (1.85+0.1,0) circle (0.15);
                \draw[photon] (aA) -- node [midway, left]{$\ell_1+q\downarrow$}  (aM);
                \draw[photon] (aM) -- node [midway, left]{$ \ell_3+q\downarrow$}  (aB);
               \draw[photon,red] (bM) -- (bL);
                \draw[photon] (bL) -- (bB);
                \draw[photon] (dA) -- node [midway, right]{$\downarrow \ell_2$}  (dM);
                \draw[photon] (dM) -- node [midway, right]{$\downarrow \ell_4$}  (dB);
                 \draw[photon] (bL) -- (bA);
                \foreach \v in {aA,bA,dA,aM,bM,dM,aB,bB,dB,bL}
                  \draw[fill] (\v) circle (.08);
              \end{tikzpicture}
\end{align*}
When this topology is reduced with \texttt{Kira}, the program detects in the topsector a single symmetry relation, corresponding to a mirror reflection of the graph along the horizontal axis---in our integral family, the relabelling $\ell_1\leftrightarrows \ell_2$, $\ell_3\leftrightarrows \ell_4$:
\begin{align*}
        \begin{tikzpicture}[baseline={([yshift=-1ex]current bounding box.south)},scale=1.5]
                \coordinate (inA)  at (0.6,0.6);
                \coordinate (outA) at (2.625,0.6);
                \coordinate (inB)  at (0.6,-0.6);
                \coordinate (outB) at (2.625,-0.6);
                \draw[dotted] (inA) -- (outA);
                \draw[dotted] (inB) -- (outB);
                \coordinate (aA) at (0.6,0.6);
                \coordinate (bA) at (1.85+0.1,0.6);
                \coordinate (cA) at (1.95,0.6);
                \coordinate (dA) at (2.625,0.6);
                \coordinate (eA) at (3.3,0.6);
                \coordinate (aM) at (0.6,0);
                \coordinate (bM) at (1.275,0);
                \coordinate (cM) at (1.95,0);
                \coordinate (dM) at (2.625,0);
                \coordinate (eM) at (3.3,0);
                \coordinate (aB) at (0.6,-0.6);
                \coordinate (bB) at (1.275,-0.6);
                \coordinate (cB) at (1.95,-0.6);
                \coordinate (dB) at (2.625,-0.6);
                \coordinate (eB) at (3.3,-0.6);
                \coordinate (bL) at (1.85+0.1,-0.35);
                \draw[photon,red] (aM) -- (dM);
                \draw[fill, white] (1.85+0.1,0) circle (0.15);
                \draw[photon] (aA) -- node [midway, left]{$\ell_1+q\downarrow$}  (aM);
                \draw[photon] (aM) -- node [midway, left]{$ \ell_3+q\downarrow$}  (aB);
               \draw[photon,red] (bM) -- (bL);
                \draw[photon] (bL) -- (bB);
                \draw[photon] (dA) -- node [midway, right]{$\downarrow \ell_2$}  (dM);
                \draw[photon] (dM) -- node [midway, right]{$\downarrow \ell_4$}  (dB);
                 \draw[photon] (bL) -- (bA);
                 \draw[righleft][out=40,in=140] (aB)to [out=-45,in=-135](dB);
                \foreach \v in {aA,bA,dA,aM,bM,dM,aB,bB,dB,bL}
                  \draw[fill] (\v) circle (.08);
              \end{tikzpicture}.
\end{align*}
Inspecting the topology more closely reveals a further symmetry which is not detected by the IBP package. The diagram contains three active propagators, each of which is connected to two potential ones. Every vertex may be moved freely along the worldlines, since they are cut, and in particular the middle arm may be swapped with the left or right arm:
\begin{align*}
        \begin{tikzpicture}[baseline={([yshift=-1ex]current bounding box.south)},scale=1.5]
                \coordinate (inA)  at (0.6,0.6);
                \coordinate (outA) at (2.625,0.6);
                \coordinate (inB)  at (0.6,-0.6);
                \coordinate (outB) at (2.625,-0.6);
                \draw[dotted] (inA) -- (outA);
                \draw[dotted] (inB) -- (outB);
                \coordinate (aA) at (0.6,0.6);
                \coordinate (bA) at (1.85+0.1,0.6);
                \coordinate (cA) at (1.95,0.6);
                \coordinate (dA) at (2.625,0.6);
                \coordinate (eA) at (3.3,0.6);
                \coordinate (aM) at (0.6,0);
                \coordinate (bM) at (1.275,0);
                \coordinate (cM) at (1.95,0);
                \coordinate (dM) at (2.625,0);
                \coordinate (eM) at (3.3,0);
                \coordinate (aB) at (0.6,-0.6);
                \coordinate (bB) at (1.275,-0.6);
                \coordinate (cB) at (1.95,-0.6);
                \coordinate (dB) at (2.625,-0.6);
                \coordinate (eB) at (3.3,-0.6);
                \coordinate (bL) at (1.85+0.1,-0.35);
                \draw[photon,red] (aM) -- (dM);
                \draw[fill, white] (1.85+0.1,0) circle (0.15);
                \draw[photon] (aA) -- node [midway, left]{$\ell_1+q\downarrow$}  (aM);
                \draw[photon] (aM) -- node [midway, left]{$ \ell_3+q\downarrow$}  (aB);
               \draw[photon,red] (bM) -- (bL);
                \draw[photon] (bL) -- (bB);
                \draw[photon] (dA) -- node [midway, right]{$\ell_2$}  (dM);
                \draw[photon] (dM) -- node [midway, right]{$ \ell_4$}  (dB);
                 \draw[photon] (bL) -- (bA);
                  \draw[righleft][out=40,in=140] (0.7,-0.1)to [out=-60,in=-180](1.75,-0.35);
                \foreach \v in {aA,bA,dA,aM,bM,dM,aB,bB,dB,bL}
                  \draw[fill] (\v) circle (.08);
              \end{tikzpicture}.
\end{align*}
This corresponds to the shifts $\ell_1\to\ell_1+\ell_2$ and $\ell_3\to\ell_3+\ell_4$, followed by $\ell_{2/4}\to-\ell_{2/4}$. It is also immediately clear that standard IBP packages cannot detect this symmetry, since it relies on the identity $\delta(\ell_1\cdot v_1)\delta((\ell_1-\ell_2)\cdot v_1)=\delta(\ell_1\cdot v_1)\delta(\ell_2\cdot v_1)$. Exploiting it is, however, crucial: it reduces the number of master integrals dramatically, and in this topology it brings the count down from 197 to 87.

\section{IBP reduction and choice of reduction basis}
We have already discussed the difficulties of an IBP reduction at 5PM-1SF order. At first self-force order, the principal obstacle was the generation of an equation system which did not exhaust the available memory; the solution of the system itself, performed over finite fields, was relatively unproblematic. At 2SF, this second step also becomes a major bottleneck.

When solving the equation system over finite fields, one solves the system multiple times modulo a prime. The number of numerical evaluations (probes) required depends on the polynomial degree of the rational functions to be reconstructed---in our case, the coefficients appearing in the IBP reduction. At 1SF we were fortunate that the coefficients were simple enough that the finite-field evaluation went through smoothly. At 2SF this is no longer the case. The total reduction time scales roughly as
\begin{align}
       \dfrac{\#\text{probes}\times \#\text{prime fields}\times \text{probe time}}{\#\text{cores}}.
\end{align}
The number of probes is set by the polynomial degree, the number of prime fields by the size of the integers entering the equations, and the probe time itself by the size of the equation system. The remaining factor, the number of cores, is fixed by the available compute resources. There have been many approaches to optimise these parameters by various methods~\cite{Klappert:2019emp,Klappert:2020aqs,Liu:2023cgs,Belitsky:2024jhe,Smirnov:2024onl}.

For the reduction of non-planar integrals, all of these parameters take very unfavourable values when one attempts a naive reduction. There are several reasons for this. First, non-planar integrals carry larger IBP vectors, so that individual equations contain more terms than in the planar case. The reason lies primarily in the presence of the denominator $(\ell_1-\ell_2-\ell_3+\ell_4)^2$: differentiating with respect to any loop momentum produces numerators involving all four loop momenta, which proliferate the terms in the IBP equations. While this cannot be avoided altogether, the size of the IBP vectors can be reduced somewhat by a good parametrisation of the family. Furthermore, more equations must be generated in order for the system to close: to reduce integrals carrying up to two scalar products, one must seed equations carrying up to two scalar products and three dots. Both of these effects drive the probe time up, and there is no further room to improve them.

The number of probes and the number of prime fields, by contrast, depend strongly on the basis to which the integrals are reduced. In the worst case, the basis introduces unphysical denominators that depend simultaneously on $d$ and $\gamma$ and do not factorise; such denominators typically inflate the size of the integers and the polynomial degrees substantially. Even when these denominators are avoided, however, some bases perform considerably better than others, and one may use algorithms such as those of refs.~\cite{Usovitsch:2020jrk,Smirnov:2020quc} to search for a good basis. We have observed in this regard that, although such algorithms can be used to construct improved bases, every basis we obtained algorithmically gave rise to higher polynomial degrees and required more prime fields than a reduction performed directly to a pre-canonical basis. By a pre-canonical basis we mean a basis as close as possible to the canonical one, but without introducing square roots or special functions.

Identifying such a basis is, of course, not straightforward. We adopt the following strategy. We first use algorithms and reductions on the maximal cut to construct a factorised basis which contains no denominators depending on both $d$ and $\gamma$ simultaneously. This basis is then used to perform the reductions required for the differential equations, with which we identify a pre-canonical basis. The resulting pre-canonical basis is in turn used for the reduction of the full impulse. A pleasant side effect of this strategy is that inserting integral results at the end becomes very straightforward, since the integral results no longer require large transformation matrices to the canonical basis.

The use of a pre-canonical basis carries a drawback: such integrals often contain multiple dots, while the integrals required for the impulse predominantly carry a high number of scalar products. To avoid an inflated equation system, we work with two bases simultaneously. The first is a Laporta basis $\vec{I}_{SP}$ built on scalar products, which yields a smaller equation system but requires substantially more time for finite-field evaluation. The second is the pre-canonical basis $\vec{I}_{pre}$, optimal for finite-field reduction. To combine the advantages of both, we perform a pre-reduction in which the Laporta basis is first reduced to the pre-canonical basis,
\begin{align}
        \vec{I}_{\text{SP}}=R_{pr}\, \vec{I}_{pre}, \label{eq:preRed}
\end{align}
where $R_{pr}$ is an $n\times n$ matrix and $n$ is the size of the basis. When we then wish to reduce the integrals $\vec J$ that enter the impulse, we set up an equation system reducing them to the Laporta basis,
\begin{align}
        \vec{J}=R\, \vec{I}_{\text{SP}},
\end{align}
where $R$ is now an $m\times n$ matrix and $m$ is the number of integrals to be reduced. Rather than evaluating $R$ over finite fields directly, we substitute~\refeq{eq:preRed},
\begin{align}
        \vec{J}=R\, R_{\text{pr}}\, \vec{I}_{\text{pre}}=R_2\, \vec{I}_{\text{pre}},
\end{align}
so that the reduction can be performed directly to the pre-canonical basis without generating any additional equations.

This also links the IBP reduction to the second main obstacle to computing the integrals: the construction of a canonical basis, which we discuss in detail in the following sections.
\section{A sector related to a K3 surface at 2SF order}

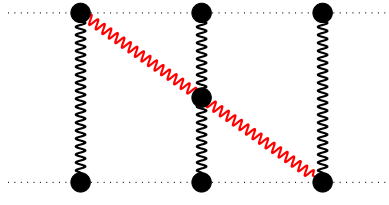
\begin{figure}[h!]
    \centering
\begin{tikzpicture}[baseline={([yshift=-1ex]current bounding box.south)},scale=1.6]
  \coordinate (inA) at (0.4,.7);
  \coordinate (outA) at (3.6,.7);
  \coordinate (inB) at (0.4,-.7);
  \coordinate (outB) at (3.6,-.7);
  \coordinate (xA) at (1,.7);
  \coordinate (xxA) at (1.825,.7) ;
  \coordinate (xxB) at (1.825,-.7) ;
  \coordinate (yA) at (1.60+0.4,.7);
  \coordinate (yyA) at (2.375,0) ;
  \coordinate (zA) at (3,.7);
  \coordinate (zzA) at (3.75,.7) ;
  \coordinate (xB) at (1,-.7);
  \coordinate (yB) at (1.6+0.4,-.7);
  \coordinate (zB) at (3,-.7);
  \coordinate (zzB) at (3.75,-.7);
  \coordinate (xM) at (1,0);
  \coordinate (yM) at (1.6+0.4,0);
  \coordinate (zM) at (3,-0);
  \draw [dotted] (inA) -- (outA);
  \draw [dotted] (inB) -- (outB) node [midway, below]{} ;
  \draw [draw=none] (xA) to[out=40,in=140] (zA);
  \draw [photon] (xA) -- (xB);
  \draw [photon] (yA) -- (yM) -- (yB);
  \draw [photon] (zA) -- (zB);
  \draw [photon,red] (xA) -- (yM) -- (zB);
  \draw [fill] (yM) circle (.08);
  \draw [fill] (xA) circle (.08);
  \draw [fill] (yA) circle (.08);
  \draw [fill] (zA) circle (.08);
  \draw [fill] (xB) circle (.08);
  \draw [fill] (yB) circle (.08);
  \draw [fill] (zB) circle (.08);
\end{tikzpicture}
    \caption{A K3 sector at 5PM-2SF}
    \label{fig:K3GW}
\end{figure}
The first interesting sector is shown in Fig.~\ref{fig:K3GW} and contains nine master integrals on the maximal cut. Its canonical form was discussed in ref.~\cite{Duhr:2025lbz}, which we will follow closely. The $9\times 9$ block decomposes, however, into a $7\times 7$ and a $2\times 2$ block, in a manner reminiscent of the symmetry split. Unlike that earlier split, the two blocks here do not decouple completely: the $2\times 2$ block receives an off-diagonal coupling from the $7\times 7$ block, as if it were a topsector. Schematically, the differential-equation matrix on the maximal cut takes the form
\begin{align}
        \left(
                \begin{array}{ccccccccc}
                 \times  & \times  &\times  &\times&\times&\times&\times  &0  &0      \\
                 \times  & \times  &\times  &\times&\times&\times&\times  &0  &0      \\
                 \times  & \times  &\times  &\times&\times&\times&\times  &0  &0      \\
                 \times  & \times  &\times  &\times&\times&\times&\times  &0  &0      \\
                 \times  & \times  &\times  &\times&\times&\times&\times  &0  &0      \\
                 \times  & \times  &\times  &\times&\times&\times&\times  &0  &0      \\
                 \times  & \times  &\times  &\times&\times&\times&\times  &0  &0      \\
                 \times  & \times  &\times  &\times&\times&\times&\times  &\times  &\times      \\
                 \times  & \times  &\times  &\times&\times&\times&\times  &\times  &\times      \\
                \end{array}
                \right).
\end{align}
The basis we choose for this splitting is
\begin{align}
        I_1 =& \frac{x^2-1  }{x^2}\text{P}^{5,2}(1,1,1,1,0,0,0,0,1,0,0,1,1,1,1,0,0,0,0,0,0,1) \, , \quad I_2 =   \partial_x I_1 \, , \quad I_3 = \partial_x^2 I_1 \, , \nonumber\\
        I_4 =& \frac1\epsilon  \frac{\left(x^2-1\right)^2  }{x^2}\text{P}^{5,2}(1,1,1,1,0,0,0,0,1,0,0,2,1,1,1,0,0,0,0,0,0,1)\, , \nonumber\\
        I_5 =& \dfrac{1+2 \epsilon}{\epsilon^2} \text{P}^{5,2}(1,1,1,1,0,0,0,0,1,0,0,1,1,1,1,0,0,0,0,0,0,2)\, \nonumber ,\\
        I_6 =& \frac1{\epsilon^2}\frac{ x^2 }{\left(x^2-1\right)^2}\text{P}^{5,2}(1,1,1,1,-1,0,0,-1,1,0,0,1,1,1,2,0,0,0,0,0,0,2)\, , \nonumber\\
        I_7 =& \frac1{\epsilon^2}\frac{x^2-1}{x} \big[ \text{P}^{5,2}(1,1,1,2,0,0,0,-1,1,0,0,2,1,1,1,0,0,0,0,0,0,1) \nonumber \\
        & +\text{P}^{5,2}(1,1,1,2,0,0,0,-1,2,0,0,1,1,1,1,0,0,0,0,0,0,1) \big]\nonumber\\
             &  + \frac4\epsilon  \frac{x^2-1  }{x^2}\text{P}^{5,2}(1,1,1,1,0,0,0,0,1,0,0,2,1,1,1,0,0,0,0,0,0,1)\,  \nonumber \displaybreak\\
             I_8 =& \frac1{\epsilon^2}\frac{x^2-1  }{x }\text{P}^{5,2}(1, 2, 2, 1, 0, 0, 0, 0, 1, 0, -1, 1, 1, 1, 1, 0, 0, 0, 0, 0, 0, 1)\nonumber\\
             I_9 =&\frac1{\epsilon} \frac{(x^2-1 )^2 }{x^2}\text{P}^{5,2}(1, 1, 1, 1, 0, 0, 0, 0, 2, 0, 0, 1, 1, 1, 1, 0, 0, -1, 0, 0, 0, 1) \, .
             \label{eq:basisK32sff}
\end{align}
A splitting of this kind appears at 2SF in many larger sectors. Its presence can often be detected by passing to a full derivative basis of the corner integral: if such a splitting exists, it will not be possible to fill out the entire basis by derivatives alone. Once the split has been performed, the $2\times 2$ block can be set aside and treated as a topsector in its own right.

We focus our attention on the $7\times 7$ block. We begin with an integrand analysis of the corner integral in a loop-by-loop Baikov representation. After two integrations we find
\begin{align}
    {\rm{LS}(\text{P}^{5,2}(1,1,1,1,0,0,0,0,1,0,0,1,1,1,1,0,0,0,0,0,0,1))}=\frac{  x^2}{x^2-1} \int \dfrac{\mathrm dt_1 \mathrm dt_2}{\sqrt{P_{4,4}(t_1,t_2)}} \,,
\label{eq:K3LS}
\end{align}
where
\begin{align}
    P_{4,4}(t_1,t_2) &= (1-t_1)(1+t_1)(1-t_2)(1+t_2)\nn \\
                    &\quad\times (1-2x(1-2t_1t_2)+x^2)(1+2x(1+2t_1t_2)+x^2) \, .
\end{align}
The polynomial $P_{4,4}$ has total degree eight in the variables $t_1, t_2$, while in either variable separately it appears only at degree four. 
This indicates an underlying K3 geometry.
The variety defined by $W=y^2-P_{4,4}(t_1,t_2)=0$, however, is singular, since $W$ and its Jacobian have common zeros. One must therefore be careful before identifying the integrand with a smooth K3 surface. One way to establish that~\refeq{eq:K3LS} does in fact correspond to a smooth K3 is to examine the Picard-Fuchs operator and check if it satisfies the properties of a third order Calabi--Yau operator. We can read the operator off the differential equations but could also compute it from the Baikov representation:
\begin{equation}
\begin{aligned}
    \mathcal{L}^{(3)}_{\text{K3,2}} &= \left(1-34 x^2+x^4\right)\theta ^3-6 x^2 \left(17-x^2\right)\theta ^2 -12 (3-x) (3+x) x^2 \theta -8 \left(5-x^2\right) x^2 \, .
\end{aligned}
\label{eq:aperyop}
\end{equation}
This operator is a third order CY operator, which was discussed in the mathematical literature~\cite{MR3890449, MR0749676}. It was used by Apéry to prove the irrationality of $\zeta(3)$~\cite{MR3363457}.
The Frobenius basis of the Apéry operator~\refeq{eq:aperyop} reads
\begin{align}
    \varpi_0(x)&=1+5 x^2+73 x^4 +1445 x^6+\mathcal{O}(x^8) \, ,\\
    \varpi_1(x)&=\varpi_0(x) \log(x)+S_1 = \varpi_0(x) \log(x) + 6 x^2+105 x^4 +2219 x^6+\mathcal{O}(x^8) \, , \nonumber\\
    \varpi_2(x)&=\dfrac{1}{2} \varpi_0(x) \log^2(x) +S_1 \log(x)+ 18 x^4 +540 x^6+\mathcal{O}(x^8) \, . \nonumber
\label{eq:frobK3BK}
\end{align}
We can connect again the corner integral, normalized by $x^2/(x^2-1)$, to the holomorphic form of the K3 geometry associated to the Apéry operator and we can choose it as our first integral in our basis. Additionally we get two more basis integrals by its first and second derivative. This gives us the basis of the pure K3 block.

We need six more integrals. We found $I_8$ and $I_9$ by demanding the first seven integrals to fully decouple from them. 
The other four integrals can be found again with a similar strategy that we used for the K3 sector at 5PM-1SF. In a ``good'' starting basis the K3 block should decouple from $I_4$-$I_7$ for $\epsilon=0$. Additionally we want the $4\times 4$ block formed by these integrals to be in $\epsilon$ form. Finally we don't want any denominators in our differential equations depending on $\epsilon$.
To make this search easier we could find one ``good'' integral by integrand analysis, 
\begin{align}
\text{P}^{5,2}(1,1,1,1,0,0,0,0,1,0,0,2,1,1,1,0,0,0,0,0,0,1)\;, 
\end{align}
which thus serves as the next starting integral. We found the other three basis elements by a list search using IBP reductions and checking if the potential candidates fulfill all criteria. This led to the choice of $I_5$-$I_7$.

The last two integrals can be ignored for now and we will treat them as a higher sector.
In this manner we arrived at the initial basis~\refeq{eq:basisK32sff}.

In this basis, $\vec I=(I_1,\ldots,I_7)^T$, the differential equation reads
\begin{align}
    \partial_x \vec I=M(x,\epsilon) \vec I \,.
\end{align}
The full matrix $M(x,\epsilon)$ is too large to display here. Of particular interest is its $\epsilon=0$ part, given by
\begin{align}
 M(x;0)=  \left(
\begin{array}{ccccccc}
 0 & 1 & 0 & 0 & 0 & 0 & 0 \\
 0 & 0 & 1 & 0 & 0 & 0 & 0 \\
 \frac{8 \left(5-x^2\right)}{x (1-34 x^2+x^4)} & -\frac{1-244x^2+19 x^4}{x^2 (1-34 x^2+x^4)} & -\frac{3 \left(1-68x^2+3 x^4\right)}{x (1-34 x^2+x^4)} & 0 & 0 & 0 & 0 \\
 0 & 0 & 0 & 0 & 0 & 0 & 0 \\
 \frac{10 x \left(17-x^2\right)}{3 (1+x) (1-x)} & -\frac{5 \left(1-34 x^2+x^4\right)}{3 (1+x) (1-x)} & 0 & 0 & 0 & 0 & 0 \\
 \frac{5 x^2 \left(5+x^2\right)}{3 (1-x)^2 (1+x)^2} & -\frac{5 x \left(1+x^2\right)}{6 (1-x) (1+x)} & 0 & 0 & 0 & 0 & 0 \\
 0 & 0 & 0 & 0 & 0 & 0 & 0 \\
\end{array}
\right) \, . \label{eq:K32sfeps0}
\end{align}
We show again the highest power of $\epsilon$ in each entry of the full matrix,
\begin{align}
    \left(
\begin{array}{ccccccc}
 -  & 0  & -  & -  & -  & -  & -  \\
 -  & -  & 0  & -  & -  & -  & -  \\
 3  & 2  & 1  & 3  & 3  & 3  & 3  \\
 -  & -  & -  & 1  & -  & -  & 1  \\
 1  & 0  & -  & 1  & 1  & 1 & 1   \\
 1  & 0  & -  & 1  & 1  & 1 & -   \\
 1  & -  & -  & 1  & 1  & -  & 1  \\
\end{array}
\right) \, ,
\label{eq:K3eps}
\end{align}
where each integer indicates the highest power of $\eps$ in the corresponding entry, and a `$-$' marks an entry that vanishes. 

Before rotating by the inverse of the semi-simple part, it is convenient to improve the integrals $I_4,\ldots,I_7$ by an intermediate rotation, in such a way that these four integrals couple at most to $I_1$ for $\epsilon=0$. This rotation has the form
\begin{align}
    T=\left(
\begin{array}{ccccccc}
 1 & 0 & 0 & 0 & 0 & 0 & 0 \\
 0 & 1 & 0 & 0 & 0 & 0 & 0 \\
 0 & 0 & 1 & 0 & 0 & 0 & 0 \\
 0 & 0 & 0 & 1 & 0 & 0 & 0 \\
 \frac{5 \left(1-34x^2+x^4\right)}{3 \left(1-x^2\right)} & 0 & 0 & 0 & 1 & 0 & 0 \\
 \frac{5 \left(1+x^2\right)x}{6 \left(1-x^2\right)} & 0 & 0 & 0 & 0 & 1 & 0 \\
 0 & 0 & 0 & 0 & 0 & 0 & 1 \\
\end{array}
\right) \, .
\end{align}
We then rotate the $3\times 3$ K3 block by the inverse of the semi-simple part of the Wronskian,
\begin{align}
    {W}^\text{SS}=\left(
\begin{array}{ccc}
 \varpi_0 & 0 & 0 \\
 \varpi_0' & \sqrt{C_2} & 0 \\
 \varpi_0'' & \sqrt{C_2}\frac{\varpi_0'}{\varpi_0} + \partial_x\sqrt{C_2} & \frac{C_2}{\varpi_0} \\
\end{array}
\right)  \; \text{with} \quad   C_2 = \frac{1}{x^2 \left(1-34 x^2+x^4\right)} \, ,
\label{eq:Grav_K3_W2}
\end{align}
and rescale by
\begin{align}
    D(\epsilon) = \text{diag}(\epsilon^2,\epsilon,1,\epsilon^2,\epsilon^2,\epsilon^2,\epsilon^2) \, .
    \label{eq:Grav_K3_D2}
\end{align}
We introduce here for the first time in our rotation a square root in the variable $x$. This is a new feature at 2SF and is especially remarkable since $\sqrt{1-34 x^2+x^4}^{-1}$ has a branch point in the physical region at $\gamma=3$.

Now we can remove terms $\sim 1/\epsilon$ and $\sim \epsilon^0$ succeedingly
by an ansatz of the form
\begin{align}
    \hat T_{i,j}=\sum_{k=-1}^0  \big(f_{i,j,k}(x)\,  \varpi^2_0(x)+g_{i,j,k}(x)\,  \varpi_0(x)\varpi'_0(x)+h_{i,j,k}(x)\,  \varpi_0(x)\big)\epsilon^k \, ,
\label{eq:rottotderiv}
\end{align}
where $f_{i,j,k}, g_{i,j,k}, h_{i,j,k}$ are algebraic functions of $x$. 
Other than for the K3 at 1SF, we cannot remove all terms with this ansatz but we have to introduce for a full canonicalisation of the differential equations four new transcendental functions.
The final rotation containing these functions has the form
\begin{align}
    \tilde{T}=
\begin{pmatrix}
 1 & 0 & 0 & 0 & 0 & 0 & 0 \\
 G_4 & 1 & 0 & 0 & 0 & 0 & 0 \\
 \tilde{T}_{3,1} & -G_4 & 1 & -64 G_2 & -\frac{896}{3} G_2 & 256 G_3 & 0 \\
 0 & 0 & 0 & 1 & 0 & 0 & 0 \\
 \frac{160 }{3}G_2 & 0 & 0 & 0 & 1 & 0 & 0 \\
 -\frac{5}{6}  G_3 & 0 & 0 & 0 & 0 & 1 & 0 \\
 0 & 0 & 0 & 0 & 0 & 0 & 1 \\
\end{pmatrix}\,,
\label{eq:K3G}
\end{align}
with
\begin{align}
    \tilde{T}_{3,1} = \frac1\epsilon G_1 -\frac{71680}{9}G_2^2-\frac{320}{3}G_3^2-\frac12 G_4^2 \, .
\end{align}
The defining first-order differential equations for the $G_i(x)$ functions read
\begin{equation}
\begin{aligned}
   G_1'(x) &= -\frac{16 x \left(1+x^2\right)p_1(x)}{9 (1-x)^3 (1+x)^3 \left(1-34 x^2+x^4\right)^2}\varpi_0(x)^2 \, ,\\
   p_1(x)  &= 115+4062 x^2-80259 x^4+1495652 x^6-80259 x^8+4062 x^{10}+115 x^{12} \, ,\\
   G_2'(x) &= \frac{x \left(1+x^2\right) }{(1-x)^2 (1+x)^2}\varpi_0(x) \, ,\\
   G_3'(x) &= \frac{\left(1+14x^2+x^4\right) }{(1-x)^2 (1+x)^2}\varpi_0(x) \, ,\\
   G_4'(x) &= \frac{1}{x \sqrt{(1-34 x^2+x^4)}} \frac{G_1(x)}{\varpi_0(x)} \, .
\end{aligned}
\end{equation}
In comparison with the K3 sectors appearing at 1SF order, we see an increase in complexity. We have to introduce four new transcendental functions and we have to introduce a square root. Nonetheless the transformation remains quite simple after choosing the right basis to start with.

After the rotation by the inverse semi-simple part of the Wronskian~\refeq{eq:Grav_K3_W2} and the $\epsilon$-\hspace{0pt}rescaling~\refeq{eq:Grav_K3_D2}, only $I_2$ and $I_3$ received corrections of the form~\refeq{eq:rottotderiv}. The integrals $I_4,\ldots,I_7$ required only corrections proportional to $I_1$. In particular, $I_4$ and $I_7$ required no corrections at all, which validates their original choice. 
We will refer to $\varpi_0$  from now on as $\varpi_{\text{K3,2}}$.

\section{A sector related to a Calabi--Yau threefold at 2SF order}
\begin{figure}[h]
    \centering
\begin{tikzpicture}[baseline={([yshift=-1ex]current bounding box.south)},scale=1.7]
  \coordinate (inA) at (0.4,.7);
  \coordinate (outA) at (3.6,.7);
  \coordinate (inB) at (0.4,-.7);
  \coordinate (outB) at (3.6,-.7);
  \coordinate (xA) at (1,.7);
  \coordinate (xxA) at (1.825,.7) ;
  \coordinate (xxB) at (1.825,-.7) ;
  \coordinate (yA) at (1.60+0.4,.7);
  \coordinate (yyA) at (2.375,0) ;
  \coordinate (zA) at (3,.7);
  \coordinate (zzA) at (3.75,.7) ;
  \coordinate (xB) at (1,-.7);
  \coordinate (yB) at (1.6+0.4,-.7);
  \coordinate (zB) at (3,-.7);
  \coordinate (zzB) at (3.75,-.7);
  \coordinate (xM) at (1,0);
  \coordinate (yM) at (1.6+0.4,0);
  \coordinate (zM) at (3,-0);
  \draw [dotted] (inA) -- (outA);
  \draw [dotted] (inB) -- (outB) node [midway, below]{};
  \draw [draw=none] (xA) to[out=40,in=140] (zA);
  \draw [photon] (xA) -- (xB);
  \draw [photon] (yA) -- (xM);
  \draw [photon] (zA) -- (zB);
  \draw [photon,red ] (xM) -- (yM) -- (zM);
  \draw [photon] (yB) -- (zM);
  \draw [fill] (xM) circle (.08);
  \draw [fill] (zM) circle (.08);
  \draw [fill] (xA) circle (.08);
  \draw [fill] (yA) circle (.08);
  \draw [fill] (zA) circle (.08);
  \draw [fill] (xB) circle (.08);
  \draw [fill] (yB) circle (.08);
  \draw [fill] (zB) circle (.08);
\end{tikzpicture}
    \caption{The sector related to a second CY threefold.}
    \label{fig:CY3GW}
\end{figure}
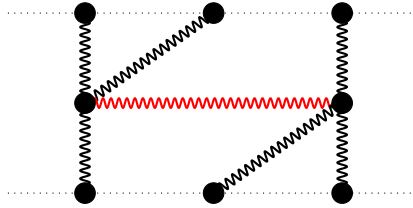

A second new geometry appears at 2SF in the form of a CY threefold in the sector depicted in Fig.~\ref{fig:CY3GW}.
This sector was discussed in ref.~\cite{Frellesvig:2023bbf}, and the geometry itself was fully classified in ref.~\cite{Klemm:2024wtd}.
Its canonical form was also discussed in ref.~\cite{Frellesvig:2024rea,Duhr:2025lbz}.
We will follow again closely ref.~\cite{Duhr:2025lbz}.

We use the starting basis
\begin{equation}
\begin{aligned}
I_1&=\dfrac{1}{x} \text{P}^{5,2}(1,1,1,1,0,0,0,0,0,0,1,0,1,1,1,0,1,0,0,1,0,1)  ,\,
I_2=\partial_x I_1       ,\,
I_3=\partial_x^2 I_1     ,\,
I_4=\partial_x^3 I_1    ,\\
I_5&= \frac{x^2-1 }{2x} \text{P}^{5,2}(1,1,1,1,0,0,0,0,0,0,2,0,1,1,1,0,1,-1,-1,1,0,1) \, .
\label{eq:startCY3}
\end{aligned}
\end{equation}
Let us motivate this choice of basis. In an integrand analysis it was shown in refs.~\cite{Frellesvig:2023bbf, Klemm:2024wtd} that the leading singularity of $I_1$ takes the form
\begin{align}
        \rm{LS}(I_1) = &\oint \dfrac{\mathrm dt_1 \mathrm dt_2 \mathrm dt_3 }{\sqrt{P}}
\end{align}
with
\begin{align}
P =&64\, t_1^2 t_2^2 t_3^2 x^2 (t_1+t_2+1)+(t_1+1)^2 (t_2+1)^2 \left(x^2-1\right)^2 \left(t_1 t_2+t_3^2\right)^2 \, .
\end{align}
This representation defines a singular CY threefold $X$ as a double cover of $\mathbb P^3$,
\begin{equation}
   X = \left\{t_4^2 = P(t_1,t_2,t_3)\right\}  \, ,
\end{equation}
in direct generalisation of the description of an elliptic curve as a double cover of $\mathbb P^1$ we discussed in Sec.~\ref{sec:geom} (see also ref.~\cite{Duhr:2024hjf} and references therein). The Picard-Fuchs operator is a CY operator of order four (see also ref.~\cite{Klemm:2024wtd} for details about the corresponding CY variety), and reads
\begin{align}
    \mathcal{L}^{(4)}_{CY3,2}=& (1-x)^3 (1+x)^3 \left(1+x^2\right)^2 \theta_x ^4-8 x^2 (1-x)^2 (1+x)^2 \left(1+x^2\right) \left(2+x^2\right)\theta ^3\nonumber \\
    &-4 x^2 (1-x) (1+x) \left(7-12 x^2-17 x^4-6 x^6\right)\theta ^2\\
\nonumber    &-8 x^2 \left(3-15 x^2+3 x^4+9 x^6+4 x^8\right) \theta -x^2 \left(9-71 x^2+39 x^4+39 x^6+16 x^8\right) \, .
\end{align}
Its Frobenius basis is given by
\begin{align}
    \varpi_{0}(x)=&1+\frac{9 x^2}{16}+\frac{1681 x^4}{4096}+\frac{21609 x^6}{65536}+\mathcal{O}(x^8) \, ,\\
    \varpi_{1}(x)=&\varpi_{0}(x) \log(x)+S_1(x) \nn \\
    =& \varpi_{0}(x) \log(x) +\frac{3 x^2}{8}+\frac{1517 x^4}{4096} +\frac{22295 x^6}{65536}+\mathcal{O}(x^8) \, , \nonumber \\
    \varpi_{2}(x)=&\dfrac{1}{2} \varpi_{0}(x) \log^2(x) +S_1(x) \log(x)+S_2(x) \nonumber \\
                &= \dfrac{1}{2} \varpi_{0}(x) \log^2(x) +S_1(x) \log(x)+ \frac{5 x^2}{32}+\frac{7289 x^4}{32768}+\frac{1123325 x^6}{4718592}+\mathcal{O}(x^8) \, , \nonumber \\
    \varpi_{3}(x)=&\dfrac{1}{6} \varpi_{0}(x) \log^3(x) +\dfrac{1}{2}S_1(x) \log^2(x)+S_2(x) \log(x) \nonumber \\
    &-\frac{5}{32}  x^2 -\frac{7865 x^4}{65536}-\frac{2448943 x^6}{28311552}+\mathcal{O}(x^8) \, . \nonumber
\end{align}

Once the integral of the first kind, $I_1$, has been identified, we can choose $I_2$-$I_4$ as the first, second and third derivative of $I_1$. 

For the fifth integral we choose again an approach to select an integral, for which, at $\epsilon=0$, the first four integrals decouple from the fifth one. Additionally we choose a normalisation such that $M_{5,5}$ is in $\epsilon$-form.
We find two integrals which are equal on the maximal cut up to normalisation. We find, however, that 
\begin{align}
        I_5&= \frac{x^2-1 }{2x}  \text{P}^{5,2}(1,1,1,1,0,0,0,0,0,0,2,0,1,1,1,0,1,-1,-1,1,0,1)
\end{align}
gives rise to simpler off-diagonal differential equations, and we therefore choose it.

In this basis, $\vec I=(I_1,\ldots,I_5)^T$, the differential equation reads
\begin{align}
    \partial_x \vec I =M(x,\epsilon) \vec I,
\end{align}
with the $\epsilon=0$ part given by
\begin{align}
 \resizebox{\textwidth}{!}{$\displaystyle
 M(x,0)=  \left(
\begin{array}{ccccc}
 0 & 1 & 0 & 0 & 0 \\
 0 & 0 & 1 & 0 & 0 \\
 0 & 0 & 0 & 1 & 0 \\
 \frac{9-71 x^2+39 x^4+39 x^6+16 x^8}{x^2(1-x)^3  (1+x)^3 \left(1+x^2\right)^2} & -\frac{1-69 x^2+202 x^4+22 x^6-123 x^8-65 x^{10}}{x^3(1-x)^3  (1+x)^3 \left(1+x^2\right)^2} & -\frac{7-76 x^2+10 x^4+116 x^6+55 x^8}{x^2(1-x)^2
   (1+x)^2 \left(1+x^2\right)^2} & -\frac{2 \left(3-8 x^2-7 x^4\right)}{x(1-x)  (1+x) \left(1+x^2\right)} & 0 \\
 -\frac{2 x \left(1-2 x^2\right)}{1+x^2} & \frac{(1-x) (1+x) \left(1-5 x^2\right)}{1+x^2} & \frac{x(1-x)^2  (1+x)^2}{1+x^2} & 0 & 0 \\
\end{array}
\right) \, .
   $ }
\end{align}
As before, it is also instructive to display the highest power of $\epsilon$ in each entry,
\begin{align}
    \left(
\begin{array}{ccccc}
 -  & 0 & -  & -  & -  \\
 -  & -  & 0 & -  & -  \\
 -  & -  & -  & 0 & -  \\
 4 & 3 & 2 & 1 & 3 \\
 2 & 1 & 0 & -  & 1 \\
\end{array}
\right) \, ,
\end{align}
where, again, integers indicate the highest power of $\epsilon$ and `$-$' marks vanishing entries. The fifth integral now couples to three integrals rather than to two, as in the K3 case.

We can try to improve our basis again by a rational transformation. 
We can remove the coupling of $I_5$ to $I_2$ and $I_3$ by rotating by
\begin{align}
    T=\left(
\begin{array}{ccccc}
 1 & 0 & 0 & 0 & 0 \\
 0 & 1 & 0 & 0 & 0 \\
 0 & 0 & 1 & 0 & 0 \\
 0 & 0 & 0 & 1 & 0 \\
 \frac{(1-x) (1+x) \left(1+6 x^2-3 x^4\right)}{4 \left(1+x^2\right)^2} & \frac{(1-x)^2 (1+x)^2 x}{2 \left(1+x^2\right)} & 0 & 0 & 1 \\
\end{array}
\right) \, .
\end{align}
The basis is then rotated by the inverse of the semi-simple part of the Wronskian, which we can get from the general form in~\refeq{eq:SeSiExplicit}
\begin{align}
    {W}^\text{SS}=\left(
\begin{array}{cccc}
 \varpi_{0} & 0 & 0 & 0 \\
 \varpi_{0}' & \frac{\varpi_{0}}{x \alpha_1} & 0 & 0 \\
 \varpi_{0}'' & \frac{2\varpi_{0}'}{x\alpha_1}-\left( \frac{1}{x^2\alpha_1} + \frac{\alpha_1'}{x\alpha_1^2} \right)\varpi_{0} & \frac{xC_3\alpha_1}{\varpi_{0}} & 0 \\
 \varpi_{0}''' &{W}^\text{SS}_{4,2}& {W}^\text{SS}_{4,3} & \frac{C_3}{\varpi_{0}} \\
\end{array}
\right) \, ,
\end{align}
where
\begin{align}
    {W}^\text{SS}_{4,2} =& \frac{2{\varpi_{0}'}^2}{x\alpha_1\varpi_{0}} - \frac{\alpha_1'\varpi_{0}'}{x\alpha_1^2} - \frac{\varpi_{0}''}{x\alpha_1} +  \frac{2 \left(1+5 x^2-7 x^4-3 x^6\right)}{(1-x)^2 (1+x)^2 \left(1+x^2\right) x^3}\frac{\varpi_{0}}{\alpha_1} \nonumber \\
            & \quad - \frac{\left(1-3 x^2\right) \left(7+5 x^2\right)}{(1-x) (1+x) \left(1+x^2\right) x^2}\frac{\varpi_{0}'}{\alpha_1} + \frac{3-8 x^2-7 x^4}{(1-x) (1+x) \left(1+x^2\right) x^2}\frac{\alpha_1'\varpi_{0}}{\alpha_1^2} \, ,\\
    {W}^\text{SS}_{4,3} =& \left(\frac{x\alpha_1\varpi_{0}'}{\varpi_{0}^2} -\frac{3-8 x^2-7 x^4}{(1-x) (1+x) \left(1+x^2\right)}\frac{\alpha_1}{\varpi_{0}}\right)C_3 \, . \nonumber
\end{align}
The structure series $\alpha_1(x)$ and the three-point coupling $C_3(x)$ are
\begin{equation}
    \alpha_1 = \frac{\varpi_{0}^2}{x(\varpi_{0}\varpi_{1}'-\varpi_{0}'\varpi_{1})} \quad\text{and}\quad C_3 = \frac{1+x^2}{(1-x)^3 (1+x)^3 x^3} \, .
\end{equation}
We rescale by
\begin{align}
    D(\epsilon) = \text{diag}(\epsilon^3,\epsilon^2,\epsilon,1,\epsilon^2) \, ,
\end{align}
where we have a reduced rescaling of $I_5$ by $\epsilon^2$. This may be explained by a weight drop of $I_5$ but without a proper integrand analysis and a proper understanding of the generalisation of third-kind differential forms to CY3 varieties, this cannot be discussed in detail here. 
For the canonicalisation, however, this poses no problem. 
We remove once again terms order by order in $\epsilon$ by an ansatz built from products of $\alpha_1(x), \varpi_{0}(x)$, their derivatives, and rational functions in $x$. We start at $1/\epsilon^2$ and remove all terms containing $\alpha_1'(x)$, $\varpi_{0}''(x)$ and $\varpi_{0}'''(x)$. 
Finally we introduce seven new transcendental functions, defined via first order differential equations in App.~\ref{app:bh}, to remove all terms left, which are not $\sim \epsilon$. The final rotation takes the form
\begin{align}
    \tilde{T}=\left(
\begin{array}{ccccc}
 1 & 0 & 0 & 0 & 0 \\
 G_6 & 1 & 0 & 0 & 0 \\
\tilde{T}_{3,1} & 20 G_4^2 & 1 & 0 & 40 G_4 \\
 \tilde{T}_{4,1} & \frac{G_2}{\epsilon }+G_7 & -G_6 & 1 & \frac{40G_1}{\epsilon}-40G_4G_6+40G_5 \\
 -\frac{G_1}{\epsilon }+G_5 & G_4 & 0 & 0 & 1 \\
\end{array}
\right),
\label{eq:finalrotCY3}
\end{align}
where
\begin{equation}
\begin{aligned}
    \tilde{T}_{3,1}=&-\frac{40 G_1 G_4}{\epsilon }+\frac{G_2}{\epsilon }+40 G_4 G_5-20 G_4^2 G_6-G_7 \, , \\
    \tilde{T}_{4,1}=&-\frac{20G_1^2}{\epsilon^2} + \frac{40G_1G_4G_6}{\epsilon} - \frac{40G_1G_5}{\epsilon} - \frac{G_2G_6}{\epsilon} + \frac{G_3}{\epsilon} \\
    &\quad + 20 G_4^2G_6^2 - 40 G_4G_5G_6 +20 G_5^2 + G_6G_7 \, .
\end{aligned}
\end{equation}
The final $\epsilon$-form is too large to print here.

We will refer to $\varpi_0$ and $\alpha_1$ from now on as $\varpi_{\text{CY3,2}}$ and $\alpha_{\text{CY3,2}}$.
Having the canonical form of the diagonal of this sector, we can now look at the full differential equation matrix, including the off-diagonal blocks.
\newpage
\section{Canonicalisation of the full DE}
We saw already at 5PM-1SF a huge increase in the number of the MIs and therefore also in the size of the differential equation matrix. At 2SF order we see again an increase in size. This is reduced somewhat, on the other hand, by the use of symmetry relations, which decrease the number of MIs. But in particular the increase of MIs in specific sectors is significant. At 1SF we had a maximum size of $7 \times 7$ blocks and at 2SF we have a maximum size of $16 \times 16$. 

Having discussed all the relevant geometries on the maximal cut, we now turn to the canonicalisation of the full system of differential equations. Our strategy is similar to the one followed at 1SF order. The principal difference between 1SF and 2SF lies in the size of the diagonal blocks, which can grow up to $16\times 16$, a regime in which algorithms such as \texttt{CANONICA} become very inefficient. We therefore adapt the procedure slightly. For most blocks, \texttt{CANONICA} is still used to find the transformation rapidly. For the larger blocks, we instead employ the Baikov representation to identify a single canonical integral on the maximal cut, and then use \texttt{INITIAL}, together with knowledge of the admissible kernels, to canonicalise the diagonal block. For the non-planar family it is sometimes not possible to find a good candidate via Baikov analysis, since the non-planarity renders the loop-by-loop approach inefficient and the resulting expressions become unmanageable. In those cases, we resort to educated guesses based on integrals carrying two scalar products on the worldline propagators and a normalisation of the form $((x^2-1)/x)^n$ for some integer $n$.

This strategy succeeds for all but one planar topsector with 16 master integrals, which requires the introduction of the square roots $\sqrt{4-x^2}$ and $\sqrt{1-4x^2}$. Since \texttt{INITIAL} becomes rather slow when square roots have to be introduced, we adopt a different strategy in this one case. We identify 14 canonical master integrals through a loop-by-loop Baikov analysis; for the remaining two integrals we choose two derivatives, and the corresponding transformation is then constructed by hand.

After the canonicalisation of all diagonal blocks, we again perform an IBP-list search to identify integrals which are canonical on the maximal cut without further transformation, in order to simplify the off-diagonals. This works well for the planar integrals, but becomes inefficient for the non-planar ones. The list-search method relies on the existence of single integrals in the family which lie close to canonical integrals, and the existence of such integrals depends sensitively on the chosen parametrisation of the integral family. With a poor parametrisation, linear combinations of integrals are generically required. The planar family admits a very natural and symmetric parametrisation, which is why the approach succeeds there. It is, however, not clear what the best parametrisation of the non-planar integrals is, and the choice of ISPs is rather arbitrary. As a consequence, the list search remains workable to some extent, but with reduced efficiency, and one must allow for linear combinations to find good candidates. These linear combinations are also more frequently bad on the subcuts. The approach nevertheless still works.

\begin{figure}[h!]
        \centering
        \begin{subfigure}{0.49\textwidth}
        \includegraphics[width=\textwidth]{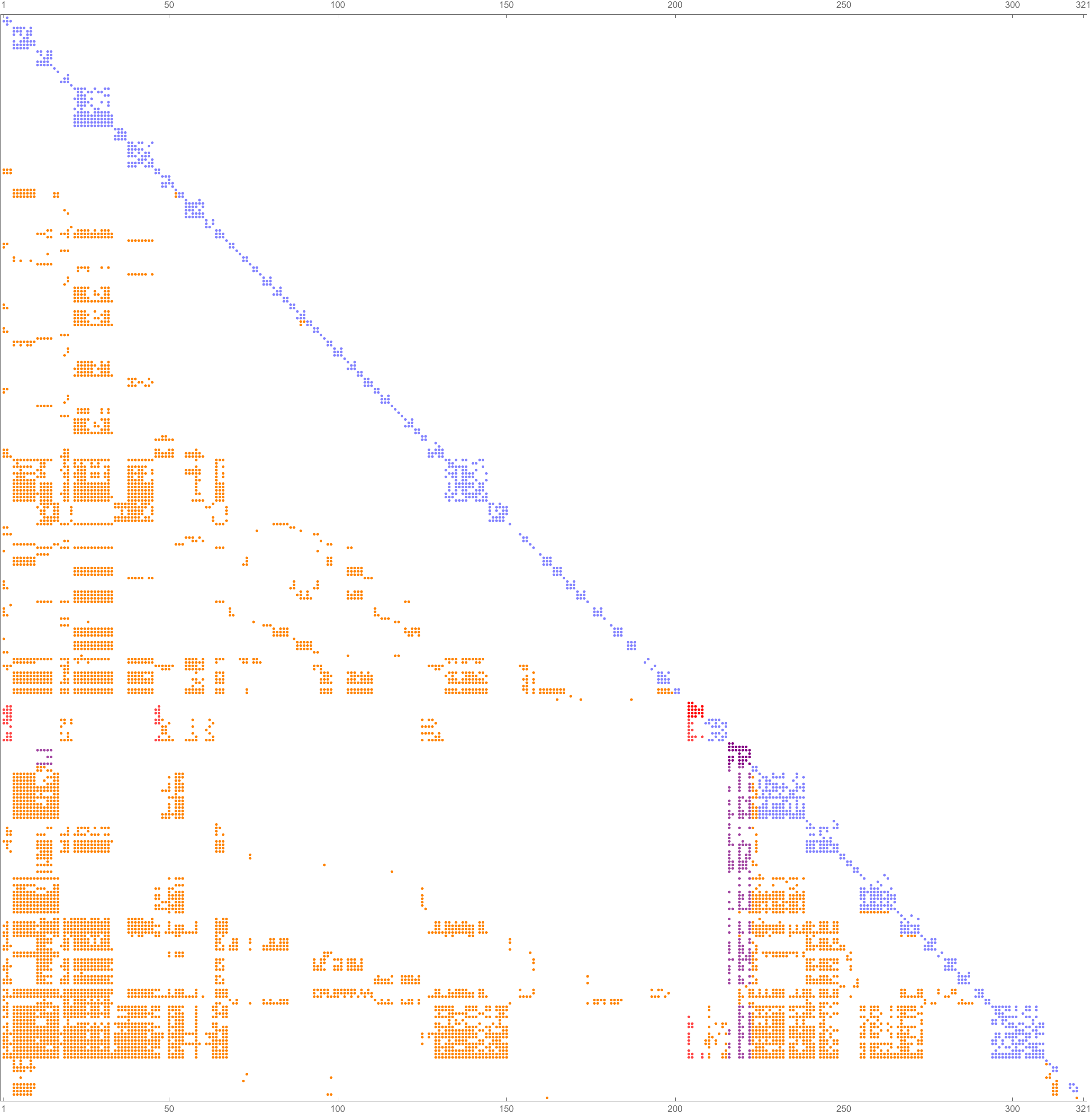}
        \caption{$\text{P}^{\text{5,2}}$ integrals.}
\end{subfigure}
        \hfill
        \begin{subfigure}{0.49\textwidth}
        \includegraphics[width=\textwidth]{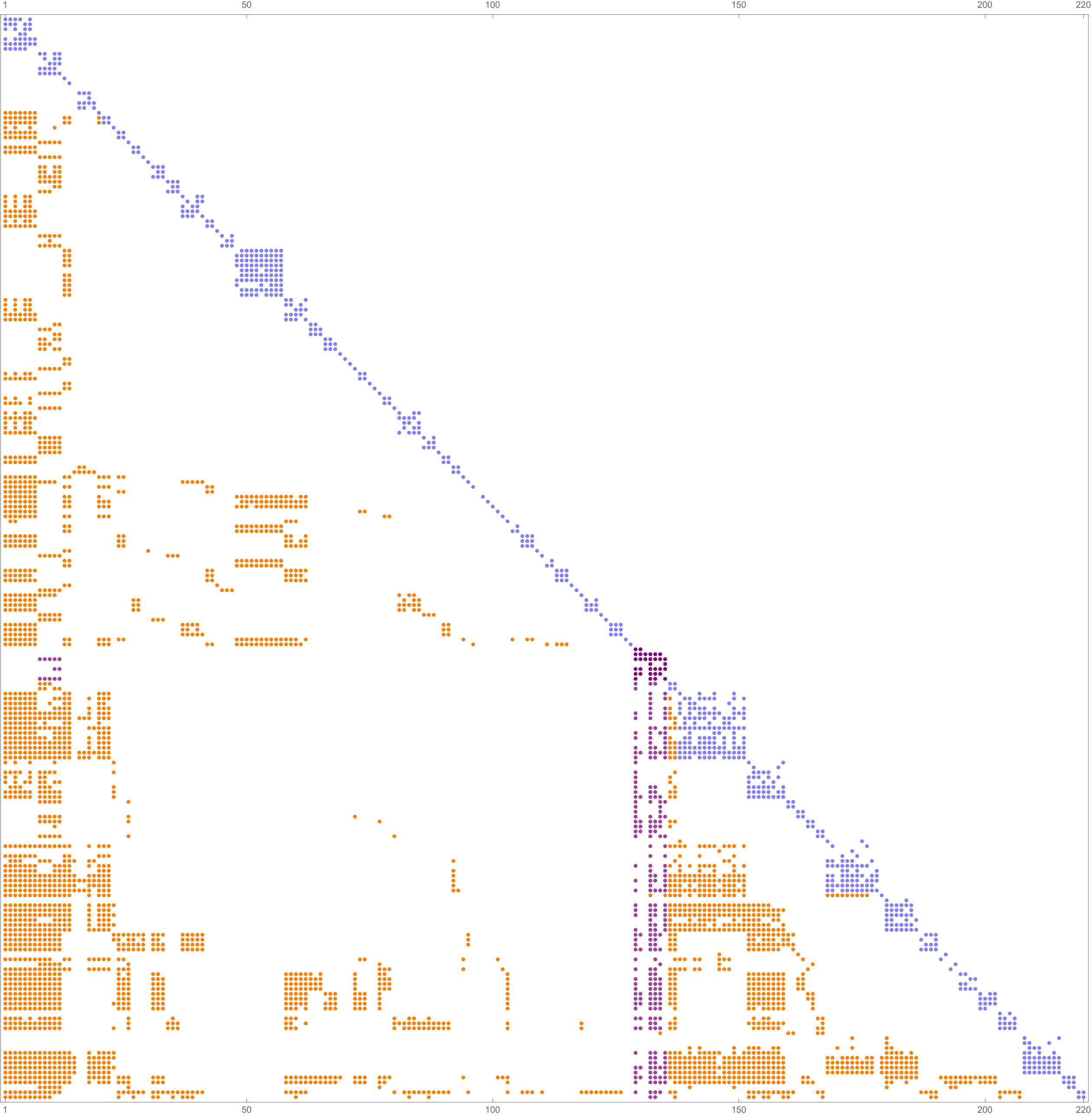}
        \caption{$\text{NP}$ integrals.}
\end{subfigure}
\hfill
        \begin{subfigure}{0.49\textwidth}
        \includegraphics[width=\textwidth]{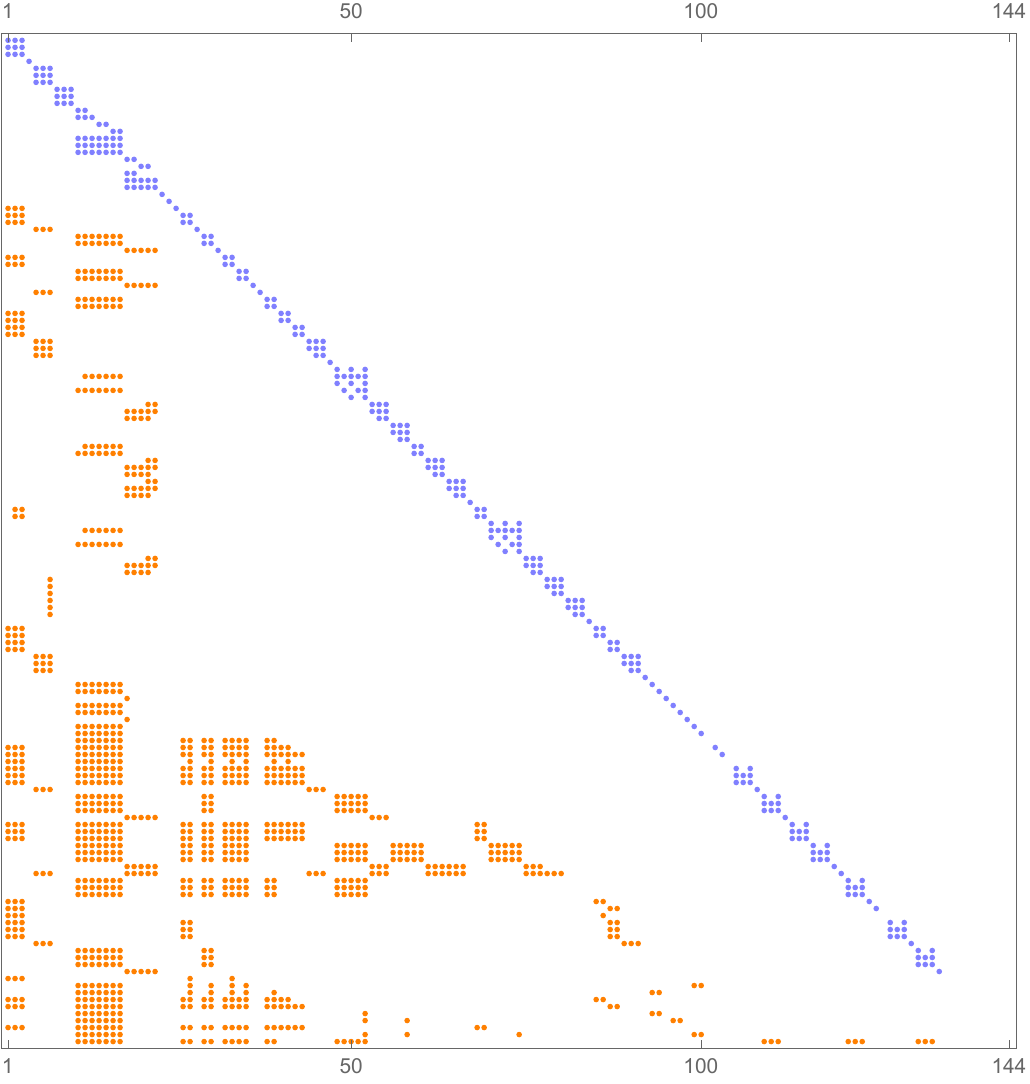}
        \caption{$\text{NX1}$ integrals.}
\end{subfigure}
\hfill
        \begin{subfigure}{0.49\textwidth}
        \includegraphics[width=\textwidth]{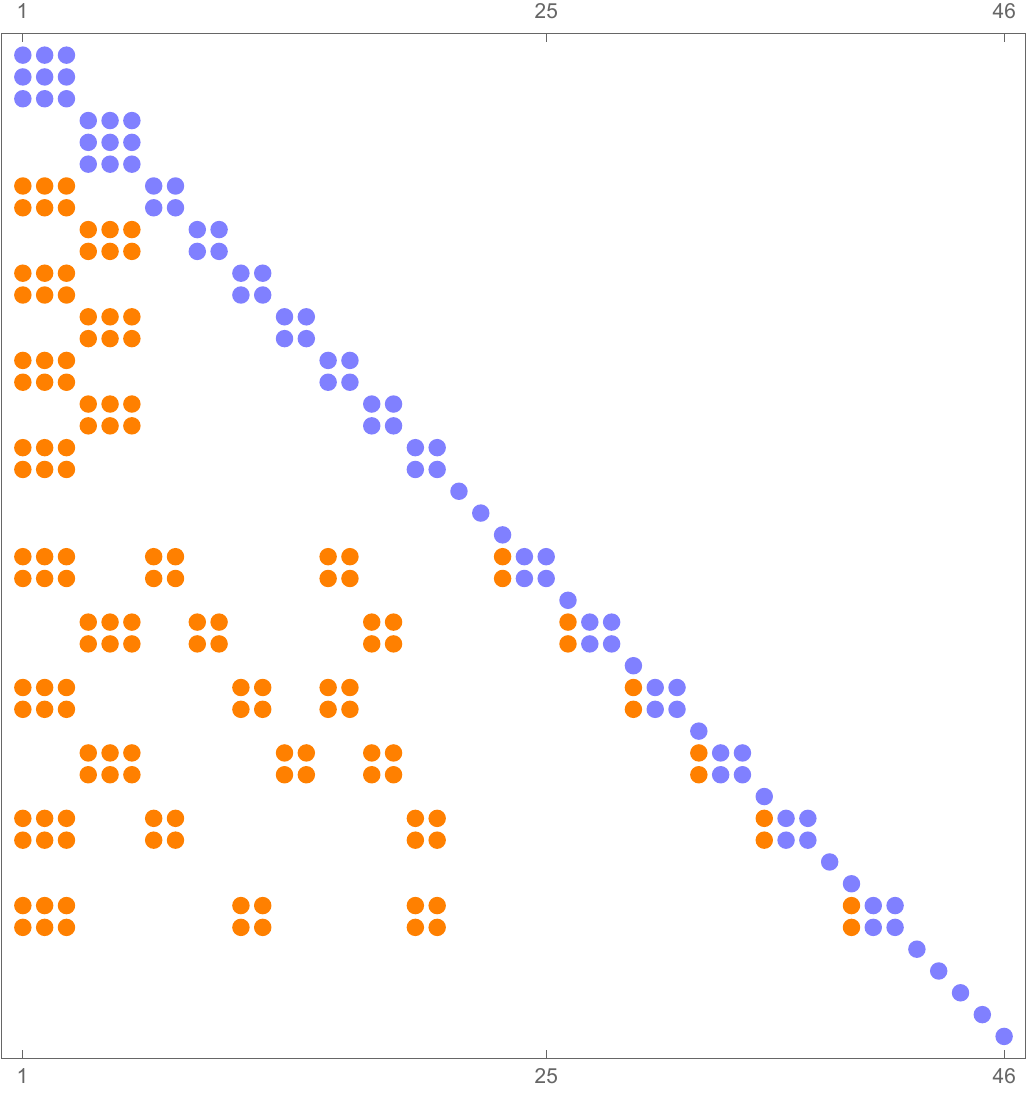}
        \caption{$\text{NX2}$ integrals.}
\end{subfigure}
\caption{The differential equation matrices for the integrals appearing at 5PM-2SF. $\d\!\log$ entries are marked blue and orange, K3 entries are marked purple and CY3 entries are marked red.}\label{fig:heatplot5PM2SF}
\end{figure}

The non-planar and planar families share a substantial overlap, since the subsectors of the non-planar topologies often become planar. One could, in principle, map these integrals onto the planar family. The mapping is, however, non-trivial because of the delta functions, and IBP packages cannot perform it automatically. We therefore do not perform such a mapping for the differential equations, but we do use the planar basis as a starting basis for the non-planar integrals, by mapping it to non-planar integrals by hand. We can see the heat plots of the matrices in Fig.~\ref{fig:heatplot5PM2SF}.

In this manner the differential equations are fully canonicalised for all four families. We omit the V-type integrals, since they are not required for the conservative dynamics.

\section{Choice of periods to fix the boundaries of the differential equations}
To fix the boundary coefficients, we have to choose again a holomorphic solution of the periods around $x=1$. 
In the case of the CY3 this choice is unique since $x=1$ is a MUM-point:
\begin{align}
        \varpi_{\text{CY3,2}}(x)=&1+\delta+\dfrac{57}{64}\delta^2+... \\
        \alpha_{\text{CY3,2}}(x)=&-\delta-\dfrac{\delta^2}{2}-\dfrac{5}{16} \delta^4+...\;.
\end{align}
For the K3, on the other hand, $x=1$ is a regular point, so that we can choose a linear combination of three different solutions. We decide to choose:
\begin{align}
        \varpi_{\text{K3,2}}(x)=&\dfrac{1}{2}+\dfrac{1}{2}\delta+\dfrac{31}{64} \delta^2+... \; ,
\end{align}
such that $(1+x^2) \varpi_{\text{K3,2}}(x)$ is an even function of $v$. 
With this choice we can do the boundary fixing as done at 1SF order.
Furthermore, we redefine the introduced square root
\begin{align}
        \sqrt{1-34 x^2 +x^4} \rightarrow i \sqrt{-(1-34 x^2 +x^4)}
\end{align}
and normalise all integrals so that all canonical integrals are purely real in an expansion around $x=1$.
\section{Conservative regions at 5PM-2SF}

The regions at 5PM-2SF are very similar to those at 1SF order. We again find at most three radiative gravitons, and for the conservative part of the calculation we only consider regions containing an even number of radiative gravitons. The planar integrals carry four active propagators which can go on-shell, of which at most three appear in the topsectors; the same holds for the extended planar families.

All relevant regions can be identified by examining two example topologies, one planar and one non-planar. The expansion of the NX families is analogous to that of the planar one,
\begin{align*}
        \begin{tikzpicture}[baseline={([yshift=-1ex]current bounding box.south)},scale=1.5]
                \coordinate (inA)  at (0.6,0.6);
                \coordinate (outA) at (2.625,0.6);
                \coordinate (inB)  at (0.6,-0.6);
                \coordinate (outB) at (2.625,-0.6);
                \draw[dotted] (inA) -- (outA);
                \draw[dotted] (inB) -- (outB);
                \coordinate (aA) at (0.6,0.6);
                \coordinate (bA) at (1.275,0.6);
                \coordinate (cA) at (1.95,0.6);
                \coordinate (dA) at (2.625,0.6);
                \coordinate (eA) at (3.3,0.6);
                \coordinate (aM) at (0.6,0);
                \coordinate (bM) at (1.275,0);
                \coordinate (cM) at (1.95,0);
                \coordinate (dM) at (2.625,0);
                \coordinate (eM) at (3.3,0);
                \coordinate (aB) at (0.6,-0.6);
                \coordinate (bB) at (1.275,-0.6);
                \coordinate (cB) at (1.95,-0.6);
                \coordinate (dB) at (2.625,-0.6);
                \coordinate (eB) at (3.3,-0.6);
                \draw[photon]     (aA) -- (aB) ;
                \draw[photon,red] (aM) -- (bM) node [midway, below]{$A_1$} ;
                \draw[photon,red] (cM) -- (bM) node [midway, below]{$A_2$} ;
                \draw[photon,red] (cM) -- (dM) node [midway, below]{$A_3$} ;
                \draw[photon] (bA) -- (bM);
                \draw[photon] (cB) -- (cM);
                \draw[photon]     (dA) -- (dB);
                \foreach \v in {aA,bA,dA,aM,bM,cM,dM,aB,cB,dB}
                  \draw[fill] (\v) circle (.08);
              \end{tikzpicture}\qquad
              \begin{tikzpicture}[baseline={([yshift=-1ex]current bounding box.south)},scale=1.5]
                 \coordinate (inA)  at (0.6,0.6);
                  \coordinate (outA) at (2.625,0.6);
                  \coordinate (inB)  at (0.6,-0.6);
                  \coordinate (outB) at (2.625,-0.6);
                  \draw[dotted] (inA) -- (outA);
                  \draw[dotted] (inB) -- (outB);
                  \coordinate (aA) at (0.6,0.6);
                  \coordinate (bA) at (1.85+0.1,0.6);
                  \coordinate (cA) at (1.95,0.6);
                  \coordinate (dA) at (2.625,0.6);
                  \coordinate (eA) at (3.3,0.6);
                  \coordinate (aM) at (0.6,0);
                  \coordinate (bM) at (1.275,0);
                  \coordinate (cM) at (1.95,0);
                  \coordinate (dM) at (2.625,0);
                  \coordinate (eM) at (3.3,0);
                  \coordinate (aB) at (0.6,-0.6);
                  \coordinate (bB) at (1.275,-0.6);
                  \coordinate (cB) at (1.95,-0.6);
                  \coordinate (dB) at (2.625,-0.6);
                  \coordinate (eB) at (3.3,-0.6);
                \coordinate (bL) at (1.85+0.1,-0.35);
                \draw[photon,red] (aM) -- (bM)node [midway, below]{$A_1$} ;
                \draw[photon,red] (bM) -- (dM)node [midway, below]{$\; \; \;\; \; \;\; \; A_2$} ;
                \draw[fill, white] (1.85+0.1,0) circle (0.15);
                \draw[photon] (aA) -- (aB);
               \draw[photon,red] (bM) -- (bL)node [midway, below,sloped]{$\hspace{-0.5cm} A_3$};
                \draw[photon] (bL) -- (bB);
                \draw[photon] (dA) -- (dB);
                 \draw[photon] (bL) -- (bA);
                \foreach \v in {aA,bA,dA,aM,bM,dM,aB,bB,dB,bL}
                  \draw[fill] (\v) circle (.08);
              \end{tikzpicture}
\end{align*}
where $A_i\in\{P,R\}$.

\subsubsection*{Potential PPP}

The potential region $(PPP)$ is in direct analogy with the 1SF case for the planar family. Diagrammatically, the expansion takes the form
\begin{align*}
        \begin{tikzpicture}[baseline={([yshift=-1ex]current bounding box.south)},scale=1.5]
                \coordinate (inA)  at (0.6,0.6);
                \coordinate (outA) at (2.625,0.6);
                \coordinate (inB)  at (0.6,-0.6);
                \coordinate (outB) at (2.625,-0.6);
                \draw[dotted] (inA) -- (outA);
                \draw[dotted] (inB) -- (outB);
                \coordinate (aA) at (0.6,0.6);
                \coordinate (bA) at (1.275,0.6);
                \coordinate (cA) at (1.95,0.6);
                \coordinate (dA) at (2.625,0.6);
                \coordinate (eA) at (3.3,0.6);
                \coordinate (aM) at (0.6,0);
                \coordinate (bM) at (1.275,0);
                \coordinate (cM) at (1.95,0);
                \coordinate (dM) at (2.625,0);
                \coordinate (eM) at (3.3,0);
                \coordinate (aB) at (0.6,-0.6);
                \coordinate (bB) at (1.275,-0.6);
                \coordinate (cB) at (1.95,-0.6);
                \coordinate (dB) at (2.625,-0.6);
                \coordinate (eB) at (3.3,-0.6);
                \draw[photon]     (aA) -- (aB) ;
                \draw[photon,red] (aM) -- (bM) node [midway, below]{$P$} ;
                \draw[photon,red] (cM) -- (bM) node [midway, below]{$P$} ;
                \draw[photon,red] (cM) -- (dM) node [midway, below]{$P$} ;
                \draw[photon] (bA) -- (bM);
                \draw[photon] (cB) -- (cM);
                \draw[photon]     (dA) -- (dB) node [midway, right]{$ \qquad\rightarrow$};
                \foreach \v in {aA,bA,dA,aM,bM,cM,dM,aB,cB,dB}
                  \draw[fill] (\v) circle (.08);
              \end{tikzpicture}\qquad
              \begin{tikzpicture}[baseline={([yshift=-1ex]current bounding box.south)},scale=1.2]
                \coordinate (inA) at (0.4,.7);
                \coordinate (outA) at (3.4,.7);
                \coordinate (inB) at (0.4,-.7);
                \coordinate (outB) at (3.4,-.7);
                \coordinate (xA) at (0.9,.7);
                \coordinate (xB) at (0.9,-.7);
                \coordinate (xM) at (1.1,0);
                \coordinate (xAM) at (0.9,.35);
                \coordinate (xBM) at (0.9,-.35);
                \coordinate (yA) at (1.9,.7);
                \coordinate (yB) at (1.9,-.7);
                \coordinate (yC) at (1.9,-1.2);
                \coordinate (xC) at (0.9,-1.2);
                \coordinate (zC) at (2.9,-1.2);
                \coordinate (yM) at (1.9,0);
                \coordinate (yAM) at (1.7,0);
                \coordinate (yBM) at (2.1,0);
                \coordinate (zA) at (2.9,.7);
                \coordinate (zB) at (2.9,-.7);
                \coordinate (zM) at (2.6,0);
                \coordinate (zAM) at (2.9,.35);
                \coordinate (zBM) at (2.9,-.35);
                \draw [dotted] (inA) --node [midway, above]{$\rightarrow \vec{e}$}  (outA);
                \draw [photon] (yB) --node [midway, right]{$\downarrow \vec{q}$} (yC);
                \draw[photon] ($(yA)!0.5!(yB)$) circle [radius=0.7,rotate=90];
                \draw [photon] (xM) -- (zM);
                \draw [photon] (yA) -- (yAM);
                \draw [photon] (yB) -- (yBM);
                \foreach \v in {yA,yB,yC,xM,zM,yAM,yBM}
                \draw[fill] (\v) circle (.08);
              \end{tikzpicture}
\end{align*}
\begin{align*}
        \begin{tikzpicture}[baseline={([yshift=-1ex]current bounding box.south)},scale=1.5]
                 \coordinate (inA)  at (0.6,0.6);
                  \coordinate (outA) at (2.625,0.6);
                  \coordinate (inB)  at (0.6,-0.6);
                  \coordinate (outB) at (2.625,-0.6);
                  \draw[dotted] (inA) -- (outA);
                  \draw[dotted] (inB) -- (outB);
                  \coordinate (aA) at (0.6,0.6);
                  \coordinate (bA) at (1.85+0.1,0.6);
                  \coordinate (cA) at (1.95,0.6);
                  \coordinate (dA) at (2.625,0.6);
                  \coordinate (eA) at (3.3,0.6);
                  \coordinate (aM) at (0.6,0);
                  \coordinate (bM) at (1.275,0);
                  \coordinate (cM) at (1.95,0);
                  \coordinate (dM) at (2.625,0);
                  \coordinate (eM) at (3.3,0);
                  \coordinate (aB) at (0.6,-0.6);
                  \coordinate (bB) at (1.275,-0.6);
                  \coordinate (cB) at (1.95,-0.6);
                  \coordinate (dB) at (2.625,-0.6);
                  \coordinate (eB) at (3.3,-0.6);
                \coordinate (bL) at (1.85+0.1,-0.35);
                \draw[photon,red] (aM) -- (bM)node [midway, below]{$P$} ;
                \draw[photon,red] (bM) -- (dM)node [midway, below]{$\; \; \;\; \; \;\; \; P$} ;
                \draw[fill, white] (1.85+0.1,0) circle (0.15);
                \draw[photon] (aA) -- (aB);
               \draw[photon,red] (bM) -- (bL)node [midway, below,sloped]{$\hspace{-0.5cm} P$};
                \draw[photon] (bL) -- (bB);
                \draw[photon] (dA) -- (dB)node [midway, right]{$ \qquad\rightarrow$};
                 \draw[photon] (bL) -- (bA);
                \foreach \v in {aA,bA,dA,aM,bM,dM,aB,bB,dB,bL}
                  \draw[fill] (\v) circle (.08);
              \end{tikzpicture}\qquad
              \begin{tikzpicture}[baseline={([yshift=-1ex]current bounding box.south)},scale=1.2]
                \coordinate (inA) at (0.4,.7);
                \coordinate (outA) at (3.4,.7);
                \coordinate (inB) at (0.4,-.7);
                \coordinate (outB) at (3.4,-.7);
                \coordinate (xA) at (0.9,.7);
                \coordinate (xB) at (0.9,-.7);
                \coordinate (xM) at (1.1,0);
                \coordinate (xAM) at (1.4,.35);
                \coordinate (xBM) at (0.9,-.35);
                \coordinate (yA) at (1.9,.7);
                \coordinate (yB) at (1.9,-.7);
                \coordinate (yC) at (1.9,-1.2);
                \coordinate (xC) at (0.9,-1.2);
                \coordinate (zC) at (2.9,-1.2);
                \coordinate (yM) at (1.9,0);
                \coordinate (yAM) at (2.3,-0.25);
                \coordinate (yBM) at (1.9,0);
                \coordinate (zA) at (2.9,.7);
                \coordinate (zB) at (2.9,-.7);
                \coordinate (zM) at (2.6,0);
                \coordinate (zAM) at (2.9,.35);
                \coordinate (zBM) at (2.9,-.35);
                \draw [dotted] (inA) --node [midway, above]{$\rightarrow \vec{e}$}  (outA);
                \draw [photon] (yB) --node [midway, right]{$\downarrow \vec{q}$} (yC);
                \draw[photon] ($(yA)!0.5!(yB)$) circle [radius=0.7,rotate=90];
                \draw [photon] (xM) -- (zM);
                \draw [photon] (yA) -- (yAM);
                \draw [photon] (yB) -- (yAM);
                \draw [photon] (yBM) -- (yAM);
                \foreach \v in {yA,yB,yC,xM,zM,yAM,yBM}
                \draw[fill] (\v) circle (.08);
              \end{tikzpicture}
\end{align*}

We see, that non-planar potential boundary integrals appear here for the first time. After IBP reduction, these can all be reduced to planar boundaries. Most of the required boundaries are familiar from 1SF, but a few additional ones must be computed.

\subsubsection*{Tail RRP+PRR}

We again find two regions involving two radiative gravitons, which give rise to tail-type boundary integrals:
\begin{align*}
        \centering
        \begin{tikzpicture}[baseline={([yshift=-1ex]current bounding box.south)},scale=1.5]
                \coordinate (inA)  at (0.6,0.6);
                \coordinate (outA) at (2.625,0.6);
                \coordinate (inB)  at (0.6,-0.6);
                \coordinate (outB) at (2.625,-0.6);
                \draw[dotted] (inA) -- (outA);
                \draw[dotted] (inB) -- (outB);
                \coordinate (aA) at (0.6,0.6);
                \coordinate (bA) at (1.275,0.6);
                \coordinate (cA) at (1.95,0.6);
                \coordinate (dA) at (2.625,0.6);
                \coordinate (eA) at (3.3,0.6);
                \coordinate (aM) at (0.6,0);
                \coordinate (bM) at (1.275,0);
                \coordinate (cM) at (1.95,0);
                \coordinate (dM) at (2.625,0);
                \coordinate (eM) at (3.3,0);
                \coordinate (aB) at (0.6,-0.6);
                \coordinate (bB) at (1.275,-0.6);
                \coordinate (cB) at (1.95,-0.6);
                \coordinate (dB) at (2.625,-0.6);
                \coordinate (eB) at (3.3,-0.6);
                \draw[photon]     (aA) -- (aB) ;
                \draw[photon,red] (aM) -- (bM) node [midway, below]{$R$} ;
                \draw[photon,red] (cM) -- (bM) node [midway, below]{$R$} ;
                \draw[photon,red] (cM) -- (dM) node [midway, below]{$P$} ;
                \draw[photon] (bA) -- (bM);
                \draw[photon] (cB) -- (cM);
                \draw[photon]     (dA) -- (dB) node [midway, right]{$ \qquad\rightarrow$};
                \foreach \v in {aA,bA,dA,aM,bM,cM,dM,aB,cB,dB}
                  \draw[fill] (\v) circle (.08);
              \end{tikzpicture}\qquad
              \begin{tikzpicture}[baseline={([yshift=-1ex]current bounding box.south)},scale=1.2]
                \coordinate (inA) at (0.3,.7);
                         \coordinate (outA) at (3.6,.7);
                         \coordinate (inB) at (0.3,-.7);
                         \coordinate (outB) at (3.6,-.7);
                         \coordinate (xA) at (0.6,.7);
                         \coordinate (yA) at (1.5,.7);
                         \coordinate (zA) at (2.4,.7);
                         \coordinate (wA) at (3.3,.7);
                         \coordinate (xB) at (0.6,-.7);
                         \coordinate (yB) at (1.5,-.7);
                         \coordinate (zB) at (2.4,-.7);
                         \coordinate (wB) at (3.3,-.7);
                         \coordinate (xM) at (0.6,0);
                         \coordinate (yM) at (1.5,1.05);
                         \coordinate (zM) at (2.4,0);
                         \coordinate (wM) at (3.3,0);
                         \coordinate (xyMB) at (1.05,-.35);
                         \coordinate (yzMB) at (1.95,-.35);
                         \coordinate (zwMB) at (2.85,-.35);
                         \coordinate (yzMA) at (1.95,.35);
                \draw [dotted] (inA) -- (outA);
                \draw [dotted] (inB) -- (outB) node [midway, below]{} ;
                \draw [draw=none] (xA) to[out=40,in=140] (zA);
                \draw [photon] (zA) -- (zB);
                \draw [photon] (zA) -- (wB);
                \draw [photon] (xA) -- (xB);
                \draw [photon] (yA) -- (yM);
                \draw [photon,red][out=40,in=140] (xA) to[out=40,in=140] (zA);
          \draw [fill] (xA) circle (.08);
          \draw [fill] (yA) circle (.08);
          \draw [fill] (yM) circle (.08);
          \draw [fill] (wB) circle (.08);
                \draw [fill] (zA) circle (.08);
                \draw [fill] (xB) circle (.08);
                \draw [fill] (zB) circle (.08);
              \end{tikzpicture}.
\end{align*}

The $RRP$ expansion---and equally the $PRR$ expansion---gives rise to the same tail topologies that we encountered at 5PM-1SF. For the non-planar integrals no analogous region is found.

\subsubsection*{Memory}

A genuinely new region appears at 5PM-2SF order, in which two propagators are again rescaled by $v$. We refer to this as the memory region. For the planar family the memory region corresponds to the $RPR$ scaling:
\begin{align*}
        \begin{tikzpicture}[baseline={([yshift=-1ex]current bounding box.south)},scale=1.5]
                \coordinate (inA)  at (0.6,0.6);
                \coordinate (outA) at (2.625,0.6);
                \coordinate (inB)  at (0.6,-0.6);
                \coordinate (outB) at (2.625,-0.6);
                \draw[dotted] (inA) -- (outA);
                \draw[dotted] (inB) -- (outB);
                \coordinate (aA) at (0.6,0.6);
                \coordinate (bA) at (1.275,0.6);
                \coordinate (cA) at (1.95,0.6);
                \coordinate (dA) at (2.625,0.6);
                \coordinate (eA) at (3.3,0.6);
                \coordinate (aM) at (0.6,0);
                \coordinate (bM) at (1.275,0);
                \coordinate (cM) at (1.95,0);
                \coordinate (dM) at (2.625,0);
                \coordinate (eM) at (3.3,0);
                \coordinate (aB) at (0.6,-0.6);
                \coordinate (bB) at (1.275,-0.6);
                \coordinate (cB) at (1.95,-0.6);
                \coordinate (dB) at (2.625,-0.6);
                \coordinate (eB) at (3.3,-0.6);
                \draw[photon]     (aA) -- (aB) ;
                \draw[photon,red] (aM) -- (bM) node [midway, below]{$R$} ;
                \draw[photon,red] (cM) -- (bM) node [midway, below]{$P$} ;
                \draw[photon,red] (cM) -- (dM) node [midway, below]{$R$} ;
                \draw[photon] (bA) -- (bM);
                \draw[photon] (cB) -- (cM);
                \draw[photon]     (dA) -- (dB) node [midway, right]{$ \qquad\rightarrow$};
                \foreach \v in {aA,bA,dA,aM,bM,cM,dM,aB,cB,dB}
                  \draw[fill] (\v) circle (.08);
              \end{tikzpicture}\qquad
              \begin{tikzpicture}[baseline={([yshift=-1ex]current bounding box.south)},scale=1.2]
                \coordinate (inA) at (0.3,.7);
                         \coordinate (outA) at (3.6,.7);
                         \coordinate (inB) at (0.3,-.7);
                         \coordinate (outB) at (3.6,-.7);
                         \coordinate (xA) at (0.6,.7);
                         \coordinate (yA) at (1.5,.7);
                         \coordinate (zA) at (2.4,.7);
                         \coordinate (wA) at (3.3,.7);
                         \coordinate (xB) at (0.6,-.7);
                         \coordinate (yB) at (1.5,-.7);
                         \coordinate (zB) at (2.4,-.7);
                         \coordinate (wB) at (3.3,-.7);
                         \coordinate (xM) at (0.6,0);
                         \coordinate (yM) at (1.5,1.05);
                         \coordinate (zM) at (2.4,0);
                         \coordinate (wM) at (3.3,0);
                         \coordinate (xyMB) at (1.05,-.35);
                         \coordinate (yzMB) at (1.95,-.35);
                         \coordinate (zwMB) at (2.85,-.35);
                         \coordinate (yzMA) at (1.95,.35);
                \draw [dotted] (inA) -- (outA);
                \draw [dotted] (inB) -- (outB) node [midway, below]{} ;
                \draw [draw=none] (xA) to[out=40,in=140] (zA);
                \draw [photon] (zA) -- (zB);
                \draw [photon] (yA) -- (yB);
                \draw [photon] (xA) -- (xB);
                \draw [photon,red][out=40,in=140] (xA) to[out=40,in=140] (yA);
                \draw [photon,red][out=40,in=140] (yA) to[out=40,in=140] (zA);
          \draw [fill] (xA) circle (.08);
          \draw [fill] (yA) circle (.08);
          \draw [fill] (yB) circle (.08);
                \draw [fill] (zA) circle (.08);
                \draw [fill] (xB) circle (.08);
                \draw [fill] (zB) circle (.08);
              \end{tikzpicture}
\end{align*}

The same region exists for the non-planar integral family, with one caveat. For the non-planar integrals one may choose $A_1=A_2=R$ and rescale the corresponding loop momenta. The active propagators are
\begin{align}
        D^{\text{NP}}_5=(\ell_1-\ell_2-\ell_3+\ell_4)^2, \; D^{\text{NP}}_6=(\ell_1-\ell_3)^2, \; D^{\text{NP}}_7=(\ell_2-\ell_4)^2 \;,
\end{align}
so that rescaling $k_1=\ell_1-\ell_3$ and $k_2=\ell_2-\ell_4$ automatically rescales $D_5=(k_1-k_2)^2$ at the same time. Despite this, the region belongs to the same class as the planar $RPR$ expansion and likewise scales as $v^{-4\epsilon}$. It is, in a sense, analogous to the $RRR$ region at 1SF, where one had integrals with only two radiative gravitons but a $v^{-6\epsilon}$ scaling. The boundary integrals from the non-planar case give the region its name: they are of memory type, familiar from the PN literature~\cite{Porto:2026fsd,Blumlein:2021txe,Foffa:2019eeb}.
\begin{align*}
        \begin{tikzpicture}[baseline={([yshift=-1ex]current bounding box.south)},scale=1.5]
                 \coordinate (inA)  at (0.6,0.6);
                  \coordinate (outA) at (2.625,0.6);
                  \coordinate (inB)  at (0.6,-0.6);
                  \coordinate (outB) at (2.625,-0.6);
                  \draw[dotted] (inA) -- (outA);
                  \draw[dotted] (inB) -- (outB);
                  \coordinate (aA) at (0.6,0.6);
                  \coordinate (bA) at (1.85+0.1,0.6);
                  \coordinate (cA) at (1.95,0.6);
                  \coordinate (dA) at (2.625,0.6);
                  \coordinate (eA) at (3.3,0.6);
                  \coordinate (aM) at (0.6,0);
                  \coordinate (bM) at (1.275,0);
                  \coordinate (cM) at (1.95,0);
                  \coordinate (dM) at (2.625,0);
                  \coordinate (eM) at (3.3,0);
                  \coordinate (aB) at (0.6,-0.6);
                  \coordinate (bB) at (1.275,-0.6);
                  \coordinate (cB) at (1.95,-0.6);
                  \coordinate (dB) at (2.625,-0.6);
                  \coordinate (eB) at (3.3,-0.6);
                \coordinate (bL) at (1.85+0.1,-0.35);
                \draw[photon,red] (aM) -- (bM)node [midway, below]{$R$} ;
                \draw[photon,red] (bM) -- (dM)node [midway, below]{$\; \; \;\; \; \;\; \; R$} ;
                \draw[fill, white] (1.85+0.1,0) circle (0.15);
                \draw[photon] (aA) -- (aB);
               \draw[photon,red] (bM) -- (bL)node [midway, below,sloped]{$\hspace{-0.5cm} R$};
                \draw[photon] (bL) -- (bB);
                \draw[photon] (dA) -- (dB)node [midway, right]{$ \qquad\rightarrow$};;
                 \draw[photon] (bL) -- (bA);
                \foreach \v in {aA,bA,dA,aM,bM,dM,aB,bB,dB,bL}
                  \draw[fill] (\v) circle (.08);
              \end{tikzpicture}\qquad
              \begin{tikzpicture}[baseline={([yshift=-1ex]current bounding box.south)},scale=1.5]
                \coordinate (inA)  at (0.6,0.6);
                \coordinate (outA) at (2.625,0.6);
                \coordinate (inB)  at (0.6,-0.6);
                \coordinate (outB) at (2.625,-0.6);
                \draw[dotted] (inA) -- (outA);
                \draw[dotted] (inB) -- (outB);
                \coordinate (aA) at (0.6,0.6);
                \coordinate (bA) at (1.6125,0.6);
                \coordinate (bAA) at (1.6125,1.2);
                \coordinate (cA) at (1.95,0.6);
                \coordinate (dA) at (2.625,0.6);
                \coordinate (eA) at (3.3,0.6);
                \coordinate (aM) at (0.6,0);
                \coordinate (bM) at (1.275,0);
                \coordinate (cM) at (1.95,0);
                \coordinate (dM) at (2.625,0);
                \coordinate (eM) at (3.3,0);
                \coordinate (aMd) at (0.6,-0.1);
                \coordinate (bMd) at (1.275,-0.1);
                \coordinate (cMd) at (1.95,-0.1);
                \coordinate (dMd) at (2.625,-0.1);
                \coordinate (eMd) at (3.3,-0.1);
                \coordinate (aMdd) at (0.6,-0.15);
                \coordinate (bMdd) at (1.275,-0.15);
                \coordinate (cMdd) at (1.95,-0.15);
                \coordinate (dMdd) at (2.625,-0.15);
                \coordinate (eMdd) at (3.3,-0.15);
                \coordinate (aMu) at (0.6,0.1);
                \coordinate (bMu) at (1.275,0.1);
                \coordinate (cMu) at (1.95,0.1);
                \coordinate (dMu) at (2.625,0.1);
                \coordinate (eMu) at (3.3,0.1);
                \coordinate (aB) at (0.6,-0.6);
                \coordinate (bB) at (1.6125,-0.6);
                \coordinate (cB) at (1.95,-0.6);
                \coordinate (dB) at (2.625,-0.6);
                \coordinate (eB) at (3.3,-0.6);
                \draw[photon]     (aA) -- (aB) node [midway,left] {} node [midway,left] {};
                \draw[photon ,red] (aA) to[out=80, in=-180] (bAA);
                \draw[photon ,red] (bAA) to[out=0, in=100] (dA);
                \draw[photon ,red] (bA) to (bAA);
                \draw[photon] (dA) -- (dB);
                \draw[photon] (bA) -- (bB);
                \foreach \v in {aA,bA,dA,aB,bB,dB,bAA}
                  \draw[fill] (\v) circle (.08);
              \end{tikzpicture}
\end{align*}

In this region we find, for the first time, two master integrals in the same sector after the IBP reduction of all boundary integrals:
\begin{align*}
        \begin{tikzpicture}[baseline={([yshift=-1ex]current bounding box.south)},scale=1.5]
                \coordinate (inA) at (0.3,.7);
                         \coordinate (outA) at (3.6,.7);
                         \coordinate (inB) at (0.3,-.7);
                         \coordinate (outB) at (3.6,-.7);
                         \coordinate (xA) at (0.6,.7);
                         \coordinate (yA) at (1.5,.7);
                         \coordinate (zA) at (2.4,.7);
                         \coordinate (wA) at (3.3,.7);
                         \coordinate (xB) at (0.6,-.7);
                         \coordinate (yB) at (1.5,-.7);
                         \coordinate (zB) at (2.4,-.7);
                         \coordinate (wB) at (3.3,-.7);
                         \coordinate (xM) at (0.6,0);
                         \coordinate (yM) at (1.5,1.05);
                         \coordinate (zM) at (2.4,0);
                         \coordinate (wM) at (3.3,0);
                         \coordinate (xyMB) at (1.05,-.35);
                         \coordinate (yzMB) at (1.95,-.35);
                         \coordinate (zwMB) at (2.85,-.35);
                         \coordinate (yzMA) at (1.95,.35);
                \draw [dotted] (inA) -- (outA);
                \draw [dotted] (inB) -- (outB) node [midway, below]{} ;
                \draw [draw=none] (xA) to[out=40,in=140] (zA);
                \draw [photon] (zA) -- (zB);
                \draw [photon] (yA) -- (yB);
                \draw [photon] (xA) -- (xB);
                \draw [photon,red][out=40,in=140] (xA) to[out=40,in=140] (yA);
                \draw [photon,red][out=40,in=140] (yA) to[out=40,in=140] (zA);
          \draw [fill] (xA) circle (.08);
          \draw [fill] (yA) circle (.08);
          \draw [fill] (yB) circle (.08);
                \draw [fill] (zA) circle (.08);
                \draw [fill] (xB) circle (.08);
                \draw [fill] (zB) circle (.08);
              \end{tikzpicture}
              \begin{tikzpicture}[baseline={([yshift=-1ex]current bounding box.south)},scale=1.5]
                \coordinate (inA) at (0.3,.7);
                         \coordinate (outA) at (3.6,.7);
                         \coordinate (inB) at (0.3,-.7);
                         \coordinate (outB) at (3.6,-.7);
                         \coordinate (xA) at (0.6,.7);
                         \coordinate (yA) at (1.5,.7);
                         \coordinate (zA) at (2.4,.7);
                         \coordinate (wA) at (3.3,.7);
                         \coordinate (xB) at (0.6,-.7);
                         \coordinate (yB) at (1.5,-.7);
                         \coordinate (zB) at (2.4,-.7);
                         \coordinate (wB) at (3.3,-.7);
                         \coordinate (xM) at (0.6,0);
                         \coordinate (yM) at (1.5,1.05);
                         \coordinate (zM) at (2.4,0);
                         \coordinate (wM) at (3.3,0);
                         \coordinate (xyMB) at (1.05,-.35);
                         \coordinate (yzMB) at (1.95,-.35);
                         \coordinate (zwMB) at (2.85,-.35);
                         \coordinate (yzMA) at (1.95,.35);
                \draw [dotted] (inA) -- (outA);
                \draw [dotted] (inB) -- (outB) node [midway, below]{} ;
                \draw [draw=none] (xA) to[out=40,in=140] (zA);
                \draw [photon] (zA) -- (zB);
                \draw [photon] (yA) -- (yB);
                \draw [photon] (xA) -- (xB);
                \draw [photon,red][out=40,in=140] (xA) to[out=40,in=140] (yA);
                \draw [photon,red][out=40,in=140] (yA) to[out=40,in=140] (zA);
          \draw [fill] (xA) circle (.08);
          \draw [fill] (yA) circle (.08);
          \draw [fill,blue!70] (1.05,.87) circle (.08);
          \draw [fill,blue!70] (1.95,.87) circle (.08);
          \draw [fill] (yB) circle (.08);
                \draw [fill] (zA) circle (.08);
                \draw [fill] (xB) circle (.08);
                \draw [fill] (zB) circle (.08);
              \end{tikzpicture}.
\end{align*}

\chapter{Observables}
\label{sec:obs}

In the previous chapters we have discussed how to compute all integrals required for the impulse up to 5PM order. The impulse encodes a number of observables, some of which we now want to discuss. The full results, especially at 5PM order, are too lengthy to be printed here, and we therefore focus on four aspects: the vectorial structure of the result, the split into conservative and dissipative pieces, the function space in which the result lives, and the physical effects contributing to the observables together and the complications they bring. We concentrate on the scattering angle and on the radiated momentum.

The conservative scattering angle $\theta_{\text{cons}}$ is obtained from the conservative impulse through
\begin{align}
        |\Delta p^\mu_{i,\rm cons}|=2p_\infty\sin(\theta_{\text{cons}}/2)\,,
\end{align}
where $p_\infty=m_1 m_2 \sqrt{\gamma^2-1}/E$, the total (conserved) energy is $E=M \Gamma =M\sqrt{1+2\nu(\gamma-1)}$, the total mass reads $M=m_1+m_2$, and the symmetric mass ratio is $\nu=m_1m_2/M^2$. The precise definition of the conservative impulse will be discussed in detail below for each PM order. The PM expansion of the scattering angle takes the form
\begin{align}
        \label{thetadef}
          \theta_{\text{cons}}=\frac{E}{M} \sum_{m\geq1}\sum_{s=0}^{\lfloor\frac{m-1}{2}\rfloor}
          \bigg(
            \frac{GM}{|b|}
            \bigg)^m
            \nu^s\,
          \theta^{(m,s)}_{\text{cons}}(\gamma)\,,
\end{align}
where $s$ counts the self-force order and $|b|=\sqrt{-b^2}$.\footnote{The self-force expansion of the scattering angle is defined slightly differently from that of the impulse. For the scattering angle, we count orders in the symmetric mass ratio~$\nu$. For the impulse, we instead organize the expansion by the powers $m_1^i m_2^j$, since these reflect which types of integrals appear.}
Additionally one can compute a relative, dissipative scattering angle~\cite{Bini:2021gat,Jakobsen:2023pvx} using the full dissipative impulse as 
\begin{align}
        \tan{(\theta_{\text{diss}})}=-\dfrac{ b \cdot \Delta p}{p_\infty |b| -  \frac{|b|}{|p|}\, p \cdot \Delta p},
\end{align}
where 
\begin{align}
        p^\mu=\dfrac{\nu}{\Gamma^2}\left( \dfrac{\gamma m_1+m_2}{m_1} p_1^\mu-\dfrac{\gamma m_2+m_1}{m_2} p_2^\mu \right).
\end{align}
The PM expansion up to 5PM of the dissipative angle is defined as
\begin{align}
                \theta_{\text{diss}}
                &=
                \Gamma
                \sum_{n=1}^5
                \bigg(\frac{GM}{|b|}\bigg)^n
                \sum_{m=0}^{\lfloor\frac{n-1}{2}\rfloor}
                \nu^m \theta_{\text{diss}}^{(n,m)}(\gamma)
                \nonumber
                \\
                &\ \ \ \ \ +
               \Gamma^{-1}
                \sum_{n=4}^5
                \bigg(\frac{GM}{|b|}\bigg)^n
                \nu^{n-2} \theta_{\text{diss}}^{(n,n-2)}(\gamma)
                +\mathcal{O}(G^6) \, .\label{eq:dissStruct}
\end{align}
The radiated momentum is defined by
\begin{align}
        P_{\text{rad}}^\mu=\Delta p_1^\mu+\Delta p_2^\mu=\sum_{n} G^n\, P_{\text{rad}}^{(n),\mu}\,.
\end{align}
Our results will be given in terms of iterated integrals
\begin{align}
	{\cal I}(\varphi_1,\varphi_2,\ldots,\varphi_n;x)
	\coloneqq\!\!
	 \int_1^x\!\!
	 \d x'
	 \varphi_1(x'){\cal I}(\varphi_2,\ldots,\varphi_n;x'),
\end{align}
with the base case ${\cal I}(;x)=1$. We will in particular be interested in which integration kernels appear and  in the maximal iteration of integration, which is the transcendental weight in the case of MPLs.

The full dissipative results can be found in \zenodoTwo; conservative results can be found in \zenodoOne.
\section{1PM and 2PM}

The impulse at 1PM and 2PM is fully given by the $0\text{SF}/\bar{0}\text{SF}$ contributions, which means that it is determined by probe motion along the geodesic trajectory in the Schwarzschild background. As discussed previously, the integrals contributing at these self-force orders can only give rise to algebraic singularities of the form $\sim \sqrt{\gamma^2-1}^{-n}$. The impulse up to 2PM order reads
\begin{align}
\Delta p_1^\mu=&
\frac{Gm_1m_2b^\mu}{|b|^2}
\left(\frac{2(2\gamma^2-1)}{\sqrt{\gamma^2-1}}+\frac{3\pi}4
\frac{(5\gamma^2-1)}{\sqrt{\gamma^2-1}}\frac{G(m_1+m_2)}{|b|}\right)\nn \\
&+
2G^2m_1m_2\frac{(2\gamma^2-1)^2}{(\gamma^2-1)|b|^2}
\left(m_{2} \check v_{1}^{\mu} - m_{1} \check v_{2}^{\mu} \right) +\cO(G^{3})
\,,\label{eq:2PMdeflection}
\end{align}
and the corresponding scattering angle is
\begin{align}
        \theta^{(1,0)}&=\frac{4 \gamma ^2-2}{\gamma ^2-1}\,, \\
        \theta^{(2,0)}&= \frac{3 \pi  \left(5 \gamma ^2-1\right)}{4 \left(\gamma ^2-1\right)}\,.
\end{align}
This agrees with the classic result obtained by Westpfahl~\cite{Westpfahl:1985tsl}. It also satisfies the mass-shell condition
\begin{equation}
p_1^2=(p_1+\Delta p_1)^2 = m_{1}^{2}.
\end{equation}
 Total momentum conservation, $\Delta p_{1}^\mu + \Delta p_{2}^\mu =0$, is also satisfied at this order, so that no momentum is radiated off in gravitational-wave emission. As we shall see, this picture changes at 3PM order.

\section{3PM}
The impulse at 3PM order has the following self-force expansion
\begin{align}
        \Delta p_1^{(3)\mu}=&m_1 m_2^3 p_1^{(3,\text{0SF})\mu}+m_1^2 m_2^2 p_1^{(3,\text{1SF})\mu}+m_1^3 m_2 p_1^{(3,\bar{0}\text{SF})\mu}\;.
\end{align}
As already discussed, we encounter at 3PM order for the first time 1SF terms, which will allow for radiation of momentum.
\subsection{Radiation reaction and the split into conservative and dissipative effects}

At 3PM order total momentum conservation no longer holds, and instead $\Delta p_1^\mu+\Delta p_2^\mu = P^\mu_{\text{rad}}\neq 0$, due to first self-force order effects. The two black holes emit radiative gravitons and back-react with them instantaneously, and this radiation reaction, which is local in time, produces a net radiated momentum for the binary system.

At the level of the integrals this is intimately connected to the appearance of active propagators in the 3PM-1SF integral family~\refeq{eq:1sfgamma}. Graviton propagators can go on-shell, so they are no longer invariant under time reversal. The use of retarded propagators plays an important role here in obtaining the physically correct result.

Our results, however, are typically not directly usable for phenomenological purposes. The measurements performed by the current detectors are of gravitational waves, not of scattering angles or radiated momenta, and the main use of our results is therefore to inform EOB models, which in turn generate gravitational-wave templates. For this purpose one usually needs to split the result into a conservative and a dissipative part in order to construct a Hamiltonian and the associated fluxes. We cannot define this split from first principles, and it is moreover not necessarily unique (see also ref.~\cite{Blanco:2026wwi} for a more general discussion).

In the PN literature it is well established to perform this split through Feynman propagators in an in-out formalism. Yet even at the PN level it matters whether one evaluates on- or off-shell quantities~\cite{Galley:2009px}. The physical intuition is that Feynman propagators are associated with vanishing boundary conditions at $t=\pm\infty$ whereas retarded propagators are associated with vanishing boundary conditions only at $t=-\infty$. Feynman propagators are in addition time symmetric, which guarantees energy conservation.

We will attempt to perform the split for PM quantities in the same spirit, although a number of subtleties will appear along the way. The guiding principle up to 4PM will be that the slow-velocity expansion of the PM result has to reproduce the conservative PN result. These subtleties begin already at 3PM, as we now discuss.

\subsubsection*{V-terms}

For the V-terms the situation is comparatively simple. As discussed in~\ref{sec:feynmanparity}, many V-type integrals are trivially zero when using a time symmetric propagator prescription. The remaining integrals must be evaluated in two different regions, the potential and the radiative one. The potential region is insensitive to any graviton $\i0$ prescription and is not affected by the choice of propagators. This part is universal and purely conservative.

The second region is the radiative region, in which one graviton can go on-shell, as discussed in~\ref{sec:regions}. If we use Feynman propagators, only one relevant topsector remains for the V-terms, since all integrals without worldlines vanish trivially:
\begin{align*}
        \begin{tikzpicture}[baseline={([yshift=-1ex]current bounding box.south)},scale=1.2]
                \coordinate (inA) at (0.4,.7);
                \coordinate (outA) at (3.4,.7);
                \coordinate (inB) at (0.4,-.7);
                \coordinate (outB) at (3.4,-.7);
                \coordinate (xA) at (0.9,.7);
                \coordinate (xB) at (0.9,-.7);
                \coordinate (yA) at (1.9,.7);
                \coordinate (yB) at (1.9,-.7);
                \coordinate (zA) at (2.9,.7);
                \coordinate (zB) at (2.9,-.7);
                \draw [dotted] (inA) -- (outA);
                \draw [dotted] (inB) -- (outB) node [midway, below]{} ;
                \draw [draw=none] (xA) to[out=40,in=140] (zA);
                \draw [photon] (xA) -- (xB);
                \draw [photon,red] (yA) -- (yB);
                \draw [photon] (zA) -- (zB);
                \draw [zUndirected] (xA) -- (yA);
                \draw [zUndirected] (zB) -- (yB);
                \draw [fill] (xA) circle (.08);
                \draw [fill] (yA) circle (.08);
                \draw [fill] (zA) circle (.08);
                \draw [fill] (zB) circle (.08);
                \draw [fill] (yB) circle (.08);
                \draw [fill] (xB) circle (.08);
              \end{tikzpicture}
\end{align*}

Upon expanding this region in the radiative region, we are left with boundary integrals of mushroom type, with Feynman propagators and worldlines. The worldlines can be removed through an IBP reduction, as shown in~\ref{sec:mushroom}, so that the boundary integrals reduce to V-type integrals without worldlines and with Feynman propagators only. These are trivially zero,
\begin{align*}
        \begin{tikzpicture}[baseline={([yshift=-1ex]current bounding box.south)},scale=1]
                \coordinate (inA) at (0.4,.7);
                \coordinate (outA) at (3.4,.7);
                \coordinate (inB) at (0.4,-.7);
                \coordinate (outB) at (3.4,-.7);
                \coordinate (xA) at (0.9,.7);
                \coordinate (xB) at (0.9,-.7);
                \coordinate (yA) at (1.9,.7);
                \coordinate (yB) at (1.9,-.7);
                \coordinate (zA) at (2.9,.7);
                \coordinate (zB) at (2.9,-.7);
                \draw [dotted] (inA) -- (outA);
                \draw [dotted] (inB) -- (outB) node [midway, below]{} ;
                \draw [draw=none] (xA) to[out=40,in=140] (zA);
                \draw [photon] (xA) -- (xB);
                \draw [photon,red] (yA) -- (yB);
                \draw [photon] (zA) -- (zB)node [midway, right]{$\qquad \longrightarrow$};
                \draw [zUndirected] (xA) -- (yA);
                \draw [zUndirected] (zB) -- (yB);
                \draw [fill] (xA) circle (.08);
                \draw [fill] (yA) circle (.08);
                \draw [fill] (zA) circle (.08);
                \draw [fill] (zB) circle (.08);
                \draw [fill] (yB) circle (.08);
                \draw [fill] (xB) circle (.08);
                \draw [fill,blue!70] (1.4,-.7) circle (.08);
              \end{tikzpicture}
              \begin{tikzpicture}[baseline={([yshift=-1ex]current bounding box.south)},scale=1]
                \coordinate (inA) at (0.4,.7);
                \coordinate (outA) at (4.4,.7);
                \coordinate (inB) at (0.4,-.7);
                \coordinate (outB) at (4.4,-.7);
                \coordinate (xA) at (0.9,.7);
                \coordinate (xB) at (0.9,-.7);
                \coordinate (yA) at (1.9,.7);
                \coordinate (yB) at (1.9,-.7);
                \coordinate (yyA) at (2.9,.7);
                \coordinate (yyB) at (2.9,-.7);
                \coordinate (zA) at (3.9,.7);
                \coordinate (zB) at (3.9,-.7);
                \draw [dotted] (inA) -- (outA);
                \draw [dotted] (inB) -- (outB) node [midway, below]{} ;
                \draw [draw=none] (xA) to[out=40,in=140] (zA);
                \draw [photon] (xA) -- (xB);
                \draw [photon,red][out=40,in=140] (yB) to[out=40,in=140] node[midway, above]{$F$}(yyB);
                \draw [photon] (zA) -- (zB)node [midway, right]{$\qquad \longrightarrow$};
                \draw [zUndirected] (xB) -- (yB);
                \draw [zUndirected] (zB) -- (yyB);
                \draw [fill] (xA) circle (.08);
                \draw [fill] (yB) circle (.08);
                \draw [fill] (zA) circle (.08);
                \draw [fill] (zB) circle (.08);
                \draw [fill] (yyB) circle (.08);
                \draw [fill] (xB) circle (.08);
                \draw [fill,blue!70] (2.4,-0.7) circle (.08);
              \end{tikzpicture}
              \begin{tikzpicture}[baseline={([yshift=-1ex]current bounding box.south)},scale=1]
                \coordinate (inA) at (0.4,.7);
                \coordinate (outA) at (3.4,.7);
                \coordinate (inB) at (0.4,-.7);
                \coordinate (outB) at (3.4,-.7);
                \coordinate (xA) at (0.9,.7);
                \coordinate (xB) at (0.9,-.7);
                \coordinate (yA) at (1.9,.7);
                \coordinate (yB) at (1.9,-.7);
                \coordinate (zA) at (2.9,.7);
                \coordinate (zB) at (2.9,-.7);
                \draw [dotted] (inA) -- (outA);
                \draw [dotted] (inB) -- (outB) node [midway, below]{} ;
                \draw [draw=none] (xA) to[out=40,in=140] (zA);
                \draw [photon] (xA) -- (xB);
                \draw [photon] (zA) -- (zB)node [midway, right]{$\qquad =0$};
                \draw [photon,red][out=40,in=140] (xB) to[out=40,in=140]node[midway, above]{$F$} (zB);
                \draw [fill] (xA) circle (.08);
                \draw [fill] (zA) circle (.08);
                \draw [fill] (zB) circle (.08);
                \draw [fill] (xB) circle (.08);
                \draw [fill,blue!70] (yB) circle (.08);
              \end{tikzpicture}
\end{align*}
so that for the V-terms the desired result is obtained automatically by using Feynman propagators. In fact, by imposing $\Delta p_1^\mu+\Delta p_2^\mu=0$ together with the mass-shell condition $(p_1+\Delta p_1)^2=p_1^2$, these terms could equivalently have been computed from lower-loop results alone.

An alternative is to construct a time-symmetric prescription as the average of the retarded and advanced propagators for the active graviton. A similar idea is used in self-force calculations~\cite{Pound:2014koa} to define a conservative result by averaging the retarded and advanced solutions. At 3PM this is equivalent to averaging the prescription on the single active propagator, and it also respects time-reversal symmetry. Since
\begin{align*}
\begin{tikzpicture}[baseline={([yshift=-1ex]current bounding box.south)},scale=1]
        \coordinate (inA) at (0.4,.7);
        \coordinate (outA) at (3.4,.7);
        \coordinate (inB) at (0.4,-.7);
        \coordinate (outB) at (3.4,-.7);
        \coordinate (xA) at (0.9,.7);
        \coordinate (xB) at (0.9,-.7);
        \coordinate (yA) at (1.9,.7);
        \coordinate (yB) at (1.9,-.7);
        \coordinate (zA) at (2.9,.7);
        \coordinate (zB) at (2.9,-.7);
        \draw [dotted] (inA) -- (outA);
        \draw [dotted] (inB) -- (outB) node [midway, below]{} ;
        \draw [draw=none] (xA) to[out=40,in=140] (zA);
        \draw [photon] (xA) -- (xB);
        \draw [photon] (zA) -- (zB)node [midway, right]{$\qquad =- $};
        \draw [photon,red][out=40,in=140] (xB) to[out=40,in=140] node[midway, above]{$\rightarrow$} (zB);
        \draw [fill] (xA) circle (.08);
        \draw [fill] (zA) circle (.08);
        \draw [fill] (zB) circle (.08);
        \draw [fill] (xB) circle (.08);
        \draw [fill,blue!70] (yB) circle (.08);
      \end{tikzpicture}
      \begin{tikzpicture}[baseline={([yshift=-1ex]current bounding box.south)},scale=1]
        \coordinate (inA) at (0.4,.7);
        \coordinate (outA) at (3.4,.7);
        \coordinate (inB) at (0.4,-.7);
        \coordinate (outB) at (3.4,-.7);
        \coordinate (xA) at (0.9,.7);
        \coordinate (xB) at (0.9,-.7);
        \coordinate (yA) at (1.9,.7);
        \coordinate (yB) at (1.9,-.7);
        \coordinate (zA) at (2.9,.7);
        \coordinate (zB) at (2.9,-.7);
        \draw [dotted] (inA) -- (outA);
        \draw [dotted] (inB) -- (outB) node [midway, below]{} ;
        \draw [draw=none] (xA) to[out=40,in=140] (zA);
        \draw [photon] (xA) -- (xB);
        \draw [photon] (zA) -- (zB);
        \draw [photon,red][out=40,in=140] (xB) to[out=40,in=140] node[midway, above]{$\leftarrow$} (zB);
        \draw [fill] (xA) circle (.08);
        \draw [fill] (zA) circle (.08);
        \draw [fill] (zB) circle (.08);
        \draw [fill] (xB) circle (.08);
        \draw [fill,blue!70] (yB) circle (.08);
      \end{tikzpicture}
\end{align*}
the averaged prescription gives the same result as the Feynman prescription.
We use here $\rightarrow$ and $\leftarrow$ to denote the causality flow for retarded and advanced propagators. Additionally we assume a causality flow on the dotted worldlines from left to right on both sides (this can usually be neglected but plays a role here because of the dot on the dotted worldline). 

\subsubsection*{B-terms}

A similar picture emerges for the B-terms. The potential region is again insensitive to any graviton propagator prescription, and in the radiative region we encounter a single boundary integral,
\begin{align*}
      \begin{tikzpicture}[baseline={([yshift=-1ex]current bounding box.south)},scale=1]
        \coordinate (inA) at (0.4,.7);
        \coordinate (outA) at (3.4,.7);
        \coordinate (inB) at (0.4,-.7);
        \coordinate (outB) at (3.4,-.7);
        \coordinate (xA) at (0.9,.7);
        \coordinate (xB) at (0.9,-.7);
        \coordinate (yA) at (1.9,.7);
        \coordinate (yB) at (1.9,-.7);
        \coordinate (zA) at (2.9,.7);
        \coordinate (zB) at (2.9,-.7);
        \draw [dotted] (inA) -- (outA);
        \draw [dotted] (inB) -- (outB) node [midway, below]{} ;
        \draw [draw=none] (xA) to[out=40,in=140] (zA);
        \draw [photon] (xA) -- (xB);
        \draw [photon] (zA) -- (zB);
        \draw [photon,red][out=40,in=140] (xB) to[out=40,in=140]  (zB);
        \draw [fill] (xA) circle (.08);
        \draw [fill] (zA) circle (.08);
        \draw [fill] (zB) circle (.08);
        \draw [fill] (xB) circle (.08);
      \end{tikzpicture}
\end{align*}

With the Feynman prescription, this integral has to be evaluated using a Feynman propagator and the real part has to be retained at the end, which yields a non-vanishing contribution. On the other hand,
\begin{align*}
\begin{tikzpicture}[baseline={([yshift=-1ex]current bounding box.south)},scale=1]
        \coordinate (inA) at (0.4,.7);
        \coordinate (outA) at (3.4,.7);
        \coordinate (inB) at (0.4,-.7);
        \coordinate (outB) at (3.4,-.7);
        \coordinate (xA) at (0.9,.7);
        \coordinate (xB) at (0.9,-.7);
        \coordinate (yA) at (1.9,.7);
        \coordinate (yB) at (1.9,-.7);
        \coordinate (zA) at (2.9,.7);
        \coordinate (zB) at (2.9,-.7);
        \draw [dotted] (inA) -- (outA);
        \draw [dotted] (inB) -- (outB) node [midway, below]{} ;
        \draw [draw=none] (xA) to[out=40,in=140] (zA);
        \draw [photon] (xA) -- (xB);
        \draw [photon] (zA) -- (zB)node [midway, right]{$\qquad = $};
        \draw [photon,red][out=40,in=140] (xB) to[out=40,in=140] node[midway, above]{$\rightarrow$} (zB);
        \draw [fill] (xA) circle (.08);
        \draw [fill] (zA) circle (.08);
        \draw [fill] (zB) circle (.08);
        \draw [fill] (xB) circle (.08);
      \end{tikzpicture}
      \begin{tikzpicture}[baseline={([yshift=-1ex]current bounding box.south)},scale=1]
        \coordinate (inA) at (0.4,.7);
        \coordinate (outA) at (3.4,.7);
        \coordinate (inB) at (0.4,-.7);
        \coordinate (outB) at (3.4,-.7);
        \coordinate (xA) at (0.9,.7);
        \coordinate (xB) at (0.9,-.7);
        \coordinate (yA) at (1.9,.7);
        \coordinate (yB) at (1.9,-.7);
        \coordinate (zA) at (2.9,.7);
        \coordinate (zB) at (2.9,-.7);
        \draw [dotted] (inA) -- (outA);
        \draw [dotted] (inB) -- (outB) node [midway, below]{} ;
        \draw [draw=none] (xA) to[out=40,in=140] (zA);
        \draw [photon] (xA) -- (xB);
        \draw [photon] (zA) -- (zB)node [midway, right]{$\qquad = $};
        \draw [photon,red][out=40,in=140] (xB) to[out=40,in=140] node[midway, above]{$\leftarrow$} (zB);
        \draw [fill] (xA) circle (.08);
        \draw [fill] (zA) circle (.08);
        \draw [fill] (zB) circle (.08);
        \draw [fill] (xB) circle (.08);
      \end{tikzpicture}
      \begin{tikzpicture}[baseline={([yshift=-1ex]current bounding box.south)},scale=1]
        \coordinate (inA) at (0.4,.7);
        \coordinate (outA) at (3.4,.7);
        \coordinate (inB) at (0.4,-.7);
        \coordinate (outB) at (3.4,-.7);
        \coordinate (xA) at (0.9,.7);
        \coordinate (xB) at (0.9,-.7);
        \coordinate (yA) at (1.9,.7);
        \coordinate (yB) at (1.9,-.7);
        \coordinate (zA) at (2.9,.7);
        \coordinate (zB) at (2.9,-.7);
        \draw [dotted] (inA) -- (outA);
        \draw [dotted] (inB) -- (outB) node [midway, below]{} ;
        \draw [draw=none] (xA) to[out=40,in=140] (zA);
        \draw [photon] (xA) -- (xB);
        \draw [photon] (zA) -- (zB)node [midway, right]{};
        \draw [photon,red][out=40,in=140] (xB) to[out=40,in=140] node[midway, above]{$F$} (zB);
        \draw [fill] (xA) circle (.08);
        \draw [fill] (zA) circle (.08);
        \draw [fill] (zB) circle (.08);
        \draw [fill] (xB) circle (.08);
      \end{tikzpicture}
\end{align*}

\noindent where the symbol $F$ denotes the use of Feynman propagators with a subsequent real-part projection. The averaged prescription therefore yields the same result as the Feynman prescription, in line with what we observed for the V-terms.

Comparing to PN results, however, one finds a mismatch in the scattering angle. Reproducing the PN result would require the contributions from the radiative region to vanish also for the B-terms. The standard remedy in the literature is therefore to evaluate these integrals only in the potential region and to set the radiative contributions to zero by hand.

Evaluating the integrals consistently in all regions, with either an averaged prescription or a real-part-of-Feynman prescription, would yield a $\Delta p_i$ that certainly implies $P_{\text{rad}}=0$, but it is not a priori clear whether the same prescription would also imply a vanishing radiated angular momentum $J_{\text{rad}}$.

These results are by no means new, but they already illustrate a problem with the split into dissipative and conservative pieces, which will be even more pronounced at higher PM orders. We currently have no way of computing conservative PM results that match conservative PN results within a single consistent prescription. The integrals must be evaluated region by region, using different prescriptions for different regions, or simply setting some of them to zero by hand. One could of course adopt a single consistent prescription throughout, although it is then not clear whether the resulting impulse implies a vanishing radiated angular momentum.

\subsection{The scattering angle}

At 3PM-1SF both the conservative and dissipative scattering angle contain for the first time non-trivial functions, namely $\log(x)$ together with rational functions of $x$. 
The iterated integrals have maximum weight one and only involve a single integration kernel,
\begin{align}
        \varphi_i(x) \in    \bigg\{ &\dfrac{1}{x}\bigg\} \label{eq:3PMRadfunct}.
\end{align}

\subsection{The radiated momentum}

Radiation appears for the first time at 3PM. The full impulse has the structure
\begin{align}
        \Delta p_1^{(3)}=&b^\mu(m_1 m_2^3 p_1^{3,b,\text{0SF}}+m_1^2 m_2^2 p_1^{3,b,\text{1SF}}+m_1^3 m_2^1 p_1^{3,b,\text{0SF}})\nn \\
        &+v_1^\mu(m_1 m_2^3 p_1^{3,v_1,\text{0SF}}+m_1^2 m_2^2 p_1^{3,v_1,\text{1SF}}-m_1^3 m_2^1 p_1^{3,v_2,\text{0SF}})\nn\\
        &+v_2^\mu(m_1 m_2^3 p_1^{3,v_2,\text{0SF}}+m_1^2 m_2^2 p_1^{3,v_2,\text{1SF}}-m_1^3 m_2^1 p_1^{3,v_1,\text{0SF}}),
\end{align}
where we use $p_1^{n,b,\text{0SF}}=p_1^{n,b,\bar{0}\text{SF}}$ and $p_1^{n,v_{1/2},\text{0SF}}=-p_1^{n,v_{2/1},\bar{0}\text{SF}}$.
 Since $\Delta p_2^{(3)}$ is obtained from $\Delta p_1^{(3)}$ by relabelling, $P_{\text{rad}}$ can only have contributions in the direction of $v_i$,
\begin{align}
        P_{\text{rad}}^{(3),\mu}=m_1^2 m_2^2\,(p_1^{v_1,\text{1SF}}+p_1^{v_2,\text{1SF}})\,(v_1^\mu+v_2^\mu)\,.
\end{align}
At leading order the radiated momentum is expressed in terms of weight-one iterated integrals built from the two kernels
\begin{align}
        \varphi_i(x) \in    \bigg\{ &\dfrac{1}{x},\, \dfrac{1}{1+x}\bigg\} \label{eq:3PMRadfunct2}.
\end{align}

\section{4PM}
The impulse at 4PM order has the following self-force expansion
\begin{align}
        \Delta p_1^{(4)\mu}=&m_1 m_2^4 p_1^{(4,\text{0SF})\mu}+m_1^2 m_2^3 p_1^{(4,\text{1SF})\mu}+m_1^3 m_2^2 p_1^{(4,\bar{1}\text{SF})\mu}+m_1^4 m_2 p_1^{(4,\bar{0}\text{SF})\mu}\;.
\end{align}
The impulse still has no 2SF contributions but gets $\bar{1}$SF contributions.

\subsection{The tail effect and the conservative--dissipative split}

As already at 3PM, the split into conservative and dissipative contributions has to be performed by hand, and is not derived from first principles. We instead try to identify a prescription that reproduces the PN result in the low-velocity expansion. At 3PM the conservative part was defined as just the potential contribution to the scattering angle. We will see that this is no longer the case at 4PM order. Computing the potential part of the scattering angle we find
\begin{align}
        \theta^{(4)}_{pot}=\dfrac{P_{4}(\gamma)\, e^{-3 \gamma_E \epsilon}}{4 \epsilon}+\theta^{(4)}_{pot,fin}\,,
\end{align}
with $P_{4}(\gamma)$ and $\theta^{(4)}_{pot,fin}$ functions of $\gamma$ or $x$ and independent of $\epsilon$.
The potential-only result is therefore not finite but carries a pole in $\epsilon$ which diverges in the limit $\epsilon \to 0$. Defining the conservative angle as this single region therefore makes no physical sense at 4PM. We also know that this pole can only be cancelled by a region which is even in velocity, which in the present case means only the tail region.

Unlike the potential region, the tail region depends on the $\i0$ prescription of the propagators. The conservative prescription should preserve energy, and we can again choose either Feynman propagators or the average of advanced and retarded propagators to achieve time-reversal symmetry and energy conservation. The two choices yield the same result, since
\begin{align*}
        \begin{tikzpicture}[baseline={([yshift=-1ex]current bounding box.south)},scale=1.5]
          \coordinate (inA) at (0.4,.7);
          \coordinate (outA) at (3.6,.7);
          \coordinate (inB) at (0.4,-.7);
          \coordinate (outB) at (3.6,-.7);
          \coordinate (xA) at (1,.7);
          \coordinate (yA) at (1.60+0.4,.7);
          \coordinate (zA) at (3,.7);
          \coordinate (xB) at (1,-.7);
          \coordinate (yB) at (1.6+0.4,-.7);
          \coordinate (zB) at (3,-.7);
          \draw [dotted] (inA) -- (outA);
          \draw [dotted] (inB) -- (outB) node [midway, below]{} ;
          \draw [draw=none] (xA) to[out=40,in=140] (zA);
          \draw [photon] (zA) -- (zB) node [midway, right]{$\qquad =$};
          \draw [photon] (xA) -- (xB);
          \draw [photon,red][out=40,in=140] (xA) to[out=40,in=140]node [midway, above]{$F$} (yA);
          \draw [photon,red][out=40,in=140] (yA) to[out=40,in=140]node [midway, above]{$F$} (zA);
          \draw [fill] (xA) circle (.08);
          \draw [fill] (yA) circle (.08);
          \draw [fill] (zA) circle (.08);
          \draw [fill] (xB) circle (.08);
          \draw [fill] (zB) circle (.08);
        \end{tikzpicture}
        \begin{tikzpicture}[baseline={([yshift=-1ex]current bounding box.south)},scale=1.5]
                \coordinate (inA) at (0.4,.7);
                \coordinate (outA) at (3.6,.7);
                \coordinate (inB) at (0.4,-.7);
                \coordinate (outB) at (3.6,-.7);
                \coordinate (xA) at (1,.7);
                \coordinate (yA) at (1.60+0.4,.7);
                \coordinate (zA) at (3,.7);
                \coordinate (xB) at (1,-.7);
                \coordinate (yB) at (1.6+0.4,-.7);
                \coordinate (zB) at (3,-.7);
                \draw [dotted] (inA) -- (outA);
                \draw [dotted] (inB) -- (outB) node [midway, below]{} ;
                \draw [draw=none] (xA) to[out=40,in=140] (zA);
                \draw [photon] (zA) -- (zB) node [midway, right]{$\qquad =$};
                \draw [photon] (xA) -- (xB);
                \draw [photon,red][out=40,in=140] (xA) to[out=40,in=140]node [midway, above]{$\leftarrow$} (yA);
                \draw [photon,red][out=40,in=140] (yA) to[out=40,in=140]node [midway, above]{$\leftarrow$} (zA);
                \draw [fill] (xA) circle (.08);
                \draw [fill] (yA) circle (.08);
                \draw [fill] (zA) circle (.08);
                \draw [fill] (xB) circle (.08);
                \draw [fill] (zB) circle (.08);
              \end{tikzpicture}
              \begin{tikzpicture}[baseline={([yshift=-1ex]current bounding box.south)},scale=1.5]
                \coordinate (inA) at (0.4,.7);
                \coordinate (outA) at (3.6,.7);
                \coordinate (inB) at (0.4,-.7);
                \coordinate (outB) at (3.6,-.7);
                \coordinate (xA) at (1,.7);
                \coordinate (yA) at (1.60+0.4,.7);
                \coordinate (zA) at (3,.7);
                \coordinate (xB) at (1,-.7);
                \coordinate (yB) at (1.6+0.4,-.7);
                \coordinate (zB) at (3,-.7);
                \draw [dotted] (inA) -- (outA);
                \draw [dotted] (inB) -- (outB) node [midway, below]{} ;
                \draw [draw=none] (xA) to[out=40,in=140] (zA);
                \draw [photon] (zA) -- (zB);
                \draw [photon] (xA) -- (xB);
                \draw [photon,red][out=40,in=140] (xA) to[out=40,in=140]node [midway, above]{$\rightarrow$} (yA);
                \draw [photon,red][out=40,in=140] (yA) to[out=40,in=140]node [midway, above]{$\rightarrow$} (zA);
                \draw [fill] (xA) circle (.08);
                \draw [fill] (yA) circle (.08);
                \draw [fill] (zA) circle (.08);
                \draw [fill] (xB) circle (.08);
                \draw [fill] (zB) circle (.08);
              \end{tikzpicture}
\end{align*}
 again with the convention that the Feynman prescription is followed by a real-part projection. We have checked this relation explicitly up to NNLO in $\epsilon$, which is all we need for the impulse. Using these boundary conditions we find
\begin{align}
        \theta^{(4)}_{tail}=- \dfrac{(\gamma-1)^{-4 \epsilon}\, P_{4}(\gamma)\, e^{-3 \gamma_E \epsilon}}{4 \epsilon}+\theta^{(4)}_{tail,fin}\,,
\end{align}
and the sum of the potential and tail contributions gives a finite total conservative angle,
\begin{align}
        \theta^{(4)}_{cons}= \log(\gamma-1)\, P_{4}(\gamma)+\theta^{(4)}_{pot,fin}+\theta^{(4)}_{tail,fin}\,.
\end{align}
Neglecting again the regions odd in $v$, this slow-velocity expansion matches exactly the known PN calculations.

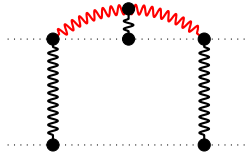
\begin{figure}[h]\centering
        \begin{tikzpicture}[baseline={([yshift=-1ex]current bounding box.south)},scale=1]
                \coordinate (inA) at (0.4,.7);
                \coordinate (outA) at (3.6,.7);
                \coordinate (inB) at (0.4,-.7);
                \coordinate (outB) at (3.6,-.7);
                \coordinate (xA) at (1,.7);
                \coordinate (yA) at (1.60+0.4,.7);
                \coordinate (zA) at (3,.7);
                \coordinate (xB) at (1,-.7);
                \coordinate (yB) at (1.6+0.4,-.7);
                \coordinate (zB) at (3,-.7);
                \coordinate (yM) at (1.6+0.4,1.1);
                \draw [dotted] (inA) -- (outA);
                \draw [dotted] (inB) -- (outB) node [midway, below]{} ;
                \draw [draw=none] (xA) to[out=40,in=140] (zA);
                \draw [photon] (zA) -- (zB);
                \draw [photon] (xA) -- (xB);
                \draw [photon] (yA) -- (yM);
                \draw [photon,red][out=40,in=140] (xA) to[out=40,in=140] (zA);
                \draw [fill] (xA) circle (.08);
                \draw [fill] (yA) circle (.08);
                \draw [fill] (yM) circle (.08);
                \draw [fill] (zA) circle (.08);
                \draw [fill] (xB) circle (.08);
                \draw [fill] (zB) circle (.08);
              \end{tikzpicture}\caption{The tail graph.}\label{fig:tail}
\end{figure}

Let us discuss what has happened physically. The expansion in regions commonly introduces divergences in individual regions which cancel against each other once all regions are summed. Because of the different $v^{m\epsilon}$ scalings of the regions, this cancellation generates a $\log(\gamma-1)$ term in the final result. Note that in the slow-velocity expansion without an $\epsilon$ expansion no $\log(\gamma-1)$ terms appear. This discontinuity is introduced by non-local tail terms coming from diagrams as depicted in Fig.~\ref{fig:tail}. At 3PM the boundary integrals describe processes in which a graviton is emitted and re-absorbed by the same worldline. Such interactions are instantaneous, since the graviton propagates at the speed of light. At 4PM the boundaries look similar, but before IBP reduction there is an additional potential graviton interacting with the radiative ones. Even though this structure is reduced by an IBP reduction, it generates terms that are non-local in time. This is the familiar tail effect~\cite{PhysRevD.37.1410, PhysRevD.46.4304, Porto:2024cwd}: the interaction of the waveform with the static background field delays the gravitational-wave signal and produces a ``tail'' of the signal. In our case it gives rise to $\log(\gamma-1)$ terms and also complicates the map to the bound case~\cite{Kalin:2019rwq,Kalin:2019inp,Cho:2021arx,Dlapa:2024cje}. Note that this would never occur in electromagnetism, since it originates entirely from the self-interactions (graviton vertices) of Einstein gravity.

\subsection{The scattering angle}

The 4PM-1SF conservative scattering angle contains not only logarithmic functions but also the period over the K3 surface in terms of squared elliptic functions. We encounter $K^2(1-x^2)$, $E^2(1-x^2)$ and $K(1-x^2)E(1-x^2)$, together with iterated integrals up to weight two. The tail effect produces for the first time the integration kernel $\frac{1}{1-x}$. All iterated integrals over K3 kernels drop out of the final result, and what remains is
\begin{align}
        \varphi_i(x) \in    \bigg\{ &\dfrac{1}{x},\, \dfrac{1}{1+x},\, \dfrac{1}{1-x},\, \dfrac{x}{1+x^2}\bigg\} \label{eq:4PMangle}.
\end{align}
The same is true for the 1SF contribution of the dissipative scattering angle $\theta^{4,1}_{\text{diss}}$. We find at 4PM in eq.~\refeq{eq:dissStruct} for the first time a $\bar 1$SF contribution $\theta^{4,2}_{\text{diss}}$, coming from the recoil. It contains no elliptic functions, but only up to twice-iterated integrals in terms of the integration kernels defined in eq.~\refeq{eq:4PMangle}.
\subsection{The radiated momentum}

The structure of the radiated momentum becomes more involved through the appearance of $\bar{1}\text{SF}$ contributions. The full impulse is given by
\begin{align}
        \Delta p_1^{(4)}=&b^\mu(m_1 m_2^4 p_1^{4,b,\text{0SF}}+m_1^2 m_2^3 p_1^{4,b,\text{1SF}}+m_1^3 m_2^2 p_1^{4,b,\bar{1}\text{SF}}+m_1^4 m_2^1 p_1^{4,b,\text{0SF}})\nn \\
        +&v_1^\mu(m_1 m_2^4 p_1^{4,v_1,\text{0SF}}+m_1^2 m_2^3 p_1^{4,v_1,\text{1SF}}+m_1^3 m_2^2 p_1^{4,v_1,\bar{1}\text{SF}}-m_1^4 m_2^1 p_1^{4,v_2,\text{0SF}})\nn\\
        +&v_2^\mu(m_1 m_2^4 p_1^{4,v_2,\text{0SF}}+m_1^2 m_2^3 p_1^{4,v_2,\text{1SF}}+m_1^3 m_2^2 p_1^{4,v_2,\bar{1}\text{SF}}-m_1^4 m_2^1 p_1^{4,v_1,\text{0SF}}),
\end{align}
so that the radiated momentum receives for the first time a contribution along $b$ from the $\bar{1}\text{SF}$, or recoil, terms,
\begin{align}
        P_{\text{rad}}^{(4)}&=b^\mu\, M^4\, \nu^2\, (m_1-m_2)\, (p_1^{4,b,\bar{1}\text{SF}}-p_1^{4,b,\text{1SF}}) \nn\\
        &+(v_1^\mu+v_2^\mu  )\, M^5\, \nu^2\,\frac{1}{2}( p_1^{4,v_1,\text{1SF}}+ p_1^{4,v_1,\bar{1}\text{SF}}+ p_1^{4,v_2,\text{1SF}}+ p_1^{4,v_2,\bar{1}\text{SF}})\nn \\
        &+(v_1^\mu-v_2^\mu  )\,M^4\,(m_2-m_1)\,\nu^2\,\frac{1}{2} ( p_1^{4,v_1,\text{1SF}}- p_1^{4,v_1,\bar{1}\text{SF}}- p_1^{4,v_2,\text{1SF}}+ p_1^{4,v_2,\bar{1}\text{SF}})\,.
\end{align}
The radiated momentum receives no contribution from the tail and contains no K3 period. The iterated integrals up to weight two are built from the kernels
\begin{align}
        \varphi_i(x) \in    \bigg\{ &\dfrac{1}{x},\, \dfrac{1}{1+x},\, \dfrac{x}{1+x^2}\bigg\} \label{eq:4PMrad}.
\end{align}

\section{5PM-1SF}
The impulse at 5PM order has the following self-force expansion
\begin{align}
        \Delta p_1^{(5)\mu}=&m_1 m_2^5 p_1^{(5,\text{0SF})\mu}+m_1^2 m_2^4 p_1^{(5,\text{1SF})\mu}+m_1^3 m_2^3 p_1^{(5,\text{2SF})\mu}+m_1^4 m_2^2 p_1^{(5,\bar{1}\text{SF})\mu}+m_1^5 m_2 p_1^{(5,\bar{0}\text{SF})\mu}\;.
\end{align}
The impulse gets for the first time contributions from the second self-force order. We will discuss first the 1SF contributions though.
\subsection{The tail-of-tail effect}

At 5PM we encounter a higher-order tail effect, the tail-of-tail, which is characterised by a repeated interaction of a radiative graviton with potential ones:
\begin{align*}
        \begin{tikzpicture}[baseline={([yshift=-1ex]current bounding box.south)},scale=1.2]
                \coordinate (inA) at (0.3,.7);
                         \coordinate (outA) at (3.6,.7);
                         \coordinate (inB) at (0.3,-.7);
                         \coordinate (outB) at (3.6,-.7);
                         \coordinate (xA) at (0.6,.7);
                         \coordinate (yA) at (1.5,.7);
                         \coordinate (zA) at (2.4,.7);
                         \coordinate (wA) at (3.3,.7);
                         \coordinate (xB) at (0.6,-.7);
                         \coordinate (yB) at (1.5,-.7);
                         \coordinate (zB) at (2.4,-.7);
                         \coordinate (wB) at (3.3,-.7);
                         \coordinate (xM) at (0.6,0);
                         \coordinate (yM) at (1.5,1.15);
                         \coordinate (zM) at (2.4,1.15);
                         \coordinate (wM) at (3.3,0);
                \draw [dotted] (inA) -- (outA);
                \draw [dotted] (inB) -- (outB) node [midway, below]{} ;
                \draw [draw=none] (xA) to[out=40,in=140] (zA);
                \draw [photon] (wA) -- (wB);
                \draw [photon] (xA) -- (xB);
                \draw [photon] (yA) -- (yM);
                \draw [photon] (zA) -- (zM);
                \draw [photon,red][out=40,in=140] (xA) to[out=40,in=140] (wA);
                \draw [fill] (xA) circle (.08);
                \draw [fill] (wA) circle (.08);
                \draw [fill] (yM) circle (.08);
                \draw [fill] (zM) circle (.08);
                \draw [fill] (wB) circle (.08);
                \draw [fill] (zA) circle (.08);
                \draw [fill] (xB) circle (.08);
                \draw [fill] (yA) circle (.08);
              \end{tikzpicture}.
\end{align*}
This effect has two consequences. For the first time, two regions that are odd in the velocity coexist, and non-trivial cancellations between them become possible. The $\text{RRR}$ and $\text{RPP}+\text{PRP}+\text{PPR}$ regions each carry a pole in $\epsilon$, and these poles cancel against each other while leaving behind $\log(\gamma-1)$ terms in the finite result, similarly to what happens at 4PM with the simple tail effect. This now occurs in the radiated momentum. We find
\begin{align}
        {P_{\text{rad}}}^{(5,1),\mu}_{R}&=\dfrac{P_{5,1}^\mu(\gamma)\, e^{-4 \gamma_E \epsilon}\, (\gamma-1)^{-2 \epsilon}}{4 \epsilon}+ {P_{\text{rad}}}^{(5,1),\mu}_{fin,R}\,, \\
       {P_{\text{rad}}}^{(5,1),\mu}_{RRR}&=-\dfrac{P_{5,1}^\mu(\gamma)\, e^{-4 \gamma_E \epsilon}\, (\gamma-1)^{-6 \epsilon}}{4 \epsilon}+ {P_{\text{rad}}}^{(5,1),\mu}_{fin,RRR}\,,\\
       {P_{\text{rad}}}^{(5,1),\mu}&={P_{\text{rad}}}^{(5,1),\mu}_{fin,RRR}+{P_{\text{rad}}}^{(5,1),\mu}_{fin,R}+ P_{5,1}^\mu(\gamma)\, \log(\gamma-1)\,. \label{eq:tailoftailterm}
\end{align}
In addition to the $\log(\gamma-1)$ term in the radiated energy, the tail-of-tail poses also a problem for the split into conservative and dissipative parts. From the PN literature one would expect a conservative contribution to the scattering angle at 5.5PN order~\cite{Bini:2020hmy}, which arises entirely from the RRR region. As we have already seen, however, the prescriptions used so far (Feynman, or the average of advanced and retarded propagators) never reproduce the conservative PN result from the single $\text{R}$ regions, and the latter have had to be set to zero by hand. Similarly both prescriptions don't reproduce for the $\text{RRR}$ region the conservative piece known from PN~\cite{Bini:2021gat,Bini:2022enm,Bini:2020hmy}.

In fact, we have not found any prescription which yields a conservative piece from the RRR region that matches the known PN result, although the full physical answer using retarded propagators is completely unambiguous and reproduces the known PN results exactly up to 5.5PN order. We also note that the would-be conservative RRR piece is non-local in time, and can therefore not be mapped to the bound case in any way; it should be removed before such a mapping is attempted~\cite{Dlapa:2025biy}.

The remaining conservative part, which is even in velocity, can be defined using Feynman prescription or by an average of retarded and advanced propagators in the tail boundaries, similarly to 4PM.

\subsection{The scattering angle}

The function space of the integrals contributing to the conservative scattering angle is exactly the same as the one encountered at 4PM. The logarithmic kernels are
\begin{align}
        \varphi_i(x) \in    \bigg\{ &\dfrac{1}{x},\, \dfrac{1}{1-x},\, \dfrac{1}{1+x},\, \dfrac{x}{1+x^2}\bigg\} \label{eq:5pm1sfangleKern}.
\end{align}
In addition we encounter the squared elliptic functions in our integrals, together with iterated integrals over kernels containing $K^2$. At 4PM we have already seen all such iterated integrals drop out, leaving only the K3 period itself and its derivatives. At 5PM-1SF all elliptic functions or iterated integrals over them drop out, so that the scattering angle is fully expressed in terms of rational functions and iterated integrals of weight at most three over the kernels in~\refeq{eq:5pm1sfangleKern}. The same is also true for the $1$SF and $\bar 1$SF contributions $\theta_{\text{diss}}^{5,1}$ and $\theta_{\text{diss}}^{5,3}$

\subsection{The radiated momentum}

The full impulse at 5PM has the form
\begin{align}
        \Delta p_1^{(5),\mu}=&b^\mu(m_1 m_2^5 p_1^{5,b,\text{0SF}}+m_1^2 m_2^4 p_1^{5,b,\text{1SF}}+m_1^3 m_2^3 p_1^{5,b,\text{2SF}}+m_1^4 m_2^2 p_1^{5,b,\bar{1}\text{SF}}+m_1^5 m_2 p_1^{5,b,\text{0SF}})\nn \\
        +&v_1^\mu(m_1 m_2^5 p_1^{5,v_1,\text{0SF}}+m_1^2 m_2^4 p_1^{5,v_1,\text{1SF}}+m_1^3 m_2^3 p_1^{5,v_1,\text{2SF}}+m_1^4 m_2^2 p_1^{5,v_1,\bar{1}\text{SF}}-m_1^5 m_2 p_1^{5,v_2,\text{0SF}})\nn\\
        +&v_2^\mu(m_1 m_2^5 p_1^{5,v_2,\text{0SF}}+m_1^2 m_2^4 p_1^{5,v_2,\text{1SF}}+m_1^3 m_2^3 p_1^{5,v_2,\text{2SF}}+m_1^4 m_2^2 p_1^{5,v_2,\bar{1}\text{SF}}-m_1^5 m_2 p_1^{5,v_1,\text{0SF}})\,,
\end{align}
which yields for the radiated momentum
\begin{align}
                P_{\text{rad}}^{(5),\mu}&=b^\mu\, M^5\, \nu^2\, (m_1-m_2)\, (p_1^{5,b,\bar{1}\text{SF}}-p_1^{5,b,\text{1SF}})  \nn\\
                &+(v_1^\mu+v_2^\mu  )\, M^6\, \nu^2\,\frac{1}{2}( p_1^{5,v_1,\text{1SF}}+ p_1^{5,v_1,\bar{1}\text{SF}}+ p_1^{5,v_2,\text{1SF}}+ p_1^{5,v_2,\bar{1}\text{SF}})\nn \\
                &+(v_1^\mu-v_2^\mu  )\,M^5\,\nu^2\,(m_1-m_2)\,\frac{1}{2} ( p_1^{5,v_1,\text{1SF}}- p_1^{5,v_1,\bar{1}\text{SF}}- p_1^{5,v_2,\text{1SF}}+ p_1^{5,v_2,\bar{1}\text{SF}})\nn \\
                &+\nu^3\,M^6\, (v_1^\mu+v_2^\mu  )(p_1^{5,v_1,\text{2SF}}+p_1^{5,v_2,\text{2SF}}-p_1^{5,v_1,\text{1SF}}-p_1^{5,v_2,\text{1SF}}-p_1^{5,v_1,\bar{1}\text{SF}}-p_1^{5,v_2,\bar{1}\text{SF}})\,.
\end{align}
The last line represents the 2SF contribution, which we neglect here. The function space of the 1SF part is more involved. 
We encounter in our integrals the same K3-period structure as in the scattering angle. In addition the CY3 sector enters and introduces a host of new functions. Some K3 sectors couple to the CY3 sector, so we also encounter integration kernels that involve both the K3 and CY3 periods. We also have the $G_i$ and $H_i$ functions, which we introduced to canonicalise our basis, and which can appear in integration kernels as well as in their prefactors.

In the final result we find almost all of these functions to be present. The different possible choices of periods have been discussed already. If we follow the choice of ref.~\cite{Driesse:2024feo}, the result contains iterated integrals of weight at most four, which breaks the pattern observed at lower orders, where at $n$-PM order at most $n-2$ iterated integrals appeared. With the choice presented in this thesis in eq.~\refeq{eq:periodchoice} the result simplifies and we find a maximum of three times iterated integrals.

The relevant integration kernels are,
\begin{align}
       & \varphi_i(x) \in \bigg\{\dfrac{1}{x},\, \dfrac{1}{1-x},\, \dfrac{1}{1+x},\, \dfrac{x}{1+x^2},\nn \\
       &\varpi_{\text{K3,1}},\,
        x \, \varpi_{\text{K3,1}},\,\frac{\varpi_{\text{K3,1}}}{x},\,\frac{\varpi_{\text{K3,1}}}{1-x},\,\frac{\varpi_{\text{K3,1}}}{1+x},\,\frac{1}{x \varpi_{\text{K3,1}}},\,\frac{1}{(1-x) \varpi_{\text{K3,1}}},\,\frac{1}{(1+x) \varpi_{\text{K3,1}}},\nn\\
          & \varpi_{\text{CY3,1}},\,
            \frac{\varpi_{\text{CY3,1}}}{x},\,
            \frac{\varpi_{\text{CY3,1}}}{x^2},\,
            \frac{\varpi_{\text{CY3,1}}}{1-x},\,
            \frac{\varpi_{\text{CY3,1}}}{(1-x)^2},\,
            \frac{\varpi_{\text{CY3,1}}}{1+x},\,
            \frac{\varpi_{\text{CY3,1}}}{(1+x)^2},\nn\\
          & \frac{1}{x\,\alpha},\,
            \frac{\alpha^2}{(1-x)\,\varpi_{\text{CY3,1}}^2},\,
            \frac{\alpha^2}{(1+x)\,\varpi_{\text{CY3,1}}^2},\,
            \frac{x\,\alpha^2}{(x^2+1)\,\varpi_{\text{CY3,1}}^2},\nn\\
          & \frac{\alpha\,\varpi_{K3,1}}{\varpi_{\text{CY3,1}}},\,
            \frac{\alpha\,\varpi_{K3,1}}{(1-x)\,\varpi_{\text{CY3,1}}},\,
            \frac{\alpha\,\varpi_{K3,1}}{(1+x)\,\varpi_{\text{CY3,1}}},\,
            \frac{\alpha\,\varpi_{K3,1}}{(x^2+1)\,\varpi_{\text{CY3,1}}},\nn\\
          & \frac{H_1}{x\,\alpha},\,
            \frac{H_1}{x},\,
            \frac{H_1}{1-x},\,
            \frac{H_1}{1+x},\nn\\
          & \frac{H_2}{x\,\alpha},\,
            \frac{H_2}{x},\,
            \frac{H_2}{1-x},\,
            \frac{H_2}{1+x},\nn\\
          & \frac{H_3}{x\,\alpha},\,
            \frac{H_3}{x},\,
            \frac{H_3}{1-x},\,
            \frac{H_3}{1+x},\nn\\
          & \frac{H_6}{x\,\alpha},\,
            \frac{H_6}{x\,\varpi_{K3,1}},\,
            \frac{H_6}{(1-x)\,\varpi_{K3,1}},\,
            \frac{H_6}{(1+x)\,\varpi_{K3,1}},\nn\\
          & \frac{H_8\,\alpha^2}{(1-x)\,\varpi_{\text{CY3,1}}^2},\,
            \frac{H_8\,\alpha^2}{(1+x)\,\varpi_{\text{CY3,1}}^2},\,
            \frac{x\,H_8\,\alpha^2}{(x^2+1)\,\varpi_{\text{CY3,1}}^2},\,
           \frac{H_8}{x\,\varpi_{K3,1}},\,
            \frac{H_8}{(1-x)\,\varpi_{K3,1}},\,
            \frac{H_8}{(1+x)\,\varpi_{K3,1}},\nn\\
          & \frac{H_9}{x\,\alpha},\,
            \frac{H_9}{x\,\varpi_{K3,1}},\,
            \frac{H_9}{(1-x)\,\varpi_{K3,1}},\,
            \frac{H_9}{(1+x)\,\varpi_{K3,1}},\nn\\
          & \frac{H_{11}\,\alpha^2}{(1-x)\,\varpi_{\text{CY3,1}}^2},\,
            \frac{H_{11}\,\alpha^2}{(1+x)\,\varpi_{\text{CY3,1}}^2},\,
            \frac{x\,H_{11}\,\alpha^2}{(x^2+1)\,\varpi_{\text{CY3,1}}^2},\,
           \frac{H_{11}}{x\,\varpi_{K3,1}},\,
            \frac{H_{11}}{(1-x)\,\varpi_{K3,1}},\,
            \frac{H_{11}}{(1+x)\,\varpi_{K3,1}}
        \bigg\}.
\end{align}
The coefficients of the iterated integrals additionally contain the periods, their derivatives and the functions $H_1, H_2, H_3, H_4, H_5, H_6, H_8, H_9, H_{11}$. All of the functions $G_1,\ldots,G_4$ disappear from the final result. 
We note that the physical meaning of the $G_i$ was less clear from the outset, whereas the remaining functions can be associated with a generalisation of integrals of the third kind familiar from the elliptic case.
We also note that the CY3 periods already appear in the pole of the tail-of-tail, e.g. the $P_{5,1}^\mu (\gamma)$ term in eq.~\refeq{eq:tailoftailterm}.

The result of the radiated energy completely agrees with results from PN in a small velocity expansion~\cite{Bini:2025vuk} up to 6PN and also shows good numerical agreement with self-force computations~\cite{Warburton:2025ymy}. Especially the latter shows the strength of this PM calculation, since it agrees well with the 1SF result, including contributions from all orders in $G$. It also shows the advantage of PM over PN, exhibiting much better agreement with the SF results at high velocities.

\section{5PM-2SF}

\subsection{The memory effect and radiation-reaction squared}

So far at first self-force order we have seen that ever higher-order tail effects arise with every new PM order. This started at 3PM with radiation reaction, which can be viewed as a tail effect of zeroth order, continued with the simple tail effect at 4PM, and reached the tail-of-tail at 5PM-1SF. They all originate from topologies of the form
\begin{align*}
\begin{tikzpicture}[baseline={([yshift=-1ex]current bounding box.south)},scale=1.2]
        \coordinate (inA) at (0.4,.7);
        \coordinate (outA) at (3.6,.7);
        \coordinate (inB) at (0.4,-.7);
        \coordinate (outB) at (3.6,-.7);
        \coordinate (xA) at (1,.7);
        \coordinate (xxM) at (1.525,0) ;
        \coordinate (yyM) at (2.475,0) ;
        \coordinate (xxB) at (1.525,-.7) ;
        \coordinate (yA) at (1.60+0.4,.7);
        \coordinate (yyB) at (2.475,-.7) ;
        \coordinate (zA) at (3,.7);
        \coordinate (xB) at (1,-.7);
        \coordinate (yB) at (1.6+0.4,-.7);
        \coordinate (zB) at (3,-.7);
        \coordinate (xM) at (1,0);
        \coordinate (yM) at (1.6+0.4,0);
        \coordinate (zM) at (3,-0);
        \draw [dotted] (inA) -- (outA);
        \draw [dotted] (inB) -- (outB) node [midway, below]{} ;
        \draw [draw=none] (xA) to[out=40,in=140] (zA);
        \draw [photon] (xA) -- (xB);
        \draw [photon] (zA) -- (zB);
        \draw [photon,red] (xM) -- (zM);
        \draw [photon] (xxM) -- (xxB);
        \draw [photon] (yyM) -- (yyB);
         \node[below=5pt] at (2, 0) { $.....$};
        \draw [fill] (xA) circle (.08);
        \draw [fill] (xxM) circle (.08);
        \draw [fill] (xxB) circle (.08);
        \draw [fill] (yyM) circle (.08);
        \draw [fill] (yyB) circle (.08);
         \draw [fill] (xM) circle (.08);
        \draw [fill] (zA) circle (.08);
        \draw [fill] (xB) circle (.08);
         \draw [fill] (zM) circle (.08);
        \draw [fill] (zB) circle (.08);
      \end{tikzpicture}
\end{align*}

As discussed before, this is the most important topology at first self-force order. The tail effects all originate from repeated interactions of a single active graviton with several potential gravitons emitted from one of the black holes of the binary system. We emphasise that we can never have a three-point graviton vertex attached to three active gravitons at 1SF.

The situation changes at second self-force order, where a new physical effect appears, the non-linear memory effect. The radiative graviton scatters off another radiative graviton. The name ``non-linear memory effect'' originates from the resulting gravitational-wave signal emitted by the binary system: the scattering on the non-static background induces a permanent change in the waveform, which does not vanish after the passage of the wave~\cite{PhysRevD.46.4304, Porto:2024cwd, Cunningham:2024dog, Georgoudis:2025vkk}. As the name suggests, the effect is driven by the non-linearity of gravity itself. In our impulse calculation it arises from topologies such as
\begin{align*}
        \begin{tikzpicture}[baseline={([yshift=-1ex]current bounding box.south)},scale=1.2]
                 \coordinate (inA)  at (0.6,0.6);
                  \coordinate (outA) at (2.625,0.6);
                  \coordinate (inB)  at (0.6,-0.6);
                  \coordinate (outB) at (2.625,-0.6);
                  \draw[dotted] (inA) -- (outA);
                  \draw[dotted] (inB) -- (outB);
                  \coordinate (aA) at (0.6,0.6);
                  \coordinate (bA) at (1.85+0.1,0.6);
                  \coordinate (dA) at (2.625,0.6);
                  \coordinate (aM) at (0.6,0);
                  \coordinate (bM) at (1.275,0);
                  \coordinate (dM) at (2.625,0);
                  \coordinate (aB) at (0.6,-0.6);
                  \coordinate (bB) at (1.275,-0.6);
                  \coordinate (dB) at (2.625,-0.6);
                \coordinate (bL) at (1.85+0.1,-0.35);
                \draw[photon,red] (aM) -- (dM);
                \draw[fill, white] (1.85+0.1,0) circle (0.15);
                \draw[photon] (aA) -- (aB);
                \draw[photon,red] (bM) -- (bL);
                \draw[photon] (bL) -- (bB);
                \draw[photon] (dA) -- (dB);
                \draw[photon] (bL) -- (bA);
                \foreach \v in {aA,bA,dA,aM,bM,dM,aB,bB,dB,bL}
                  \draw[fill] (\v) circle (.08);
        \end{tikzpicture}
\end{align*}
in which two active gravitons interact and back-react with the binary system. Interestingly this effect produces the first non-planar integrals encountered, so that the non-planar topology can be directly tied to this physical effect. It also makes clear that the memory effect is a primary driver of the increased complexity of 2SF versus 1SF, and it gives rise to a new integration region containing two radiative modes.

In addition we have planar topologies such as
\begin{align*}
          \begin{tikzpicture}[baseline={([yshift=-1ex]current bounding box.south)},scale=1.2]
            \coordinate (inA)  at (0.6,0.6);
            \coordinate (outA) at (2.625,0.6);
            \coordinate (inB)  at (0.6,-0.6);
            \coordinate (outB) at (2.625,-0.6);
            \draw[dotted] (inA) -- (outA);
            \draw[dotted] (inB) -- (outB);
            \coordinate (aA) at (0.6,0.6);
            \coordinate (bA) at (1.275,0.6);
            \coordinate (cA) at (1.95,0.6);
            \coordinate (dA) at (2.625,0.6);
            \coordinate (aM) at (0.6,0);
            \coordinate (bM) at (1.275,0);
            \coordinate (cM) at (1.95,0);
            \coordinate (dM) at (2.625,0);
            \coordinate (aB) at (0.6,-0.6);
            \coordinate (bB) at (1.275,-0.6);
            \coordinate (cB) at (1.95,-0.6);
            \coordinate (dB) at (2.625,-0.6);
            \draw[photon]     (aA) -- (aB) ;
            \draw[photon,red] (aM) -- (dM);
            \draw[photon] (bA) -- (bM);
            \draw[photon] (cB) -- (cM);
            \draw[photon]     (dA) -- (dB);
            \foreach \v in {aA,bA,dA,aM,bM,cM,dM,aB,cB,dB}
              \draw[fill] (\v) circle (.08);
          \end{tikzpicture}
        \end{align*}

\noindent which introduce another new effect: the active graviton back-reacts with potential gravitons emitted by both black holes simultaneously. This generates the same region as the memory effect but corresponds physically to radiation-reaction-squared effects~\cite{Porto:2024cwd}. This is again similar to 4PM, where the tail effect and the repeated radiation-reaction were part of the same region.

\subsection{Conservative prescriptions}

At 1SF order we found a (more or less) working prescription, in which we either use Feynman propagators and take the real part of the result or we use the average of the advanced and retarded solutions. Both of these prescriptions fail at second self-force order. Since they only worked anyway for the regions even in velocity, we will examine only the potential, tail and memory regions. The potential region is again completely unambiguous, since it is insensitive to any graviton prescription. It carries a pole in $\epsilon$ coming from the tail, just as at 4PM, and in addition it has a singular point in the physical region around $\gamma=3$,
\begin{align}
        \theta^{(5,2)}_{pot}&=\dfrac{P_{5,2}(\gamma)\, e^{-4 \gamma_E \epsilon}}{4 \epsilon}+\theta^{(5,2)}_{pot,fin}\,,\\
        \theta^{(5,2)}_{pot,fin}&=\dfrac{n_1}{\sqrt{3-\gamma}^{15}}+\mathcal{O}\!\left((3-\gamma)^{-13/2}\right),
\end{align}
with $n_1$ some number.
Two poles thus have to cancel for the result to be physically sensible. The $\epsilon$ pole should again be cancelled by the tail, and the $\gamma-3$ pole has to be cancelled as well. Not only the pole but every half-integer power of $(\gamma-3)$ must drop out, otherwise the conservative angle would diverge or become complex for $\gamma>3$. We should remind ourselves that the conservative angle is not a physical quantity but an auxiliary construction, and as such it is to some degree a matter of convention what exactly is called ``conservative''. It is clear, however, that energy must be conserved by any sensible such prescription.

We can next examine the tail region, where we find
\begin{align}
        \theta^{(5,2)}_{tail}&=-\dfrac{P_{5,2}(\gamma)\, e^{-4 \gamma_E \epsilon}\,(\gamma-1)^{-4 \epsilon}}{4 \epsilon}+\theta^{(5,2)}_{tail,fin}\,,\\
        \theta^{(5,2)}_{tail,fin}&=\mathcal{O}\!\left((3-\gamma)^{0}\right).
\end{align}
The $\epsilon$ pole of the potential region is again cancelled by the tail region, but the tail region carries no $\gamma-3$ pole. Once again Feynman propagators and the average of advanced and retarded propagators for the gravitons both work as prescriptions and yield the same result, since
\begin{align*}
        \begin{tikzpicture}[baseline={([yshift=-1ex]current bounding box.south)},scale=1.2]
                \coordinate (inA) at (0.3,.7);
                         \coordinate (outA) at (3.6,.7);
                         \coordinate (inB) at (0.3,-.7);
                         \coordinate (outB) at (3.6,-.7);
                         \coordinate (xA) at (0.6,.7);
                         \coordinate (yA) at (1.5,.7);
                         \coordinate (zA) at (2.4,.7);
                         \coordinate (wA) at (3.3,.7);
                         \coordinate (xB) at (0.6,-.7);
                         \coordinate (zB) at (2.4,-.7);
                         \coordinate (wB) at (3.3,-.7);
                \draw [dotted] (inA) -- (outA);
                \draw [dotted] (inB) -- (outB) node [midway, below]{} ;
                \draw [draw=none] (xA) to[out=40,in=140] (zA);
                \draw [zUndirected] (wA) -- (zA);
                \draw [photon] (zA) -- (zB);
                \draw [photon] (wA) -- (wB)node [midway, right]{$\qquad =$};
                \draw [photon] (xA) -- (xB) node [midway, left]{\hspace*{-2cm}$2 \times$};
                \draw [photon,red][out=40,in=140] (xA) to[out=40,in=140]node [midway, above]{$F$}  (yA);
                \draw [photon,red][out=40,in=140] (yA) to[out=40,in=140]node [midway, above]{$F$}  (zA);
                \draw [fill] (xA) circle (.08);
                \draw [fill] (yA) circle (.08);
                \draw [fill] (wB) circle (.08);
                \draw [fill] (zA) circle (.08);
                \draw [fill] (wA) circle (.08);
                \draw [fill] (xB) circle (.08);
                \draw [fill] (zB) circle (.08);
              \end{tikzpicture}
              \begin{tikzpicture}[baseline={([yshift=-1ex]current bounding box.south)},scale=1.2]
                \coordinate (inA) at (0.3,.7);
                         \coordinate (outA) at (3.6,.7);
                         \coordinate (inB) at (0.3,-.7);
                         \coordinate (outB) at (3.6,-.7);
                         \coordinate (xA) at (0.6,.7);
                         \coordinate (yA) at (1.5,.7);
                         \coordinate (zA) at (2.4,.7);
                         \coordinate (wA) at (3.3,.7);
                         \coordinate (xB) at (0.6,-.7);
                         \coordinate (zB) at (2.4,-.7);
                         \coordinate (wB) at (3.3,-.7);
                \draw [dotted] (inA) -- (outA);
                \draw [dotted] (inB) -- (outB) node [midway, below]{} ;
                \draw [draw=none] (xA) to[out=40,in=140] (zA);
                \draw [zUndirected] (wA) -- (zA);
                \draw [photon] (zA) -- (zB);
                \draw [photon] (wA) -- (wB)node [midway, right]{$\qquad +$};
                \draw [photon] (xA) -- (xB);
                \draw [photon,red][out=40,in=140] (xA) to[out=40,in=140]node [midway, above]{$\leftarrow$} (yA);
                \draw [photon,red][out=40,in=140] (yA) to[out=40,in=140]node [midway, above]{$\leftarrow$} (zA);
          \draw [fill] (wA) circle (.08);
          \draw [fill] (xA) circle (.08);
          \draw [fill] (yA) circle (.08);
          \draw [fill] (wB) circle (.08);
                \draw [fill] (zA) circle (.08);
                \draw [fill] (xB) circle (.08);
                \draw [fill] (zB) circle (.08);
              \end{tikzpicture}
              \begin{tikzpicture}[baseline={([yshift=-1ex]current bounding box.south)},scale=1.2]
                \coordinate (inA) at (0.3,.7);
                         \coordinate (outA) at (3.6,.7);
                         \coordinate (inB) at (0.3,-.7);
                         \coordinate (outB) at (3.6,-.7);
                         \coordinate (xA) at (0.6,.7);
                         \coordinate (yA) at (1.5,.7);
                         \coordinate (zA) at (2.4,.7);
                         \coordinate (wA) at (3.3,.7);
                         \coordinate (xB) at (0.6,-.7);
                         \coordinate (zB) at (2.4,-.7);
                         \coordinate (wB) at (3.3,-.7);
                \draw [dotted] (inA) -- (outA);
                \draw [dotted] (inB) -- (outB) node [midway, below]{} ;
                \draw [draw=none] (xA) to[out=40,in=140] (zA);
                \draw [zUndirected] (wA) -- (zA);
                \draw [photon] (zA) -- (zB);
                \draw [photon] (wA) -- (wB);
                \draw [photon] (xA) -- (xB);
                \draw [photon,red][out=40,in=140] (xA) to[out=40,in=140]node [midway, above]{$\rightarrow$} (yA);
                \draw [photon,red][out=40,in=140] (yA) to[out=40,in=140]node [midway, above]{$\rightarrow$} (zA);
          \draw [fill] (wA) circle (.08);
          \draw [fill] (xA) circle (.08);
          \draw [fill] (yA) circle (.08);
          \draw [fill] (wB) circle (.08);
                \draw [fill] (zA) circle (.08);
                \draw [fill] (xB) circle (.08);
                \draw [fill] (zB) circle (.08);
              \end{tikzpicture}
\end{align*}
where we again implicitly take the overall real part of the integral when Feynman propagators are used. We have checked this up to NLO in $\epsilon$, which is again all we need for the impulse.

The remaining task is then to cancel the $\gamma-3$ pole, which can only come from the memory region, since the K3 sector that introduces this pole has only the potential and the memory region as even-in-velocity regions.

In the memory region we have to evaluate the two boundary integrals
\begin{align*}
        \begin{tikzpicture}[baseline={([yshift=-1ex]current bounding box.south)},scale=1.2]
                \coordinate (inA) at (0.3,.7);
                         \coordinate (outA) at (3.6,.7);
                         \coordinate (inB) at (0.3,-.7);
                         \coordinate (outB) at (3.6,-.7);
                         \coordinate (xA) at (0.6,.7);
                         \coordinate (yA) at (1.5,.7);
                         \coordinate (zA) at (2.4,.7);
                         \coordinate (xB) at (0.6,-.7);
                         \coordinate (yB) at (1.5,-.7);
                         \coordinate (zB) at (2.4,-.7);
                \draw [dotted] (inA) -- (outA);
                \draw [dotted] (inB) -- (outB) node [midway, below]{} ;
                \draw [draw=none] (xA) to[out=40,in=140] (zA);
                \draw [photon] (zA) -- (zB);
                \draw [photon] (yA) -- (yB);
                \draw [photon] (xA) -- (xB);
                \draw [photon,red][out=40,in=140] (xA) to[out=40,in=140] (yA);
                \draw [photon,red][out=40,in=140] (yA) to[out=40,in=140] (zA);
          \draw [fill] (xA) circle (.08);
          \draw [fill] (yA) circle (.08);
          \draw [fill] (yB) circle (.08);
                \draw [fill] (zA) circle (.08);
                \draw [fill] (xB) circle (.08);
                \draw [fill] (zB) circle (.08);
              \end{tikzpicture}
              \begin{tikzpicture}[baseline={([yshift=-1ex]current bounding box.south)},scale=1.2]
                \coordinate (inA) at (0.3,.7);
                         \coordinate (outA) at (3.6,.7);
                         \coordinate (inB) at (0.3,-.7);
                         \coordinate (outB) at (3.6,-.7);
                         \coordinate (xA) at (0.6,.7);
                         \coordinate (yA) at (1.5,.7);
                         \coordinate (zA) at (2.4,.7);
                         \coordinate (xB) at (0.6,-.7);
                         \coordinate (yB) at (1.5,-.7);
                         \coordinate (zB) at (2.4,-.7);
                \draw [dotted] (inA) -- (outA);
                \draw [dotted] (inB) -- (outB) node [midway, below]{} ;
                \draw [draw=none] (xA) to[out=40,in=140] (zA);
                \draw [photon] (zA) -- (zB);
                \draw [photon] (yA) -- (yB);
                \draw [photon] (xA) -- (xB);
                \draw [photon,red][out=40,in=140] (xA) to[out=40,in=140] (yA);
                \draw [photon,red][out=40,in=140] (yA) to[out=40,in=140] (zA);
          \draw [fill] (xA) circle (.08);
          \draw [fill] (yA) circle (.08);
          \draw [fill,blue!70] (1.05,.87) circle (.08);
          \draw [fill,blue!70] (1.95,.87) circle (.08);
          \draw [fill] (yB) circle (.08);
                \draw [fill] (zA) circle (.08);
                \draw [fill] (xB) circle (.08);
                \draw [fill] (zB) circle (.08);
              \end{tikzpicture}.
\end{align*}

Two problems arise when we attempt to evaluate these integrals. With the Feynman prescription we encounter in this region a $\tfrac{1}{\epsilon^2}$ pole that cannot be cancelled by any other contribution. On the other hand, it is no longer clear how to build an averaged advanced/retarded prescription for these integrals. Looking at the two boundaries above, there are in principle two distinct possibilities. We should also recall that we have boundary integrals with three radiative gravitons, which we IBP-reduced to integrals with only two radiative gravitons. For these boundaries there is just one way of writing down a time-symmetric combination involving only two terms,
\begin{align*}
        \begin{tikzpicture}[scale=1.2]
                \coordinate (inA)  at (0.6,0.6);
                \coordinate (outA) at (2.625,0.6);
                \coordinate (inB)  at (0.6,-0.6);
                \coordinate (outB) at (2.625,-0.6);
                \draw[dotted] (inA) -- (outA);
                \draw[dotted] (inB) -- (outB);
                \coordinate (aA) at (0.6,0.6);
                \coordinate (bA) at (1.6125,0.6);
                \coordinate (bAA) at (1.6125,1.6);
                \coordinate (dA) at (2.625,0.6);
                \coordinate (aB) at (0.6,-0.6);
                \coordinate (bB) at (1.6125,-0.6);
                \coordinate (dB) at (2.625,-0.6);
                \draw[photon]     (aA) -- (aB) node [midway,left] {} node [midway,left] {\mbox{}};
                \draw[photon ,red] (aA) to[out=80, in=-180]node [midway, above,sloped]{$\longleftarrow$} (bAA);
                \draw[photon ,red] (bAA) to[out=0, in=100]node [midway, above,sloped]{$\longrightarrow$} (dA);
                \draw[photon ,red] (bA) to node [midway, left]{$\downarrow$} (bAA);
                \draw[photon] (dA) -- (dB);
                \draw[photon] (bA) -- (bB);
                \foreach \v in {aA,bA,dA,aB,bB,dB,bAA}
                  \draw[fill] (\v) circle (.08);
              \end{tikzpicture}\, ,
 \begin{tikzpicture}[scale=1.2]
  \coordinate (inA)  at (0.6,0.6);
  \coordinate (outA) at (2.625,0.6);
  \coordinate (inB)  at (0.6,-0.6);
  \coordinate (outB) at (2.625,-0.6);
  \draw[dotted] (inA) -- (outA);
  \draw[dotted] (inB) -- (outB);
  \coordinate (aA) at (0.6,0.6);
  \coordinate (bA) at (1.6125,0.6);
  \coordinate (bAA) at (1.6125,1.6);
  \coordinate (dA) at (2.625,0.6);
  \coordinate (aB) at (0.6,-0.6);
  \coordinate (bB) at (1.6125,-0.6);
  \coordinate (dB) at (2.625,-0.6);
  \draw[photon]     (aA) -- (aB) node [midway,left] {} node [midway,left] {\mbox{}};
  \draw[photon ,red] (aA) to[out=80, in=-180]node [midway, above,sloped]{$\longrightarrow$} (bAA);
  \draw[photon ,red] (bAA) to[out=0, in=100]node [midway, above,sloped]{$\longleftarrow$} (dA);
  \draw[photon ,red] (bA) to node [midway, left]{$\uparrow$} (bAA);
  \draw[photon] (dA) -- (dB);
  \draw[photon] (bA) -- (bB);
  \foreach \v in {aA,bA,dA,aB,bB,dB,bAA}
    \draw[fill] (\v) circle (.08);
\end{tikzpicture}.
\end{align*}

\begin{figure}[h!]
        \includegraphics[width=0.9\textwidth]{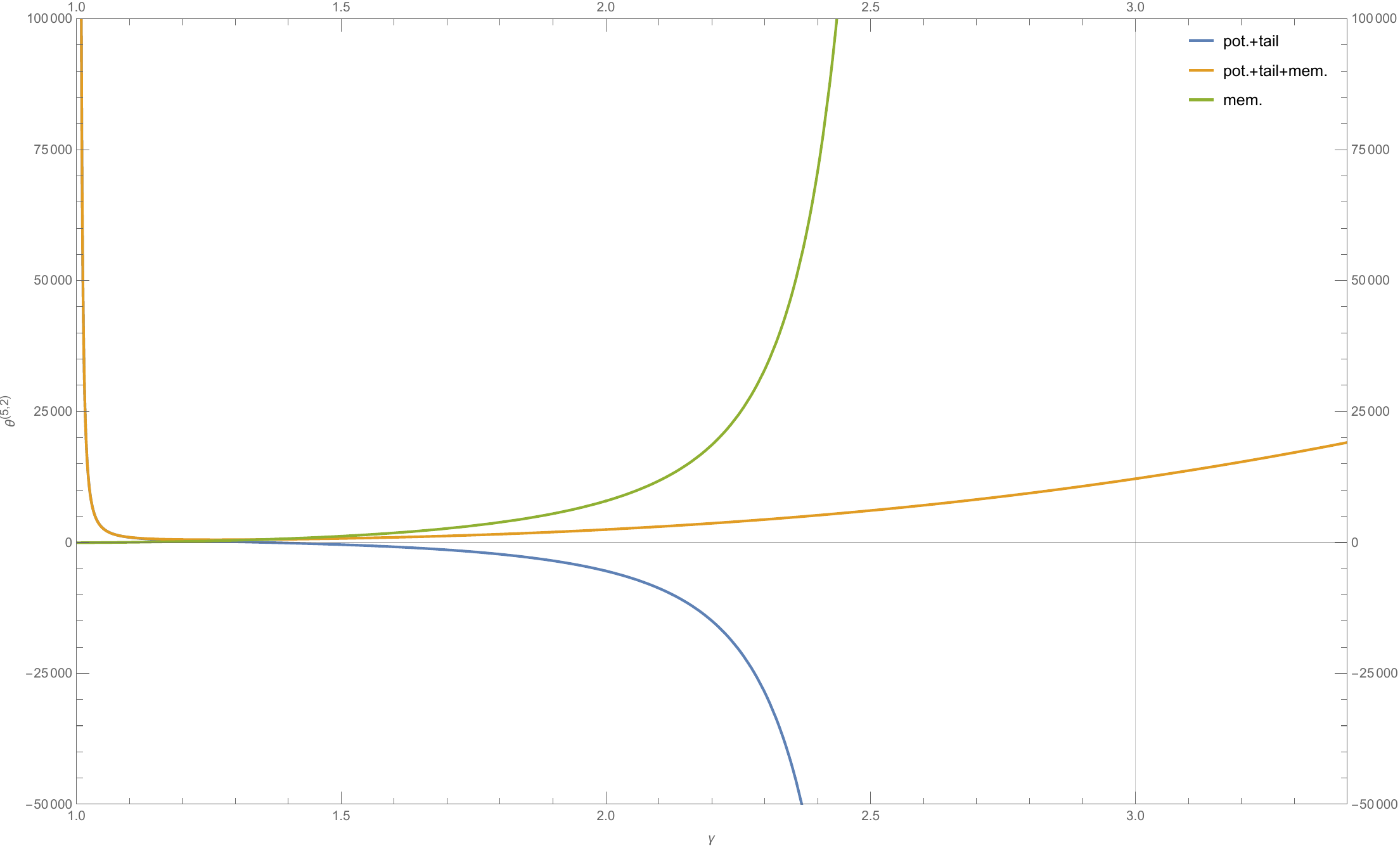}
        \centering
        \caption{The potential/tail contributions and memory contributions to the 2SF component of the 5PM scattering angle have a divergence at $\gamma=3$. The sum of both is finite over the whole physical region without any singular points from $\gamma=1$ to $\gamma=\infty$.}
        \label{fig:g3plot}
        \end{figure}

\noindent We refer to the average of those two prescriptions as the $\gamma$-3 prescription, which we apply only to the memory region. Crucially, this prescription enjoys the same symmetries as a Feynman prescription, including the $S_3$ symmetry permuting the three subdiagrams that join at the three-graviton interaction. Using this prescription we find, as depicted in Fig.~\ref{fig:g3plot}, a well-defined, real result over the entire physical region,
\begin{align}
        \theta^{(5,2)}_{mem}&=\theta^{(5,2)}_{mem,fin}\,,\\
        \theta^{(5,2)}_{mem,fin}&=-\dfrac{n_1}{\sqrt{3-\gamma}^{15}}+\mathcal{O}\!\left((3-\gamma)^{-13/2}\right),\\
        \theta^{(5,2)}&=\log(\gamma-1)\, P_{5,2}(\gamma)+\theta^{(5,2)}_{pot,fin}+\theta^{(5,2)}_{tail,fin}+\theta^{(5,2)}_{mem,fin}\,,\\
        \theta^{(5,2)}&=\mathcal{O}\!\left((\gamma-3)^{0}\right).
\end{align}
We note that a region-by-region prescription for the conservative part is in fact necessary already at lower PM orders. It is also worth noting that the self-force approach itself currently only provides a clear-cut notion of conservative dynamics at 1SF order, so it does not yet supply guidance for the 2SF case~\cite{Pound:2014koa}.

The conservative prescriptions used so far were designed to reproduce the known PN results, and we now perform this comparison region by region. The contributions from the potential and tail regions match the known PN data exactly up to 4PN order~\cite{Bini:2020hmy,Blumlein:2020pog,Bini:2021gat,Bini:2022enm,Bini:2025vuk}. The comparison to 5PN results~\cite{Porto:2026fsd} shows a different picture. The result does not agree with the PN calculation performed using the standard Feynman computation, nor with the purely conservative local prescription introduced in ref.~\cite{Porto:2026fsd}. However, the difference compared to the local prescription amounts only to a sign in front of the terms coming from the memory region. It is also shown that the PN calculation does agree with the PM calculation when the $\gamma$-3 prescription is used. From a PN perspective the $\gamma$-3 pole is invisible, since one expands around $\gamma=1$, so the PN prescription differing by a sign would not cancel the $\gamma$-3 pole. From the PM perspective it is therefore more natural to use the $\gamma$-3 prescription. On the other hand, one might also argue that, for the analytic continuation to the bound case, one has $\gamma<1$, so that this pole is not problematic. We emphasise that this entire discussion is an artefact of the definition of a conservative part of a physical observable; the full dissipative result has no spurious poles and is unambiguous. We leave the definition of what the best conservative prescription might be for future work.

\subsection{The conservative scattering angle}
\begin{figure}[h!]
        \includegraphics[width=0.49\textwidth]{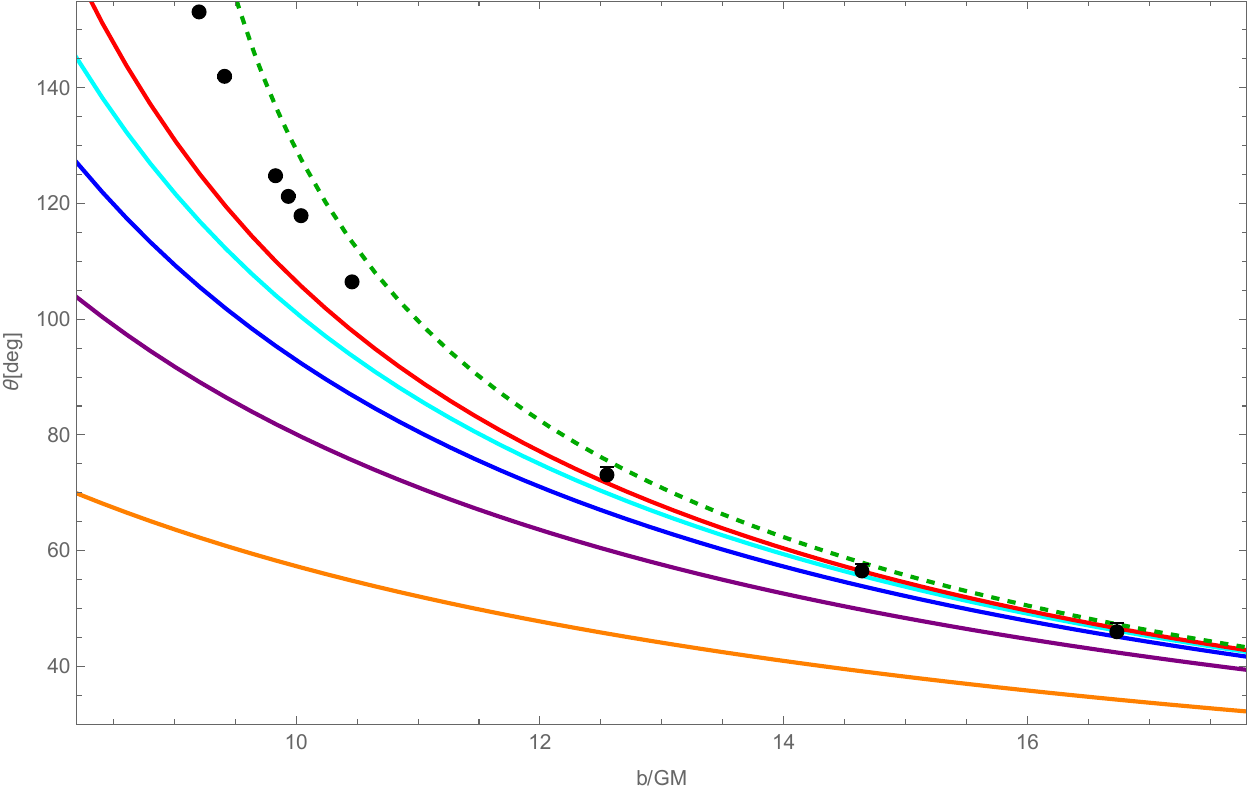}
        \includegraphics[width=0.49\textwidth]{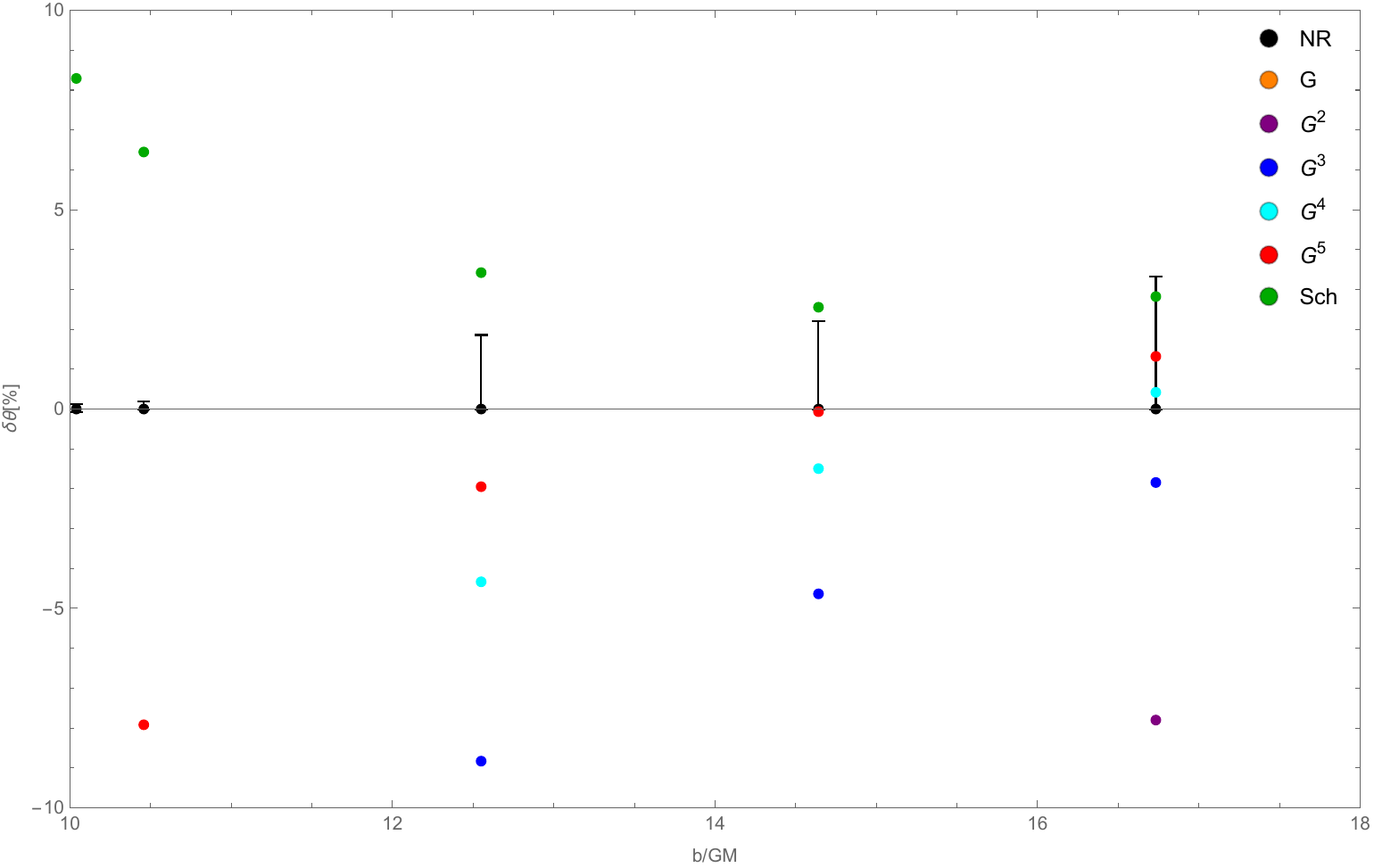}
        \centering
        \caption{The scattering angle $\theta$ up to 5PM as a function of the impact parameter, which is given in units of the Schwarzschild radius $GM$. The initial relative velocity is  $v=0.5125$ and the symmetric mass ratio $\nu=0.25$.
        The black dots show numerical relativity simulations~\cite{Rettegno:2023ghr}. The 5PM contribution includes the full dissipative 5PM-1SF contributions and the 5PM-2SF conservative result.
        On the right hand side we show the error $\delta \theta =(\theta_{\text{PM}}-\theta_{\text{NR}})/\theta_{\text{NR}}$ of the PM results $\theta_{\text{PM}}$ compared to the NR result $\theta_{\text{NR}}$.
        Our result agrees with the NR data within their error bars for $b/GM>12.5$. The successive $G^{n}$ corrections decrease monotonically, which provides an intrinsic error estimate for our $G^{5}$ result; by this measure it is more precise than the NR data for $b/GM>14$.}
        \label{fig:anglePlot}
        \end{figure}
We compute only the conservative part of the scattering angle. The function space of the 2SF integrals again contains a K3 and a CY3 sector, together with a larger set of logarithmic kernels. New kernels involving the square roots $\sqrt{4-x^2}$ and $\sqrt{-1+4x^2}$ also appear and would introduce spurious poles, but they drop out after IBP reduction. All contributions from the CY3 sector likewise drop out (see also ref.~\cite{Bern:2026oqp}), even in $d$ dimensions. The kernels surviving in the physical observable are
\begin{align}
        \varphi_i(x) \in	\bigg\{ &\dfrac{1}{x},\, \dfrac{1}{1+x^2},\, \dfrac{1+x}{x(x-1)},\, \dfrac{-1+x}{x(x+1)},\, \dfrac{1-x^2}{x+x^2+x^3},\, \dfrac{1-x^2}{x-x^2+x^3}, \nonumber \\
	&\quad \dfrac{-1+x^2}{x+x^3},\, \dfrac{(1+x^2)\,\varpi_\text{K3}(x)}{x},\, \dfrac{1}{x \,r(x)\,\varpi_\text{K3}(x)}\bigg\}
\end{align}
with the square root $r(x) = \sqrt{-1+34x^2-x^4}$. Note that the individual iterated integrals still possess spurious poles around $\gamma=3$; their contributions only cancel exactly against each other in the expansion around $\gamma=3$. We find at most three iterated integrals built from these kernels.

This is the last contribution of the scattering angle that we compute. We compare all these contributions in Fig.~\ref{fig:anglePlot} to numerical relativity computations and also to the resummed 0SF Schwarzschild solution.

\chapter{Conclusions and outlook}
In this thesis we have computed the impulse on a scattering binary up to fifth Post-Minkowskian order and second self-force order, and have given a self-contained review of the impulse PM computation as a whole. The only piece still missing for the complete 5PM result is the dissipative sector at second self-force order.

We began in Ch.~\ref{sec:wqft} and Ch.~\ref{sec:inin} with the worldline description of black hole scattering and the worldline quantum field theory formalism. Chapter~\ref{sec:intred} reviewed modern integrand-reduction techniques and analysed the structural features of the Post-Minkowskian integrand specific to this kinematic regime, while Ch.~\ref{sec:struct} laid out the integrand at the various self-force orders and the way it organises itself perturbatively. We then revisited the 3PM calculation in Ch.~\ref{sec:methdiff} as a worked example of the differential-equations method combined with the method of regions, before showing in Ch.~\ref{sec:geom} how to construct a canonical basis for a much broader class of Feynman integrals and how these integrals can be associated with their underlying geometries. These techniques were applied at fourth and fifth Post-Minkowskian order in Ch.~\ref{sec:4pm}--Ch.~\ref{sec:5pm2sf}, where we discussed in detail how Calabi--Yau two- and threefolds appear in the integrals and how to handle them when a direct integrand-level analysis becomes infeasible at four loop order. We also showed how the tail, tail-of-tail and non-linear memory effects appear as distinct integration regions and how they contribute to the impulse.

Our central physical results, derived from the impulse in Ch.~\ref{sec:obs}, are the first analytic expressions for the scattering angle and the radiated energy and recoil at fifth Post-Minkowskian and first self-force order, together with the conservative scattering angle at second self-force order. The most striking feature of these results is that periods of a Calabi--Yau threefold enter the radiative sector and contribute to the radiated momentum --- to our knowledge the first appearance of Calabi--Yau threefold periods in a classical physical observable. Additionally we saw in the conservative scattering angle at 4PM and 5PM order two different periods over a K3 manifold contributing. Comparing the scattering angle to numerical-relativity simulations (Fig.~\ref{fig:anglePlot}), we find percent-level agreement throughout the perturbative regime.

A number of independent consistency checks support these results. All intermediate $1/\epsilon$ divergences cancel in the limit $d\to 4$ and the impulse satisfies the mass-shell condition $p_i\cdot\Delta p_i = -\tfrac12\Delta p_i^2$ order by order. At lower orders our results reproduce the established 1PM--4PM literature, at 5PM the results agree with known PN results~\cite{Bini:2020hmy,Blumlein:2020pog,Bini:2021gat,Bini:2022enm,Bini:2025vuk,Porto:2026fsd} and the potential region agrees with the independent amplitude-based computation of ref.~\cite{Bern:2025wyd}. Taken together, these checks make us confident in the new fifth-order data.

Beyond the explicit results, the analytic approach taken here permitted a structural analysis of the observables themselves. We found that the transcendental weight of the iterated integrals in the final result is always $n-2$ at $n$PM order, even when Calabi--Yau periods contribute at 5PM-1SF, and that several of the periods present in intermediate expressions drop out of the physical observables entirely. 

We also emphasise the assumptions underlying our computation. We have treated the bodies as spinless Schwarzschild black holes, neglecting both spin and finite-size (tidal) effects, and we have worked with a specific prescription for separating conservative from dissipative dynamics.

The techniques developed here should suffice to obtain the still-missing dissipative part of the impulse at 5PM 2SF order without further conceptual complications.
Furthermore the inclusion of spin and tidal effects at four loop order should be possible without conceptual issues. The tensor structure for spin effects is more involved, which makes the IBP reduction more difficult, though it should still be feasible, at least at linear order in spin.

The situation is quite different at 6PM. The IBP reduction at that order is beyond the reach of current technology, and new ideas will be required before a complete sixth-order calculation becomes feasible. 

It is also clear, that our results here still need to be transformed into waveform models. The most straightforward approach for this, would be the usage of our results for new EOB model development. 

A long-standing problem is the mapping of scattering results to the bound problem. The non-local-in-time pieces of the scattering observables cannot be transferred to the bound case with the maps currently in use~\cite{Kalin:2019rwq,Kalin:2019inp,Cho:2021arx,Gonzo:2023goe,Gonzo:2024zxo}. Closing this gap will require either a more general analytic continuation or direct calculations in the bound-orbit setting.

Another issue, which becomes particularly visible at 5PM, is the ambiguity of the conservative--dissipative split. Such a split is necessary in order to feed the conservative part into EOB resummation, but the way in which it is performed is not dictated by first principles, and different prescriptions give different results. Further work will be needed in order to identify the split that combines best with the EOB framework. Another possibility would be to find a different resummation technique, which does not rely on this split.

On the mathematical side, our results invite a more systematic study of the geometries underlying Post-Minkowskian integrals: a classification of which Calabi--Yau varieties appear at each order and of the period cancellations in the observables.

Looking beyond gravitational scattering, it will be interesting to apply the integration methods~\cite{Gorges:2023zgv,Duhr:2025lbz} to the multi-scale Feynman integrals appearing in particle-physics processes. The present work has shown that techniques going beyond the $\d\!\log$ and elliptic cases are not only viable in principle but robust enough to serve as building blocks in a high-loop computation with a large number of master integrals. The advances achieved for the IBP reduction in the course of this four-loop calculation~\cite{Driesse:2024feo,Driesse:2024xad,Driesse:2026qiz,Bern:2024adl,Bern:2025wyd,Bern:2025zno} should likewise find immediate application in reduction problems from collider physics. 

Finally, looking beyond the theory, we expect the results presented in this thesis to play an important role in connecting the theory of general relativity to experimental data when the next generation of gravitational-wave detectors comes online. 

\appendix
\chapter{Conventions}
\label{App:Conventions}

We work with a mostly minus metric, $\eta_{\mu\nu}= \text{diag}(1,-1,\ldots,-1)$ and dimensionally regularise
in $d=4-2\epsilon$ dimensions.
The Riemann tensor is defined as
\be
R^{\lambda}{}_{\mu\nu\kappa} =\partial_{\nu}\Gamma^{\lambda}{}_{\kappa\mu}- \partial_{\kappa}\Gamma^{\lambda}{}_{\nu\mu} 
+\Gamma^{\lambda}{}_{\eta \kappa}\Gamma^{\eta}{}_{\nu\mu} -\Gamma^{\lambda}{}_{\eta \nu}\Gamma^{\eta}{}_{\kappa\mu}\, .
\ee
The  Ricci tensor, Ricci scalar  and affine connection are defined by
\begin{align} \begin{aligned}
R_{\mu\nu}& = R^\alpha{}_{\mu \nu \alpha}\,, 
\qquad R= R_{\mu\nu}g^{\mu\nu}\, ,
\\
\Gamma^{\rho}{}_{\mu\nu}&= \sfrac{1}{2} g^{\rho\kappa} \left ( \partial_{\mu}g_{\nu\kappa} + \partial_{\nu}g_{\mu\kappa}
-\partial_{\kappa}g_{\mu\nu}
\right ) \, ,
\end{aligned} \end{align}
We write for the loop integrations
\begin{align}
        \int_{l_1,\dots,l_n} =\int \dfrac{\d^d l_1}{(2 \pi )^d}\dfrac{\d^d l_2}{(2 \pi )^d}\cdots \dfrac{\d^d l_n}{(2 \pi )^d}.
\end{align}
We use the variables $x,\gamma, v, \delta$, which are related by
\begin{align}
        \gamma&=v_1 \cdot v_2, \\
        \gamma&=(1-v^2)^{-1/2}, \\
        \gamma&=1/2(x+x^{-1}),\\
        \delta&=1-x.
\end{align}
The dual velocities are defined as
\begin{align}
        \check v_1^\mu =\dfrac{\gamma v_2^\mu-v_1^\mu}{\gamma^2-1},\;\text{and}\;\check v_2^\mu =\dfrac{\gamma v_1^\mu-v_2^\mu}{\gamma^2-1}.
 \end{align}
 We use for the integrals, when dropping the $\i0$ prescription the notation
 \begin{align}
        {{\cal P}\;^{\text{mPM-jSF}}}_{\{n\}}^{\{\sigma\}}=\text{P}^{m,j}(n).
\end{align}
\chapter{The semi-simple and unipotent parts of the Wronskian}
\label{semisimple}
In this appendix we follow the discussion in~\cite{Duhr:2025lbz} about the splitting of the Wronskian associated to a Calabi--Yau operator of order $n+1$,
\begin{equation}
        \mathcal L_0^{(n+1)} = \partial_z^{n+1} + p_n(z) \partial_z^n + \hdots + p_0(z) \,,
\end{equation}
restricting our attention to the cases $n\in \{1,2,3\}$ relevant for the present thesis. 
We can use the first two solutions of the Frobenius basis at a MUM-point to define
\begin{equation}
        t(z) = \frac{\varpi_1(z)}{\varpi_0(z)} = \log(z) + \ord(z)\,, \quad q(z) = e^{t(z)} = z + \ord(z^2) \, ,
    \label{eq:canvar}
\end{equation}
The variable $t(z)$ extends the elliptic $\tau$ parameter of eq.~\refeq{eq:tau_def} to higher-dimensional Calabi--Yau families and provides a canonical local coordinate on moduli space. Its inverse is the \emph{mirror map},
\begin{equation}
        z(q) = q + \ord(q^2) \, .
    \label{eq:mirrormap}
\end{equation}
Two further ingredients of the construction are the \emph{structure series} $\alpha_i(z)$ and the associated \emph{$Y$-invariants}, see e.g.\ refs.~\cite{BognerCY, BognerThesis}. The structure series, for $i=0,1,\hdots, n$, is built recursively from the periods via $\alpha_i = u_{i,i}(z)^{-1}$, with
\begin{equation}
        u_{i,j}(z) = \theta_z \left( \frac{u_{i-1,j}(z)}{u_{i-1,i-1}(z)} \right)\,,\quad u_{0,j}(z) = \varpi_j(z) \, ,
    \label{eq:defu}
\end{equation}
where $\theta_z=z\partial_z$ is the logarithmic derivative. For $i=1,2$ we can write $\alpha_0(z) = 1/\varpi_0(z)$ and $\alpha_1(z)=(\theta_z t(z))^{-1}$. The $Y$-invariants are given by
\begin{equation}
        Y_i(z) = \frac{\alpha_1(z)}{\alpha_{i+1}(z)}\, , \quad i=1,2, \hdots, n-2 \, ,
\end{equation}
and we set $Y_0=1$ by convention. Because of the symmetry properties of the structure series, $\alpha_i =  \alpha_{n-i+1}$ we find
\begin{align}
        \label{eq:Y_self-duality}
    Y_i(z) = Y_{n-i+1}(z)\,.
\end{align}
With these ingredients in place, the splitting of the Wronskian reads
\begin{align}\label{eq:W_Wss_Wu}
        W_n(z) = W_n^{\textrm{ss}}(z)\,W_n^{\textrm{u}}(z)\,.
\end{align}
where the unipotent factor for a Calabi--Yau variety satisfies the differential equation~\cite{Broedel:2019kmn,Ducker:2025wfl,Duhr:2025lbz}
\begin{align}
        \theta_q W^\text{u}_n(q) = N_n(q) \, W^\text{u}_n(q)\,,
\end{align}
where $N_n(q)$ is the nilpotent matrix
\begin{equation}\label{eq:N_n(q)}
    N_n(q) = \begin{pmatrix}
        0 & 1 & 0 &\ldots & & &  \\
        & 0 & Y_1(q) & 0 &\ldots & &  \\
        & \ldots& 0 & Y_2(q) & 0 &\ldots &  \\
        & & \ldots& 0 & \ddots & 0 &\ldots  \\
        & & & \ldots& 0 & Y_1(q) & 0  \\
        & & & & \ldots& 0 & 1  \\
        & & & & & \ldots& 0  \\
    \end{pmatrix} \, .
\end{equation}
Because $N_n$ is strictly upper triangular, the equation is solved by direct iteration, and the solution takes the form of iterated integrals~\cite{Duhr:2022dxb}
\begin{align}
        \frac{\varpi_i(q)}{\varpi_{0}(q)} = I(Y_0,Y_1,\ldots,Y_{i-1};q)\,,\qquad i=1,\ldots,n\,,
\end{align}
where we have introduced the iterated integrals
\begin{align}
        I(f_1,\ldots,f_p;q) = \int_0^{q} \frac{d q'}{q'}f_1(q')\,I(f_2,\ldots,f_p;q')\,,
\end{align}
with the recursion starting from $I(;q)=1$. The integrals can diverge at $q=0$, and the lower integration boundary should be interpreted as a tangential base point, cf.\ ref.~\cite{Brown:mmv}. 
We can use them to rewrite the unipotent part of the period matrix as
\begin{equation}
    W^\text{u}_n(q) = \begin{pmatrix}
1 &I(1;q) & I(1,Y_1;q) &I(1,Y_1,Y_2;q)& \hdots &I(1,Y_1\ldots,Y_1;q) &  I(1,Y_1\ldots,1;q)\\
0& 1 & I(Y_1;q) &I(Y_1,Y_2;q) & \hdots & I(Y_1\ldots,Y_1;q) & I(Y_1\ldots,1;q) \\
0& 0&1 &I(Y_2;q) & \hdots &I(Y_2\ldots,Y_1;q) & I(Y_2\ldots,1;q) \\
& \vdots& &\ddots& &  \vdots &\vdots   \\
0& 0& 0&0& \ldots&  1 & I(1;q)  \\
0& 0& 0& 0& \ldots& 0& 1  \\
    \end{pmatrix} \, .
\end{equation}
In this form one can show that the unipotent part fulfills the quadratic identity~\cite{Duhr:2022dxb}
\begin{align}\label{eq:quad_rel_Wu}
        W^\text{u}_n(q)\,\Sigma_n\,W^\text{u}_n(q)^T = \Sigma_n\,.
\end{align}
We can use this relation to also get an analogue quadratic relation for the semi-simple part,
\begin{align}\label{eq:qual_rel_ss}
        W_n^{\textrm{ss}}(z)\,\Sigma_n\,W_n^{\textrm{ss}}(z)^T = Z_n(z)\,.
\end{align}
The bilinear relation~\refeq{eq:qual_Rel_wronskian} therefore decouples into two independent quadratic constraints on $W_n^{\textrm{ss}}$ and $W_n^{\textrm{u}}$, and the rational matrix $Z_n(z)$ enters only the first of them.
One can compute now $Z_n(z)$ by using \refeq{eq:defnptcoupling} and by demanding that the derivatives of \refeq{eq:qual_rel_ss} obey \refeq{eq:qual_rel_ss} itself and the Picard-Fuchs equation. We find for $n=1,2,3$:
\begin{equation}
        Z_1 = C_1(z)\begin{pmatrix}
            0 & 1 \\
            -1 & 0
        \end{pmatrix} \, .
    \end{equation}
    \begin{equation}
        Z_2 = C_2(z)\begin{pmatrix}
            0 & 0 & 1 \\
            0 & -1 & \frac{1}{3}p_2(z) \\
            1 & \frac{1}{3}p_2(z) & \frac{1}{9}p_2(z)^2+\frac{1}{3}\partial_z p_2(z)-p_1(z)
        \end{pmatrix}\, ,
    \end{equation}
    \begin{equation}
        Z_3 = C_3(z)\begin{pmatrix}
            0 & 0 & 0 & 1 \\
            0 & 0 & -1 & \frac{1}{2}p_3(z) \\
            0 & 1 & 0 & \frac{1}{2} p_3(z)+\partial_z p_3(z)-2p_2(z)  \\
            -1 & \frac{1}{2}p_3(z) & \frac{1}{2} p_3(z)+\partial_z p_3(z)-2p_2(z) & 0
        \end{pmatrix} \, .
    \end{equation}
We can furthermore use these relations to write down expressions for the semi-simple factor $W_n^{\textrm{ss}}(q)$ in the three cases of interest,
\begin{align}
     W^\text{SS}_1(z) &= \begin{pmatrix}
        \varpi_0 & 0 \\
        \varpi_0' & \frac{C_1(z)}{\varpi_0}
    \end{pmatrix} \, , \\
    W^\text{SS}_2(z) &= \begin{pmatrix}
        \varpi_0 & 0 & 0 \\
        \varpi_0' & {\scriptstyle\sqrt{C_2(z)}} & 0 \\
        \frac{\varpi_0'^2}{2\varpi_0}+\frac{2p_2(z)^2+3 p_2'(z)-9p_1(z)}{18}\varpi_0 - \frac{p_2(z)}{3}\varpi_0' \ \ & \left(\frac{\varpi_0'}{\varpi_0} - \frac{p_2(z)}{3} \right){\scriptstyle \sqrt{C_2(z)}} \ \ & \frac{C_2(z)}{\varpi_0}
    \end{pmatrix} \, , \\
     W^\text{SS}_3(z) &= \begin{pmatrix}
        \varpi_0 & 0 & 0 & 0 \\
        \varpi_0' & \frac{\varpi_0}{z\alpha_1} & 0 & 0 \\
        \varpi_0'' & \frac{2\varpi_0'}{z\alpha_1}-\left( \frac{1}{z^2\alpha_1} + \frac{\alpha_1'}{z\alpha_1^2} \right)\varpi_0 & \frac{zC_3(z)\alpha_1}{\varpi_0} & 0 \\
        \varpi_0''' & \frac{2\varpi_0'^2}{z\varpi_0\alpha_1}
        - \frac{\varpi_0''}{z\alpha_1}
        + v_1(z)\varpi_0-v_2(z)\varpi_0' \ &
        \left(\frac{z\varpi_0'\alpha_1}{\varpi_0^2}
        -\frac{zp_3(z)\alpha_1}{2\varpi_0}\right) {\scriptstyle C_3(z)}\ & \frac{{\scriptstyle C_3(z)}}{\varpi_0}
    \end{pmatrix} \, ,
    \end{align}
with
\begin{align}
     v_1(z) &\,= \frac{p_3(z)\alpha_1'}{2z\alpha_1^2} + \frac{2p_3(z)+zp_3(z)^2+2zp_3'(z)-4zp_2(z)}{4z^2\alpha_1} \, , \\
     v_2(z) &\,= \frac{\alpha_1'}{z\alpha_1^2} + \frac{1+zp_3(z)}{z^2\alpha_1} \, .
\end{align}
In deriving these expressions we have made repeated use of the following identities, which follow from the quadratic relations in eq.~\refeq{eq:qual_rel_ss} for $n=2$ and $n=3$ respectively,
\begin{align}
 \nonumber  \varpi_0'' &\,= \frac{\varpi_0'^2}{2\varpi_0}+\frac{2p_2(z)^2+3 p_2'(z)-9p_1(z)}{18}\varpi_0 - \frac{p_2(z)}{3}\varpi_0' \qquad &&\text{for} \ n=2 \, , \\
  \alpha_1'' &\,= (zp_3(z)-2)\frac{\varpi_0'\alpha_1}{z\varpi_0}-2\frac{\varpi_0'\alpha_1'}{\varpi_0}+4\frac{\varpi_0''\alpha_1}{\varpi_0}-2\frac{\varpi_0'^2\alpha_1}{\varpi_0^2}+2\frac{\alpha_1'^2}{\alpha_1}  && \\
        &\quad -\frac{2zp_3(z)+z^2p_3(z)^2+2z^2p_3'(z)-4z^2p_2(z)-8}{4z^2}\alpha_1 - \frac{zp_3(z)-4}{2z}\alpha_1' \qquad &&\text{for} \ n=3\nonumber \, .
\end{align}

\chapter{Finite-field evaluation and rational reconstruction}
\label{App:FiniteField}

Large symbolic computations often face the following situation: the input data and the final result have moderate algebraic complexity, but the intermediate expressions encountered along the way swell tremendously in size. One of the most important examples in which this occurs is the IBP reduction of Feynman integrals~\cite{vonManteuffel:2014ixa}. The coefficients of the master integrals at the end of the reduction have moderate complexity, but the intermediate steps of the Gaussian elimination can become arbitrarily complicated. Keeping track of those intermediate expressions in memory and in CPU time can make a computation expensive, and in some cases altogether impossible. Finite-field methods bypass these intermediate steps. The idea is simple: instead of carrying out the computation symbolically, one carries it out many times at different probe points over a finite field $\mathbb Z_p$ for some prime $p$, so that every number stays bounded by $p$ and the intermediate results cannot swell. The rational answer is then reconstructed from a certain number of such runs. We follow refs.~\cite{Peraro:2016wsq,Weinzierl:2022eaz} to present the method, which is by now standard in state-of-the-art loop calculations.

\section*{Calculations in $\mathbb Z_p$}

The integers modulo a prime $p$ form the finite field $\mathbb Z_p$ with $p$ elements. Addition, subtraction and multiplication are inherited from $\mathbb Z$ followed by reduction modulo $p$. The multiplicative inverse, on the other hand, requires a little more care. We define the inverse of a non-zero element $a\in\mathbb Z_p$ through
\begin{equation}
        \left(a\, a^{-1}\right)\mod p = 1\,,
\end{equation}
and compute it efficiently through the extended Euclidean algorithm applied to the pair $(a,p)$, see e.g.~\cite{Weinzierl:2022eaz} for a detailed discussion. A rational number $q=a/b\in\mathbb Q$ whose denominator is coprime to $p$ is then represented in $\mathbb Z_p$ by
\begin{equation}
        q\mod p = \left(a\, b^{-1}\right)\mod p\,.
\end{equation}
A symbolic linear system whose coefficients are rational functions can therefore be turned into a purely numerical system over $\mathbb Z_p$ by choosing a probe point for every kinematic variable and mapping the resulting rational coefficients into $\mathbb Z_p$. The row reduction is then carried out numerically and at a fraction of the cost, with no large symbolic intermediate expressions appearing at any step.

\section*{Reconstruction of the analytic numbers in $\mathbb Q$}

The non-trivial step is the reverse one, that is, the analytic reconstruction. We begin with the recovery of rational numbers. After the computational step, for instance a row reduction, we are left with a number $\hat q\in\{0,1,\ldots,p-1\}$ that represents the rational answer modulo $p$. To reconstruct the rational itself one uses Wang's algorithm~\cite{Wang:1981,Wang:1982}, which is applied iteratively to the pair $(\hat q, p)$ until one obtains a pair $(a,b)$ satisfying
\begin{equation}\label{eq:rational-recon-bound}
|a|\,,\;|b|\;<\;\sqrt{p/2}\,.
\end{equation}
This pair fulfils $\hat q = a\, b^{-1}\!\mod p$ and yields the rational number $\hat q = a/b\in\mathbb Q$. The reconstruction is correct provided the numerator and denominator of the true answer are well below $\sqrt{p/2}$. In practice we use $64$-bit primes, and the reductions in this thesis occasionally produce rational numbers that exceed this bound. The solution is then to use several primes through the Chinese remainder theorem.

\section*{Chinese remainder theorem}

We cannot recover numerators or denominators larger than $\sqrt{p/2}$ from a single prime number. For the reconstruction of these numbers we have to use the Chinese remainder theorem (CRT), which allows us to combine several runs over different prime fields. 
If we repeat the finite-field reduction over $k$ pairwise coprime primes $p_1,p_2,\ldots,p_k$, we get residues $\hat q_i\equiv\hat q\pmod{p_i}$, and the CRT yields the unique residue
\begin{equation}
        \hat Q\equiv\hat q\pmod{P}\,,\qquad P=p_1p_2\cdots p_k\,,
\end{equation}
to which Wang's algorithm is then applied. 
We combine multiple prime numbers effectively to a larger prime number.
The bound for reconstructable numbers get enlarged to $\sqrt{P/2}$, and the procedure can be terminated once adding a further prime no longer changes the reconstructed rational.

\section*{Rational functions}

We are ultimately interested in reconstructing not just rational numbers but rational functions. The same strategy applies, almost unchanged. We sample the function at many probe points and reconstruct each of the numerical residues by the procedure above. In the univariate case the resulting samples are fitted to a rational function via Thiele's continued-fraction interpolation, and an analogous recursive procedure handles the multivariate case~\cite{Peraro:2016wsq,Peraro:2019svx,Klappert:2019emp,Klappert:2020aqs}. The number of probes required scales with the polynomial degree of numerator and denominator, rather than with the size of the intermediate expressions.

This last observation makes plain the two main advantages of the approach. The swelling of intermediate expressions is avoided entirely, and any structural information we have about the result, such as a lower polynomial degree of the coefficients, directly translates into a reduction in the number of probes and primes. In the IBP context this last point is particularly relevant, since we are free to choose the basis of master integrals: a basis that yields more compact coefficients leads, automatically, to a faster reduction.

\chapter{Definition of the 0SF-families}\label{App:0SFFam}
\subsection*{4PM}
The 4PM-0SF planar family is defined as follows:
\begin{align}
        {{\cal P}^{\text{4PM-0SF}}}_{\{n\}}^{\{\sigma\}}
        =
        \int_{\ell_1\ell_2 \ell_3}
        \frac{
          \prod_{i=1}^3\dd^{(n_i-1)}(\ell_i\cdot v_{1})
          }{
          \prod_{i=1}^{12}D_{i}^{n_{i+3}}
          }\,,
      \end{align}
with
      \begin{align}
        D_{1/2/3}=&\ell_{1/2/3}\cdot v_{2}+\sigma_i \i0\,,
        D_{4/5}=(\ell_{1/2}-\ell_3)^2
        \,,
         D_{6}=(\ell_{1}-\ell_2)^2
        \,,\nn\\
       & D_{7/8/9}=\ell_{1/2/3}^2
        \,,
        D_{10/11/12}=(\ell_{1/2/3}-q)^2
        \,.
      \end{align}
      \subsection*{5PM}
      The 5PM-0SF planar family is defined as follows:
      \begin{align}
              {{\cal P}^{\text{5PM-0SF}}}_{\{n\}}^{\{\sigma\}}
              =
              \int_{\ell_1\ell_2 \ell_3 \ell_4}
              \frac{
                \prod_{i=1}^4\dd^{(n_i-1)}(\ell_i\cdot v_{1})
                }{
                \prod_{i=1}^{18}D_{i}^{n_{i+4}}
                }\,,
            \end{align}
           with
            \begin{align}
              D_{1/2/3/4}=\ell_{1/2/3/4}&\cdot v_{2}+\sigma_i \i0\,,
              D_{5/6/7}=(\ell_{1/2/3}-\ell_4)^2
              \,,
               D_{8/9}=(\ell_{1}-\ell_{2/3})^2
              \,,D_{10}=(\ell_{2}-\ell_{3})^2
              \,,\nn\\
             & D_{11/12/13/14}=\ell_{1/2/3/4}^2
              \,,
              D_{15/16/17/18}=(\ell_{1/2/3/4}-q)^2
              \,.
            \end{align}

\chapter{Additional functions for the canonicalisation of the CY3-sector at 1SF order} \label{app:cy31G}
The defining first-order differential equations of the functions needed for the canonicalisation of the full DE matrix at 5PM-1SF order:
\begin{align}
        H_1'=& \frac{\varpi_0}{(x-1)^2}\, ,\\
        H_2'=& \frac{\varpi_0}{x}\, ,\\
        H_3'=& \frac{\varpi_0}{(x+1)^2}\, ,\\
        H_4'=& \frac{\varpi_{K3,1} \left(3 x^6 \varpi_{0}''+9 x^5 \varpi_{0}'-10 x^4 \varpi_{0}''+3 x^4 \varpi_{0}-10 x^3 \varpi_{0}'+3 x^2 \varpi_{0}''-2 x^2 \varpi_{0}-3 x \varpi_{0}'+3 \varpi_{0}\right)}{32 x^2} \, ,\\
        H_5'=& \frac{\varpi_{K3,1} \left(3 x^6 \varpi_{0}''+9 x^5 \varpi_{0}'+3 x^4 \varpi_{0}-4 x^3 \varpi_{0}'-3 x^2 \varpi_{0}''+3 x \varpi_{0}'-3 \varpi_{0}\right)}{8 x^2}\, ,\\
        H_6'=& \frac{2 H_4}{(x-1) x (x+1) \varpi_{K3,1}}\, ,\\
        H_7'=& \frac{\left(3 x^4-10 x^2+3\right) G_1 \alpha \varpi_{K3,1}}{32 \left(x^4-1\right) \varpi_{0}}\nn \\
        &+G_3  \frac{\varpi_{K3,1} \left(x \left(-3 x^4+10 x^2-3\right) \varpi_{0} \alpha'+\alpha \left(6 \left(x^4-1\right) \varpi_{0}+2 x \left(3 x^4-10 x^2+3\right) \varpi_{0}'\right)\right)}{32 x^2 \alpha^2}\nn \\
        &+G_3 \frac{H_4}{x \alpha}+\frac{\left(2 x^4-32 x^2+2\right) H_4}{3 x-3 x^5}\nn\\
        &+\frac{\varpi_{K3,1} x \left(x^{12}-44 x^{10}+101 x^8-101 x^4+44 x^2-1\right) \varpi_{0}'}{16 x^2 \left(x^4-1\right)^2}\nn\\
        &+\frac{\varpi_{K3,1} \left(x^{12}+2 x^{10}-73 x^8+236 x^6-73 x^4+2 x^2+1\right) \varpi_{0}}{16 x^2 \left(x^4-1\right)^2}\, ,
\end{align}
\begin{align}
        H_8'=& \frac{\varpi_{K3,1} \left(x \left(-3 x^4+10 x^2-3\right) \varpi_{0} \alpha'+\alpha \left(6 \left(x^4-1\right) \varpi_{0}+2 x \left(3 x^4-10 x^2+3\right) \varpi_{0}'\right)\right)}{32 x^2 \alpha^2}-\frac{H_4}{x \alpha}\, ,\\
        H_9'=& -\frac{2 H_5}{x \varpi_{K3,1}-x^3 \varpi_{K3,1}}\, ,\\
        H_{10}'=& \frac{3 G_1 \alpha \varpi_{K3,1}}{8 \varpi_{0}}\nn \\
        &+G_3 \left(\frac{H_5}{x \alpha}-\frac{\varpi_{K3,1} \left(\alpha \left(6 x \left(x^4-1\right) \varpi_{0}'+\left(6 x^4-4 x^2+6\right) \varpi_{0}\right)-3 x \left(x^4-1\right) \varpi_{0} \alpha'\right)}{8 x^2 \alpha^2}\right)\nn \\
        &+\frac{\left(2 x^4-32 x^2+2\right) H_5}{3 x-3 x^5}\nn \\
        &+\frac{\varpi_{K3,1} \left(x \left(x^8-10 x^6+10 x^2-1\right) \varpi_{0}'+\left(x^8-12 x^6-2 x^4-12 x^2+1\right) \varpi_{0}\right)}{4 x^2 \left(x^4-1\right)}\, ,\\
        H_{11}'=& \frac{\varpi_{K3,1} \left(\alpha \left(6 x \left(x^4-1\right) \varpi_{0}'+\left(6 x^4-4 x^2+6\right) \varpi_{0}\right)-3 x \left(x^4-1\right) \varpi_{0} \alpha'\right)}{8 x^2 \alpha^2}-\frac{H_5}{x \alpha}\, ,
\end{align}
where $\varpi_0=\varpi_{\text{CY3,1}}$.
\chapter{Additional functions for the canonicalisation of the CY3-sector at 2SF order }
\label{app:bh}
The defining first-order differential equations of the functions needed for the canonicalisation of the CY3 appearing at 2SF order are:
\begin{align}
     x\,G_1'(x)&= \frac{x^2\, \left(3-x^2\right) \left(1-3 x^2\right)}{\left(1+x^2\right)^3} \varpi_0(x) \, ,\\
     x\,G_2'(x)&= 20 \frac{G_1(x)^2}{ \alpha_1 (x)} 
     + \frac{20 x^2 (1-x^2)^2 \left(1-8 x^2+x^4\right) }{\left(1+x^2\right)^4}\frac{\varpi_0(x)^2}{\alpha_1(x)} \, ,\\
    x\,G_3'(x)&= 
    -\frac{40 \left(1-48 x^2+x^4\right)}{(1-x^2) \left(1+x^2\right) }G_1(x)^2 
    + \frac{1+x^2}{(1-x^2)^3 }\frac{G_2(x)^2\alpha_1(x)^2}{\varpi_0(x)^2} \\
    &\quad + \frac{1280 \left(2-11 x^2+2 x^4\right)}{\left(1+x^2\right)^3 }G_1(x)\varpi_0(x)  \nonumber\\
    &\quad -\frac{4 x^2 \left(41-271 x^2-1412 x^4+2606 x^6-1343 x^8-51 x^{10}-2 x^{12}\right)}{\left(1+x^2\right)^5} \varpi_0(x)^2 \nonumber \\
    &\quad + \frac{2 x^2 (1-x^2)  \left(1+152 x^2-198 x^4+152 x^6+x^8\right)}{\left(1+x^2\right)^3}\varpi_0'(x)^2 \, ,\nonumber\\
    x\,G_4'(x)&= \frac{G_1(x)}{ \alpha_1(x)} \, , \\
    x\,G_5'(x)&= -\frac{1-48 x^2+x^4}{(1-x^2) \left(1+x^2\right)}G_1(x) + \frac{1+x^2}{(1-x^2)^3 }\frac{G_2(x) G_4(x) \alpha_1(x)^2}{\varpi_0(x)^2} \\
            & \quad + \frac{16 \,x^2 \left(2-11 x^2+2 x^4\right) }{\left(1+x^2\right)^3} \varpi_0(x) \, , \nonumber\\
    x\,G_6'(x)&= \frac{1+x^2}{(1-x^2)^3 }\frac{G_2(x)\alpha_1(x)^2}{\varpi_0(x)^2} \, , \\
\end{align}
\begin{align}
    x\,G_7'(x)&= \frac{40\, G_1(x)}{\alpha_1(x)} \Big( G_5(x) - G_4(x)G_6(x)  \Big)
    - \frac{G_3(x)}{\alpha_1(x)} \nonumber \\
      &\quad + \frac{2 (1-x^2) \left(1+152 x^2-198 x^4+152 x^6+x^8\right)}{\left(1+x^2\right)^3}\frac{\varpi_0(x)\varpi_0'(x)}{\alpha_1(x)} \, . 
\end{align}
\chapter{Fourier Integrals}\label{App:Integrals}

The Fourier transform in $d$-dimensional Euclidean space reads
\be
\int \frac{d^{d} \mathbf{q}}{(2\pi)^{d}}e^{\I \mathbf{q}\cdot \mathbf{x}} 
\frac{1}{[\mathbf{q}^{2}]^{a}} = \frac{\Gamma(d/2-a)}{4^{a}\pi^{d/2} \Gamma(a)}
\left ( \mathbf{x}^{2} \right )^{a-d/2}
\ee
while the Lorentzian variant is
\be
\int \frac{d^{d} {q}}{(2\pi)^{d}}e^{\I {q}\cdot {x}} 
\frac{\I}{[{q}^{2}+\i0]^{a}} = \frac{\Gamma(d/2-a)}{4^{a}\pi^{d/2} \Gamma(a)}
\frac{(-1)^{a+1}}{(-x^{2}-\i0)^{d/2-a}}\,.
\ee
The central Fourier integral reads
\be
\int_{q}e^{\I q\cdot b} \dd(q\cdot v_1)\, \dd(q\cdot v_2) |q|^{n} = \frac{1}{\sqrt{\gamma^{2}-1}} \frac{2^{n}}{\pi^{(d-2)/2}}
\frac{\Gamma(\frac{d-2+n}{2})}{\Gamma(-\frac{n}{2})} | b|^{2-d-n} \label{eq:FT}
\ee
with  $|q|:=\sqrt{-q^{2}}$.

\chapter{Graviton Gauge Fixing}\label{sec:gaugefixing}

In order to minimise the number of terms in the three- and four-point graviton vertex we rewrite the Einstein-Hilbert action via a total derivative and by adding a gauge-fixing term
\begin{align}\label{gfactionbulk}
 \mathcal{S}_{\text{bulk}} &= -\frac{1}{16\pi G} \int \mathrm d^{d}x \Big [
 \sqrt{-g} g^{\alpha\beta} g^{\gamma\delta} g^{\mu\nu} ( \Gamma_{\alpha\gamma\delta}
 \Gamma_{\mu\beta\nu} - \Gamma_{\alpha\gamma\mu}  \Gamma_{\delta\beta\nu} )
 +\sfrac{1}{2} G^{\mu}G^{\nu}\eta_{\mu\nu} \Bigr ] \,.
\end{align}
The gauge fixing vector $G^{\mu}$ with the graviton field $h_{\mu\nu}$ and its trace $h=h_{\mu}{}^{\mu}$ takes the form
\begin{align}
\begin{aligned}
G^{\mu} = & \, \partial_{\nu_{1}}h^{\mu\nu_{1}} - \sfrac{1}{2} \partial^{\mu}h 
+\sfrac{1}{2} \partial_{\nu_1}h\,  h^{\mu\nu_{1}}
+\sfrac{1}{2} \partial^{\mu}h_{\nu_1\nu_2}\, h^{\nu_1\nu_2} 
-\partial^{\nu_1}h^{\mu \nu_2} h_{\nu_1\nu_2} 
-\sfrac{1}{4}  \partial^{\mu}h\, h \\ &
+ \partial_{\nu_1}h^{\nu_3\mu}   h^{\nu _1\nu_2}\, h_{\nu_2\nu _3} 
+\sfrac{1}{2} \partial_{\nu_1} h^{\nu_1\nu _2}\, h_{\nu_2\nu_3}\, h^{\nu_3\mu}
-\sfrac{1}{4} \partial^{\nu_2} h\, h_{\nu_2\nu_3} \, h^{\nu_3\mu}
+\partial_{\nu_1}h_{\nu_2\nu_3}\,  h^{\nu_1\nu_2} h^{\nu_3\mu} \\ &
-\sfrac{1}{2} \partial^{\nu_2} h_{\nu_1\nu_2}\,  h^{\mu \nu _1}\, h
+\sfrac{1}{4} \partial^{\mu}h_{\nu_1\nu_2}\, h^{\nu_1\nu_2} h
-\sfrac{1}{2} \partial^{\nu_1} h^{\mu\nu_2}\, h_{\nu_1\nu_2} h
+\sfrac{1}{8} \partial_{\nu_1}h^{\mu\nu_1}\, h^{2}\, .
\end{aligned}
\end{align}
The quadratic-in-$h_{\mu\nu}$ term yields the standard graviton propagator in De-Donder gauge in eq.~\refeq{eq:hPropagator},
and the non-linear gauge fixing terms are engineered in such a fashion as 
to maximally simplify the cubic and quartic graviton vertices. 

\bibliographystyle{elsarticle-num-names} 
\bibliography{thesis}

\end{document}